%% file: thesis.tex
\pdfoutput=1
\documentclass[11pt]{report}
\usepackage[a4paper,left=4cm,right=2cm,top=2.5cm, bottom=3cm]{geometry}
\usepackage{ditthesis}
\usepackage[numbers]{natbib}
\usepackage{tikz}
\usetikzlibrary{positioning}
\usetikzlibrary{shapes.geometric}
\usetikzlibrary{shadows}
\usepackage{booktabs}
\usepackage{listings}
\usepackage{notoccite}
\usepackage{url}
\usepackage{amsmath}
\usepackage{datatool}

\usepackage[acronym,toc]{glossaries}

\usepackage{amsmath}
\usepackage{fixltx2e}
\usepackage{pgfplots}
\usepackage{filecontents}
\usepackage{subcaption}

\usepackage{bchart}

\usetikzlibrary{shapes,arrows}

\usepackage{algorithm}
\usepackage{algpseudocode}
\usepackage[hang]{footmisc}
\usepackage{pgfplots}
\usepackage{pgfplotstable}
\usepackage[]{color}

\usetikzlibrary{calc}
\usepackage[titletoc]{appendix}
\usepackage{layout}
\usepackage{longtable}
\usetikzlibrary{shapes,intersections,shapes.multipart}

\usepackage[T1]{fontenc}

\pgfplotsset{}

\newglossaryentry{ccd}{name={CCD},description={Charge-Coupled Devices}}
\newglossaryentry{cmos}{name={CMOS},description={Complementary Metal Oxide Semiconductor}}
\newglossaryentry{cpu}{name={CPU},description={Central Processing Unit}}
\newglossaryentry{lsst}{name={LSST},description={Large Synoptic Survey Telescope}}
\newglossaryentry{acn}{name={ACN},description={Astronomical Compute Node}}
\newglossaryentry{jpl}{name={JPL},description={Jet Propulsion Laboratory}}
\newglossaryentry{pmt}{name={PMT},description={Photomultiplier Tube}}
\newglossaryentry{ska}{name={SKA},description={Square Kilometre Array}}
\newglossaryentry{cte}{name={CTE},description={Charge Transfer Efficiency}}
\newglossaryentry{fits}{name={FITS},description={Flexible Image Transport Systems}}
\newglossaryentry{opus}{name={OPUS},description={Operational Pipeline Unified System}}
\newglossaryentry{stsi}{name={STScI},description={Space Telescope Science Institute}}

\newglossaryentry{evalso}{name={EVALSO},description={Enabling Virtual Access to Latin-American Southern Observatories}}
\newglossaryentry{hpc}{name={HPC},description={High Performance Computing}}
\newglossaryentry{tdrss}{name={TDRSS},description={Tracking and Data Relay Satellite System}}
\newglossaryentry{hst}{name={HST},description={Hubble Space Telescope}}
\newglossaryentry{mast}{name={MAST},description={Mikulski Archive for Space Telescopes}}
\newglossaryentry{eso}{name={ESO},description={European Southern Observatory}}
\newglossaryentry{sdss}{name={SDSS},description={Sloan Digital Survey Systems}}
\newglossaryentry{lbto}{name={LBT},description={Large Binocular Telescope}}
\newglossaryentry{gtc}{name={GTC},description={Gran Telescopio Canarias}}
\newglossaryentry{stsdas}{name={STSDAS},description={Space Telescope Science Data Analysis System }}
\newglossaryentry{iraf}{name={IRAF},description={Image Reduction and Analysis Facility}}
\newglossaryentry{jwst}{name={JWST},description={James Webb Space Telescope}}
\newglossaryentry{dsn}{name={DSN},description={Deep Space Network}}
\newglossaryentry{soc}{name={SOC},description={Space Operations Center}}
\newglossaryentry{ec2}{name={EC2},description={Elastic Compute Cloud}}
\newglossaryentry{mpt}{name={MPT},description={Message Passing Toolkit}}
\newglossaryentry{openmp}{name={OpenMP},description={Open MultiProcessing}}
\newglossaryentry{noao}{name={NOAO},description={National Optical Astronomy Observatory }}
\newglossaryentry{nhpps}{name={NHPPS},description={NOAO High-Performance Pipeline System }}
\newglossaryentry{cpl}{name={CPL},description={ESO Common Pipeline Library}}

\newglossaryentry{nist}{name={NIST},description={National Institute of Standards and Technology}}
\newglossaryentry{jms}{name={JMS},description={Java Message Service}}
\newglossaryentry{bco}{name={BCO},description={Blackrock Castle Observatory}}
\newglossaryentry{emccd}{name={EMCCD},description={Electron Multiplying Charge Coupled Device}}

\newglossaryentry{vo}{name={VO},description={Virtual Observatory}}
\newglossaryentry{s3}{name={S3},description={Simple Storage Service}}

\newglossaryentry{ittd}{name={ITTD},description={Institute of Technology Tallaght in Dublin}}
\newglossaryentry{nfs}{name={NFS},description={Network File System}}
\newglossaryentry{hdfs}{name={HDFS},description={Hadoop Distributed File System}}

\newglossaryentry{sqs}{name={SQS},description={Simple Queue Service}}
\newglossaryentry{ami}{name={AMI},description={amazon machine image}}
\newglossaryentry{utc}{name={UTC},description={Coordinated Universal Time}}
\makeglossaries

\begin{document} 

\thesistitlepage
\thesisdeclarationpage

\begin{thesisacknowledgments}
The completion of this thesis has been possible  through the support of my family, friends and colleagues who have helped and encouraged me over the last four years. I would like to acknowledge the support and guidance of my supervisors who remained ever confident that this journey would be enlightening and fulfilling;  my  children Connor, Ois\'{i}n and Cillian, who shared their hugs of encouragement, and my wife Orla who gave me understanding, space and time to finish, which was no small sacrifice. To my colleagues and friends who generously gave their time and opinions I can assure you that this thesis was very much better for your contribution. Special thanks to my friends and colleagues at HEAnet, the Blackrock Castle Observatory, the Cork Institute of Technology, and the Institute of Technology Tallaght who provided services and support which were fundamental to the research performed within this thesis. This research was also supported in part by Amazon Web Services Ireland who provided extensive online virtual infrastructure resources.  Finally I would like to express my deepest thanks to my parents who first set me on the path to a higher education. I will endeavour to pass the same values on to my own children. 
\end{thesisacknowledgments}

\thesisabstract

Astronomical photometry is the science of measuring the flux of a celestial object. Since its introduction in the 1970s the CCD has been the principle method of measuring flux to calculate the apparent magnitude of an object. Each CCD image taken must go through a process of cleaning and calibration prior to its use. As the number of research telescopes increases the overall computing resources required for image processing also increases. As data archives increase in size to Petabytes, the data processing challenge requires image processing approaches to evolve to continue to exceed the growing data capture rate. 
\par Existing processing techniques are primarily sequential in nature,  requiring increasingly powerful servers, faster disks and faster networks  to process data. Existing High Performance Computing solutions involving high capacity data centres are both complex in design and expensive to maintain, while providing resources primarily to high profile science projects. 
\par This research describes three distributed pipeline architectures, a virtualised cloud based IRAF, the Astronomical Compute Node (ACN), a private cloud based pipeline, and NIMBUS,  a globally distributed system. The ACN pipeline processed data at a rate of 4 Terabytes per day  demonstrating data compression and upload to a central cloud storage service at a rate faster than data generation. The  primary contribution of this research however is NIMBUS, which is rapidly scalable, resilient to failure and capable of processing CCD image data at a rate of hundreds of Terabytes per day. This pipeline is implemented using a decentralised web queue to control the compression of data, uploading of data to distributed web servers, and creating web messages to identify the location of the data. Using distributed web queue messages, images are downloaded by computing resources distributed around the globe. Rigorous experimental evidence is presented verifying the horizontal  scalability of the system which has demonstrated a processing rate of 192 Terabytes per day with clear indications that higher processing rates are possible.

\tableofcontents 
\listoffigures
\listoftables
\lstlistoflistings

\begin{thesispublications}

	The publications that are related to this thesis are listed below:
	\begin{enumerate}
		\item Doyle, P; Mtenzi, F; Smith, N; Collins, A and O'Shea, B "Significantly reducing the processing times of high-speed photometry datasets using a distributed computing model", Proc. SPIE 8451, Software and Cyber infrastructure for Astronomy II, 84510C (September 24, 2012); doi: 10.1117/12.924863; 
		\item Doyle, P.; Deegan, M.; Markey, D.; Tinabo, R.; Masamila, B.; Tracey, D. Case Studies In Thin Client Acceptance. Ubiquitous Computing and Communication Journal. Special Issue on ICIT 2009 conference - Applied Computing, 2009.		
		\item Doyle, P.; Deegan, M.; O'Driscoll, C.; Gleeson, M.; Gillespie, B.; , "Ubiquitous desktops with multi-factor authentication," Digital Information Management, 2008. ICDIM 2008. Third International Conference, vol., no., pp.198-203, 13-16 Nov. 2008 doi: 10.1109/ICDIM.2008.4746797 		

	\end{enumerate}

\end{thesispublications}

\include{chapter1}

\include{chapter2}


\include{chapter4}   

\include{chapter5}

\include{chapter6}

\include{chapter7}


\include{appendixA}
\clearpage 
\addcontentsline{toc}{chapter}{Bibliography}

\bibliography{thesis}
\bibliographystyle{unsrt}

\end{document}

%% file: chapter1.tex

\chapter{Introduction}

\par Photometry is defined as the branch of science that deals with measuring the intensity of the electromagnetic radiation, or flux of a celestial object \citep{merriamw:2014}. The word photometry is derived from the Greek word photos, meaning light, and m\'{e}tron, meaning measure \citep{sterken1992astronomical}.  This science can be traced as far back as 130 BC to Hipparchus \citep{1977ccp..book.....N}, who devised the first measurement system categorising objects' apparent brightness from brightest to faintest. While the initial Hellenistic method provided only six classifications, the sophistication and sensitivity of the tools used in measurements evolved dramatically throughout the ages, with the current system of measurement of apparent brightness allowing for fractional measurements, both positive and negative, which have no specific upper or lower limits.

\par Modern photometry has evolved and been revolutionised by the use of  \gls{ccd}. CCDs are light sensitive integrated circuits that are used in imaging and signal processing. Information is represented as packets of electric charge which are stored in an array of closely space capacitors which can be moved from one capacitor to another. Charges can be moved systematically to a location on the device where the charge is converted to a digital value representing the light intensity for each pixel to form an image. Since their first introduction to astronomy, CCDs have received considerable attention from the astronomical community \citep{1975ccdt.proc...65M}, and revolutionised this field of science providing levels of sensitivity beyond the capability of photographic plates, extending the detection range into the infrared spectrum, providing immediate results with a linear response, and allowing for software to compensate for CCD defects.  When a CCD digital image is recorded it contains a digital count of the electrical charge of each of the pixels on the CCD array.  The electrical charge per cell is converted to a digital pixel value by first transferring the charge to a corner of the array and then using an analog to digital conversion to record its value.  This digital image contains a number of different artefacts introduced by the process of recording and reading, which must be removed. These and other sources of noise require a computation operation to be performed across the image  in order to quantify the signal to noise ratio. So for each image taken there is a computational overhead required before scientific analysis can be performed. As the number of images increases, so does this computation cost. 

\par Over the last few decades the number of ground based astronomical research observatories has continued to increase and currently stands at approximately 400 sites globally \cite {esondelmott:2007} with each site using some form of CCD device. While CCD devices, which have improved both in terms of resolution and capture rates  are still the primary capture device in use, \gls{cmos} devices  which offer the potential for faster imaging capture rates with lower power utilisation and higher resolutions \cite {Janesick:2003hm} are increasing in popularity. It is the combination of these developments, faster image capture rates, higher resolutions and more telescope observatories  that drives the increase in image processing based computing requirements. Even this requirement pales in comparison to the potential of highly distributed robotic telescope projects being developed \cite {2004SPIE.5496..302W} which could, over time, increase the number of CCD or CMOS images being produced to billions per second, raising the issue of data processing to terabytes and petabytes per day. Within the last few years astronomers have voiced serious concerns about the growing data tsunami with Berriman  \cite{Berriman:2011fw} predicting 60PB of data being available as an archive by the year 2020. The International Digital Coporation's $7^{th}$ annual Digital Universe Study echoes this concern with the issues of data growth and the subsequent data generation, storage, search and  processing highlighted, identifying exponential growth in digital data expanding from 4.4 to 44  zettabytes\footnote{A zettabyte is one billion terabytes, or $10^{21}$ bytes } by the year 2020 \cite{Gantz:2012wp}. As with astronomical data, much of the cause for this growth is identified as falling costs of capture devices and the increase in digital versus analogue data collection technology.   

\par Many of the existing approaches to CCD image data processing are still primarily sequential in nature with software tools written for single  \gls{cpu} cores using single threads. This approach was reasonably successful given that for many years Moore's Law continued to hold true, and faster machines could be purchased to speed up the overall system performance every one to two years without dramatically altering the underlying software applications used to process raw CCD images.  However, given that Moore's law  stated that there would be a doubling of the number of transistors on an affordable CPU every two years, and that this has led to multi-core CPU chip designs, the law, while still technically true will no longer provide significant performance enhancements for single threaded sequential processing applications. Programs must become multi-threaded and software must evolve to take advantage of multiple CPU cores. 

\par Other limitations also appear once the volume of data becomes sufficiently large. Network bandwidth becomes a limiting factor when large volumes of data are centralised, and even single server multi-core CPU systems may not provide enough of a performance enhancement to processing rates. As it becomes cheaper to generate  data, more sophisticated processing techniques are required which can easily utilise large arrays of computing resources.  A more sensible redesign of data processing software should consider exploiting additional processing and networking resources beyond a single infrastructure and embrace a distributed processing approach.  

\par In order to address the issue of cleaning and preparing terabytes or even petabytes of CCD based astronomical photometry images  per day,   a distributed elastic cloud based computing model to perform standard image data processing is required. A processing pipeline has been designed which demonstrates a working CCD image reduction pipeline, which incorporates an elastic data processing model where resources can join or leave a swarm of distributed computing workers which communicate via a distributed web based messaging queue.  Furthermore, taking advantage of the fact that CCD images can be cleaned in isolation from each other, image data is distributed for parallel processing to eliminate sequential image processing bottlenecks.    

\par To ensure that access is provided to all data sources, experimental results and source code used within this thesis, the following URL is provided as an access point to all material \url{http://www.dit.ie/computing/staff/pauldoyle/research/}.

\section{Background}

\subsection{Definitions}

\par To ensure clarity of the terms used within the context of astronomical photometry, the following definitions are provided for reference.
\newpage

	\begin{description}
	\item [Apparent Magnitude]  \hfill \\
	The apparent magnitude of a source is based on its apparent brightness as seen on earth, adjusted for atmosphere. The brighter the source appears, the lower the apparent magnitude value. The apparent magnitude of an object is measured using a standard reference as shown in this equation where $m_{2}$ and $F_{2}$ are reference magnitude and reference flux values and $F_{1}$ is the observed flux. $m_{1} - m_{2} = 2.5 log \frac{F_{1}}{F_{2}}$
		\item [Absolute Magnitude]  \hfill \\
	The absolute magnitude is a measure of a stars brightness as seen from a distance of 10 parsecs (32.6 light years) from the observer.  The absolute magnitude $M$ of an object can be calculated  given the apparent magnitude $m$ and luminosity distance $D$ which is measured in parsecs. $M = m - 5 ((log_{10}D_{L}) - 1)$
	\item [Instrumental Magnitude]  \hfill \\
	The instrumental magnitude is an uncalibrated measure of the apparent magnitude of an object which is only used for comparison with other magnitude values on the same image. This is calculated using the following formula where $f$ is the measure of the intensity value of the source image. $m = 2.5log_{10}(f)$
	\item [Luminosity]  \hfill \\
	The Luminosity of an object is a measure of the total energy emitted by a star or other celestial body per unit of time and is independent of distance and is measured in watts. The luminosity of a star $L$ is related to temperature $T$ and the radius $R$ of the star and is defined as follows where $\sigma$ is the Stefan-Boltzmann constant.  $L=4\pi R^{2}\sigma T^{4}$
	\item [Flux]  \hfill \\
	The flux is a measure of the apparent brightness of a source which is inversely proportional to the square of the distance and is measured in watts per square meter $W/m^{2}$. How bright a source appears is based on the distance from the object and the Luminosity of the object which can be defined as follows where $L$ is the Luminosity and $d$ is the distance to the source. $F=\frac{L}{4\pi d^2}$
	\end{description}

\subsection{Photometry}

\par There have been many methods devised to estimate apparent magnitude values within astronomical photometry with each new system or advancement aimed to reduce the error margin and increase repeatability for each measurement, but until relatively recently, it was the skill of the observer that ultimately determined the accuracy of the recorded magnitude value \cite{Milone:2011td}. A brief look at the origins of photometry and the evolution of magnitude measurements will help to explain the magnitude scale.

\par
\begin{figure} [ht]
\centering
  \includegraphics[width=0.4\textwidth] {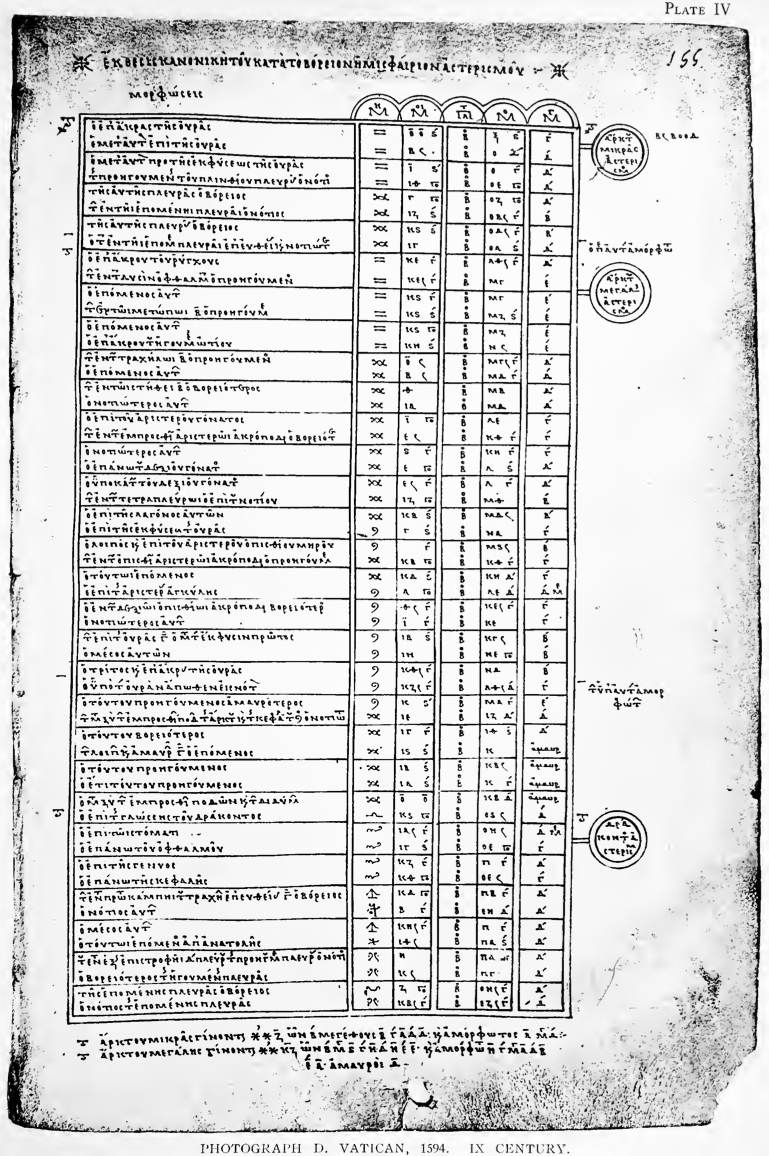}
    \caption{Claudius Ptolemy Star Catalogue. Alexandria, 2nd century. \cite {Ptolemy:2010wu} }
  \label{fig:ptolemy}
\end{figure}

 The earliest measurement of a star's magnitude is credited to Hipparchus (190 --120 BC), a Greek astronomer and mathematician who it is believed made his observations in the 2nd century, BCE. While his original work did not survive, it was referenced by Ptolemy in the Almagest (which is dated at approximately 147 AD, see Figure  \ref{fig:ptolemy} ), which contains a star catalogue of just over 1000 stars referenced by positions within a constellation and their apparent brightness or magnitude. The original magnitude scale used by Hipparchus was a six point system where the brightest stars were designated as $m=1$ (First Magnitude) and the faintest stars are designated $m=6$ (Sixth Magnitude). An increase from magnitude 1 to 2 for example represents more than a halving of the light visible from an object. While Ptolemy claimed to have observed all of the stars himself, it has been argued that the data was based at least partially on the Hipparchus observations (almost 300 year earlier) \cite {1977ccp..book.....N}. The reason for the increasing number for less brilliant stars is most likely based on the division of the twilight into 6 equal parts and stars which became visible within each segment were assigned a magnitude value, hence more faint stars were visible later and received a high magnitude numerical number.  This system of magnitude calculation stood for almost 2000 years and it is only relatively recently that the precision of magnitude calculations has increased significantly from these ancient times.

\par
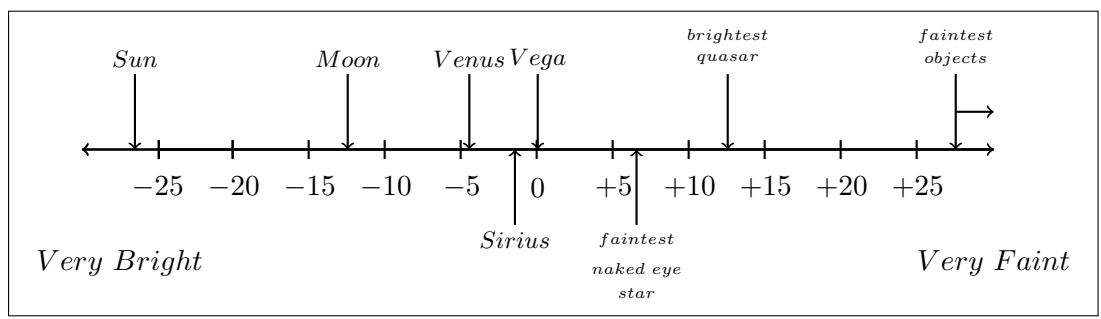
\begin{figure}  [ht]
\fbox{ 
\begin{tikzpicture}
	\draw (-0.3,-0.3) node {\footnotesize$Sun$};
	\draw[<-,thick] (-0.3,-1.5) -- (-0.3,-0.5);		
	\draw (2.5,-0.3) node {\footnotesize$Moon$};
	\draw[<-,thick] (2.5,-1.5) -- (2.5,-0.5);
	\draw (4.1,-0.3) node {\footnotesize$Venus$};
	\draw[<-,thick] (4.1,-1.5) -- (4.1,-0.5);	
	\draw (5,-0.3) node {\footnotesize$Vega$};
	\draw[<-,thick] (5,-1.5) -- (5,-0.5);
	\draw (4.7,-2.7) node {\footnotesize$Sirius$};
	\draw[->,thick] (4.7,-2.5) -- (4.7,-1.5);
	\draw (6.3,-2.7) node {\tiny$faintest$};
	\draw (6.3,-3.1) node {\tiny$naked \  eye$};
	\draw (6.3,-3.4) node {\tiny$star$};
	\draw (7.5,0) node {\tiny$brightest$};
	\draw (7.5,-0.3) node {\tiny$quasar$};
	\draw[<-,thick] (7.5,-1.5) -- (7.5,-0.5);	
	\draw (10.5,0) node {\tiny$faintest$};
	\draw (10.5,-0.3) node {\tiny$objects$};
	\draw[<-,thick] (10.5,-1.5) -- (10.5,-0.5);
	\draw[->,thick] (10.5,-1) -- (11,-1);	
	\draw[->,thick] (6.3,-2.5) -- (6.3,-1.5);				
	\draw (-0.5,-3) node {$Very\ Bright$};
	\draw (11,-3) node {$Very\  Faint$};	
	\draw[<-,thick] (-1,-1.5) -- (11,-1.5);	
	\draw[|-|,thick] (0,-1.5) -- (1,-1.5);
	\draw  (0,-2) node {$-25$};
	\draw  (1,-2) node {$-20$};
	\draw  (2,-2) node {$-15$};
	\draw  (3,-2) node {$-10$};
	\draw  (4,-2) node {$-5$};
	\draw  (5,-2) node {$0$};
	\draw  (6,-2) node {$+5$};
	\draw  (7,-2) node {$+10$};
	\draw  (8,-2) node {$+15$};
	\draw  (9,-2) node {$+20$};
	\draw  (10,-2) node {$+25$};
	\draw[<-,thick] (-1,-1.5) -- (11,-1.5);	
	\draw[|-|,thick] (0,-1.5) -- (1,-1.5);
	\draw[-|,thick] (1,-1.5) -- (2,-1.5);
	\draw[-|,thick] (2,-1.5) -- (3,-1.5);
	\draw[-|,thick] (3,-1.5) -- (4,-1.5);
	\draw[-|,thick] (4,-1.5) -- (5,-1.5);
	\draw[-|,thick] (5,-1.5) -- (6,-1.5);
	\draw[-|,thick] (6,-1.5) -- (7,-1.5);	
	\draw[-|,thick] (7,-1.5) -- (8,-1.5);	
	\draw[-|,thick] (8,-1.5) -- (9,-1.5);	
	\draw[-|,thick] (9,-1.5) -- (10,-1.5);	
	\draw[->,thick] (10,-1.5) -- (11,-1.5);	
\end{tikzpicture}
}
  \vskip -0.8em
  \caption{Apparent brightness of a selection of objects using the magnitude system }
   \label{fig:magnitudes}	

\end{figure}

 Hearnshaw \cite {2005mest.book.....H} provides an excellent account of this progression pointing out that while many systems existed, the ability to combine and standardise star catalogues was a consistent concern. When telescopes were introduced to astronomy it became possible to see even fainter stars than those observed with the naked eye, and the magnitude scale moved beyond the value of 6. With current state of the art instrumentation we are entering an era of observing objects as faint as magnitude 30 \cite {Zackrisson:2010kr} and possibly beyond. To determine where a star is on the magnitude scale, a reference star is chosen and allocated a standard number. At one point Polaris was assigned the magnitude value of 2.0, but this star is a variable star (apparent magnitude changes over time) so this was not an appropriate reference star. Vega was finally selected and assigned the value of 0. Using the star Vega as a reference point for magnitude 0, the table of magnitude values for the Moon, Planets and the Sun requires the magnitude scale to enter negative values as shown in Figure \ref {fig:magnitudes}.

\par What is being measured during the photometric process is the apparent brightness (or apparent magnitude) of an object and not its actual magnitude. To illustrate the difference in actual versus apparent magnitude consider the apparent brightness of a 40 watt bulb as seen from 10 meters versus 10 kilometres. In both cases the light bulb retains the same luminosity, but the apparent brightness is dramatically different due to the distance between the observer and the light bulb.

\par Photometry measurements from the time of Ptolemy remained relatively unchanged for about fifteen centuries with photometry estimations not improving until William Herschel (1738-1822) produced the first reliable naked eye star catalog using a telescope. Herschel used a system of estimating the difference between objects, which was later formalised by Friedrich Argelander (1799-1875) who established the step method. John Herschel continued his father's work achieving an estimated error of $\pm0.12$ magnitude which is close to the practical limit of visual photometry, $\pm0.1$ mag. The visual photometer, which appeared in the mid 19th century used prisms to project two objects into the field of view of the observer who would then equalise the apparent brightness of each object through a series of adjustments.  The relative difference in the calibration process contributed to the calculation of the magnitude difference between the objects. 

\par The next major advancement in photometry was the introduction of the photographic plate in 1871. By comparing existing stars within reference catalogues using visual inspection, large photographic surveys were undertaken. By 1930 it was estimated that measurements could now be made which were at a precision of $\pm0.02$ mag. 

\par William Henry Stanley Monck (1839-1915), made the first electrical measurement of light 1892 but it was not until 1907 however, that Joel Stebbins (1878-1966) used the photoconductive cell in conjunction with the photocell achieving a photometric accuracy of $\pm0.023$ mag, surpassing existing accuracy levels of photographic photometry. Frequent technical issues ensured that photographic plates remained in place for the first half of the 20th century. 

\par The glass photomultiplier tube (\emph{pmt}) became the primary photometry measurement system in the mid 20th century due to to its quantum efficiency of about 10\%-30\% compared to just 1\% for photographic systems. These systems would make way for the CCD by the early 1970's which provided two dimensional array detectors, and ultimately higher levels of precision.  A good summary of the history of the photometric evolution is given by Richard Miles, 2007\cite{miles2007light}.

\subsection {Charge Couple Devices}

\par  It is only quite recently that our ability to measure and record highly precise magnitude values for faint objects has developed, and this ability can be attributed to the introduction and use of the CCD in the 1970s \cite{howell2006handbook}.  With highly accurate photometry measurements based on CCD technology and careful observation it is possible to detect fluctuations in apparent magnitude, which may be used, among other things, in the identification of extra-solar planets orbiting distant stars \cite {Barnes:2009wi}.

\par
\begin{figure} [htbp]
\centering
  \includegraphics[width=0.6\textwidth] {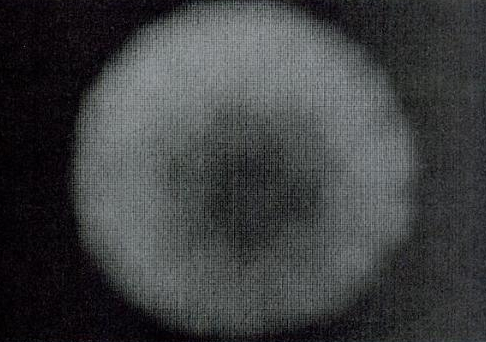}
  \vskip -0.8em
    \caption{First CCD image of Uranus taken in 1976 at Mount Lemmon \cite {janesick2001scientific}. }
  \label{fig:uranus}
\end{figure}

 The CCD was invented in 1969 at Bell Labs \cite {2005mest.book.....H} by William Boyle and George Smith and was initially conceived as a memory module, but it was only four years later that a program was initiated within the NASA \gls{jpl} to work on an imaging device of greater than 100x100 pixel resolution for space based probes using this technology. By 1974 Texas Instruments, under contract from NASA, produced a report outlining the feasibility of such a device \cite {Antcliffe:1975wu}, which would later be used in the Galileo mission to Jupiter (1989) and the Hubble telescope (1990), among others. By the mid 1970s it was decided that in order to engage the scientific community in this new CCD technology (the primary technology in use at the time was photographic film and vidicons) was to create a mobile CCD device, which was brought to a number of ground based telescopes.  According to Janesick, new scientific discoveries were  made each time the camera system visited a new site \cite {Janesick:1992ww},  and it was these trips and a high quantum efficiency rating (photon to electron conversion rate) which quickly led to a dramatic increase in the demand for CCD devices. Figure \ref{fig:uranus} shows the first ever image of Uranus \cite {janesick2001scientific} taken using a 400x400pixel CCD sensor on Mount Lemmon's 154cm telescope. Since then the CCD has evolved  and  become the dominant device used by professional astronomers.

\par
\begin{figure} [ht]
\centering
\fbox{   \includegraphics[width=0.6\textwidth] {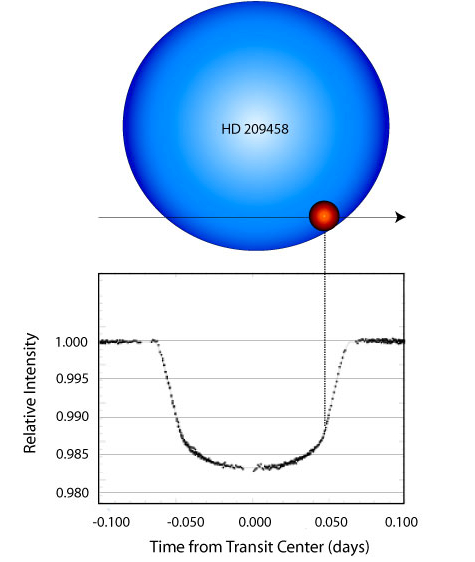}}
  \vskip -0.8em
    \caption{Detection of the first planetary transit of the star HD 209458. \cite {Charbonneau:2000fh}. }
  \label{fig:transit}
\end{figure}

 There are many reasons for requiring accurate magnitude calculations and an interesting and topical example of their use is in the hunt for extra-solar planets \cite {Barnes:2009wi}. One such method is referred to as the photometric transit method \cite {1984Icar...58..121B} where stars with planetary bodies that rotate around the star in the same plane cause a reduction in the apparent magnitude of the star. The transit is detected using a photometric light curve as shown in Figure \ref{fig:transit} for star HD 209458, which was the first planetary transit of a star identified using this method. 

\par The most well known example of this method in use, is the recent Kepler Mission, which had the specific aim of detecting Earth-class extra-solar planets around stars with a detection method determined to be viable based on two factors, "that the stellar variability on the time scale of a transit be less than the decrease in brightness caused by the transit and that the photometric instrumentation has sufficient precision to detect the transit". \cite {Koch:1998ck}. The instrumentation referred to is the CCD of which there are 42 devices. Each 50x25 mm CCD has 2200x1024 pixels. The level of change to the brightness of the object is as small as 1/10,000 (100 parts per million, ppm) for a terrestrial sized planet, which provides some indication of the need for high precision photometry. 

\par
\begin{figure} [ht]
\centering
\fbox{   \includegraphics[width=0.6\textwidth] {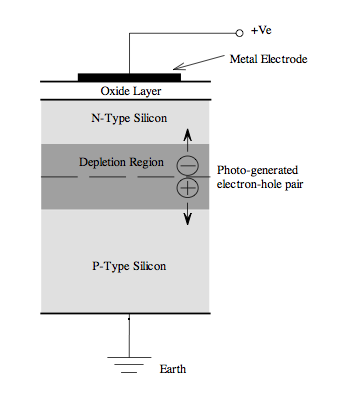}}
  \vskip -0.8em
    \caption{Cross section through a CCD pixel \cite {vikdhillon}. }
  \label{fig:ccd-diagram}
\end{figure}

 The CCD has contributed dramatically to photometry within the astronomical community. It is the primary instrument of most, if not all, of the large-scale optical astronomical survey systems currently in existence. As stated, the Kepler spacecraft uses 42 CCD devices, the Sloan Digital Survey comprises of 30 CCD devices of approximately 2048x2048 pixels in size, and the Hubble telescope initially used 8 low resolution CCD chips. A CCD chip is built on a single wafer and is made up of a two dimensional array of pixels. As a photon of light hits the silicon surface of a light sensitive CCD pixel the energy is absorbed raising some electrons to a higher energy state and releasing them, allowing them to flow towards the  n-type silicon layer as shown in Figure \ref{fig:ccd-diagram} where an electrical charge accumulates which is directly related to the level of incident light. This potential well exists for each pixel. After a period of time (the exposure time) the  accumulated charge for the pixel, is moved towards the readout point by transferring the charge across the device.  Once the parallel shift of pixel charges is performed, the pixels at the edge of the device are transferred using a serial shift to transfer the charge to the measurement electronics. Using an analog to digital converter, the charge is converted to a digital numerical value for the specific pixel charge. This is used as the raw digital image value from the exposure and is transferred to the computer. 

\subsection {The Data Processing Challenge}

\par When a single CCD detector records an image, the size of the digital image is usually dependant on the number of pixels on the device and the number of bytes used to store the value for the pixel. The size of the data-set generated by an array of CCDs is dependant on the size of each digital image, the image capture rate (ranging from milliseconds to minutes), the time period over which images are taken, and the number of CCDs in the array. While a small telescope may use a single CCD, larger telescopes may employ an array of CCDs, and robotic telescope farms may use an array of telescopes each with its own CCD array. With megapixel CCD arrays already in use, and with frame rates per second increasing, the tsunami of data production is already beginning.  Indeed, Graham \cite{Graham:2009vs} refers to the \emph{data avalanche, tsunami and explosion} of data, predicting that by 2020 the scale of the problem will be apparent as not just optical, but radio telescopes generate petabytes of data on a nightly basis. Ferguson et al \cite{Ferguson:2009ts}, looking to the next decade of data reduction and analysis sees the three major challenges as follows: 

	\begin{enumerate}
	\item Data rates growing rapidly as CPU processing rates level off.
	\item Industry trends in computing hardware leading to major changes in astronomical algorithms and software.
	\item Computationally demanding analysis techniques becoming more essential with increasing pressure on computing resource.
	\end{enumerate}

\par New sky survey systems in development, such as the \gls{lsst}, will produce up to 20 terabytes of data per day in the very near future. Supercomputers/high performance computing is proposed as a primary requirement with parallelism being the natural development to address challenge one above. With this level of data production, issues beyond processing must also be considered. Concerns about storage, input/output and processing have been in the published literature for many years. Shaw et al in 1995, in a short paper \cite{Shaw:1995ur} described the growing issue of large databases of data, the possibility of moving to lossless compression, and stating that vast data arrays will \emph{"tax networks, I/O systems, and storage capabilities, so compression techniques will be crucial"}.  These concerns remain the same today. Murtagh et al \cite{Murtagh:2002uv} summarised the issue as follows going on to discuss the requirement for some form of image compression strategy for data movement. \emph{"The quantity of astronomical data is rapidly increasing. This is partly owing to large digitized sky surveys in the optical and near infrared ranges. These surveys, in turn, are due to the development of digital imaging arrays such as charge coupled devices (CCDs). The size of digital arrays is also increasing, pushed by astronomical research demands for more data in less time."} 
 
 \par It is only when considering the combination of these challenges that the extent of the problem of large dataset production and processing can be fully appreciated. The factors which contribute to large dataset generation are summarised as follows. 

	\begin{itemize}
\item Resolution: 		Number of pixels captured per image.
\item Capture Rate: 		Number of images taken per second.
\item Capture Period:		The length of time over which images can be taken. 
\item Device Count: 	The number of capture devices operating at one time.
\item Capacity:		Ability to read and store data generated.
	\end{itemize}

\par When a CCD image is created, the pixel value is a combination of both signal and noise.  The process of performing image reduction and preparing the image for use in photometry is an essential step in all CCD image based pipelines and is often referred to as CCD reduction. There are  typically three calibration frames used in the reduction of raw images, which are bias, dark and flat field frames \cite {Gilliland:1992uk}.  In addition to these basic reductions, further image processing is required to complete the calculation of an accurate magnitude value for a series of reference objects within each image and all of these steps are precursors to the production of light curves from the CCD image. As the number of images produced increases, so does the processing requirement as this reduction process is applied to each image.

\subsection {Research Scope }

\par The data processing of CCD images is restricted in this research to pixel level calibration and basic photometric analysis. Figure \ref{fig:pcal-phot} provides a summary of the operations performed by the NIMBUS pipeline which stops short of performing any actual science on extracted magnitude values from images. To ensure that the ability to analyse magnitude values can be done in real-time, PCAL and PHOT should process data at the same rate as data is being generated and supplied to the pipeline. Just-in-time processing must be completed within a 24 hour period which would mean data processing must be no less than 3 times slower than data acquisition before a bottleneck is created, assuming an 8 hour image capture period per day. 
\tikzstyle {box} = [draw,rectangle,node distance=5cm,rounded corners=0.7ex,minimum height=7em, minimum width=8em]
\tikzstyle {box1} = [draw,rectangle,node distance=5cm,rounded corners=0.7ex,minimum height=4em, minimum width=8em]
\tikzstyle {newbox1} = [draw,rectangle,node distance=5cm,rounded corners=0.1ex,minimum height=4em, minimum width=11cm]

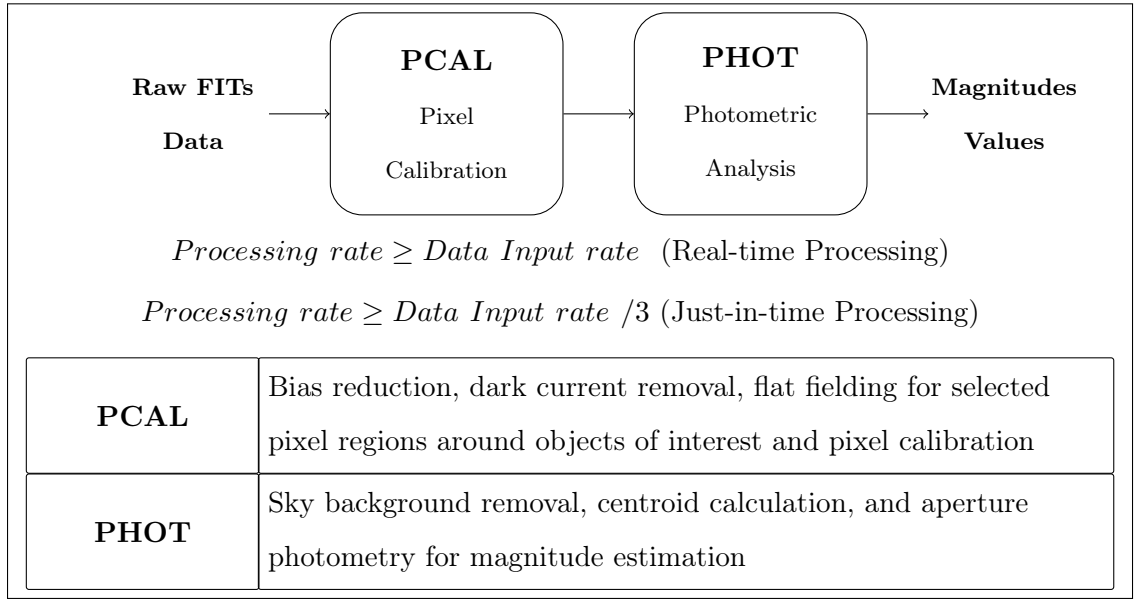
\begin{figure}[htbp]
 \begin{center}
\fbox { 
  \begin{tikzpicture}
           \node (rawdata) [box, rounded corners=12pt, draw, align=center] at (-4,0) {\textbf{PCAL}\\ \footnotesize{Pixel}\\\footnotesize{Calibration}};
           \node (caldata) [box, rounded corners=12pt, draw,align=center] at (0,0) {\textbf{PHOT}\\ \footnotesize{Photometric}\\\footnotesize{Analysis}};
           \draw  ($(rawdata.west)+(-18mm,-0mm)$)                     node [align=center]              {\textbf{\footnotesize{Raw FITs}}\\\textbf{\footnotesize{Data}}};
           \draw  ($(caldata.east)+(18mm,-0mm)$)                     node [align=center]           {\textbf{\footnotesize{Magnitudes}}\\ \textbf{\footnotesize{Values}}};

           \draw[->] (rawdata) -- (caldata) ;
            \draw[->] ($(rawdata.west)+(-8mm,-0mm)$)  -- (rawdata) ;
            \draw[->] ($(caldata.east)$)  -- ($(caldata.east)+(+8mm,-0mm)$) ;
            
            \draw  ($(caldata.south)+(-25mm,-5mm)$)                     node [align=center]              {\textbf{$ Processing\ rate \ge Data\ Input\ rate \  $}   (Real-time Processing)};
             \draw  ($(caldata.south)+(-25mm,-13mm)$)                     node [align=center]           {\textbf{$ Processing\ rate \ge Data\ Input\ rate \ /3 $}   (Just-in-time Processing)};

               \node (PVAL) [box1, rounded corners=1pt, draw, align=left] at (-8,-4) {\textbf{PCAL}};
               \node (VAL) [newbox1, rounded corners=1pt, draw, align=left,text width=11cm] at (-0.85,-4) {Bias reduction, dark current removal,  flat fielding for selected pixel regions around objects of interest and pixel calibration};
               \node (PHOT) [box1, rounded corners=1pt, draw, align=left] at (-8,-5.54) {\textbf{PHOT}};
               \node (VAL2) [newbox1, rounded corners=1pt, draw, align=left,text width=11cm] at (-0.85,-5.54) {Sky background removal, centroid calculation, and aperture \\ photometry for magnitude estimation};

   \end{tikzpicture}
  }

  \caption{An overview of calibration and photometric analysis performed on raw CCD images within the NIMBUS pipeline. }
  \label{fig:pcal-phot}
 \end{center} 
\end{figure} 

\section {Research Hypothesis}

\par The research hypothesis proposed within this thesis was to determine if a globally distributed astronomical CCD image reduction pipeline can process data at a rate which exceeds existing data processing requirements and is scalable such that it can meet future data processing challenges.  

\par To support scalable growth, a pipeline would be required to allow horizontal scaling of all components relying on parallel processing of data in a robust reliable manner. Work orchestration requires the communication of thousands of computing processes allowing for nodes to be added or removed without interfering with the running of the system.  Image processing must be suitable to parallelised processing. The pipeline should also not be restricted to the specifics of CCD photometry, but be flexible enough to process other data products, which can also be processed in parallel.

\section {Thesis Contributions}

\par A distributed pipeline was conceived which was validated to determine how processing rates upwards of 200 terabytes of data per day could be achieved. The NIMBUS pipeline, developed as the primary contribution of this thesis, accomplishes this by enabling the use of global computing resources  to easily and seamlessly contribute to image processing. The key enabling features of the architecture are distributed web queuing for message based communications, self configuring workers which allow for multiple science payloads to be processed, system resilience which allows running workers to join or leave the pipeline seamlessly, parallel processing of images, and decentralised storage and processing. 
	
The main contributions of this thesis can be identified as follows:
	\begin{enumerate}
		\item NIMBUS, a globally distributed data processing architecture that can process hundreds of terabytes of data per day which is also scalable beyond this point.
		\item A self configuring, balancing system that is scalable and resilient to failure.
		\item A dynamically reconfigurable pipeline that has wider applications than astronomical image processing. 
		\item A real-time pipeline, which in this context may involve a small processing latency in the order of one to two minutes, but this latency would be small enough to enable a feedback response to the telescope, or observatory, to allow them to react to recently captured and processed scientific data.		
		
	\end{enumerate}

\newpage 
\section{Structure of this Thesis}

This thesis is structured as follows. 
	\begin{description}
	\item [Chapter 1]  \hfill \\
This chapter provides context for the thesis and introduces the challenge posed by the growing volume of image data within astronomy. An overview of how this thesis proposes to address this problem is presented through the globally distributed NIMBUS pipeline, and the contributions are clearly identified. 
	\item [Chapter 2]  \hfill \\
This chapter reviews the literature on the processing of astronomical photometry, describing the primary noise sources within a CCD and the processes of reducing them. Data generation sources within the astronomical community are identified in addition to reviewing existing data processing  techniques,  processing rates and the data volume challenges facing the astronomical community. Distributed computing techniques and how these are being used within the scientific community at present within astronomy and other scientific disciplines are also reviewed.
	\item [Chapter 3]  \hfill \\
Chapter 3 gives an overview of the experimental methodology used in all experiments and  discusses a series of experimental designs based on distributed computing techniques and how they can be used. The experimental setup for Chapters 5 and 6 are also presented with a review of the technology used in these experiments. 
	\item [Chapter 4]  \hfill \\
	Chapter 4 presents the design, implementation, results and evaluation of the \gls{acn} experimental model. This model uses a distributed hybrid cloud, which relies on a private \gls{nfs} locking mechanism to communicate with workers available data for processing. The results and findings from these experiments are presented within the chapter. 
	\newpage
	\item [Chapter 5]  \hfill \\
	In Chapter 5 the design, implementation, results and evaluation of the NIMBUS experimental module are described. NIMBUS uses a global processing cloud controlled incorporating Amazon Web Services for web based message queuing, computing instances and data storage. The results and findings from these experiments are presented within the chapter. 
	\item [Chapter 6]  \hfill \\
	Chapter 6 provides an analysis of the findings of this thesis, reviews and summarises the work presented, drawing conclusions and identifying possible future work.

	\item [Appendicies]  \hfill \\
	Appendix A through D provide additional data and material referenced within the main chapters of this thesis.

	\end{description}

%% file: chapter2.tex

\chapter{Astronomical Photometry Data Processing}
\label{chapter2}

\section{Introduction}

\par In this chapter the literature for standard reduction techniques, basic photometry, the source of data and existing data processing practices is reviewed. These four sections are essential to demonstrate that the data processing techniques implemented within the NIMBUS pipeline are consistent with standard data processing operations performed on astronomical CCD images. It is also required to provide context for the NIMBUS pipeline against existing technologies and current state of the art practices.  

\par CCD and CMOS imaging systems have well understood reduction processing steps designed to calibrate a raw image. The accuracy of photometric measurement is based on these well defined cleaning techniques which are discussed in more detail in this chapter.   

\par Aperture photometry techniques provide a clear process for the estimation of apparent magnitude of objects, using a standard reference scale.  Finding the centre of objects, estimating the sky background and calculating the flux of an object for a range of aperture sizes are all well defined procedures. 

\par The sources of CCD data production are also considered, including existing and future data processing techniques and challenges. The most commonly  referenced projects, given as examples of the growing data challenge within astronomy, are the Large Synoptic Survey Telescope (LSST) \cite{Howell:2009vz} which is primarily optical in nature, and the \gls{ska} \cite{Dewdney:2009uz} which is radio based. The LSST (expected to come on-line in 2019) is predicting data acquisition sizes in the region of 20-30 terabytes per night, while the SKA (2024) has a variety of predicted data capture rates depending on the implementation size of the array, although most agree this has the potential for generating one terabytes per second \cite{Jones:vk}.  Data processing requirements vary extensively and the following categories are reviewed in terms of their data generation capabilities.

\par Observatories large and small, space based and ground based all employ some form of software processing on images. Well established file formats, a range of data reduction pipelines and an extensive set of software packages and technologies ensures there are a mix of approaches in existence for performing data reduction.

\section{Standard Image Reduction Techniques}
\par When a CCD instrument is used, the recorded file output stored on the computer contains a measure of the source signal in addition to unwanted random contraptions for various sources. The noise introduces an error into the measurement. In this section the sources of noise in CCD image reading are described along with the techniques used to deal with them. These techniques are incorporated into the NIMBUS system.

\subsection{Noise Sources}

\par If a CCD  recorded a single electron for every photon striking a pixel and this was the readout value obtained from the CCD then it could be considered to be a \emph {perfect CCD}. In reality this is not the case, and a number of factors contribute to the introduction of noise to the process.  Noise is the introduction of unwanted variations to the image, distorting the readings in some way.  If a CCD pixel has a well depth of 100,000 electrons (the total amount of charge that can be stored in a pixel) and the average noise can be determined to be approximately 40 electrons per pixel then the SNR (Signal to Noise Ratio) is 100,000/40  or 2,500. If the amount of noise can be reduced then the SNR is increased. The process of reducing the level of noise in an image is critical to performing high precision photometry. The standard equation for SNR is given in equation \ref{eq:SNR} and is often unofficially referred to as the CCD Equation \cite{fowler1981evaluation}. 

\begin{equation}
\label{eq:SNR}
SNR = \frac {N_{*}}{\sqrt{N_{*} + n_{pix}(N_{S} + N_{D} + N_{R}^{2}) }}  
\end{equation}

	\begin{itemize}
	\item $N\textsubscript{*}$ represents the total number of photons collected from an object of interest which can be either 1 or more pixels.  
	\item  $n\textsubscript{pix}$ represents the total number of pixels considered.  
	\item  $N\textsubscript{S}$ represents the total number of photons per pixel from the sky background.  
	\item  $N\textsubscript{D}$ represents the total number of dark current electrons per pixel.  
	\item  $N\textsubscript{R}$\textsuperscript{2} represents the total number of electrons per pixel from read noise.  

	\end{itemize}

	\par The main contributions to noise within a CCD are described here and the method for reducing them is expanded upon in the following subsections :
	\begin{itemize}
		\item Dark current is a thermal noise source build-up within a pixel. Longer exposures exhibit cumulative effects as it is exposure time dependant. Cooling the CCD is an effective strategy to reduce this noise in addition to the use of shorter exposures or through the use of \emph{Dark Frames}.
		\item Pixel non-uniformity refers to the variation in pixel sensitivity when exposed to the same levels of light. While standards are exacting, differences in pixel sensitivity exist. This noise is eliminated to some degree by \emph{Flat Fielding} which also eliminates other optical variations or dust.   
		\item Read noise is an additive noise source introduced during the reading of values from the CCD during the conversion from an analog to a digital number. It is primarily removed using a \emph {Bias Frame} as the amount of noise is independent of the exposure time. There have been dramatic improvements in this noise reduction over time. Initially this value was as high as 500 electrons per readout in early astronomical CCD imagers, but this value has been reduced to as low as 3 electrons per readout in CCDs and as low as 1 in CMOS devices. 
		\item  Charge transfer efficiency. As pixel charge values are moved across the CCD towards the readout point some electrons may get dropped or lost during the transfers. A charge may be transferred thousands of times (for example in a 1024x1024 pixel array the maximum a charge will be transferred is 2048 time). The CTE is not usually an issue with a typical efficiency rating of 99.99997\% not being uncommon. This relates to approx. 2.5 electrons out of every 85,0000 being left behind or lost. 
		\item Cosmic rays are particles which travel at high velocity and may dramatically increase the electron count for a pixel. The energy released by the particle releases many electrons which are then recorded as bright spots on the image.  If this value is used, it has the ability to distort calculations such as star magnitudes, bias pixel values or flat field pixel values.
	\end{itemize}
	
\par The noise within the CCD can be characterised using equation \ref{eq:ccd1} \cite {corl:2004:Online}. When a CCD image is taken, a two dimensional digital representation of the image is created, which we can reference using $x$ for the column position and $y$ for the row position. The values recorded in a file are the digital counts of the charge generated by the electrons detected by each pixel (the signal) and additional charges relating to unwanted sources (noise).   	
	\begin{equation}\label{eq:ccd1}
	s(x,y)=B(x,y) + tD(x,y) + tG(x,y)I(x,y) + random noise \  	
	\end{equation}
	
where 	\begin{itemize}
		\item $(x,y)$ represents the pixel row and column position on the image.
		\item $s(x,y)$ is a raw pixel digital count recorded on the CCD for an integration time of $t$.
		\item $t$ is the integration time of the exposure in seconds.
		\item $B(x,y)$ is the \emph{bias} digital count of each pixel for a $0$ length exposure.
		\item $D(x,y)$ is the \emph{dark current} digital count of each pixel for an exposure length of $t$ seconds.
		\item $G(x,y)$ is the \emph{sensitivity}  of each pixels.
		\item $I(x,y)$ is the digital count of the light flux received by the pixel.
		\end{itemize}

\subsection {Bias Frames}

\par
\begin{figure} 
\centering
 \fbox{  
\begin{tikzpicture}[scale=.7,every node/.style={minimum size=1cm}]
   \begin{scope}[
           yshift=-60,every node/.append style={
           yslant=0.5,xslant=-1},yslant=0.5,xslant=-1
           ]
       \draw[step=4mm, black,thick] (0,0) grid (5,5); 
       \draw[black,line width=0.7mm] (0,0) rectangle (5,5);
       \fill[black] (1.25,0.75) rectangle (1.55,0.45); 
   \end{scope}

     \begin{scope}[
           yshift=-40,every node/.append style={
           yslant=0.5,xslant=-1},yslant=0.5,xslant=-1
           ]
       \fill[white,fill opacity=1] (0,0) rectangle (5,5);
       \draw[step=4mm, black] (0,0) grid (5,5); 
       \draw[black,line width=0.7mm] (0,0) rectangle (5,5);

       \fill[black] (1.25,0.75) rectangle (1.55,0.45); 
   \end{scope} 

     \begin{scope}[
           yshift=-20,every node/.append style={
           yslant=0.5,xslant=-1},yslant=0.5,xslant=-1
           ]
       \fill[white,fill opacity=1] (0,0) rectangle (5,5);
       \draw[step=4mm, black] (0,0) grid (5,5); 
       \draw[black,line width=0.7mm] (0,0) rectangle (5,5);

       \fill[black] (1.25,0.75) rectangle (1.55,0.45); 
   \end{scope}

   \begin{scope}[
           yshift=0,every node/.append style={
           yslant=0.5,xslant=-1},yslant=0.5,xslant=-1
           ]
       \fill[white,fill opacity=1] (0,0) rectangle (5,5);
       \draw[step=4mm, black] (0,0) grid (5,5); 
       \draw[black,line width=0.7mm] (0,0) rectangle (5,5);

       \fill[black] (1.25,0.75) rectangle (1.55,0.45); 
   \end{scope}
   
   	\draw (7,2.5) node {$Bias\  Frame\ 1 $};
	\draw (7,1.75) node {$Bias\  Frame\ 2$};
	\draw (7,1) node {$Bias\  Frame\ 3$};

	\draw (7,0.25) node {$Bias\  Frame\ 4 $};
	\draw (7,-3.5) node {$Average\ pixel \  value\  calculated$};
	\draw[-latex,orange,line width=0.3mm](6.7,-3)node[below]{}  to[out=90,in=20] (0.8,1);
	\draw[-latex,orange,line width=0.3mm](6.7,-3)node[below]{} to[out=90,in=20] (0.8,0.25);
	\draw[-latex,orange,line width=0.3mm](6.7,-3)node[below]{}  to[out=90,in=20] (0.8,-0.5);
	\draw[-latex,orange,line width=0.3mm](6.7,-3)node[below]{}  to[out=90,in=20] (0.8,-1.25);
	\draw (6.7,-4) node {$for\ use\ in\ Master\ Bias\ frame$};

\end{tikzpicture} }

 \vskip -0.8em
    \caption{Master bias frame created using multiple bias frames. }
  \label{fig:bias-grid}
\end{figure}
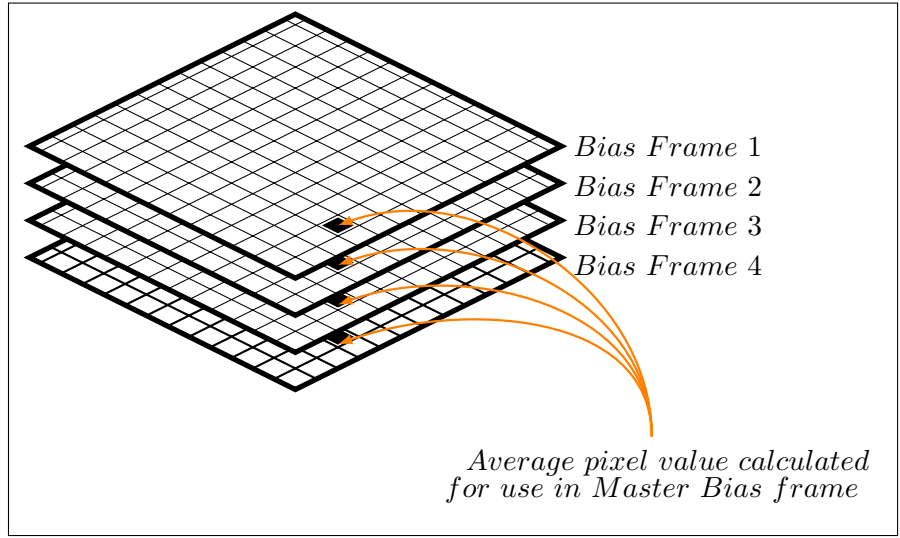

 A bias frame has a dark frame with an exposure time of zero and is a measure a pixel's read-noise. This value is usually caused by a low level spatial variation caused by the on-chip CCD amplifiers. Read-noise from a CCD is an additive noise source that is introduced during the pixel read process which does not vary with exposure time. This is a systematic  noise source which must be removed.  A master bias frame is created through the combination of multiple bias frames using the average pixel values seen across each frames as shown in Figure \ref{fig:bias-grid}.  An average value is considered acceptable given that the CCD should not be exposed to cosmic rays since there was no exposure of the CCD sensors. The master bias frame can then be used in cleaning data images by subtracting the master bias value for each pixel.

\par Given that a bias frame is taken with a time interval of $t=0$ equation \ref{eq:ccd1} can be reduced to equation \ref{eq:bias}, where $b(x,y)$ is the bias value recorded in the pixel $s(x,y)$. 

\begin{equation}
\label {eq:bias}
s(x,y) = b(x,y) = B(x,y) + noise
\end{equation}

\par While a simple estimate of the bias value $B$ of a pixel can be obtained by simply using the value $b(x,y)$  a better estimate of \emph{B} can be obtained using the average value of the pixel across  multiple bias images where $N$ is the number of bias images used as shown in equation \ref{eq:gmb}.  
\begin{equation}\label{eq:gmb}
\widehat{B}(x,y)=\frac {1}{N} \sum_{i=1}^{N}b_{i}(x,y)
\end{equation}

\begin{figure} [htbp]
\centering
  \includegraphics[width=0.5\textwidth] {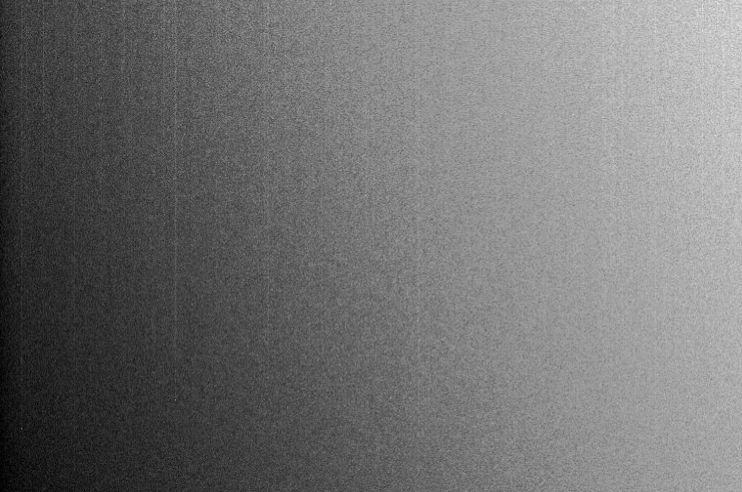}
  \vskip -0.8em
    \caption{Example of a master bias frame.}
  \label{fig:masterbias}
\end{figure}

\subsection {Dark Current}
\par A dark frame is used to determine the level of noise introduced by thermal events in the CCD device. Most devices are now cooled and thus reduce the amount of noise however it is still useful to understand this process. While a bias frame is taken with the shutter closed and the minimum exposure time, this dark frame is taken with the shutter closed, but for a specific period of time. The noise from thermal interference is time dependant, so the length of time of the exposure is important as the noise level is related to it.  

\par The formula for a dark current frame with an exposure time $t_{dark}$ is given in equation \ref{eq:dark1} when the closed shutter sets $I=0$. 

\begin{equation}
\label{eq:dark1}
s(x,y) = d(x,y) = B(x,y) + t_{dark}D(x,y) + noise
\end{equation}

\par An estimate of dark current for a pixel is found by subtracting the bias and dividing by the exposure time.    

\begin{equation}
\widehat{D}(x,y)=\frac {d(x,y) - \widehat{B}(x,y)}{t_{dark}}
\end{equation}

\par A more accurate estimate can be obtained by averaging a number of dark current frames $M$ and eliminating the bias from each value. This gives a bias reduced master dark current frame for time interval $t_{dark}$

\begin{equation}
\widehat{D}(x,y)=\frac {1}{t_{dark}}  \frac {1}{M} \sum_{i=1}^{M}d_{i}(x,y) - \widehat{B}(x,y)
\end{equation}

\par A Master Dark Image can then be expressed using equation \ref{eq:dmaster} for specific time intervals such as $t_{data}$ to suit different levels of image exposures. 

\begin{equation}\label{eq:dmaster}
\widehat{D}_{M}^{'}(x,y)=\widehat{D}(x,y) t_{data}
\end{equation}

\subsection {Flat Fielding}

\par A flat field image is taken when the CCD has been evenly illuminated by a light source.  Flat fielding is used to compensate for differences in pixel to pixel variations of the CCD response to illumination when the same amount and spectrum of light is illuminated across each pixel on the CCD. This technique also helps remove the effects of dust which can cause dark spots on an image and uneven illumination  caused by vignetting in the optical system. 

\par
\begin{figure} [ht]
\centering
  \includegraphics[width=0.5\textwidth] {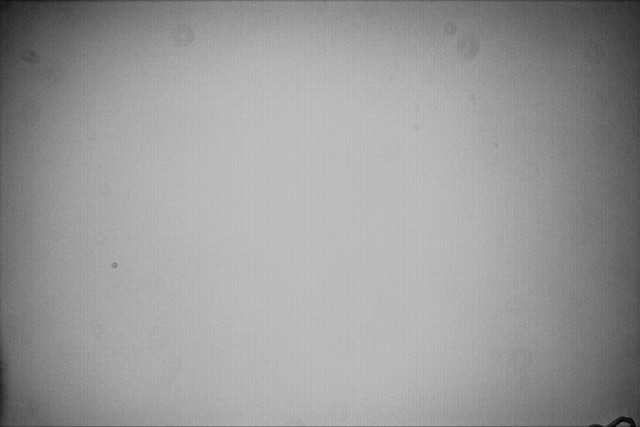}
  \vskip -0.8em
    \caption{Example of a flat field Image. }
  \label{fig:flat}
\end{figure}

 The flat field value is used to modify image pixel values to account for these variations.  There are varying opinions on the best method to create a good flat field image, such as the use of an illuminated painted screen inside the telescope dome \cite{Massey:1992vz}. The difficulty with any technique is  finding a means to illuminate  the CCD with a flat distribution of light which is representative of the wavelength of the light expected during actual image recording of objects of interest. Techniques range from closing the telescope dome and using a specially treated surface for illumination, to using an image of the evening sky prior to data capture where there is still enough light available to illuminate the entire CCD sufficiently. Howell provides an excellent overview of many of these approaches \cite{howell2006handbook}. A typical flat field is shown in Figure \ref{fig:flat}.

\par Typically there are a number of flat field images taken and combined into a Master Flat Field Frame. There are similarities to the generation of a Master Bias, however due to the exposure of the CCD to a light source the possibility of encountering cosmic rays is increased, so  median values, as opposed to average values, are used.  

\par This can be expressed as the flat field value for a pixel with an exposure time of $t_{flat}$ from Equation \ref{eq:ccd1} using Equation \ref{eq:ff1} where $L$ is the flux illuminated across all pixels equally. 

\begin{equation}
\label{eq:ff1}
s(x,y) = f(x,y)= B(x,y) + t_{flat} D(x,y) + t_{flat} G(x,y)L + noise
\end{equation}

\par As already described in equation \ref{eq:dmaster} a master dark can be created for the flat field image 
 $\widehat{D}_{M}^{F}(x,y)=\widehat{D}(x,y) t_{flat}$ 
 by using the same time integration as the flat field.

\par The noise of the flat field value can be further reduced by obtaining a $median$ value for each pixel position using a number of flat field images as per equation \ref{eq:ffmedian}. 

\begin{equation}
\label{eq:ffmedian}
\widetilde{f}(x,y)= Median f(x,y)  
\end{equation}

\par To find a median value, the values for a pixel position are read into an array and sorted with the median value found using the following formulas.

	\begin{description}
	\item [Medians with odd number of data points]  \hfill \\
	Sort data points and pick the middle data point and use its value. Where $n$ is the number of data points you must read the value at the following index point in the sorted list $(n+1)2$\footnote {for C  where the array starts at 0,  adjust this index as follows $(n-1)/2$}.
	\item [Medians with even number of data points]  \hfill \\
	Sort data points and pick the two middle data point and use the average of the two values. Where $n$ is the number of data points\footnote {for C  where the array starts at 0, adjust this index as follows $value1$ = $(n/2)-1$ and $value 2$ = $n/2$}  $value1$ = $\frac{n}{2}$ and $value 2$ = $\frac{n}{2} + 1$  giving $Median = (value1 + value 2)/2 $

	\end{description}

\par By subtracting the master dark frame  which used the same time interval as the flat frames and subtracting the master bias a corrected flat field value is obtained for each pixel which gives a bias reduced master flat field frame as per equation \ref{eq:cff}.  

\begin{equation}
\label{eq:cff}
f^{'}(x,y) = \widetilde{f}(x,y) - D^{F}_{M}(x,y) =  t_{flat} G(x,y)L 
\end{equation}

\par The final step is to create a normalised flat field which has an average value of 1. First calculate $\overline{F}$ the average value of all values in the flat field  where $n$ is the number of elements in the flat field image in equation \ref{eq:norm1}.  

\begin{equation}
\label{eq:norm1}
\overline{F} = \frac {\sum_{i=1}^{n}f^{'}_{i}(x,y) }{n}
\end{equation}

\par Normalising  each of the pixels in the master flat field can be performed by dividing each pixel by the average value $\overline{F}$ in equation \ref{eq:normff}.

\begin{equation}\label{eq:normff}
G(x,y) = \frac {f^{'}(x,y) }{\overline{F}}
\end{equation}

\subsection {Image Reduction}

\par The process of characterising the level of noise within a CCD pixel  has been presented in equation \ref{eq:ccd1}. Using the estimation techniques identified, a basic image calibration process designed to reduce noise from the CCD raw images, a necessary process in preparing the CCD images for analysis, can be summarised.  To simplify the process it can be assumed that the dark current frames were taken with the same time integration as the flat field frames and the data images. All images are stored using the \gls{fits} \cite{Wells:1981wn} format unless otherwise stated. 

	\begin{itemize}
	\item Multiple CCD bias frames are captured with time integration of $0$.  
	\item Multiple CCD flat Field frames are captured for integration time $t_{data}$.  
	\item Multiple CCD dark current frames for integration time $t_{data}$.  
	\item Generate a master bias frame using equation \ref{eq:gmb}. 
	\item Generate a bias reduced master dark current frame for integration time $t_{data}$ using equation \ref{eq:dmaster}.
	\item Generate a normalised bias reduced flat field master using equation \ref{eq:normff}. 
	\item Capture raw CCD images frames.  
	\end{itemize}

\par Following the capture of raw CCD images,  equation \ref{eq:ccd1} can be used to estimate the value of flux for a pixel for a time interval $t_{data}$ using equation \ref{eq:intensity}. 
\begin{equation}
\label{eq:intensity}
\frac {s(x,y) - \widehat{B}(x,y)  - \widehat{D}_{M}(x,y) }{G(x,y)} =    \widehat{I}(x,y)t_{data}  
\end{equation}

\par A pixel value on a CCD frame has the bias and dark current removed and is then adjusted for the calculated responsiveness of the pixel relative to all other pixels. This calculation must be performed on all pixels which are ultimately used in the calculation of magnitude values. A new version of the image can then be created containing the calibrated pixel values.  The creation of the master bias, flat field or dark frames is often done once for each night of observation and are then used in the calibration of pixels for that night. 
\section {Photometry using CCD images}

The general steps in classical photometry using a cleaned digital image are usually identified as follows.

\begin {enumerate}
\item Image centring, the process of finding the centre of an object.
\item Estimation of the sky background for the purpose of removing it from the flux intensity value. 
\item Flux value intensity calculation for an object for a specific aperture size. 
\item Magnitude calculation for an object for a specific aperture size taking into account the sky background. 
\end {enumerate}

Multiple magnitudes can be generated based on variations in the software aperture size used in the calculation of the flux intensity. Each of these steps are described below, identifying the basic techniques and formulas as appropriate.

\subsection {Centroid Algorithm}

\par
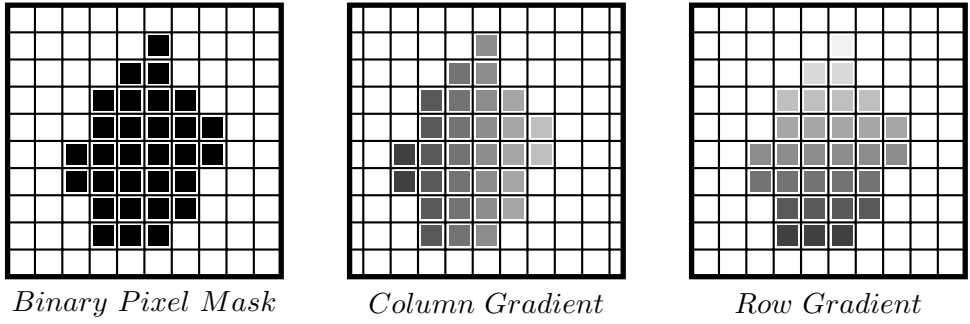
\begin{figure} [ht]
\centering
 
 \definecolor{light-gray1}{gray}{0.25}
  \definecolor{light-gray2}{gray}{0.35}
 \definecolor{light-gray3}{gray}{0.45}
 \definecolor{light-gray4}{gray}{0.55}
 \definecolor{light-gray5}{gray}{0.65}
 \definecolor{light-gray6}{gray}{0.75}
 \definecolor{light-gray7}{gray}{0.85}
 \definecolor{light-gray8}{gray}{0.95}

\begin{tikzpicture}[scale=.9,every node/.style={minimum size=1cm}]
   \begin{scope}[
           yshift=-60,every node/.append style={
           yslant=0,xslant=0},yslant=0,xslant=0
           ]
       \draw[step=4mm, black,thick] (0,0) grid (4,4); 
       \draw[black,line width=0.7mm] (0,0) rectangle (4,4);
       \fill[black] (1.25,0.75) rectangle (1.55,0.45); 
       \fill[black] (1.65,0.75) rectangle (1.95,0.45); 
       \fill[black] (2.05,0.75) rectangle (2.35,0.45); 
       
       \fill[black] (1.25,1.15) rectangle (1.55,0.85); 
       \fill[black] (1.65,1.15) rectangle (1.95,0.85); 
       \fill[black] (2.05,1.15) rectangle (2.35,0.85); 
       \fill[black] (2.45,1.15) rectangle (2.75,0.85); 

        \fill[black] (0.85,1.55) rectangle (1.15,1.25); 
        \fill[black] (1.25,1.55) rectangle (1.55,1.25); 
       \fill[black] (1.65,1.55) rectangle (1.95,1.25); 
       \fill[black] (2.05,1.55) rectangle (2.35,1.25); 
        \fill[black] (2.45,1.55) rectangle (2.75,1.25); 

       \fill[black] (0.85,1.95) rectangle (1.15,1.65); 
       \fill[black] (1.25,1.95) rectangle (1.55,1.65); 
       \fill[black] (1.65,1.95) rectangle (1.95,1.65); 
       \fill[black] (2.05,1.95) rectangle (2.35,1.65); 
       \fill[black] (2.45,1.95) rectangle (2.75,1.65); 
       \fill[black] (2.85,1.95) rectangle (3.15,1.65); 

       \fill[black] (1.25,2.35) rectangle (1.55,2.05); 
       \fill[black] (1.65,2.35) rectangle (1.95,2.05); 
       \fill[black] (2.05,2.35) rectangle (2.35,2.05); 
       \fill[black] (2.45,2.35) rectangle (2.75,2.05); 
       \fill[black] (2.85,2.35) rectangle (3.15,2.05); 

       \fill[black] (1.25,2.75) rectangle (1.55,2.45); 
       \fill[black] (1.65,2.75) rectangle (1.95,2.45); 
       \fill[black] (2.05,2.75) rectangle (2.35,2.45); 
       \fill[black] (2.45,2.75) rectangle (2.75,2.45); 
     
       \fill[black] (1.65,3.15) rectangle (1.95,2.85); 
       \fill[black] (2.05,3.15) rectangle (2.35,2.85); 

       \fill[black] (2.05,3.55) rectangle (2.35,3.25); 

   \end{scope}

      \begin{scope}[
           yshift=-60,every node/.append style={
           yslant=0,xslant=0},yslant=0,xslant=0
           ]
       \draw[step=4mm, black,thick] (10,0) grid (14,4); 
       \draw[black,line width=0.7mm] (10,0) rectangle (14,4);

       \fill[light-gray2] (6.05 ,0.75) rectangle (6.35 ,0.45); 
       \fill[light-gray3] (6.45,0.75) rectangle (6.75,0.45); 
       \fill[light-gray4] (6.85,0.75) rectangle (7.15,0.45); 
      
       \fill[light-gray2] (6.05,1.15) rectangle (6.35,0.85); 
       \fill[light-gray3] (6.45,1.15) rectangle (6.75,0.85); 
       \fill[light-gray4] (6.85,1.15) rectangle (7.15,0.85); 
       \fill[light-gray5] (7.25,1.15) rectangle (7.55,0.85); 

        \fill[light-gray1] (5.65,1.55) rectangle (5.95,1.25); 
        \fill[light-gray2] (6.05,1.55) rectangle (6.35,1.25); 
        \fill[light-gray3] (6.45,1.55) rectangle (6.75,1.25); 
       \fill[light-gray4] (6.85,1.55) rectangle (7.15,1.25); 
       \fill[light-gray5] (7.25,1.55) rectangle (7.55,1.25); 

       \fill[light-gray1] (5.65,1.95) rectangle (5.95,1.65); 
       \fill[light-gray2] (6.05,1.95) rectangle (6.35,1.65); 
       \fill[light-gray3] (6.45,1.95) rectangle (6.75,1.65); 
       \fill[light-gray4] (6.85,1.95) rectangle (7.15,1.65); 
       \fill[light-gray5] (7.25,1.95) rectangle (7.55,1.65); 
       \fill[light-gray6] (7.65,1.95) rectangle (7.95,1.65); 

       \fill[light-gray2] (6.05,2.35) rectangle (6.35,2.05); 
       \fill[light-gray3] (6.45,2.35) rectangle (6.75,2.05); 
       \fill[light-gray4] (6.85,2.35) rectangle (7.15,2.05); 
       \fill[light-gray5] (7.25,2.35) rectangle (7.55,2.05); 
       \fill[light-gray6] (7.65,2.35) rectangle (7.95,2.05); 

       \fill[light-gray2] (6.05,2.75) rectangle (6.35,2.45); 
       \fill[light-gray3] (6.45,2.75) rectangle (6.75,2.45); 
       \fill[light-gray4] (6.85,2.75) rectangle (7.15,2.45); 
       \fill[light-gray5] (7.25,2.75) rectangle (7.55,2.45); 

         \fill[light-gray3] (6.45,3.15) rectangle (6.75,2.85); 
       \fill[light-gray4] (6.85,3.15) rectangle (7.15,2.85); 

          \fill[light-gray4] (6.85,3.55) rectangle (7.15,3.25); 

   \end{scope}
   
         \begin{scope}[
           yshift=-60,every node/.append style={
           yslant=0,xslant=0},yslant=0,xslant=0
           ]
       \draw[step=4mm, black,thick] (5,0) grid (9,4); 
       \draw[black,line width=0.7mm] (5,0) rectangle (9,4);

       \fill[light-gray1] (11.25,0.75) rectangle (11.55,0.45); 
       \fill[light-gray1] (11.65,0.75) rectangle (11.95,0.45); 
       \fill[light-gray1] (12.05,0.75) rectangle (12.35,0.45); 
       
       \fill[light-gray2] (11.25,1.15) rectangle (11.55,0.85); 
       \fill[light-gray2] (11.65,1.15) rectangle (11.95,0.85); 
       \fill[light-gray2] (12.05,1.15) rectangle (12.35,0.85); 
       \fill[light-gray2] (12.45,1.15) rectangle (12.75,0.85); 

        \fill[light-gray3] (10.85,1.55) rectangle (11.15,1.25); 
        \fill[light-gray3] (11.25,1.55) rectangle (11.55,1.25); 
       \fill[light-gray3] (11.65,1.55) rectangle (11.95,1.25); 
       \fill[light-gray3] (12.05,1.55) rectangle (12.35,1.25); 
        \fill[light-gray3] (12.45,1.55) rectangle (12.75,1.25); 

       \fill[light-gray4] (10.85,1.95) rectangle (11.15,1.65); 
       \fill[light-gray4] (11.25,1.95) rectangle (11.55,1.65); 
       \fill[light-gray4] (11.65,1.95) rectangle (11.95,1.65); 
       \fill[light-gray4] (12.05,1.95) rectangle (12.35,1.65); 
       \fill[light-gray4] (12.45,1.95) rectangle (12.75,1.65); 
       \fill[light-gray4] (12.85,1.95) rectangle (13.15,1.65); 

       \fill[light-gray5] (11.25,2.35) rectangle (11.55,2.05); 
       \fill[light-gray5] (11.65,2.35) rectangle (11.95,2.05); 
       \fill[light-gray5] (12.05,2.35) rectangle (12.35,2.05); 
       \fill[light-gray5] (12.45,2.35) rectangle (12.75,2.05); 
       \fill[light-gray5] (12.85,2.35) rectangle (13.15,2.05); 

       \fill[light-gray6] (11.25,2.75) rectangle (11.55,2.45); 
       \fill[light-gray6] (11.65,2.75) rectangle (11.95,2.45); 
       \fill[light-gray6] (12.05,2.75) rectangle (12.35,2.45); 
       \fill[light-gray6] (12.45,2.75) rectangle (12.75,2.45); 
     
       \fill[light-gray7] (11.65,3.15) rectangle (11.95,2.85); 
       \fill[light-gray7] (12.05,3.15) rectangle (12.35,2.85); 

       \fill[light-gray8] (12.05,3.55) rectangle (12.35,3.25); 

   \end{scope}
   
   	\draw (2,-2.5) node {$Binary\ Pixel\ Mask $};
	\draw (7,-2.5) node {$Column\ Gradient $};
	\draw (12,-2.5) node {$Row\ Gradient $};

\end{tikzpicture} 

 \vskip -0.8em
    \caption{Centroid detection using a gradient technique }
  \label{fig:centroid}
\end{figure}

 Once an image has been reduced, the first step is the calculation of the geometric centre of an object of interest, which must be precisely determined. There are a number of different  algorithms available \cite{adams1980stellar}. A gradient based technique is presented in Figure \ref{fig:centroid} along with the corresponding algorithm,  Algorithm \ref{alg:centroid}, given in section 4.2.3.4 of Chapter 4. The first step using this method is to clip a region of the image where the point source is located and apply a binary mask where all pixels above a chosen threshold are set to the value $1$ and all pixels below the threshold are set to $0$. The $X$ position of the centre is found by using a column gradient where each pixel within the mask  has its value set to the column number that it is in. The $Y$ position of the centre is found by using a row gradient where each pixel within the mask  has its value set to the row number that it is in. Where $N = number\ of\ pixels\ in\ the\ mask$,  $M_{C}(x,y)$ representing a pixel within the column gradient image, and $M_{R}(x,y)$ representing a pixel in the row  gradient image,  the centroid $X$ and centroid $Y$values using Equation \ref{eq:centroid} can be calculated. 

\begin{equation}
\label{eq:centroid}
CentroidX =   \frac {\sum{M_{C}(x,y)}  }  {N}   \ \ \ \  CentroidY =   \frac {\sum{M_{R}(x,y)}  }  {N} 
\end{equation}

\subsection {Sky Background Estimation}

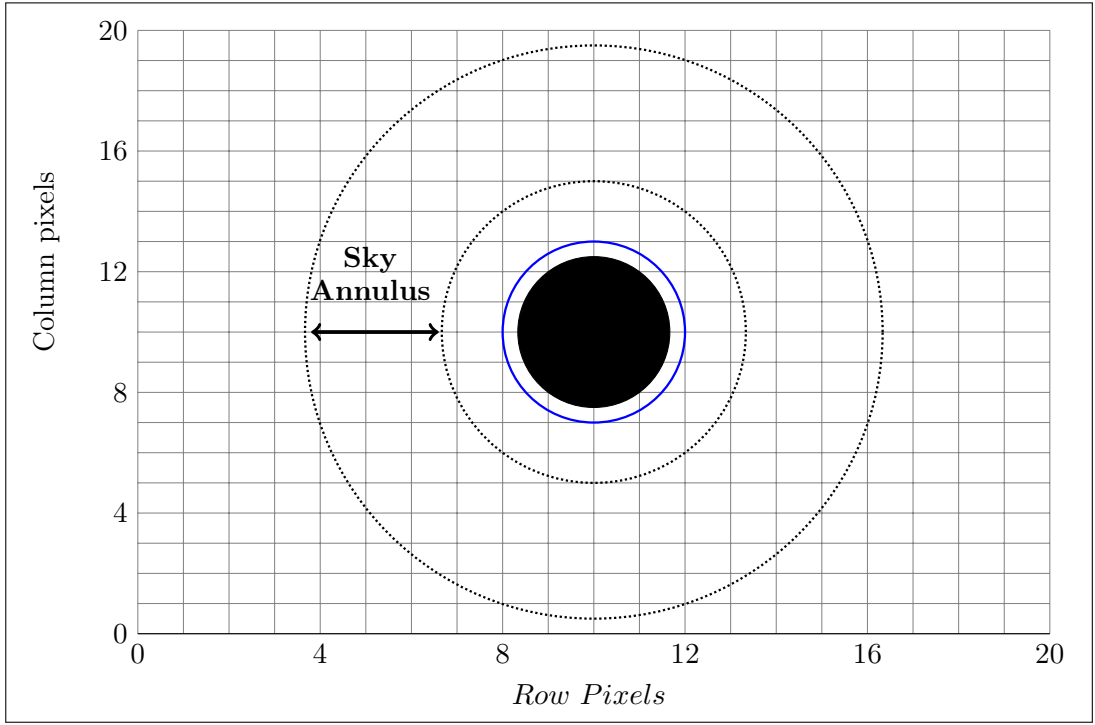
\begin{figure} 
\centering
 \fbox{  

\begin{tikzpicture}[x=0.6cm,y=0.4cm]

  \def\xmin{0}
  \def\xmax{20}
  \def\ymin{0}
  \def\ymax{20}

  \draw[style=help lines, ystep=1, xstep=1] (\xmin,\ymin) grid
  (\xmax,\ymax);


  \draw (10,\ymin -2 ) node {$Row\  Pixels\ $};

  \draw[-] (\xmin,\ymin) -- (\xmax,\ymin) node[below] {};
  \node[label=below:\rotatebox{90}{Column pixels}] at (\xmin -2 ,16) {};

  \foreach \x in {0,4,...,20}
    \node at (\x, \ymin) [below] {\x};
  \foreach \y in {0,4,...,20}
    \node at (\xmin,\y) [left] {\y};

\filldraw [black] (10,10) circle (1cm);
\draw [blue,line width=0.3mm] (10,10) circle (1.2cm);
\draw [black,densely dotted,line width=0.3mm](10,10) circle (2cm);
\draw [black,densely dotted, line width=0.3mm](10,10) circle (3.8cm);

\draw[<->,black,line width=0.5mm](3.8,10)  to (6.6,10);
\draw (5.1,12.4 ) node {{\textbf{Sky}}};
\draw (5.1,11.4 ) node {{\textbf{Annulus}}};

\end{tikzpicture}

}
 \vskip -0.8em
    \caption{Representation of a star, with the aperture around the star shown in blue and the sky annulus shown between the dashed lines.\cite{howell2006handbook} }
  \label{fig:annulus}
\end{figure}

Using a calibrated image and the centre point of an object, the next step is magnitude calculation for specific objects on an image.   Figure \ref{fig:annulus}  shows the software aperture (solid line), around the star and the sky annulus (two dashed line circles). A simple estimate of the background sky level is to calculate the per pixel average value of the pixels within the  sky annulus. These pixels also contain noise in addition to photons from the background sky which needs to be removed. More accurate estimates use the median value $B_{M}$ and exclude values which are  plus or minus 3 standard deviations form the median which will exclude cosmic rays and possibly other light sources.  A buffer exists between the aperture and the sky annulus to ensure that the background is far enough away from the object as to be representative of the background.  

\par
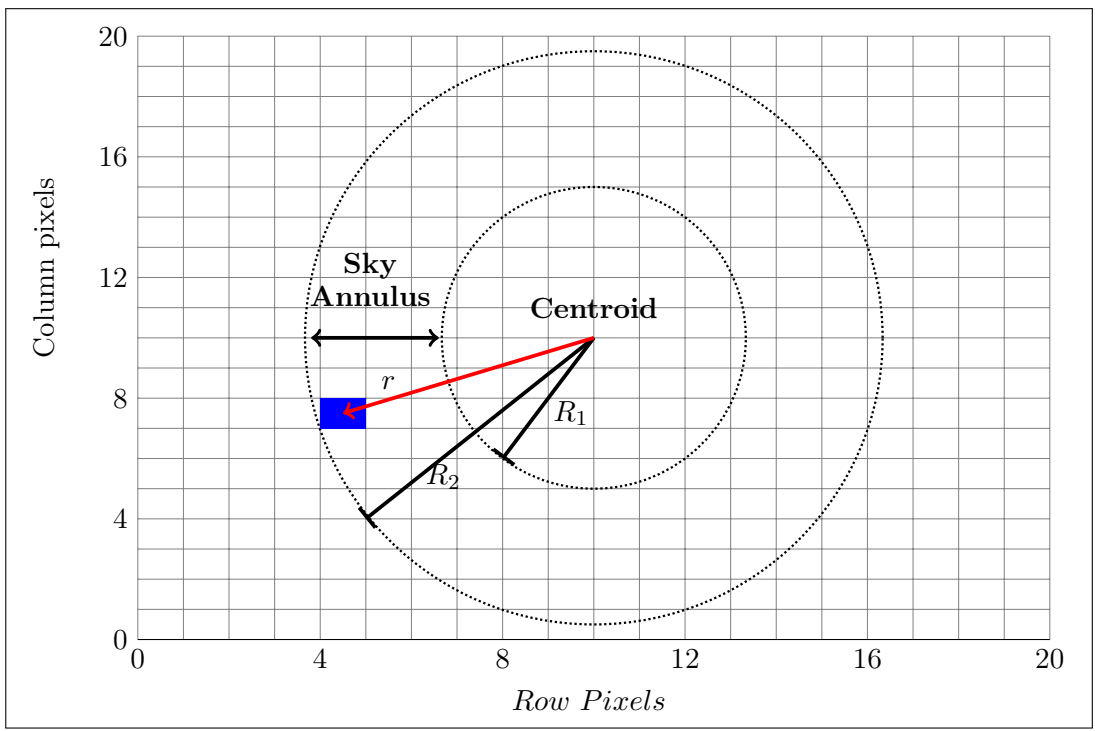
\begin{figure} 
\centering
 \fbox{  

\begin{tikzpicture}[x=0.6cm,y=0.4cm]

  \def\xmin{0}
  \def\xmax{20}
  \def\ymin{0}
  \def\ymax{20}

  \draw[style=help lines,gray,  ystep=1, xstep=1] (\xmin,\ymin) grid
  (\xmax,\ymax);


  \draw (10,\ymin -2 ) node {$Row\  Pixels\ $};

  \draw[-] (\xmin,\ymin) -- (\xmax,\ymin) node[below] {};
  \node[label=below:\rotatebox{90}{Column pixels}] at (\xmin -2 ,16) {};

  \foreach \x in {0,4,...,20}
    \node at (\x, \ymin) [below] {\x};
  \foreach \y in {0,4,...,20}
    \node at (\xmin,\y) [left] {\y};

\draw [black,densely dotted,line width=0.3mm](10,10) circle (2cm);
\draw [black,densely dotted, line width=0.3mm](10,10) circle (3.8cm);

\draw[<->,black,line width=0.5mm](3.8,10)  to (6.6,10);

\draw (10,11 ) node {{\textbf{Centroid}}};
\draw (5.5,8.5 ) node {{\textbf{$r$}}};
\draw (6.7,5.4 ) node {{\textbf{$R_{2}$}}};
\draw (9.5,7.5 ) node {{\textbf{$R_{1}$}}};
\fill[blue] (4,7) rectangle (5,8); 

\draw (5.1,12.4 ) node {{\textbf{Sky}}};
\draw (5.1,11.4 ) node {{\textbf{Annulus}}};

\draw[-|,black,line width=0.5mm](10,10)  to (8,6);
\draw[-|,black,line width=0.5mm](10,10)  to (5,4);
\draw[->,red,line width=0.5mm](10,10)  to (4.5,7.5);

\end{tikzpicture}

}
 \vskip -0.8em
    \caption{Determining if a pixel is within the sky annulus by ensuring the distance to the centre of the pixel is between $R_{1}$ and $R_{2}$ .}
  \label{fig:euclidian}
\end{figure}

 The sky background $B$ is an estimate of the amount of light which should be removed from the final flux value of a star and can be considered to be a photon based noise level. Typically the background calculation only includes pixels which are fully within the sky annulus and excludes partial pixels as shown in \ref{fig:euclidian}. By sorting these pixel values and obtaining a median value,  the sky background $\widehat{B}$ per pixel can be estimated.  The Euclidian distance is used to determine if a pixel is within the sky annulus. If $R_{1}$ is the distance between the centroid and the inner part of the sky annulus and $R_{2}$ is the distance to the outer part of the sky annulus, the distance from the centroid to a pixel $r$ must be between these two values. The distance between the centroid $Cx,Cy$ and the pixel position $x,y$  can be calculated using equation \ref{eq:euclid}. A pixel is considered within the sky annulus if $r$ is greater than $R_{1}+0.5$  and less than $R_{2}-0.5$.

\begin{equation}\label{eq:euclid}
r =   \sqrt{  (Cx - x)^2  +  (Cy - y)^2}
\end{equation}

\subsection {Calculating Flux Intensity values}

The next step is to calculate the total flux value recorded for an object by summing all of the record pixel electron counts $I$ within the software aperture. This is a total count of the unit values stored within each pixel within a specific aperture range. A more accurate flux total $F_{T}$ is then obtained by including all pixels within the aperture and partial pixel counts. Using a similar method to the sky background calculation, pixels can be determined to be either fully inside the aperture, on the line, partially inside the aperture or outside of the aperture.  If $N$ represents the total number of pixels within the aperture (counting partial and full pixels) then the total flux can be estimated and the sky background subtracted using equation \ref{eq:estbackground}.

\begin{equation}\label{eq:estbackground}
\widehat{F}_{T} =  \sum_{i=1}^{N}{I_{i} } - N \widehat{B}
\end{equation}

\subsection {Calculating Instrumental Magnitude}

Given an estimate of the total flux for the object the instrumental magnitude $m$ calculation using a standard equation \cite{Klotz:1921tk} can be completed. 

\begin{equation}\label{eq:mag}
m = -2.5\ log_{10} (\widehat{F}_{T}) 
\end{equation}

\section{Data Sources}

\begin{figure} [htbp]
\centering
  \includegraphics[width=0.5\textwidth] {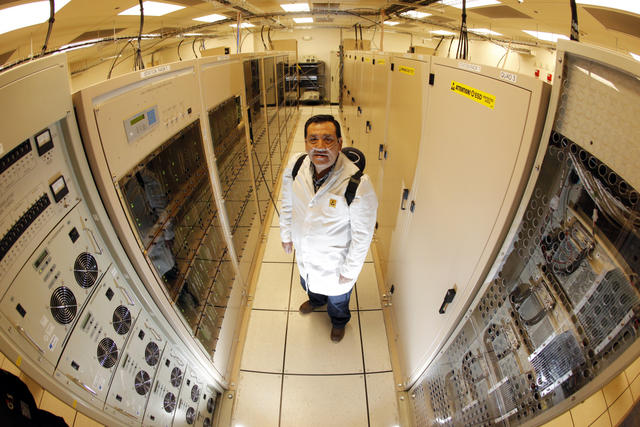}  
  \vskip -0.8em
    \caption{Technician breathing oxygen at the Alma Correlator, the world's newest and highest high performance computing system with over 134 million processors \cite{Anonymous:J1ETgENg}. }
  \label{fig:alma}
\end{figure}

With an understanding of CCD calibration and magnitude calculations it is important to consider the context within which these operate.  For any world-class scale project (space or ground based) significant investment is required in IT. Data products are produced, pre-processed to a pre-defined level, and made available to a Principle Investigator, supporting institutes or potentially to the public, either directly via download servers or via the \gls{vo} \cite{Hanisch:2003uo}. For large projects, data capture, transfer, calibration and reduction, basic processing, archiving and access are considered as part of the observatory capabilities and bespoke solutions are often implemented. Smaller institutes often capture less data due to the capabilities of their instruments but investment in IT is still required, although more modest computing resources may be sufficient. Researchers and institutes will have varying requirements and capabilities either in data processing and/or data capture and it is the ability to match computing resources to large and potentially varying data acquisition rates that is of interest. As smaller research groups have the capacity to generate larger volumes of data, a gap in processing capabilities emerges. As the pressure for data generation rates goes up, there should be pressure on bringing the IT costs in line so that smaller institutes take advantage of instrument improvements. Data processing costs cannot be allowed to grow linearly with data acquisition. Projects such as the ALMA (Atacama Large Millimeter/submillimeter Array) Correlator\footnote{Capable of 17 quadrillion operations per second. At an altitude of 5,000 meters on the Chajnantor Plateau, oxygen levels are only half of what they are at sea level.}, Figure \ref{fig:alma}, which locate computing resources physically close to the capture devices due to remoteness of location or bandwidth restrictions, fail to avail of global resources and cannot be replicated for all observatories. For the purpose of categorising and reviewing existing sources of CCD image data the following general classifications are used.

	\begin{description}
	\item [Space-based telescopes]  \hfill \\
Space-based telescopes operate with the significant advantage of being free of atmospheric conditions but have been limited in terms of data processing and data transfer. Bandwidth for sending data to Earth has been a bottleneck with transmission rates generally below 1Mbps \cite{mudgway2000uplink}, although recent tests of the Lunar Laser Communication Demonstration potentially paves the way for significant increases in bandwidth in future missions \cite{grein2014fiber}. Bespoke and evolving data processing pipelines are often used per mission to process data, although reuse is becoming evident by the \gls{opus} pipeline operated by the \gls{stsi} \cite{2008eic..work..177L}. 
	\item [Ground based telescopes - Large]  \hfill \\
Large ground based observatories involving often large consortiums or national funding providing survey and project based data. Until relatively recently many of these observatories have been required to physically move some or all of the science data to data centres for processing. Projects such as \gls{evalso} have significantly enhanced network connectivity and bandwidth and in 2010 EVALSO provided a 1Gbps connection between South America and Europe,  dramatically reducing the bandwidth bottleneck. Due to these bandwidth restrictions, often due to the remoteness of the observatory location, \gls{hpc}  centres are typically paired with these observatories and often use bespoke data processing solutions.  Smaller observatories or institutes tend to have less computational resources available, lacking significant investment in IT and either build bespoke HPC solutions or use shared HPC resources when available. 
	\item [Survey telescopes]  \hfill \\
Observatories  both large and small designed to continually survey the sky have the ability to generate large volumes of data. Data management and processing is a key factor in these systems.  Sample rates, image resolution and number of devices capturing data have the potential to exceed data bandwidth capabilities requiring local resources to have the capacity to constantly store and potentially process data. 
 	\end{description}

\subsection{Optical Space Telescopes}

Due to the restrictions on data transfer all space-based telescopes will attempt some form of optimisation at the capture point to reduce the amount of data sent to Earth. Kepler for example performs pixel selection on captured images using a number of specific criteria \cite{Bryson:2010wp}. Ultimately however all data needs to be sent to Earth for processing. The communication mechanism used by US based space crafts since the early 1960's \cite {Renzetti:1975wu} is the NASA \gls{dsn}, a collection of Earth based antenna in three primary locations, Goldstone, Canberra and Madrid all of which connect directly with the Deep Space Operations Centre, Pasadena, California. The radio link to space crafts is a point to point system using different frequencies and ultimately different data transfer rates.  For telescopes in Earth's orbit the \gls{tdrss} offers higher bandwidth communications with recent generations ranging between  300Mbits/s and 800Mbits/s depending on the microwave band used. Communications from more distant crafts tend to require higher power which typically results in reduced bandwidth.  Further details are provided based on some of the high profile optical telescopes in space. 

\subsubsection{Hubble Space Telescope and the James Webb Space Telescope}

\par The \gls{hst} was launched in 1990, produces approximately 120 gigabytes of data per week and communicates from low Earth orbit with the TDRSS \cite {teles1995overview}, a data relay and service designed to facilitate communications between Earth and orbiting space crafts. The TDRSS downloads its data to the ground station at White Sands in New Mexico where it is transferred to the Space Telescope Science Institute (STScl)  in Balimore for processing. The data pipeline used is the Operational Pipeline Unified Systems (OPUS) \cite{rose1995opus} which was designed specifically for the HST, but is now used for other programs.  OPUS does not actually perform image calibration but stages data for processing and then takes the processed data and stages it for inclusion into the \gls{mast} archive. OPUS uses a blackboard architecture of communication using file names providing a communication layer between processes which allows for distributed and parallel processing. Once data is staged for processing by OPUS, the calibration is performed by the \gls{stsdas} \cite{leitherer1995}, which is an \gls{iraf} based system. 

\par The successor to the HST is the \gls{jwst} due to be parked at L2 (Lagrange point 2) approximately 1.5 million kilometers from Earth and is expected to launch within the next 5 years. This spacecraft will also communicate with Earth via the DSN and transfer data to the STScl data centre in Baltimore Maryland for processing. The JWST and Kepler will be subject to the limits of the DSN bandwidth (the Ka-band is used for scientific data download, operating at 26 gigahertz) and while Kepler downloads approximately 23 GB of data per month, JWST may generate in the region of 30 GB per day based on an upgrade of the DSN infrastructure \cite{Johns:2008fb}. The limits of data transmission using the DSNs Ka-band effectively constrain the problem of data processing to a manageable level at a maximum of 20-30 GB per day.

\newpage
\subsubsection{Kepler Mission}

\par
\begin{figure} [ht]
\centering
  \includegraphics[width=0.6\textwidth] {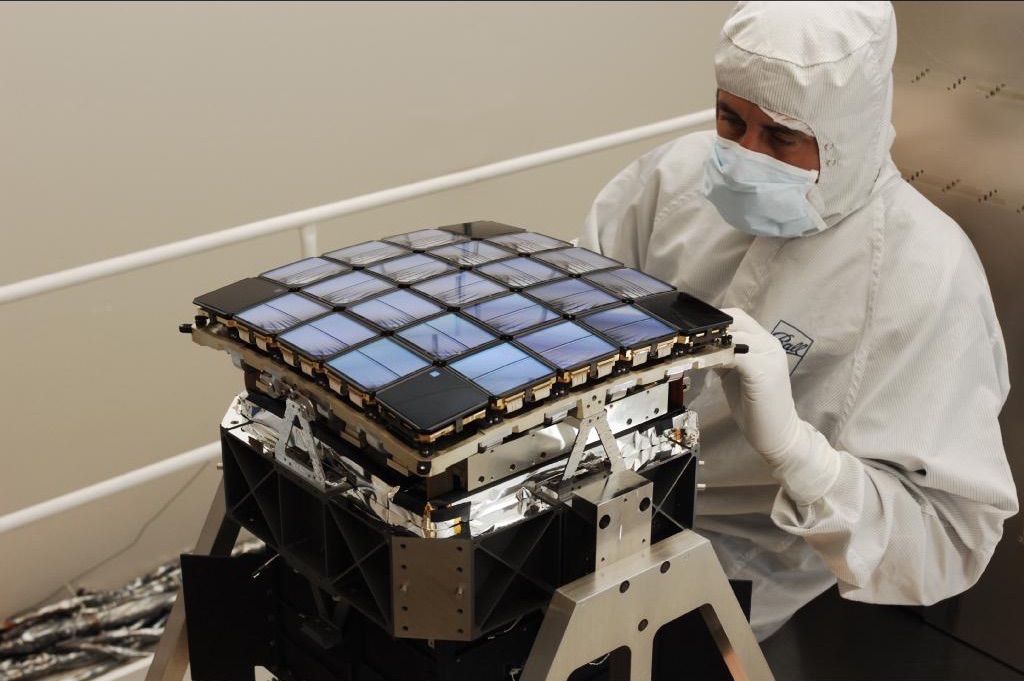} 
  \vskip -0.8em
    \caption{Kepler array consists of 42 charge coupled devices (CCDs). Credit: NASA and Ball Aerospace. } 
  \label{fig:kep}
\end{figure}

 The Kepler spacecraft (designed to search for Earth sized planets in other solar systems) sits in a heliocentric orbit (centred on the Sun) and is trailing the Earth by over 10 million kilometres staying within communications range via NASA's DSN using X-band for twice weekly command and status, and monthly Ka-band contact for data download at a speed of 4.33 Mbps \cite{Haas:2010vy} which takes 6 hours. Launched in 2009, Kepler monitored approximately 170,000 stars and downloaded approximately 23 GB of pixel data per month \cite{Klaus:2010et}, which is approximately 6 million pixels captured every 30 minutes of operation. Due to bandwidth limitations, ~5\% of pixels are downloaded each month \cite{Hall:2010er}. The detector is made up of 42 CCD arrays, which operate on 30-minute capture cycles \cite {van2009kepler}, see Figure \ref{fig:kep}. Each CCD is 2.8 by 3.0 cm with 1024 by 1100 pixels. The entire focal plane contains 95 mega pixels.  

\par
\begin{figure} [htbp]
\centering
\fbox{   \includegraphics[width=0.9\textwidth] {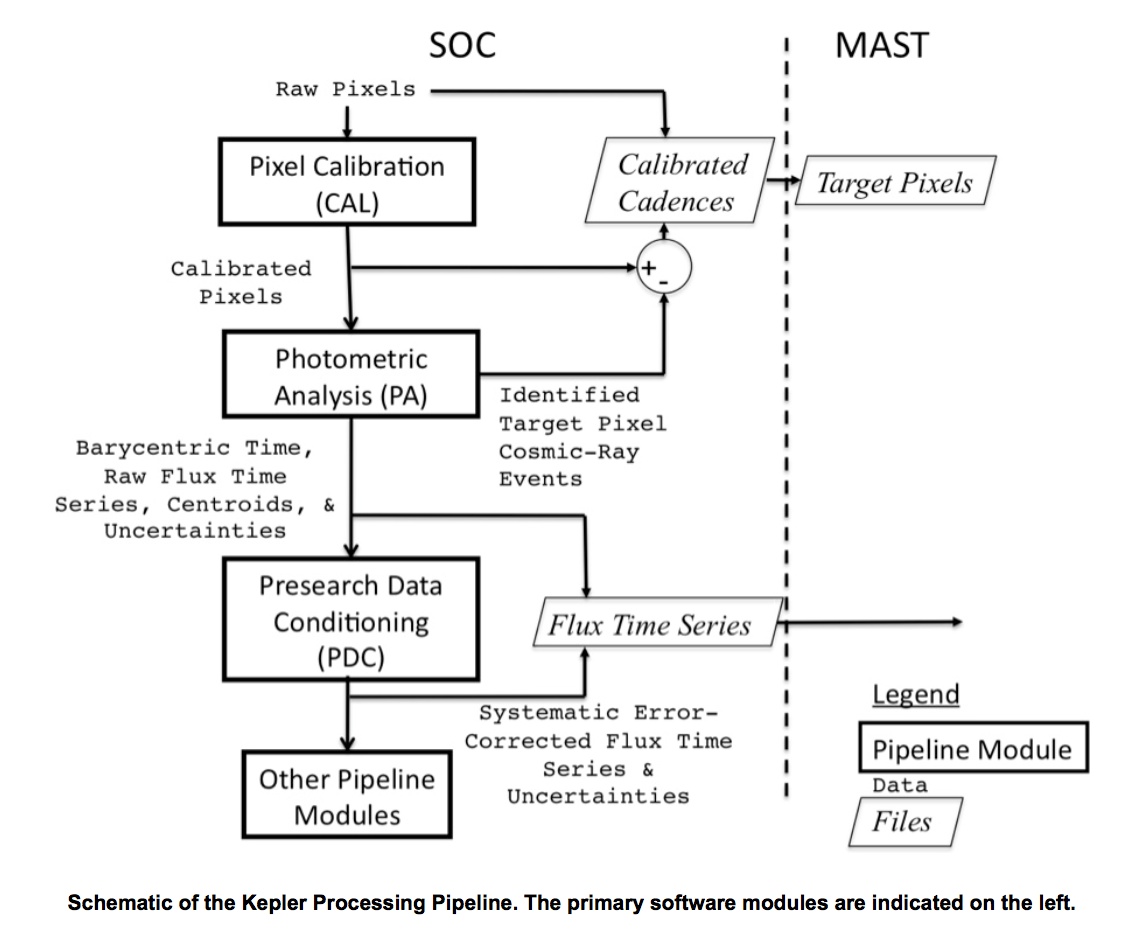}  }
  \vskip -0.8em
    \caption{Kepler science data flow in the AMES Science Operations Centre (SOC). Credit: NASA }
  \label{fig:cal}
\end{figure}

 Data from the DSN is eventually routed to the STScl and packaged into FITS files before being sent to the \gls{soc} at  the AMES research centre, Moffet Field, California where data calibration and photometric analysis is performed. The Kepler pipeline, which uses a Java framework to create units of work, can process some data in parallel by allowing some modules to be run in a customised sequences per pipeline. An instance of the framework can run for each of the 42 CCD detectors, which is parallel processing on the dataset at a very high level. Within each instance of the pipeline, operations are run more sequentially for the CAL module \cite{Quintana:2010bk} which puts images through a series of science algorithms as shown in the Figure \ref{fig:cal}. Each module is composed of a number of procedures written in Matlab. 

\par In 2011, the volume of data being generated by the Kepler mission and the associated processing requirements was sufficiently large for the pipeline to require porting to the Pleiades cluster at the NASA advanced Supercomputing Division. So while the data pipeline for Kepler was described as highly parallel, the architecture's ability to scale in the face of a growing data set was evidently limited.  This need to fully reprocess all of the raw data as part of the constant reviewing of analysis techniques and parameters was a major driving factor  \cite{Klaus:2013tc}. Porting to the new pipeline required significant investment demonstrating the lack of expandability by the existing pipeline. At the time a comparative analysis was performed against the Amazon \gls{ec2} service which was ultimately rejected not on expandability, but on raw network performance \cite{Mehrotra:2012dl}.

\par
\begin{figure} [ht]
\centering
\fbox{   \includegraphics[width=0.9\textwidth] {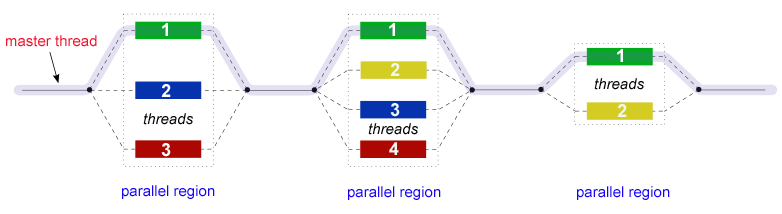}  }
  \vskip -0.8em
    \caption{Fork-Join processing }
  \label{fig:fork}
\end{figure}

 The NASA Pleiades supercomputer is an example of a HPC used for computing intensive workloads. Data must be transferred into the system and jobs must be written and submitted to the central control system for processing. The Pleiades HPC is operated by the Ames research centre in California USA, was brought online in 2008, and is a collaborative effort with SGI. This computer cluster ranks in the top 20 supercomputers and is constantly being expanded.  It supports a number of processing environments and is designed to assist NASA with a variety of processing requirements such as simulation and modelling. The Kepler mission migrated to this platform to complete processing of light curves for the hundreds of thousands of stars monitored by the Kepler space craft. The programming environment is Linux  with jobs being batched for execution. The development environment is C, C++ or Fortran and the parallel processing components are supported via the SGI \gls{mpt} . \gls{openmp}, is an application programming interface (API) designed to allow control and execution of code that takes advantage of multiprocessing, shared memory which is also supported by Pleiades. Like many systems, OpenMP uses a fork-join approach to parallelism as shown in Figure \ref{fig:fork}. 

\subsubsection{Global Astrometric Interferometer for Astrophysics}
The Global Astrometric Interferometer for Astrophysics (GAIA) is a European Space Agency survey mission launched December 19th 2013 which aims to provide a three dimensional map of the Milky Way using measures of star position and movement. The spacecraft is positioned at the L2 Lagrange point approximately 1.5 million kilometres from the Earth, in the same location planned for the James Webb Space Telescope. The mission is designed to survey appoximately 1\% of the  100 billion stars in our Galaxy. The spacecraft contains an array of 106 CCDs and the data transfer from the spacecraft will be at a rate of approximately 5 Mbps \cite{doi:10.111712.2056402} for eight hours per day, which will delivery over the lifetime of the project approximately 100 TB of raw uncompressed data. The total final storage requirements are expected to expand to 1 PB including data backup and provisional data processing steps. The final data set available to standard researchers when processed will consist of approximately 20 TB. 

\par
\begin{figure} [ht]
\centering
\fbox{   \includegraphics[width=0.9\textwidth] {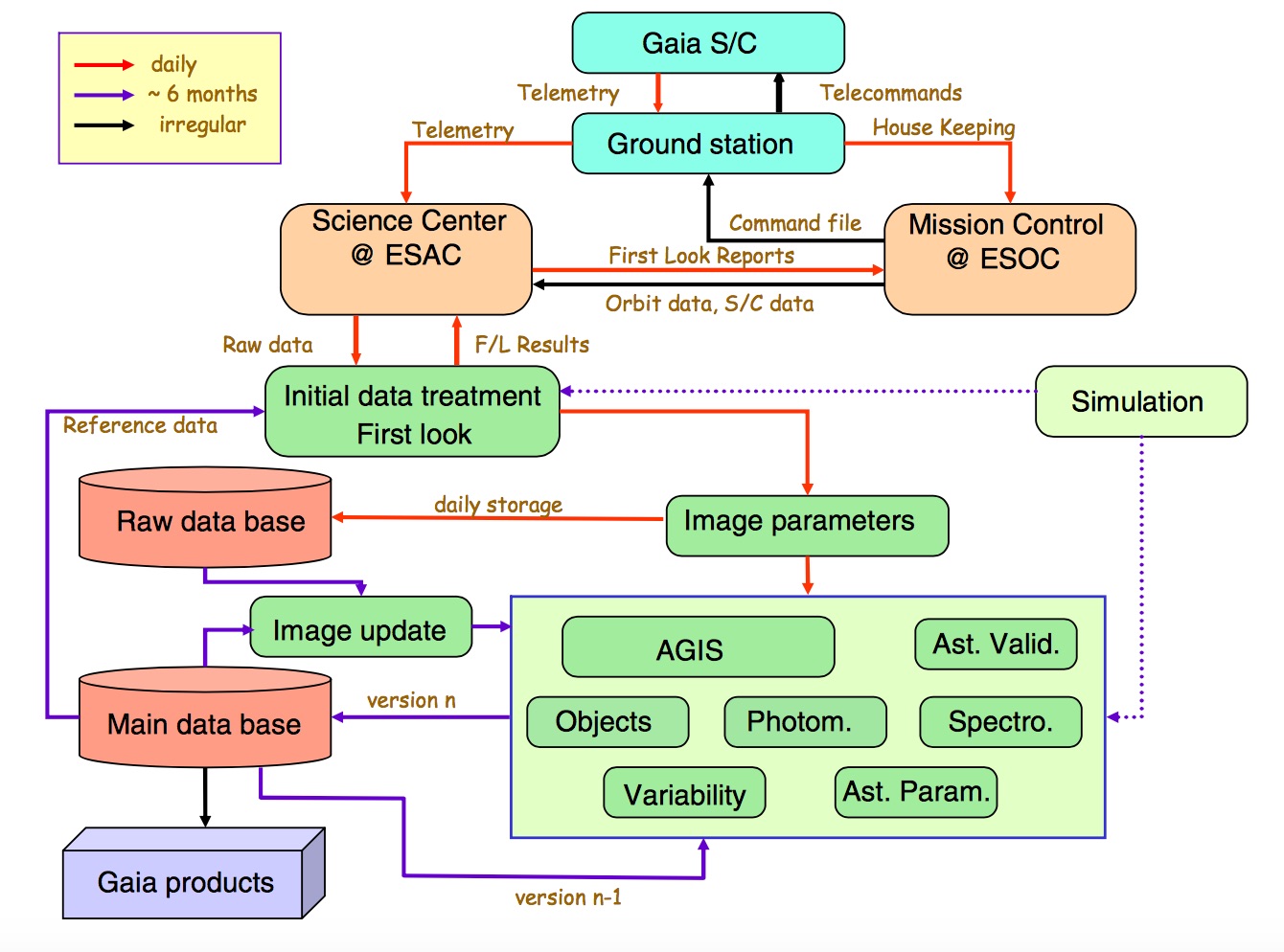}  }
  \vskip -0.8em
    \caption{Main structure of the data flow in the processing of the Gaia raw data \cite{IAU:1930144} }
  \label{fig:gaia}
\end{figure}

 Due to the format used to capture the data, significant data processing is required to reconstruct the images. Two reasons for the complexity of the data are identified by Mignard et al \cite{IAU:1930144} due to the large number of computations required to process the data, and the level of interconnections between different subsets of the data. The basic data flow for processing the data is shown in Figure \ref{fig:gaia}.

\par Data processing will be performed by a European consortium, the Data Processing and Analysis Consortium (DPAC) which is comprised of contributors from 24 European countries. 

\subsection{Large Ground-Based Telescopes}

Large ground based telescopes overcome atmospheric distortion, to some extent, by being located in the highest, and driest locations on Earth. As the number of instruments continues to increase, their data processing requirements also increase.  There are a number of other larger ground based telescopes under construction or in planning at present, such as the Giant Magellan Telescope which is a 25 meter telescope expected to generate terabytes of data daily, The European Extremely Large Telescope is planned to have a 40 meter mirror, again expected to have a large volume of data daily for processing \cite{2010jena.confE..28P}.

\par The \gls{gtc} is currently the largest operational optical telescope in the world with a 10.4 meter aperture \cite{CepaJordi98}. Observations are scheduled according to a predetermined schedule with raw data provided to participating principle investigators for data processing via FTP. Data is collected on two 2k by 4k red optimised CCD devices giving a total of 4k by 4k pixels \cite{CepaJ2007}.   Flat field and bias files are provided along with the raw data. Full resolution image readout can run at a variety of rates, with single image readouts taking between 7.8 and 21 seconds depending on the readout mode. Data generation rates are in the region of 20 GB per hour assuming 32bit pixel storage while operating at 7.8 second capture time.

\par The Keck Observatory operates two 10m telescopes each containing 36 hexagonal mirrors which operate as a single reflective unit. The observatory is based on the summit of the Mauna Kea volcano in Hawaii and uses adaptive optics to overcome the effects of the atmosphere. Observation time is allocated to partner institutes with raw data made available for download over FTP, SCP or other similar transfer protocols. The LRIS red and blue detectors are comprised of 2 x 2k x 4k CCDs with a minimum readout time of 42 seconds. Raw data capture rates are roughly equivalent to the GTC. Image processing is performed by the participating principle investigator, files are usually available in the FITS format. 

\par
\begin{figure} [ht]
\centering
\fbox{   \includegraphics[width=0.6\textwidth] {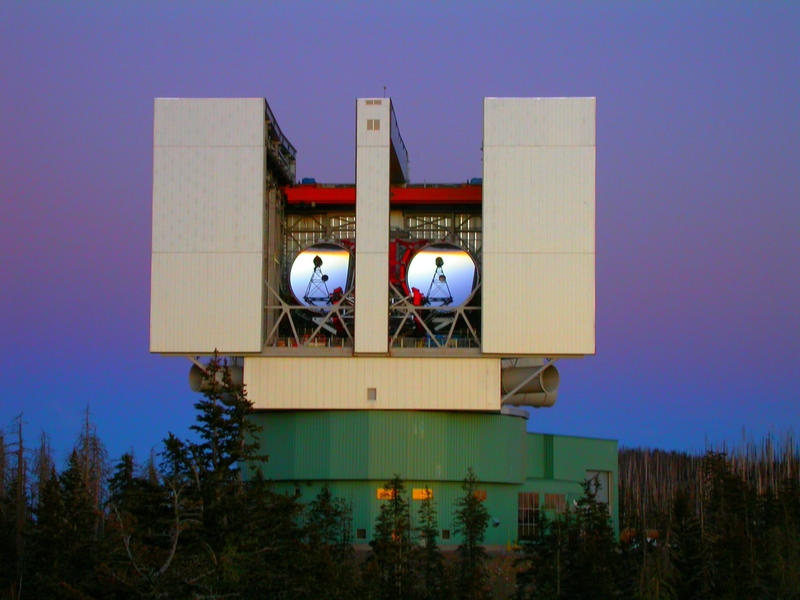}  }
  \vskip -0.8em
    \caption{LBT Large Binocular Telescope   Credit: Aaron Ceranski/LBTO }
  \label{fig:lbto}
\end{figure}

 The \gls{lbto} (Figure \ref{fig:lbto}) has two wide field cameras which can operate in tandem, one of which is blue optimised and the other red optimised. The LBT instrument consists of  4 x 2k x 4k CCDs per mirror \cite{2003SPIE.4837..140H}.  Exposure times are given as 0.3 seconds minimum for accurate photometry, with 30 minutes being the maximum observing times before the instrument focus is rechecked. Data is made available for download using a mounted NFS directory from which the observer can copy the data and proceed to process it themselves.  While the telescope has the potential to generate 42 Terabytes of data per day, this is not a constant data rate, and is unlikely to consistently hit this maximum data generating rate.

\subsection{Survey Projects}

\par In addition to telescopes which have large apertures there are projects which focus not on specific targets but rather sweep large areas of the sky to provide survey data. Surveys are expensive as data constantly streams into the instrument and must be captured, stored and processed. Surveys have the capacity to generate terabytes or petabytes of data. The LSST for example is a multi-million dollar consortium expecting to generate terabytes of data per day. 

\par The \gls{sdss} \cite {1996AJ....111.1748F}, which is currently in operation using a 2.5 metre telescope, reported a maximum data capture rate in the region of 160GB per day in 2004 \cite{Ivezic:2004bn}. This is probably closer to 200GB per day at present. The photometric data processing pipeline was written at Princeton University, and the data processing is performed at Fermilab using their HPC data centre to process data. The telescope uses a 30 x 2k x 2k CCD array, with approximately 54 seconds between exposures \cite{Gunn:1998en}.  The imaging pipeline ultimately produces files in the FITS format, but internal formats are used within the Serial Stamp Collecting pipeline (SSC) with a custom written C implementation which processes images in sequence. 

\par Building on the success of the SDSS is the LSST \cite{Anonymous:jSqSPrBn} which has an 8.4 single mirror system which will be online by about 2019 and is expected to generate in the region of 20 terabytes of data per night with a resolution of 3.2 billion pixels per image sustaining a data capture rate of 330Mbytes per second \cite {Anonymous:jCnKYjyL}. The LSST plans to move tens of terabytes of data per day over high-speed fibre network from the Chilean site to the U.S. Processing is focused on high-speed cores close to the datasets and the belief is that on going advances in server technology will allow a traditional data centre HPC approach to data processing \cite{Howell:2009vz}. The data reduction pipeline is still in development and is most likely a bespoke solution. Interestingly they rejected the \gls{eso} Common Pipeline as an option, given the volume of data to be processed and due to concerns over its stability and extensibility.

\par Extensive archives for astronomical data are being produced and made available to researchers from multiple sources. Diverse data archives exist which are consolidations of optical data from multiple observations or projects such as  the ESO Science Archive Facility which contains approximately 1/3 of a petabyte of data and is growing by terabytes per month \cite{romaniello2011eso}. Similarly the \gls{noao}  provides access to data from its primary facilities such as Kitt Peak, and limited access to other data products from partnerships such as the Keck Observatory with plans to include future data products from the LSST and the Giant Magellan Telescope. The NASA National Space Science Data centre provides approximately 230TB of digital data covering 5,500  distinct data collections, while MAST provides data from the HST and other space based telescopes. There has been a concerted effort to create a central repository of astronomical data by the International Virtual Observatory Alliance (IVOA) which was formed in 2002 \cite{Quinn:2004tj}. The concept is for astronomical data to be made available from multiple sources/instruments to scientists around the world in a centralised, standardised way. In 2003 it was reported that ESO would have greater than 1 petabyte of compressed data available by 2012 \cite {Hanisch:2003uo}. A distributed architecture to support multiple repositories of data was discussed by Hanisch in the year 2000 where it is conceded that a single data archive cannot hold all of the data products being produced. \cite {2000ASPC..216..201H}. A distributed data archive would seem to logically require a distributed data processing solution using a similar argument. No single data centre could be big enough to process all of the data available. 

\subsection{Radio Astronomy}
\par The issues associated with large data set generation is not limited to optical telescopes or CCD photometry. Additional projects exists which offer the potential to generate Petabytes of data, such as the ASTRON initiated Low Frequency Array for radio astronomy (LOFAR) project \cite{butcher2004lofar} and the Square Kilometre Array (SKA) \cite{Dewdney:2009uz}. The LOFAR project requires raw data to be transferred from each of the distributed array nodes to a Central Processing System (CEP) which recombines the data for distribution to offline user based processing.  Data recombination was initially performed using an IBM Blue Gene/P supercomputer which provided at its peak, 34 TFlops of processing power before it is transported to the Long Term Archive (LTA) were 20 PB of data is expected to be stored over the next 5 years \cite{begeman2011lofar}.  Processing is currently performed by a GPU cluster. For researchers to use the data products produced by such projects, large scale computing resources are required to access, download and process the data. Such resources are often costly  to construct and difficult to maintain.  Challenges for data processing and management also exist within an Irish context as the I-LOFAR 
\url{http://www.lofar.ie} consortium aims to join the European wide LOFAR initiative, helping to extend the east-west baseline of the array to just over 1400 kilometres. 

\par The SKA will provide even further data processing challenges with data capture rates measured in the 100's of Gbps. Construction is expected to begin in 2018  with a completion date of 2024. The Science Data Processor consortium will be responsible for focusing on the software, hardware and algorithms required to process the raw data into a series of data products.  These challenges are expected to be in excess of anything existing within the field of science at present. Using the LOFAR project as a reference example, the IBM/ASTON Dome project will also review state of the art computing technologies in an effort to build an exascale computing system to process and store data.  Further technical challenges will also be presented to researchers seeking to use these data products in terms of data management and data processing and the technology accessible by researchers will require significant advancement in line with that of the SKA itself.

\section{Data Reduction Software}

The astronomical community does not suffer from a poor selection of data processing tools. The variety of low-level tools emphasizes the lack of standards in this area. In the 2007 ESO Instrument Calibration Workshop \cite{Kaufer:2008tv} the session on Data Flow and Data Reduction Software reviewed the history of pipelines, and looked at some of the pipelines proposed across various sites. A brief summary of these technologies is provided in Section \ref{FitsFormatsandAIS}.

\subsection{Fits Formats and APIs}
\label {FitsFormatsandAIS}
The FITS format (the Flexible Image Transport System) is a standard astronomical image format endorsed by the International Astronomical Union (IAU) and NASA, which was approved in 1981. FITSIO and CFITSIO are C and FORTRAN libraries supported by NASA to access and manipulate the FITS file format. There are many other languages with interfaces to these libraries allowing reduction software to be written in a multitude of languages including Python, Java, Perl, MATLAB and C++. Most data produced for researcher to consume are provided using this format. 

\subsection{IRAF}

\par Image Reduction and Analysis Facility is a standard application used throughout the astronomical community which has been in development since the mid to early 1980s \cite{1986SPIE..627..733T}, and is a Linux based software package.  It provides the closest thing to a standard for data reduction and analysis within much of the astronomy community. Many of the tools referenced by data archives are IRAF based. It is well documented and available on a variety of platforms. Much of the code is implemented in FORTRAN and data processing of files is sequential using batch type processing. Researchers interested in processing data from data archives will often use IRAF as the basic reduction software package.

\subsection{NHPPS}

\par
\begin{figure} [ht]
\centering
\fbox{   \includegraphics[width=0.6\textwidth] {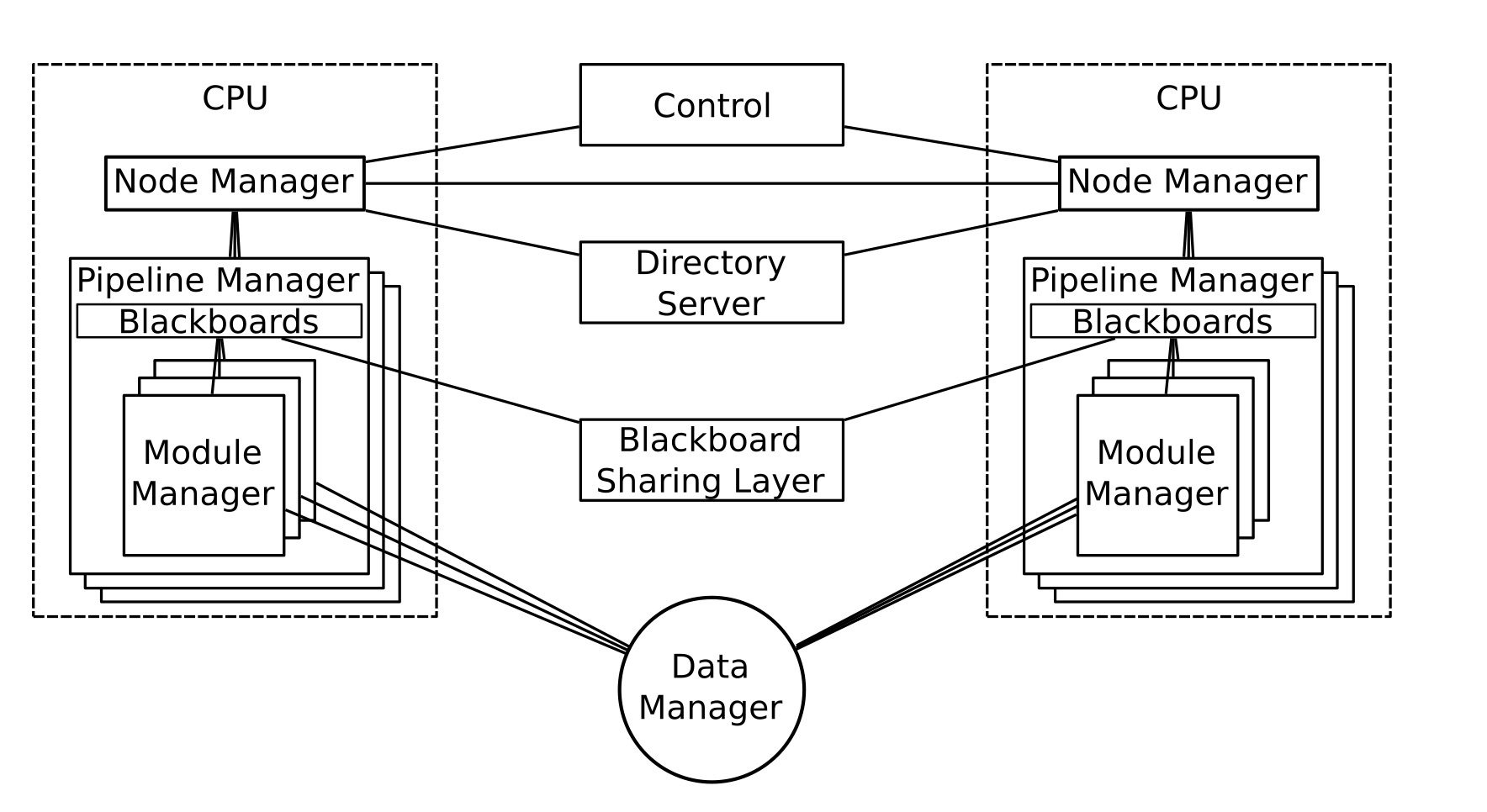}  }
  \vskip -0.8em
    \caption{NOAO High Performance Pipeline System Architecture \cite {Valdes:2006vr}. }
  \label{fig:nhpps}
\end{figure}

 The \gls{nhpps} is a python based pipeline which can operate with local processing clusters of software nodes. Connections to nodes are NFS based with SSH access required between each node. Each node must be password-less and have the the NHPPS software installed. Originally based on the OPUS system \cite {Scott:2007vk} the  NHPPS uses the blackboard architecture for communication across a multi-node distributed environment which is a multi-queue based system which contains queues of work which can be assigned to pipelines and queues which contain available datasets.  Parallelisation within the pipeline is referred to as coarse grain parallelisation, which splits the data into chunks which can be processed independently and runs multiple processes on single servers to reduce CPU idle time. The distributed features of the pipeline are based on a directory server requiring the CPU nodes to be connected via sockets as shown in Figure \ref{fig:nhpps}.  Limitations identified with the pipeline include the requirement for controlled shutdown of the computing nodes in case of data loss due to partial completion of data processing by a node.  \cite {Valdes:2006vr}. 

\subsection{ESO: Common Pipeline Library (CPL)}
The VLT instrument pipelines are based on the \gls{cpl}, a C based technology closely coupled to the FITS format, which is used as the basis of a number of pipelines, the first of which was the GIRAFFE pipeline \cite{Banse:2004va}. CPL was designed to consolidate different software implementations that existed within various ESO instruments to provide a common technology to assist with rapid data reduction development. Currently all VLT pipelines are either using CPL or are being converted to it. CPL based applications will require faster processors to run pipelines faster and does not provide multi-threaded support although work is on-going to enable multi-threading applications using a thread safe version of CPL \cite{deBilbao:2010wk}.  CPL runs on Linux based operating systems, primarily Scientific Linux, and relies on NASAs CFITSIO libraries. 

\subsection{OPUS}
In a review of NASAs experience with Data Reduction pipelines over the last 30 years  given by Don Lindler { \cite{Kaufer:2008tv}} the initial Hubble pipeline implementation was considered a step backwards from previous pipelines and the Space Telescope Science Institute, responsible for data processing quickly moved to development of the OPUS pipeline, which was later used in Chandra X-Ray Observatory, the Spitzer Space Telescope and the Far Ultraviolet Spectroscopic Explore. The new pipeline quickly proving to be quite versatile. Initially a VAX/VMS solution, it was moved to Sun/Solaris platform based on an IRAF implementation. Its legacy will continue into the JWST although Mac OSX is a more likely platform with Python/PyRAF as the implementation software.  

\subsection {IUE} 
\par Prior to OPUS, one of the first and most successful reduction pipelines was developed to support the International Ultraviolet Explorer launched in 1978 and designed to run for 3 years. It was 19 years later however that the IUE pipeline was eventually shutdown (despite the satellite being in sound working order). To maintain the accessibility of the data, the pipeline was ported to IRAF in 1998 \cite{shaw1998iraf}. The HST system, based on Sun workstations is having similar longevity issues with hardware being difficult to replace and repair. 

\subsection{Other pipelines}
\par In addition to the STScI use of OPUS and the ESO focus on CPL there are some primary technologies, which should be briefly mentioned. There are many other examples of pipeline software development within the astronomical community. In 2005 Ó Tuairisg \cite{Tuairisg:2005vm} described a distributed computing model using a GRID system allowing images to be processed in parallel. Having data on a shared file system helped reduce data transfer cost and facilitates processing times.  This approach builds a large cluster around the data to be processed. 

\par The Apsis package on the other hand is a more traditional system developed in Python to process the early release images from the Advanced Camera for Surveys in 2002.  Images and tables were processed via Pyraf and Pyfits \cite{Blakeslee:2002wy}.

\section{Distributed Computing}

The concept of distributing computing tasks to multiple machines is a well-established discipline within computer science dating back as early as 1969 to ARPNET. Instead of a single processor processing data in sequence, work is processed on distributed computing nodes at the same time.  There are many issues with this approach and much research into optimising the process. Broadly speaking there are two areas to consider, the first is how work can be partitioned into smaller jobs, and the second is the practical management of those jobs across a distributed system. A good summary of the pitfalls of working within a distributed system were initially formalised in 1994 by Peter Deutsch, a fellow at Sun Microsystems, when identifying the \emph{seven} fallacies of distributed systems. James Gosling, also of Sun extended these into the \emph{eight} fallacies (shown below) which were later explained succinctly in a white paper by Rotem-Gal-Oz \cite{RotemGalOz:2006tw}. 

	\begin{enumerate}
		\item The network is reliable
		\item Latency is zero
		\item Bandwidth is infinite
		\item The network is secure
		\item Topology doesn't change
		\item There is one administrator
		\item Transport cost is zero
		\item The network is homogeneous		
	\end{enumerate}

\par The Cloud Computing model provides an example of a form of distributed computing, enabling ubiquitous, convenient, on-demand network access to a shared  pool of configurable computing resources (e.g., networks, servers, storage, applications, and services) that  can be rapidly provisioned and released with minimal management effort or service provider interaction.  A definition of cloud computing was put forward by the \gls{nist} \cite{mell2011nist}  followed by a special publication in 2012 with a series of recommendations \cite{badger2012cloud}. The report identified five essential characteristics, three service models and four deployment models as shown in Table \ref{tab:nist}. 

 \begin{table}[H]

\centering
 \begin{tabular}{p{13.7cm}}
  \hline
  \hline

\emph{\small{The cloud model is composed of five essential characteristics, three service models, and four deployment 
models.}}\\

\end{tabular}

\begin{tabular}{p{5cm} p{8cm}}
\hline
\small{Essential Characteristics   }& \small{On-demand self-service }\\ 
\ &  \small{Broad network access }\\
\ &  \small{Resource pooling }\\
\ &  \small{Rapid Elasticity }\\
\ &  \small{Measured Service }\\

\hline
\small{Deployment Models   }& \small{Private Cloud}\\
\ &  \small{Community Cloud }\\
\ &  \small{Public Cloud }\\
\ &  \small{Hybrid Cloud }\\
\hline

\small{Service Models   }& \small{Software as a Service (SAAS)}\\
\ &  \small{Platform as a Service (PAAS }\\
\ &  \small{Infrastructure as a Service (IAAS) }\\
\hline
\hline

\end{tabular}
\caption{NIST definition of Cloud Computing.}
 \label{tab:nist}
\end{table}

\par For infrastructure as a service (IAAS), the distributed nature of a solution is based purely on the architecture used to build the system. Resources are provided, such as the Simple Queue Service, the Elastic Compute Cloud, with architectural decisions on their implementation left to the user. Platform as a service (PAAS) provides implementations of Map-Reduce, a programming model designed to allow a cluster of computer nodes to perform highly parallel operations on large datasets. Based largely on the LISP functions of similar names, the technique allows for robust parallel operations to be performed inside a cluster of machines. The system aims at being fault-tolerant, providing automatic parallelization and distribution to worker nodes while offering status monitoring.  Large volumes of data are processed in parallel by distributed CPU nodes using a distributed file system. The open-source standard for this implementation is Hadoop running on the Hadoop Distributed File System (HDFS)  \cite{Borthakur:2007tw}. The technique should allow scaling to thousands of CPU nodes with the right type of problem. While initial reference examples were text search based, additional material has been published demonstrating successful implementation of the Map-Reduce technique in scientific data processing environments. This system provides a programming paradigm that builds private HPC style solutions within a cloud infrastructure. 

\par Each of the service models outlined by NIST exist as commercial services such as Amazon's AWS \cite{huckman2008amazon} an IAAS solution. Clouds can also be constructed using open source technology such as OpenStack \cite{sefraoui2012openstack}. Lenk provides a useful list of vendors and technologies used in cloud construction \cite{Lenk:2009ht}.

\par Distributed computing is currently in use by the scientific community and a considerable amount of literature exists in reviewing the suitability, cost and performance of clouds and other techniques used for scientific data processing \cite{Taylor:2010bx}, \cite{Juve:2009tw} \cite{Wang:2008hi} \cite{Keahey:2008ub} \cite{Berriman:2012bo} \cite{Fan:2012gh} \cite{Wilson:2012wy}. In December 2011 the Natural Sciences and Engineering Research Council of Canada (NSERC) published a detailed report on the potential role of cloud computing in science based mid-range computational and data intensive workloads \cite{Ramakrishnan:2011dr}. A summary of these findings is provided below. 

	\begin{enumerate}
		\item The elastic nature of the cloud is a significant advantage, allowing for elastic provisioning primarily through the use of virtualisation technologies.
\item There is a potential issue in the level of work required for porting existing approaches to the cloud model, including considerable levels of skills required to do so. This upfront cost should be considered as part of the economic analysis when deciding to potentially move this model.  
\item Significant gaps exist in managing data within the cloud environment and the process is neither simple nor easily accessible. Scientific workflows are not specifically catered for, and there is an inherent difficulty in exploiting the features of technologies such as Map-Reduce. Other problems include the lack of bootstrap starting points and complex management of cloud environments.  
\item Economic benefits come from consolidating resources to improve system utilisation (which it was felt exist in the US Department of Energy).  Incorporating aspects of the cloud model into existing data centres is a worthy objective. Private clouds should be considered first before the use of commercial clouds avoiding issues of security, data management and performance of public clouds. 
\item Scientific applications have specific requirements that require cloud solutions tailored to their needs.
	\end{enumerate}

\par Various reactions to this report \cite{Holland:2012vr} would suggest that the DOE struggle with the adoption of public clouds on the basis of a paradigm shift from HPC to Cloud being non trivial, and the business model of pay-as-you-go not being fully compatible with the scientific requirement of open-ended need for resources. Below is a brief overview of some high profile projects, which are actively engaged in processing datasets using distributed or cloud computing.

\subsection{Scientific Projects Overiew}

\par The Kepler Project (not to be confused with the Kepler spacecraft mission) uses a Map-Reduce \cite{Dean:2008fi} programming model and demonstrates early results in processing biometric scientific data for large HADOOP clusters \cite{Wang:2009hq}. One of the concerns being addressed within this paper is that the Map-Reduce model still offers a layer of complexity, which they believe excludes many from gaining access to its potential benefits. 

\par Wiley \cite{wiley2011astronomy} demonstrated the use of Map-Reduce for image co-addition using the Sloan Digital Sky Survey imaging database to produce a single image from multiple image files with improved signal to noise ratios (SNR). The basic Map function processed each file as a single job, determining if it should be included in the co-added image, and if so, a bitmap was passed on for the Reduce function to take and include in the new image. Of interest is that 100,000 FITS files were processed using this technique and while many files were not required for the final image, the research demonstrates the feasibility of FITs file processing in some form. 

\par The Kepler CCD data is downloaded from the spacecraft and processed on Linux based clusters running 64 nodes with 512 CPU cores. Data is chopped into parallel jobs for processing on worker nodes and a \gls{jms}  used to distribute jobs across 4 clusters \cite{Klaus:2010uz} \cite{Klaus:2010et}. With worker nodes capable of being added, job definition is flexible and the examples provided are single image in and calibrated image out the other end of the pipeline, or multiple images in, and light curve out. Much of the processing is done using MATLAB libraries and Oracle. Similar in many ways to formal cluster computing, distributed elements and pipelines are controlled within a local cluster.

\subsection{SETI@home}

\par Public resource computing uses the spare CPU cycles of computers to perform data processing. Early research in the use of unused processor capacity included \emph{The worm programs} as early as 1978 at Xerox PARC using a small set of 100 machines to measure Ethernet performance \cite{Shoch:1982jo}. But it is the SETI@home project \cite{Anderson:2002gb} which stands out as the best example of a public resource project utilising spare CPU cycles from millions of online users. Radio data is recorded at source and physically shipped every few days to its headquarters in Berkeley (approximate 2 terabytes of data every few days) where the data is split into work-units, which are accessed by clients across the world. Notable within this approach is that this is a computationally intensive problem, each 350k of data requiring multiple hours of processing. Low bandwidth requirements ensure that even users with modest connections to the Internet can contribute to the overall project. Challenges identified by the Berkeley team have centred on the requirement to issue duplicated work-units (3 sets) to combat malicious users, and the challenges in running and maintaining the server infrastructure to support their data distribution model.   

\par
\begin{figure} [ht]
\centering
 \includegraphics[width=0.6\textwidth] {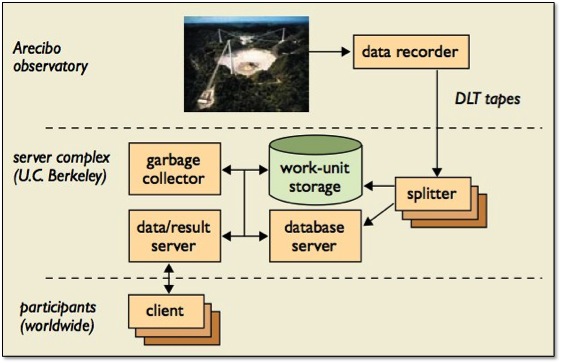}  
  \vskip -0.8em
    \caption{Distribution of radio data using SETI@Home (2002)}
  \label{fig:set1}
\end{figure}

 The SETI@home initiative demonstrated a distributed solution for astronomical data processing as shown in Figure \ref{fig:set1}. The SETI pipeline splits work into parallel jobs, which are processed by clients around the world in a distributed manner. A key factor, as already mentioned, in the approach taken by the SETI project is that the I/O rates are low and the computational requirements were high. For kilobytes of data, multiple hours of computation are potentially required \cite{Anderson:2002gb}. In 2001 the estimated processing completed by the distributed system was approximately 437,000 years of CPU, leading to the claim of being the largest supercomputer in the world (at the time). \cite{Korpela:2001gj}

\par With most examples presented from the literature so far,  existing models for large astronomical data, processing is usually performed on centralised data centres. SETI@home offers an alternative distributed approach but it stresses that the computation to data ratio should be high. 

\par This research seeks to address the key question of whether a distributed model can be created when the computation to data ratio is low while allowing for tens of terabytes of data to be processed. The distributed model potentially offers a cost advantage to the smaller institute/facility while providing a powerful processing network.

\section{The data challenge}

A dataset was provided by the \gls{bco}, a research facility engaged in high-speed photometry research.  The reference dataset contained 3262 cubed FITS\footnote{FITS is the Flexible Image Transport System digital format used for storing and processing scientific images}  files, each containing 10 images and each approximately 512x512 pixels in resolution (0.7MB per image) and the total size of the dataset was 26GB. This data was replicated to simulate a multi-terabyte data. The dataset was generated on September 22nd 2003 at Calar Alto, targeting S5 0716+71 as part of an engineering equipment test of a new hardware/software stack using an Andor CCD device. 

\par A reference image processing speed from BCO was in the order of 1 image processed per second, which is 0.7MB of data processed per second, or roughly 60GB of data per day. This processing pipeline was sequential in nature and used a fixed number of computing devices, which could not be expanded during image processing.

\par Researchers at the Blackrock Castle Observatory in 2003, using a high-speed \gls{emccd} detector, operating at 10 images per second, with a detector resolution of 0.2 megapixels, generated approximately 7 megabytes of data per second, equivalent to 200 gigabytes in an 8-hour period.  Recent CMOS detectors (Andor Zyla sCMOS 5.5) with a resolution of 5.5 megapixels are under test at that facility and are capable of capturing 100 frames per second. With each pixel value stored as a 32 bit number (8 bytes), the size of a dataset can be calculated as per equation \ref{eq:sizepix}, where  $N_{pix}$ is the number of pixels on the detector, $p$ is the numeric precision used to store the pixel value (typically 8 bytes per pixel), $t_{sec}$ is the time in seconds for the data capture period, and $f_{ps}$ is the number of frames recorded per second.  
 \begin{equation}\label{eq:sizepix}
Total_{bytes} = N_{pix} * p * f_{ps} * t_{sec}
\end{equation}

\par
\begin{figure} [ht]
\centering

\begin{tikzpicture}
    \begin{axis}[width=0.6\textwidth,
    legend style={at={(0.3,0.95)},anchor=north,legend cell align=left},
        xlabel=$Frame Rates$,
        ylabel=$Terabytes$]
    \addplot plot coordinates {
        (0,0) (1,0.6) (5,3.02)  (10,6.04) (25,15.11)    (50,30.21)  (100,60.42) };
    \addplot
        plot coordinates {
        (0,0)        (1,0.44)        (5,2.2)        (10,4.39)        (25,10.99)       (50,21.97)       (100,43.95)        };       
    \addplot
        plot coordinates {
        (0,0)        (1,.22)        (5,1.1)        (10,2.2)       (25,5.49)        (50,10.99)        (100,21.97)        };              
     \addplot
        plot coordinates {
        (0,0)        (1,0.11)       (5,0.55)        (10,1.1)        (25,2.75)        (50,5.49)        (100,10.99) };     
     \addplot
        plot coordinates {
        (0,0)        (1,0.03)        (5,0.14)        (10,0.29)        (25,0.72)        (50,1.44)        (100,2.88)};    
       \legend{$5.5\ MegaPixel$\\$4\ MegaPixel$\\$2\ MegaPixel$\\$1\ MegaPixel$\\$0.24\ MegaPixel$\\}   
    \end{axis}

\end{tikzpicture} 
  \vskip -0.8em
    \caption{Data generation rates per 8 hours for varying camera resolutions running at various frame rates. See Table \ref{tab:longtableccd} }
  \label{fig:ccdsizes}
\end{figure}
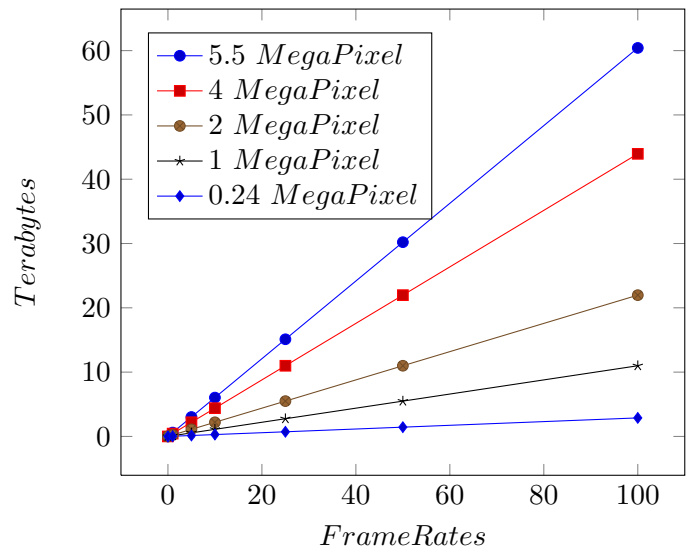

 Using the formula in Figure \ref{fig:ccdsizes} the data production rate in terabytes per 8 hours can be plotted for varying CCD or CMOS pixel resolutions and frame rates. While it may not be feasible to operate  at 100 frames per second continuously, it can be seen that at lower capture rates the dataset generated within 8 hours is in the order of terabytes. With capture rates in the order of  terabytes  per night, large dataset generation is clearly well within the capability of smaller observatories. With the development of robotic farms, even at the lower capture rates, these rates are clearly in the realm of big data.   

\subsection{Sequential versus Distributed Data Processing}

\par The approaches to processing large datasets are largely dependent on the performance requirement of the task and the volume of data. It is perfectly reasonable to use a brute force approach to solving a problem when the problem is sufficiently small, or computing resources are sufficiently powerful. In these cases results can be produced within a reasonable amount of time so there is no need to process data using any specific method other than sequential. As the volume of data increases, and traditional approaches start to incur unreasonable delays in processing time, further thought is required to address the problem of performance and processing efficiency. Within a typical pipeline, the cleaning and reduction process is a two-step sequential pipeline (Figure \ref{fig:seqpipeline}). The first step is reading in a raw image, calibrating all the pixels in the image and writing a cleaned image file. This is performed on all image files. The second step is reading the cleaned image file and calculating a series of magnitude values for each star (or light source) and writing out a file containing magnitude values. In this pipeline, work is typically performed on a single but powerful server. A sequential processing pipeline must have the capability of processing data at the capture rate to ensure that the pipeline does not back up. If run over 24 hours with a capture period of 8 hours then the slowest speed of the pipeline must be $3x$ the capture rate. As resolutions or frame rates increase there is a race for processing rates to keep pace.  With the slowing down of processor rate improvements and the end of Moore's law  in sight, alternative processing approaches are required.  

\tikzstyle{decision} = [diamond, draw,fill=white,text width=4em, text badly centered, minimum height=2em, inner sep=0pt]
\tikzstyle{block} = [rectangle, draw,text width=5em, text centered, rounded corners, minimum height=4em]
\tikzstyle{block1} = [rectangle, draw,fill=white,,rounded corners,text width=8em, text centered, minimum height=1em]
\tikzstyle{worker} = [rectangle, draw,fill=blue!10, text width=10em, text centered, rounded corners=8ex, minimum height=16em]
\tikzstyle{disk} =   [cylinder, draw,fill=white, text width=1em, shape border rotate=90, shape aspect=0.5, inner sep=0.3333em,  minimum width=2cm, minimum height=1em]

\tikzstyle{line} = [draw, very thick, color=black!50, -latex']
\tikzstyle{cloud} = [draw, ellipse, fill=white,node distance=2.5cm, minimum height=2em]
\tikzstyle {data} = [draw,trapezium,trapezium left angle=70,trapezium right angle=-70,node distance=2.5cm,minimum height=1em]

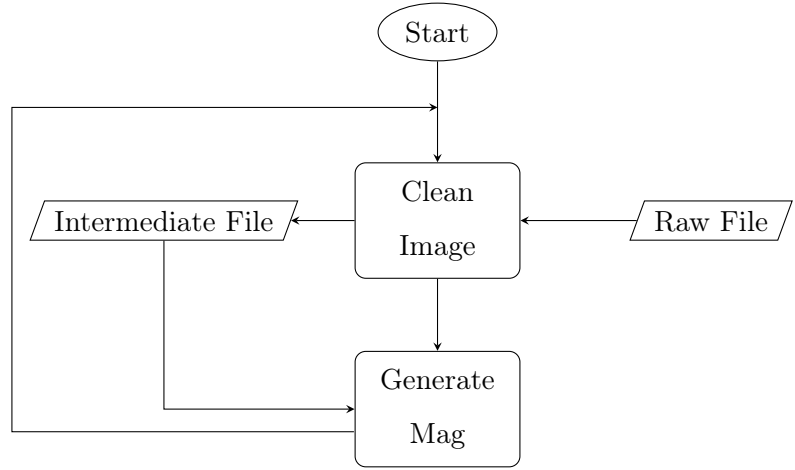
\begin{figure}[ht]
 \begin{center}

  \begin{tikzpicture} [node distance=2.5cm, auto, >=stealth]
   \node[cloud] (a)                                     {Start};
   \node[block] (b)  [below of=a]                       {Clean Image};
   \node[data] (c)  [right of=b, node distance=3.6cm]                       {Raw\  File};
   \node[data] (d)  [left of=b, node distance=3.6cm]                       {Intermediate File};
   \node[block] (e)  [below of=b]                       {Generate Mag};

   \draw[->] (a) -- (b);
   \draw[<-] (b) -- (c);
   \draw[<-] (d) -- (b);
   \draw[->] (b) -- (e);

   \draw[->,color=black]  ($ (e) - (11mm,3mm) $)  -- +(-40mm,0mm)  -| ($ (a) - (56mm,10mm) $)  -- ($(a) -(0mm,10mm)$)  ; 
   \draw[->,color=black]  (d.south)   |- (e.west)    ;


  \end{tikzpicture} 
  \caption{A sequential processing pipeline. A raw file is read and has bias, flat field and dark current master frames applied, creating an intermediate file from which instrument magnitude values are calculated before the next file is read. Files are processed in a sequential order.}
  \label{fig:seqpipeline}
 \end{center}
\end{figure}

\par
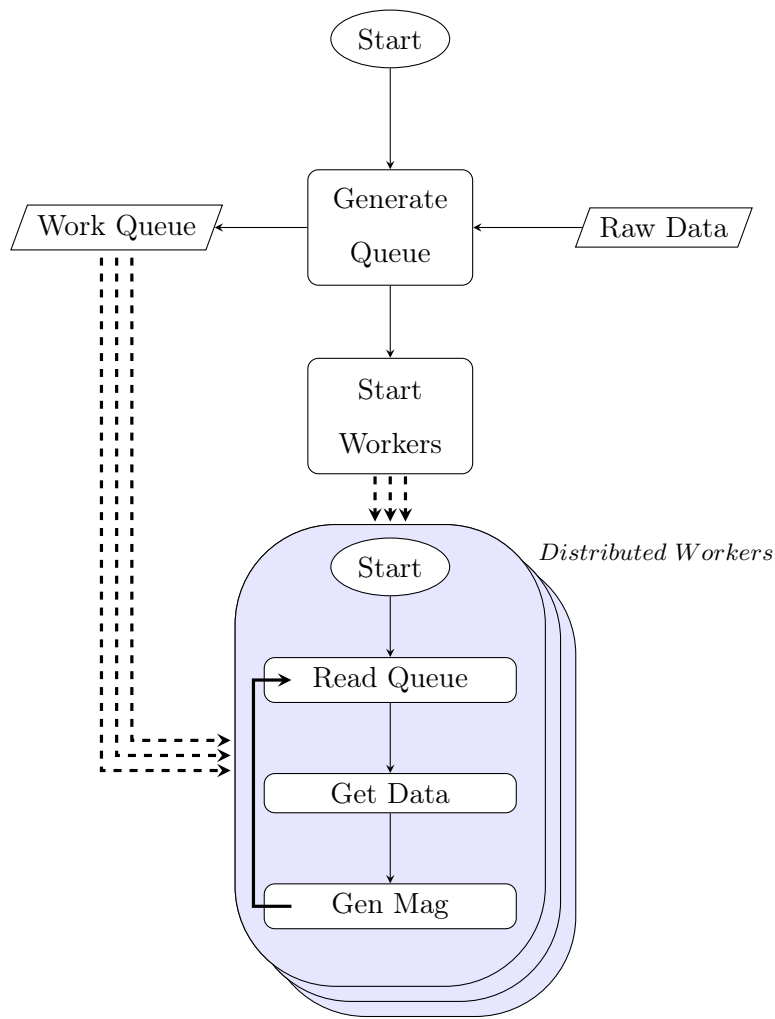
\begin{figure}[ht]
 \begin{center}

  \begin{tikzpicture}[node distance=2.5cm, auto, >=stealth]
   \node[cloud] (a)                                     {Start};
   \node[block] (b)  [below of=a]                       {Generate  Queue};
   \node[data] (c)  [right of=b, node distance=3.6cm]                       {Raw\  Data};
   \node[data] (d)  [left of=b, node distance=3.6cm]                       {Work Queue};
   \node[block] (e)  [below of=b,node distance=2.5cm]                       {Start Workers};
   \node[worker] (f)  [below of=e,node distance=4.5cm]                       {Distributed Workers};
   \node[worker] (h)  [below of=f,node distance=0.2cm, right of=f,node distance=0.2cm]                       {Distributed Workers};
   \node[worker] (h)  [below of=f,node distance=0.4cm, right of=f,node distance=0.4cm]                       {Distributed Workers};
   \node[worker] (h)  [below of=f,node distance=0.2cm, right of=f,node distance=0.2cm]                       {Distributed Workers};
   \node[worker] (f)  [below of=e,node distance=4.5cm]                       {};
   
   \node[cloud] (st)        at (0,-7)                             {Start};
   \node[block1] (rq) at (0,-8.5) {Read Queue};
   \node[block1] (gd) [below of=rq,node distance=1.5cm]    {Get Data};
   \node[block1] (gm) [below of=gd,node distance=1.5cm]    {Gen Mag};


   \draw[->] (a) -- (b);
   \draw[<-] (b) -- (c);
   \draw[->] (b) -- (d);
    \draw[->] (b) -- (e);
    
       \draw[->] (st) -- (rq);
       \draw[->] (rq) -- (gd);
       \draw[->] (gd) -- (gm);

     \draw[->,very thick,color=black]  ($ (gm) - (13mm,0mm) $)  -- +(-5mm,0mm)  -| ($ (rq) - (18mm,0mm) $)  -- ($(rq) -(13mm,0mm)$)  ; 
     
     \draw[->,very thick,dashed, color=black]  ($ (e) - (0mm,8mm) $)  -- +(-0mm,-6mm)    ; 
     \draw[->,very thick,dashed, color=black]  ($ (e) - (2mm,8mm) $)  -- +(0mm,-6mm)    ; 
     \draw[->,very thick,dashed, color=black]  ($ (e) - (-2mm,8mm) $)  -- +(0mm,-6mm)    ; 

     \draw[->,very thick,dashed, color=black]  ($ (d) - (0mm,4mm) $)   |- ($ (f) - (21mm,0mm) $)   ; 
     \draw[->,very thick,dashed, color=black]  ($ (d) - (-2mm,4mm) $)   |- ($ (f) - (21mm,-2mm) $)   ; 
     \draw[->,very thick,dashed, color=black]  ($ (d) - (2mm,4mm) $)   |- ($ (f) - (21mm,2mm) $)   ; 
	\draw (3.5,-6.8) node {\footnotesize$Distributed \ Workers$};



  \end{tikzpicture} 
  \caption{Distributed processing pipeline. A queue of work is created of available raw files. Once distributed worker nodes are activated, they use the queue to get work in parallel.  }
  \label{fig:distribpipeline}
 \end{center}
\end{figure}

 A distributed processing approach has the advantage of potentially employing large numbers of resources. To distribute the processing of data in a meaningful way, the data must be parallelised to some extent. If the data must be processed in a sequence then distributed computing may not be very relevant. Astronomical CCD data however can be reduced in parallel once the calibration frames are provided with each image.  As in Figure \ref{fig:distribpipeline} a distributed pipeline would take in blocks of raw data which can be processed independently and have them queued waiting for a distributed CPU node to  process them. Ideally limits would not be imposed by the communications protocols between a work queue and the CPU process.  The NHPPS, Kepler and OPUS pipelines use a form of message exchange to allow for multiple CPU nodes to communicate and receive notifications of work to be performed.

\section{Conclusions}

\par In this chapter a summary of the history of photometry has been presented along with an overview of the current process involving the use of CCDs for taking images.  Since its introduction in the mid 1970's, CCD technology has provided astronomers with the tools to measure the flux from stars in an increasingly precise manner facilitating a significant increase in the precision of magnitude calculation  since the early work of Hipparchus.

\par To improve the accuracy of these measurements  an understanding of the sources of noise within the CCD instrument is required and the techniques required to minimise them. The sources of noise and the steps required to reduce their impact on the final magnitude calculation have been outlined. As the number of images increases and the resolution of the images goes up, the processing required to generate magnitude values is also increasing. 

\par While space based telescopes and large survey systems employ large high performance computing solutions for initial data processing, this data, both raw and reduced is presented to researchers for analysis. The online virtual observatory now provides access to almost petabytes of data world-wide. Without software tools which easily facilitate large scale distributed computing for image processing, the volume of data will become a barrier to performing science. As the number of sources for data increase, possibly even moving to robotic farms, the tsunami of data will overwhelm most researchers and institutes. 

\par NIMBUS, a globally distributed pipeline is described in this thesis  as an alternative approach to the data processing techniques reviewed. This requires that the images be processed in parallel with as little work performed as necessary without compromising the quality of the data. Using the analysis of magnitude calculations, it can be shown  that data can be safely processed in parallel with the same outcome as a sequential pipeline as is done in some existing pipelines. The methods used to allow the NIMBUS pipeline to scale should ensure that the distribution of computing nodes can truly reach global levels and not be restricted to local network domains. The following chapter discusses the approach taken within the NIMBUS pipeline demonstrating through experimentation the capability of a globally distributed system.

%% file: chapter4.tex
\chapter{Research Methodology}
\label{chapter3}

\par The purpose of this research was to determine if a globally distributed network can process terabytes of astronomical CCD image data per day and this chapter reviews the methodology used to make that determination. 

\par The research performed is quantitative, iterative, and  experimental based. With the use of distributed non-homogenous resources, operating on shared network environments it was felt that a theoretical analysis would not identify real world system performance limitations. A pipeline was developed in the initial stages of this research as a pilot system against which experiments  were performed to ensure the accuracy of the core processing software against a reference BCO pipeline \cite{collins2006high}. The name given to this pipeline was \emph{FEBRUUS}, so named after the Roman god of purification. This pilot calibrated the cleaning software and provided an image cleaning rate baseline against which the remaining experiments would be compared. A list of the architectural designs which form the basis of this thesis is shown in Table \ref{tab:experiment-designs}. 

  \begin{table}[ht]
\centering
\begin{tabular}{p{3cm} p{10cm}  }

  \toprule
Experiment Name & Description \\
  \midrule
\small{FEBRUUS Pilot}& \small{Pilot system designed to demonstrate the basic principles of data image cleaning and to validate the accuracy of the reduction against the existing BCO Matlab pipeline}   \\
\small{IRAF Cloud}& \small{System designed to consider the possible implementation of a series of IRAF virtual instances distributed within a cloud infrastructure }  \\
\small{ACN Pipeline}& \small{Pipeline designed to test the effectiveness of data compression and distribution in a private cloud using an NFS queuing model.}  \\
\small{NIMBUS Pipeline}& \small{Pipeline designed to test a global processing pipeline which is dynamically reconfigurable and which can deal with processing nodes joining and leaving without impacting the integrity of the pipeline. }  \\

\hline
  \bottomrule
\end{tabular}
  \caption{Experimental designs and pipelines discussed within this chapter}
\label{tab:experiment-designs}
\end{table}

\par The initial pilot was performed within the Dublin Institute of Technology but later experiments were performed across multiple locations in Ireland. The final experiment was run using globally distributed resources. 

\par A number of architectures were conceived for the purpose of testing the hypothesis that a global distributed network can perform high speed data processing for large astronomical CCD images. Existing pipelines and technologies have been reviewed and it has been shown that the emphasis on parallel or distributed processing has been primarily confined to HPC systems typically within large data centres. Kepler provides a clear example of a message based distributed system which lacked the capability to scale as the data sets increased, requiring a full system port to the Pleiades supercomputer. SETI@Home while a globally distributed architecture, has an underlying principle that there is a high CPU to I/O ratio. In this chapter multiple architectures which were central to the iterative process used within this thesis are presented and discussed. 

\par In this chapter,  the data used within this thesis is described and the two principle distributed designs which form the basis of the research performed, the ACN pipeline and the NIMBUS pipeline are introduced. The initial pilot study, FEBRUUS, is also presented showing the core algorithms used for pixel calibration.

\section{Dataset}

\par
\begin{figure} [ht]
\centering
 \includegraphics[width=0.9\textwidth] {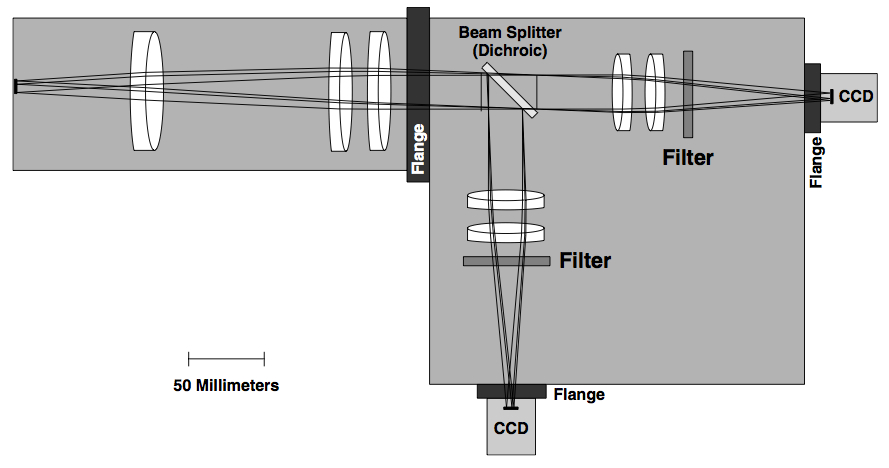}  
  \vskip -0.8em
    \caption{The optical layout of the $TO\phi CAM$}
  \label{fig:toffeecam}
\end{figure}

 Initial contact with the Blackrock Castle Observatory, Cork, in September of 2009 led to a series of discussions which explored an existing reduction pipeline system in use by the BCO research team to process raw CCD image data. BCO is engaged in high-speed photometry research \cite{smith2008emccd}  and generates datasets which contain multiple images per second. The BCO team had implemented a MATLAB based pipeline which had its science data output compared and verified against an IRAF implementation of the same algorithms. The MATLAB system provided, among other things,  faster processing rates than the IRAF solution. The processing rate of this pipeline was approximately 1 image per second, while the capture rate of the CCD device was about 10 images per second. Limitations of the pipeline included the inability to take advantage of additional computing resources, and difficulties in transferring data to alternative systems. A faster data processing pipeline was required.  The BCO are also developing a new science instrument, which has the potential of generating terabytes of data. The $TO\phi CAM$ \cite{2010jena.confE..28P}  (Two-Channel Optical Photometric Imaging Camera, pronounced toffee-cam, see figure \ref{fig:toffeecam}) uses two CCD97 EMCCDs from Andor Technologies each capable of generating 34 frames/s, approximately 68 Mbytes per second or nearly two terabytes per 8 hours observing. Existing processing rates using the MATLAB pipeline of 1 image/s would require almost 23 days of processing for a full night of observation.  BCO are involved in High time-resolution astrophysics (HTRA) and the science objective of generating high-speed photometry images was part of a research project to perform Point Spread Function fitting (PSF) photometry to an estimate accuracy over a timeframe of about one hour \cite{smith2008emccd}. 

\par  A 26 Gigabyte dataset was supplied by BCO consisting of 36,820 images stored in data cubes of 10 data images per file, with data frame integration times of $0.08$ seconds per image. This data set was the primary source of CCD images used in all experiments. The dataset was generated on September 22nd 2003 at the Calar Alto Observatory in southern Spain, targeting S5 0716+71 as part of an engineering equipment test of a new hardware/software stack using an Andor CCD device. This dataset acts as a clear reference when discussing existing processing techniques ensuring that future systems deliver the same science output. This also provided a point of qualitative and quantitative comparison for new architectures. 

\par
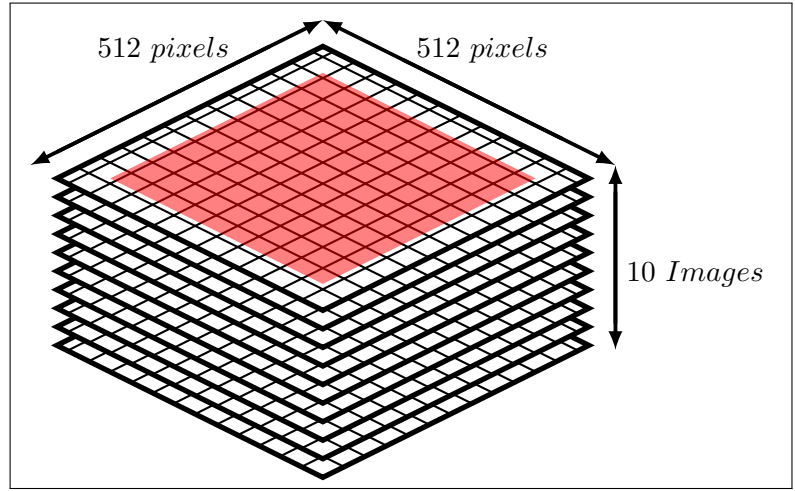
\begin{figure} 
\centering
 \fbox{  
\begin{tikzpicture}[scale=.7,every node/.style={minimum size=1cm}]
   \begin{scope}[yshift=-60,every node/.append style={yslant=0.5,xslant=-1},yslant=0.5,xslant=-1]
       \fill[white,fill opacity=1] (0,0) rectangle (5,5);
       \draw[step=4mm, black,thick] (0,0) grid (5,5); 
       \draw[black,line width=0.7mm] (0,0) rectangle (5,5);
   \end{scope}
   \begin{scope}[yshift=-50,every node/.append style={yslant=0.5,xslant=-1},yslant=0.5,xslant=-1]
       \fill[white,fill opacity=1] (0,0) rectangle (5,5);
       \draw[step=4mm, black,thick] (0,0) grid (5,5); 
       \draw[black,line width=0.7mm] (0,0) rectangle (5,5);
   \end{scope}
   \begin{scope}[yshift=-40,every node/.append style={yslant=0.5,xslant=-1},yslant=0.5,xslant=-1]
       \fill[white,fill opacity=1] (0,0) rectangle (5,5);
       \draw[step=4mm, black,thick] (0,0) grid (5,5); 
       \draw[black,line width=0.7mm] (0,0) rectangle (5,5);
   \end{scope}
   \begin{scope}[yshift=-30,every node/.append style={yslant=0.5,xslant=-1},yslant=0.5,xslant=-1]
       \fill[white,fill opacity=1] (0,0) rectangle (5,5);
       \draw[step=4mm, black,thick] (0,0) grid (5,5); 
       \draw[black,line width=0.7mm] (0,0) rectangle (5,5);
   \end{scope}
   \begin{scope}[yshift=-20,every node/.append style={yslant=0.5,xslant=-1},yslant=0.5,xslant=-1]
       \fill[white,fill opacity=1] (0,0) rectangle (5,5);
       \draw[step=4mm, black,thick] (0,0) grid (5,5); 
       \draw[black,line width=0.7mm] (0,0) rectangle (5,5);
   \end{scope} 
      \begin{scope}[yshift=-10,every node/.append style={yslant=0.5,xslant=-1},yslant=0.5,xslant=-1]
       \fill[white,fill opacity=1] (0,0) rectangle (5,5);
       \draw[step=4mm, black,thick] (0,0) grid (5,5); 
       \draw[black,line width=0.7mm] (0,0) rectangle (5,5);
   \end{scope}
   \begin{scope}[yshift=0,every node/.append style={yslant=0.5,xslant=-1},yslant=0.5,xslant=-1]
       \fill[white,fill opacity=1] (0,0) rectangle (5,5);
       \draw[step=4mm, black,thick] (0,0) grid (5,5); 
       \draw[black,line width=0.7mm] (0,0) rectangle (5,5);
   \end{scope}
   \begin{scope}[yshift=10,every node/.append style={yslant=0.5,xslant=-1},yslant=0.5,xslant=-1]
       \fill[white,fill opacity=1] (0,0) rectangle (5,5);
       \draw[step=4mm, black,thick] (0,0) grid (5,5); 
       \draw[black,line width=0.7mm] (0,0) rectangle (5,5);
   \end{scope}
   \begin{scope}[yshift=20,every node/.append style={yslant=0.5,xslant=-1},yslant=0.5,xslant=-1]
       \fill[white,fill opacity=1] (0,0) rectangle (5,5);
       \draw[step=4mm, black,thick] (0,0) grid (5,5); 
       \draw[black,line width=0.7mm] (0,0) rectangle (5,5);
   \end{scope}
   \begin{scope}[yshift=30,every node/.append style={yslant=0.5,xslant=-1},yslant=0.5,xslant=-1]
       \fill[white,fill opacity=1] (0,0) rectangle (5,5);

       \draw[step=4mm, black,thick] (0,0) grid (5,5); 
       \draw[black,line width=0.7mm] (0,0) rectangle (5,5);
       \fill[red,fill opacity=0.5] (0.5,.5) rectangle (4.5,4.5);
       
          \draw[-latex,black,line width=0.5mm] (5.5,4) -- (5.5,0);  
          \draw[-latex,black,line width=0.5mm] (5.5,1) -- (5.5,5.5);  

          \draw[-latex,black,line width=0.5mm] (4.5,5.5) -- (0,5.5);  
          \draw[-latex,black,line width=0.5mm] (0.5,5.5) -- (5.5,5.5);  
          
           \draw[-latex,black,line width=0.5mm] (5,-0.5) -- (2,-3.5);  
           \draw[-latex,black,line width=0.5mm] (2.5,-3) -- (5.5,0);

   \end{scope} 
      	\draw (3,6) node {$512\ pixels $};
	\draw (-3,6) node {$512\ pixels $};

	\draw (7,1.75) node {$10\ Images $};

\end{tikzpicture} }

 \vskip -0.8em
    \caption{Visualisation of the raw BCO data set with SUBRECT region shown in red. The unique data set contains 3682 data cubes, each containing 10 raw images.}
  \label{fig:visualdata}
\end{figure}

 The raw data images are stored in an uncompressed FITS (Flexible Image Transport System) file format and are approximately 7MB in size using 32bit integer values. The image borders of approximately 20 pixels have been removed from each raw file reducing them in size. Details of the raw image FITS header file is shown in Table \ref{tab:rawfits} giving the precise FITS SUBRECT values, which define a clipped region of the CCD image. In addition to the raw data frames, 200 bias frames stored in 20 data cubes of 10 images each, and 111 single image flat field files were provided. No dark current frames were taken. The flats and bias frames were taken without any SUBRECT which required the correct alignment of master files against raw data image file as shown in Figure \ref{fig:visualdata}. The total number of raw data pixels within this data set is approximately 6.6 billion pixels. This dataset has already been processed \cite{2008ASSL..351..257S} and it is for this reason that it was considered a good reference data set allowing for calibration of the master bias and flat images and comparison of reduced images against the BCO pipeline. Given the short integration times of the data, and the fact that the data processing was focused on differential photometry, the dark current master was not deemed a requirement by the BCO team. While the dark current process has already been described in Chapter \ref{chapter2}, dark current removal has not been performed on the data.

  \begin{table}[H]
\centering
\begin{tabular}{p{3cm} p{12cm}}
  \toprule
Keyword & Value \\
  \midrule
\small{SIMPLE}& \small{= T / file does conform to FITS standard}\\
\small{BITPIX}& \small{= -32 / number of bits per data pixel}\\
\small{NAXIS}& \small{= 3 / number of data axes}\\
\small{NAXIS1}& \small{= 428 / length of data axis 1}\\
\small{NAXIS2}& \small{= 426 / length of data axis 2}\\
\small{NAXIS3}& \small{= 10 / length of data axis 3}\\
\small{EXTEND}& \small{= T / FITS dataset may contain extensions}\\
\small{COMMENT}& \small{FITS (Flexible Image Transport System) format defined in Astronomy and Astrophysics Supplement Series v44/p363, v44/p371, v73/p359, v73/p365}\\
\small{HEAD}& \small{='DV887   '           / Head model}\\
\small{ACQMODE}& \small{= 'Kinetics'           / Acquisition mode}\\
\small{ACT}& \small{= 1.304200E-01 / Integration cycle time}\\
\small{KCT}& \small{= 1.304200E-01 / Kinetic cycle time}\\
\small{NUMACC}& \small{= 1 / Number of integrations}\\
\small{NUMKIN}& \small{= 10 / Series length}\\
\small{READMODE}& \small{= 'Image   '           / Readout mode}\\
\small{IMGRECT}& \small{= '1, 512, 512, 1'     / Image format}\\
\small{HBIN}& \small{= 1 / Horizontal binning}\\
\small{VBIN}& \small{= 1 / Vertical binning}\\
\small{SUBRECT }& \small{= '30, 457, 453, 28'   / Subimage format }\\
\small{ DATATYPE}& \small{ = 'Counts  '           / Data type}\\
\small{XTYPE }& \small{  = 'Pixel number'       / Calibration type}\\
\small{XUNIT }& \small{= 0 / Type of system}\\
\small{ TRIGGER}& \small{= 'Internal'           / Trigger mode }\\
\small{CALIB }& \small{= '0,1,0,0 '           / Calibration }\\
\small{EXPOSURE }& \small{= 8.000000E-02 / Total Exposure Time}\\
\small{TEMP }& \small{=        -6.500000E+01 / Temperature }\\
\small{READTIME }& \small{=         1.000000E-06 / Pixel readout time }\\
\small{OPERATN }& \small{ =                    4 / Type of system}\\
\small{DATE }& \small{ = '2003-09-22T02:04:34' / file creation date (YYYY-MM-DDThh:mm:ss UTC) }\\

  \bottomrule

\end{tabular}
\caption{Raw Data Fits Header}
\label{tab:rawfits}
\end{table}

\par The 26 GB dataset was sufficient for the initial pilot and the ACN processing pipeline, however for the NIMBUS pipeline,  a much larger data set was required. To accomplish this the BCO data was replicated a number of times and stored on multiple storage devices. This was considered acceptable due to the fact the focus of the experimentation was to review processing speed and did not require terabytes of unique data frames. The amount of processing for duplicated image data was identical to processing of unique data. To ensure this, data files when duplicated required unique names to eliminate the possibility of web servers or worker nodes caching the image data and artificially reducing the processing time of the system.

\subsection{Performance Analysis}

\par The calibration time of an individual CCD frame in most modern computing environments is typically measured in seconds or fractions of a second. Given that images are often taken over multiple seconds, many existing applications process astronomical images using software not specifically optimised for performance. MATLAB-based custom applications and of course the ubiquitous IRAF application offer reasonably easy access to image reduction and processing for scientists, while the CPL and other frameworks are in use by larger centres where specialised calibration workflows are required. However an issue arises, as the number of CCD images increases and the processing time becomes a function of the number of images to process. Data transfer, storage/backup, and retrieval also require careful consideration due to the number of times these operations are performed. Apparently trivial decisions relating to the use of intermediate files when cleaning an image, or the use of compression can have a dramatic influence on overall system performance and resource utilisation.  For example, using a two step reduction process where the first step uses Master Bias and Master Flat to clean the pixels, and the second step calculates magnitudes, doubles storage requirements through the use of intermediate files. 
 
\par
\begin{figure} [ht]
\centering
  \includegraphics[width=0.5\textwidth] {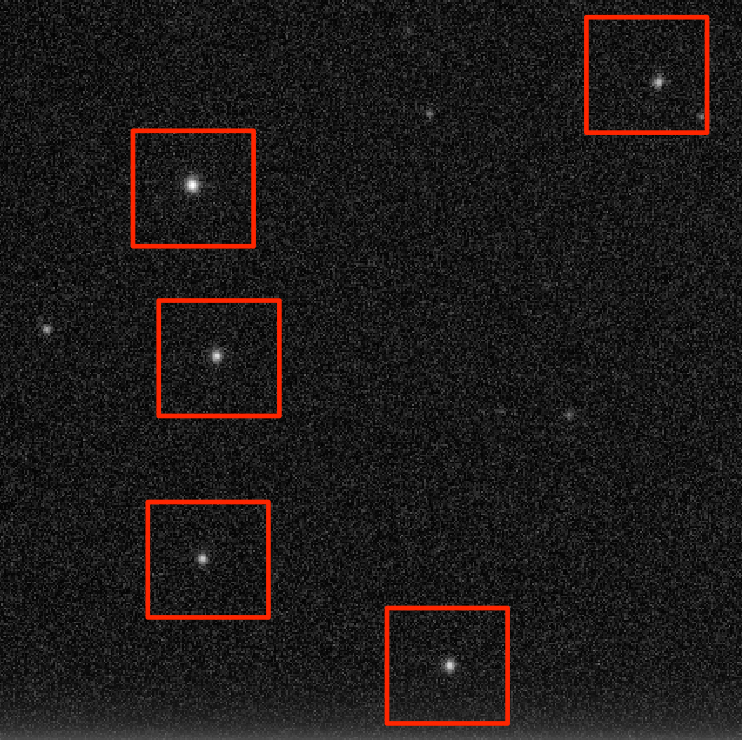}  
  \vskip -0.8em
    \caption{Clip regions on a CCD frame   Credit: BCO. }
  \label{fig:clip2}
\end{figure}

 In many cases much of the data captured within an image is not used in photometric measurements, and generic workflows that calibrate this data result in work being performed which does not contribute to the accuracy of a calculated magnitude. (E.g. Pixel cleaning is performed on all pixels within the image, including pixels not used in reference object magnitude calculations). It should be possible to only clean a subset of the image and only calibrate pixels required for magnitude calculations, eliminating a high percentage of work from the workflow as shown in Figure \ref{fig:clip2}. 

\par One reason why many infrastructures are not easily expanded is that most data reduction tools such as IRAF and CPL  are  sequential in nature, processing files interactively or in a batch sequence relying on high performance hardware devices to ensure that the data reduction process is kept within a reasonable timeframe. 

\par As more data is captured and uploaded to archives and made available through the Virtual Observatory the amount of data available for research is also increasing. When processing archive raw data, calibration is often performed, and researchers need to find resources for this data processing. Depending on the size of the dataset this could require significant computing resources.

\subsection{Parallel Data Processing}

\par The first step in working out how data should be logically grouped it is necessary to understand what data is relevant to the operations being performed. If pixels can be determined as non-contributory to the generation of magnitude values then by not including them in the data pipeline both the file IO and the CPU requirements may be reduced. 

\par CCD image calibration is often performed as a series of steps within a pipeline. In many cases this is done for flexibility within the pipeline framework allowing for a modular approach to software development or image processing as shown in the Kepler Science dataflow pipeline in Figure \ref{fig:cal}. It is worth reiterating that this research is focused on the equivalent processing performed within Kepler's CAL and the PA modules. As is evident from the Kepler pipeline there is a clear sequence that must be followed and which cannot be performed in parallel. This is true for some logical portion of data however and not necessarily true for the whole dataset, or for portions of that dataset.  In the case of Kepler, each CCD has its data processed within a parallel pipeline.

\par
\begin{figure} [ht]
\centering
 \fbox{  

\pgfdeclarelayer{bottom} \pgfdeclarelayer{top}
\pgfsetlayers{bottom,main,top}   
\begin{tikzpicture}[scale=.9,every node/.style={minimum size=1cm},on grid]       
\begin{pgfonlayer}{bottom}

        \begin{scope}[  
        yshift=-20,every node/.append style={
            yslant=0.5,xslant=-1,rotate=-10},yslant=0.5,xslant=-1,rotate=-10
          ]
        \fill[white,fill opacity=0.9] (0,0) rectangle (5,5);
        \draw[step=1.5mm, gray!70] (0,0) grid (5,5);
        \draw[step=1.5mm, gray] (2,2) grid (3,3);
        \draw[black,thick] (0,0) rectangle (5,5);
        
    \end{scope}
    
        \begin{scope}[  
        yshift=-10,every node/.append style={
            yslant=0.5,xslant=-1,rotate=-10},yslant=0.5,xslant=-1,rotate=-10
          ]
        \fill[white,fill opacity=0.9] (0,0) rectangle (5,5);
        \draw[step=1.5mm, gray!70] (0,0) grid (5,5);
        \draw[step=1.5mm, gray] (2,2) grid (3,3);
        \draw[black,thick] (0,0) rectangle (5,5);
        
    \end{scope}
    
    \begin{scope}[  
        yshift=0,every node/.append style={
            yslant=0.5,xslant=-1,rotate=-10},yslant=0.5,xslant=-1,rotate=-10
          ]
        \fill[white,fill opacity=0.9] (0,0) rectangle (5,5);
        \draw[step=1.5mm, gray!70] (0,0) grid (5,5);
        \draw[step=1.5mm, darkgray] (2,2) grid (3,3);
        
        \draw[gray, thick] (2,2) rectangle (3,3);
        \draw[black,thick] (0,0) rectangle (5,5);
        \node[name=B,draw,scale=0.9,black, thick,text width=0.95,text height=0.95,inner sep=0pt,] at (2.525,2.525) {};
    \end{scope}

\end{pgfonlayer}

  \begin{scope}[  
       yshift=80,every node/.append style={
        yslant=0.5,xslant=-1,rotate=-10},yslant=0.5,xslant=-1,rotate=-10
                     ]

        \fill[white,fill opacity=.9] (1,1) rectangle (4,4);
        \draw[step=3.33mm, gray] (1,1) grid (4,4);
        \node[name=A,scale=.9,draw,black,very thick,text width=3cm,text height=3cm,inner sep=0pt] at (2.5,2.5) {};
        \foreach \i in {north east,north west,south west,south east}
          \draw (B.\i) -- (A.\i);
        \fill[white,fill opacity=.9] (1,1) rectangle (4,4);
                \draw[step=3.33mm, gray] (1,1) grid (4,4);

        \node[name=C,scale=.1,draw,black,thick,text width=3cm,text height=3cm,inner sep=0pt] at (3.8,3.8) {};


	\filldraw [black] (2.5,2.5) circle (0.3cm);
	\draw [black,densely dotted,line width=0.3mm](2.5,2.5) circle (1cm);


    \end{scope}

        \begin{scope}[  
        yshift=160,every node/.append style={
        yslant=0.5,xslant=-1,rotate=-10},yslant=0.5,xslant=-1,rotate=-10
                     ]
          \fill[white,fill opacity=.9] (2.6,2.3) rectangle (4,3.7);
          \node[name=D,scale=.2,draw,black,very thick,text width=3cm,text height=3cm,inner sep=0pt] at (3,2.5) {};


        \foreach \i in {north east,north west,south west,south east}
          \draw (D.\i) -- (C.\i);
          \node[name=D,scale=.2,draw,fill=gray!20,very thick,text width=3cm,text height=3cm,inner sep=0pt] at (3,2.5) {};


    \end{scope}

	\draw (7.8,2.5) node {$Images$};
	\draw (7,5.5) node {$Clip\ Regions$};
	\draw (6,8.5) node {$Pixel$};
	
		\draw[<-,black,line width=0.3mm](6.5,2) {}  to  (9,2);
		\draw[<-,black,line width=0.3mm](4.5,5) {}  to  (9,5);
		\draw[<-,black,line width=0.3mm](2.5,8) {}  to  (9,8);

\end{tikzpicture}  }

 \vskip -0.8em
    \caption{Identifying parallel processing opportunities}
  \label{fig:parallel}
\end{figure}
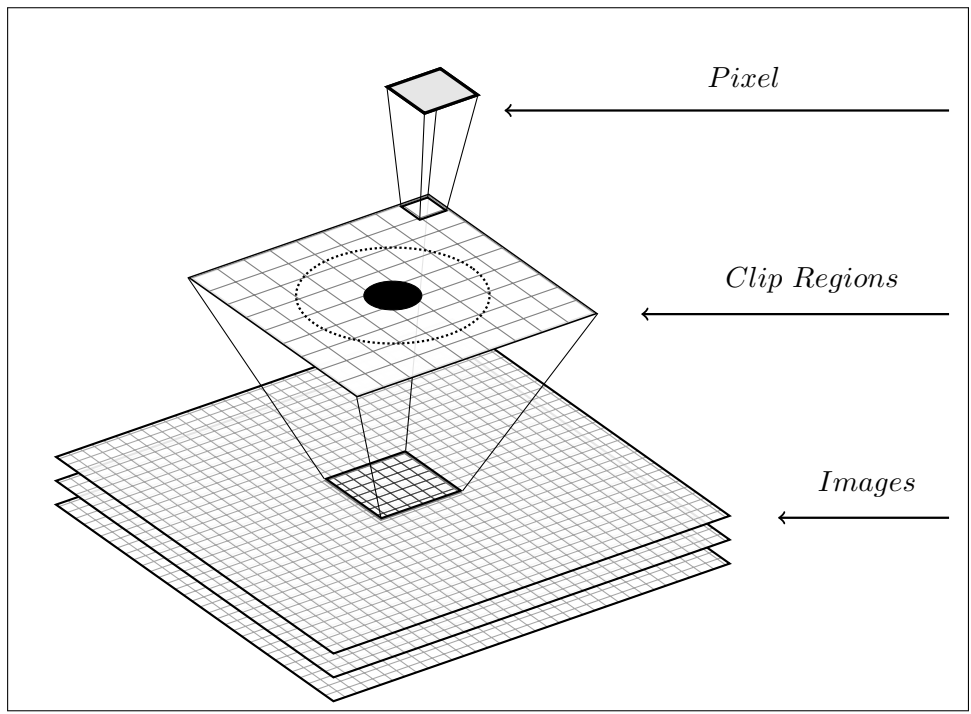

 This research considers what work can be run in parallel and how this impacts performance. As an example, looking at Figure \ref{fig:parallel}, if the output goal of a pipeline is to calibrate a pixel then all pixels can be treated as independent pieces of data which can be processed in parallel without any issue. If the output goal was the magnitude calculation of a star, a larger logical grouping of pixel data into a clipped region of the image is required. Multiple images can be grouped into data cubes, but once the timestamp is preserved then these images can be processed independently and then reassembled into light curves using the time sequence. There are other considerations such as the grouping of images in data cubes.  Existing raw data from BCO has a FITS data cube containing 10 images but there are other configurations possible. In many cases the reality of the file I/O costs of the systems will have an impact on data processing.

\section{System Designs}
\par In addition to the FEBRUUS pilot study, three primary designs have been considered and reviewed with two of them fully implemented. Experiments were devised for the implemented designs to test their limits and capabilities, with lessons learned helping improve the next evolution.  The purpose of these experiments was to test the performance, elasticity and flexibility of the architectures. Three primary designs were considered, and the final design incorporated the key components of the other three.

\begin {itemize}
\item IRAF Virtualisation.  Use of standard IRAF installations, virtualised and deployed into a cloud environment, using torrents as a distribution and replication technique for CCD image data. A torrent is a file sent via the BitTorrent protocol which is initially incomplete. The file continues to download form multiple computers using a torrent client which locates additional copies of the file on different computers.   
\item The ACN Pipeline: A distributed private cloud using commodity servers, with a lightweight data processing appliance and a private centralised queue to advertise work.  
\item The NIMBUS Pipeline: A distributed public cloud based on virtualised and physical servers using a distributed web queue to advertise work. 
\end {itemize}

\par The key reason for using a distributed model was to allow resources from multiple locations to participate in the pipeline. This requires that data is accessible to processing nodes, which obtain work from a central queue, download data, process it, and upload results. The ACN and the NIMBUS pipelines were built to facilitate experimentation. 

\subsection{Pixel Calibration - FEBRUUS Pilot }

\par The aim of the pilot system was to implement a series of programs designed to clean raw CCD image data using the identified formulas presented in Chapter \ref{chapter2}. The output of this system was a series of calibrated images, which used Master Bias and Master Flat images. A summary of the key objectives of this pipeline are presented below. 

\begin {itemize}
\item Write a series of CFITSIO based programs to generate Master BIAS, and Master FLAT frames 
\item	Write a lightweight program to perform pixel calibration of raw CCD images using the Master Flat and Master Bias frames
\item	Develop a series for tools for comparing FITS files to calibrate the results with results from BCO 
\item	Determine the performance of a CFITSIO program compared to the MATLAB pipeline from BCO
\item	Provide a benchmark for future processing pipelines
\item	Learn the process of image pixel calibration. 
\end {itemize}

By implementing the standard pixel calibration algorithms it was possible to become more familiar with the process to ensure decisions regarding what could be performed in parallel and what could be excluded from magnitude calculations. By focusing on basic calibration it was possible to obtain a basic benchmark for image cleaning. By using a CFITSIO library within a C program, the speed of imaging cleaning within a sequential environment could be estimated as a basis for comparison within a distributed environment. 

\par In total seven CFITSIO based tools were built to support the pilot. Details of these tools are provided below in Table \ref{tab:pilot-table}

  \begin{table}[ht]
\centering
\begin{tabular}{p{2cm} p{12cm}}
  \toprule
Program & Function \\
  \midrule
\small{gmb.c}& \small {Generate the master bias frame through the combination of 200 bias frames using an average pixel value per pixel coordinate}\\
\small{gmf.c}& \small {Generate master flat through the combination of 111 flat frames using median pixel values per pixel coordinate}\\
\small{bmf.c}& \small {Bias reduce the master flat by subtracting the master bias values for each pixel coordinate}\\
\small{nmf.c}& \small {Normalise the values within the bias reduced master flat frame}\\
\small{rrf.c}& \small {Reduce raw file by performing pixel calibration using the master bias and normalised, master flat frames}\\
\small{cfd.c}& \small {Compare two fits files by checking all pixel values to see if they are identical}\\
\small{lde.c}& \small {List all of the pixel values or a subset of pixel values within a FITs file}\\

\hline
  \bottomrule
\end{tabular}
  \caption{Basic tools developed to calibrate against the BCO MATLAB pipeline}
\label{tab:pilot-table}
\end{table}

\subsubsection{Generate Master Bias }

\par The purpose of this program is to create a single two-dimensional image, which contains an accurate representation of the systematic wide additive values, present in all pixels. The program takes in as input a directory location, which contains a number of BIAS files and the name of the required output master bias file. Debugging is offered at different levels to have various amounts of data generated to the screen, which can be redirected and reviewed. The output of this code is validated against a file supplied by BCO. While initial differences were found when comparing the output of the \emph{gmb.c} program and the BCO supplied file, this was found to be related to a difference in precision for the stored values during the average calculation. When the pipeline system performs calculations is uses DOUBLE\_IMG when setting the precision settings on the output MasterBIAS file.  

\par Unless the corresponding file from BCO uses similar precision, the following differences in values can occur. BCO Data: 233 compared to Pilot Data: 232.8800000000. Since this is an additive source for noise it may well make more sense to store this in an integer value within the Master Bias, however the primary aim is to retain as much high precision processing as possible.

\par Algorithm \ref{alg:algorithm1} describes the operation of the gmb.c program and how a master bias frame is created from a stack of bias images. The master bias is created one row at a time by summing pixels from the same position across multiple images and then dividing them by the number of images. This process is repeated for all pixels on a row by row basis. The flowchart for this program is  listed in Appendix \ref{app:chapter3}.

\begin{algorithm} [ht]

  \caption{Generate Master Bias}
  \label{alg:algorithm1}
  \begin{algorithmic}[1]
     \Procedure{GMB}{$row,col,img$}\Comment{row and columns width and no. of images}
     \State {$ImageRow[col]$}
     \State {$PixVal[col]$}
     \State {$BiasVal[col]$}

        \For{$i = 1$ \textbf{to} $row$}
        \For {$j = 1$ \textbf{to}  $img$}  
           \State {$ImageRow \gets readframerow [i,j] $} 
        \For {$k = 1$ \textbf{to}  $col$}  
 
            \State \texttt{$PixVal[k] \gets PixVal[k] + imageRow[k]$} \Comment{sum all values for each pixel}
            \EndFor
            \EndFor
        \For {$k = 1$ \textbf{to}  $col$}  \Comment{calculate avg value for each pixel in the row}
            \State {$BiasVal[k] \gets PixVal[k]/img$} 
            \EndFor
            \State \texttt{writerow(BiasVal)} \Comment{write out 1 row of master bias values }  
      \EndFor \Comment{go to the next row}

    \EndProcedure
  \end{algorithmic}
\end{algorithm}

\subsubsection{Generate Master Dark}

The data provided by BCO did not require dark frame cleaning due to the specification of the CCD device used, the image exposure time and the fact that differential photometry was being performed.  For completeness Algorithm \ref{alg:gmd} is presented which describes the process of creating a bias reduced dark current image master. 

\begin{algorithm} [ht]
  \caption{Generate Master Dark }
  \label{alg:gmd}
  \begin{algorithmic}[1]
     \Procedure{GMD}{$row,col,img$}\Comment{row and columns width and no. of images}
     \State {$ImageRow[col]$}
     \State {$PixVal[col]$}
     \State {$DarkVal[col]$}

        \For{$i = 1$ \textbf{to} $row$}
        \For {$j = 1$ \textbf{to}  $img$} 
        
           \State \texttt{$ImageRow \gets readframerow [i,j] $} 
        \For {$k = 1$ \textbf{to}  $col$}  \Comment{sum all values for each pixel}

            \State \texttt{$PixVal[k] \gets = PixVal[k] + imageRow[k]$} 
                        \EndFor
            \EndFor
        \For {$k = 1$ \textbf{to}  $col$} 
            \State {$DarkVal[k] \gets PixVal[k]/img - BiasVal[i,k]$} 
            \EndFor
            \State \texttt{writerow(DarkVal)} 
      \EndFor 

    \EndProcedure
  \end{algorithmic}
\end{algorithm}

\subsubsection{Generate Master Flat}

\par This process is similar to \emph{gmb.c} in that there are multiple images which combine to form a single image. There is an important difference however in that the flat files are created by exposing the CCD to light which introduces the possibility of cosmic rays striking a pixel causing its value to dramatically increase. The use of an average value across images which includes these values could affect the overall result. Instead pixel values for a specific x,y coordinate across all images are sorted to obtain the median value. 

\par The median value for a pixel coordinate is not the final step in creating a master flat file as it must be cleaned by eliminating the bias (using the master bias file) and then have the pixel values normalised. For the purpose of calibration with the BCO system these step were broken down into distinct steps. 

\par The output from the \emph{gmf.c} program was compared to the reference file provided by BCO to ensure that resulting values were identical. Due to the fact that a median is required and that the source data is a whole number, there is no need for higher precision than INT, although a LONG data type was used. The \emph{bmf.c} program uses the output of the gmf.c and subtracts the bias value associated with the specific pixel coordinate from the master bias frame. Sample output from this process is shown in Appendix \ref{app:chapter3}.

\par The discussion on pixel to pixel variation in sensitivity to light earlier in Chapter \ref{chapter2} looks at the quantum efficiency of a pixel for a particular wavelength of light. Exposing a CCD to a flat field of light should, in theory, produce the same value in each pixel after subtracting the bias value. In reality the same value is not seen, which may be due to a number of reasons such as the use of a non-uniform light source,  the quantum efficiency is actually variable across the CCD or dust particles which create shadows or patterns across the CCD. If on average a pixel collected 1000 electrons during an exposure, then to normalise a specific pixel the value needs to be divided by the average (in this case 1000). The result of this calculation is a normalisation value which can then apply to actual readings in object files either reducing the recorded value (because this specific pixel collects higher than average electrons) or increasing the recorded value (because this specific pixel collect lower than average electrons). The flowcharts for these programs are  listed in Appendix \ref{app:chapter3}.

\par Algorithm \ref{alg:gmf} is a combination of these programs that creates a final master flat frame from a stack of flat field images.

\begin{algorithm}[htbp]
  \caption{Generate Master Flat }
  \label{alg:gmf}
  \begin{algorithmic}[1]
  
     \Procedure{GMF}{$row,col,img$}\Comment{row and columns width and no. of images}
     \State $ImageRow[col]$
     \State $PixVal[col][img]$
          \State $Medians[col]$
     \State $Normalise[col]$

		 \State \Comment{Generate Median pixel values from multiple Flat Frames}

      \For{$i = 1$ \textbf{to} $row$}
                        \For {$j = 1$ \textbf{to}  $img$} 
             	\State {$ImageRow \gets readRow [i,j] $} \Comment{read in a row from each image}

                  \For {$k = 1$ \textbf{to}  $col$} 
            		\State {$PixVal[j][k] \gets PixVal[j][k] + imageRow[k]$} 
                 \EndFor
	   \EndFor   
            \For{\texttt{(k = 1 to col)}} 
          		\State {$Median[k]  \gets MEDIAN (QSORT (PixVal[k])) - BiasVal[i,k] $} 
		        \State {$writeRow(Medians)$}         
 
 	    \EndFor   

        \EndFor \Comment{ Go to next Row}

        \For{$i = 1$ \textbf{to} $row$} \Comment{Generate AVG pixel value}
 	\For {$k = 1$ \textbf{to}  $col$} 
             		\State {$Total \gets Total + ReadMedianRow[i][k]$} \Comment{total all Flat Fids values}
           	 \EndFor
        \EndFor   
        \State \texttt{AVG=Total/(row*col)}\Comment{Calculate avg value for Master Frame}

      \For{$i = 1$ \textbf{to} $row$}
	 		\For {$k = 1$ \textbf{to}  $col$} 
             			\State {$NormaliseVal[i][k] \gets ReadMedianRow[i][k]/AVG$} 
           	 	\EndFor
		  \State \texttt{writeRow(Normalise[i])} \Comment{write out 1 row of master flat values }         
           	 \EndFor

    \EndProcedure
  \end{algorithmic}
\end{algorithm}

\par The following three primary steps are incorporated within the algorithm. 

	\begin{itemize}
	\item Calculate the median pixel value across multiple flat field frames  
	\item Remove the bias value for that pixel using the master bias frame 
	\item  Normalise the pixel values across the bias reduced image
	\end{itemize}

\par The master flat frame is created a row at a time by obtaining the pixels value from the same location across multiple images, sorting them, and identifying the median value. This process is repeated for all pixels on a row by row basis. The purpose of using a median is to eliminate from the calculation any effects from a cosmic ray which could significantly distort an average value. To obtain a median value, all values must be sorted, and odd and even numbers of images must be accounted for within the algorithm.  Sample output from this process is shown in Appendix \ref{app:chapter3}.

\subsubsection{Pixel Cleaning Image Files}
The \emph{rrf.c} program is the primary appliance in the pilot system as it is responsible for cleaning the raw data images using the Master Bias and the Master Flat files. The formula used by this program has already been described in Chapter 2.  For all pixels in a raw object frame the corresponding pixel in the same location in the Master Bias frame is subtracted. This value is then divided by the corresponding pixel in the same location in the Master Flat frame, which adjusts the value up or down depending on the relative quantum efficiency for the pixel as compared to the average efficiency across the CCD. This final value is then written to the output file as the newly cleaned value. 

Algorithm \ref{alg:rrf}  demonstrate the simplicity of the actual calibration operation once master frames have been created and pixel alignment between the master frames and the raw images has been achieved. Where the raw images and the master frames are different in size, as with the BCO dataset, clipping is required on the master frames, or the use of an index into the master files to ensure pixel alignment with the raw images is correctly performed.  

\begin{algorithm} [ht]
  \caption{Calibrate raw image}
  \label{alg:rrf}
  \begin{algorithmic}[1]
     \Procedure{Calibrate}{$row,col,img$}\Comment{row and columns width}
     \State {$file\ raw$}\Comment{ file handle for raw image}
     \State {$file\ cal$}\Comment{ file handle for new calibrated image}

     \State {$file\ mb$}\Comment{ file handle for Master Bias}
     \State {$file\ mf$}\Comment{ file handle for Master Flat}
     \State {$file\ md$}\Comment{ file handle for Master Dark Current}
     \State {$CalPix[col]$}\Comment{ 1 dimensional array to store calibrated pixels}

      \For{$i = 1$ \textbf{to} $row$}
                  \For {$k = 1$ \textbf{to}  $col$} 
            \State {$biasVal \gets ReadPixel(mb,i,k)$}
            \State {$flatVal \gets ReadPixel(mf,i,k)$ }
            \State {$darkVal \gets ReadPixel(md,i,k)$} 
            \State {$rawVal \gets ReadPixel(raw,i,k)$ }
           \State {$CalPix[k] \gets (rawVal- biasVal - darkVal) / flatVal $} \Comment{clean pixel}
            \EndFor
            \State {$writerow(cal,CalPix)$} \Comment{write out 1 row of new values}  
            
            \EndFor 

    \EndProcedure
  \end{algorithmic}
\end{algorithm}

\subsubsection{Supporting Tools}

\par Pixel level comparisons were performed on the pilot master bias, master flat and reduced raw images agains the BCO equivalents to ensure that the basic processes were correctly implemented. The full data set provided by BCO was processed using the \emph{rrf.c} program from a local NFS disk on a Linux system to determine the maximum processing rate using a sequential processing approach.  The \emph{rrf.c} program held the master frames in memory and opened and processed each image data cube in sequence until all were processed. 

\subsection{Virtual IRAF instances - Design 1 }

The initial design concept for cloud based distributed computing was for a virtualisation of existing reduction pipelines, which ran on instances in a cloud environment as shown in Figure \ref{irafvirtual}. It was considered that an IRAF virtual machine appliance could be constructed and copies of the virtualisation instance could be run within a variety of cloud infrastructures. Using pre-fabricated images, IRAF based virtual machines could be created on any supported hypervisor.

\tikzstyle{iraf} = [draw, circle,fill=white,node distance=2.5cm, minimum height=1em]  
\begin{figure}[htbp]
 \begin{center}
\fbox { 

\begin{tikzpicture}
  \node [shape=cloud,name path=cloud,cloud puffs=20,aspect=2,draw,fill=white,minimum height=5cm,minimum width=6cm](cloud) at (2,2) {\strut};

  \draw node [fill,circle,inner sep=0pt,minimum size=1pt] {};
    \node[iraf] (a)    at (2.7,3)  {IRAF};
        \node[iraf] (a)    at (3.7,1.5)  {IRAF};

    \node[iraf] (a)    at (1.4,1)  {IRAF};

    \node[iraf] (a)    at (1,3.2)  {IRAF};

    \node[draw, cylinder, shape aspect=1.5, inner sep=0.3333em, fill=white,
    rotate=90, minimum width=1cm, minimum height=1cm] (cyl) at (-4,-4) {};
    
    \node[draw, cylinder, shape aspect=1.5, inner sep=0.3333em, fill=white,
    rotate=90, minimum width=1cm, minimum height=0.2cm] (cyl) at (-4,-4.2) {};

    \node[draw, cylinder, shape aspect=1.5, inner sep=0.3333em, fill=white,
    rotate=90, minimum width=1cm, minimum height=0.2cm] (cyl) at (-4,-3.8) {};


    \node[draw, cylinder, shape aspect=1.5, inner sep=0.3333em, fill=white,
    rotate=135, minimum width=2cm, minimum height=0.2cm] (cyl) at (-3.6,-0.4) {};
    \node[draw, ellipse , shape aspect=1.5, inner sep=0.3333em, fill=white,
    rotate=80, minimum width=1cm, dashed,minimum height=0.1cm] (cyl) at (-5.7,1.8) {};

     \draw[->,black,line width=0.2mm](-8,3)  to (-7,2);
    \draw[->,black,line width=0.2mm,dashed](-6.5,1.5)  to (-4.3,-0.7);
    \draw[->,black,line width=0.2mm,dashed](-4.3,-0.7) to (-5.9,1.7);
    \draw[->,black,line width=0.2mm,dashed](-5.9,1.7) to (-5.3,2.3);

     \draw[->,black,line width=0.2mm](-7,4)  to (-6,3);
    \draw[->,black,line width=0.2mm,dashed](-5.5,2.5)  to (-3.3,0.3);
    \draw[->,black,line width=0.2mm,dashed](-3.3,0.3) to (-5.5,1.7);
    \draw[->,black,line width=0.2mm,dashed](-5.5,1.7) to (-5.3,2.1);

    \node[draw, cylinder,fill opacity=.1,shape aspect=1.5, fill=white,
    rotate=135, minimum width=2cm, minimum height=4cm] (cyl) at (-5,1) {};
    
    \node[draw, cylinder,fill opacity=1,shape aspect=1, fill=white,
    rotate=45, minimum width=0.5cm, minimum height=0.8cm] (cyl) at (-5,2.4) {};
    
     \draw[->, >=latex, line width=10pt]  (-4, -1.25) to (-4,-3);
     \draw[->, >=latex, line width=10pt]  (-3, -3) to (0,0);

     \draw (-5,4 ) node {{\textbf{Data\ Source}}};
     \draw (-6,-3.5 ) node {{\textbf{Data\ Storage}}};
     \draw (4,-1 ) node {{\textbf{Cloud\ Instances}}};

\end{tikzpicture} }

  \caption{IRAF virtual instances in the cloud}
  \label{irafvirtual}
 \end{center}
\end{figure}
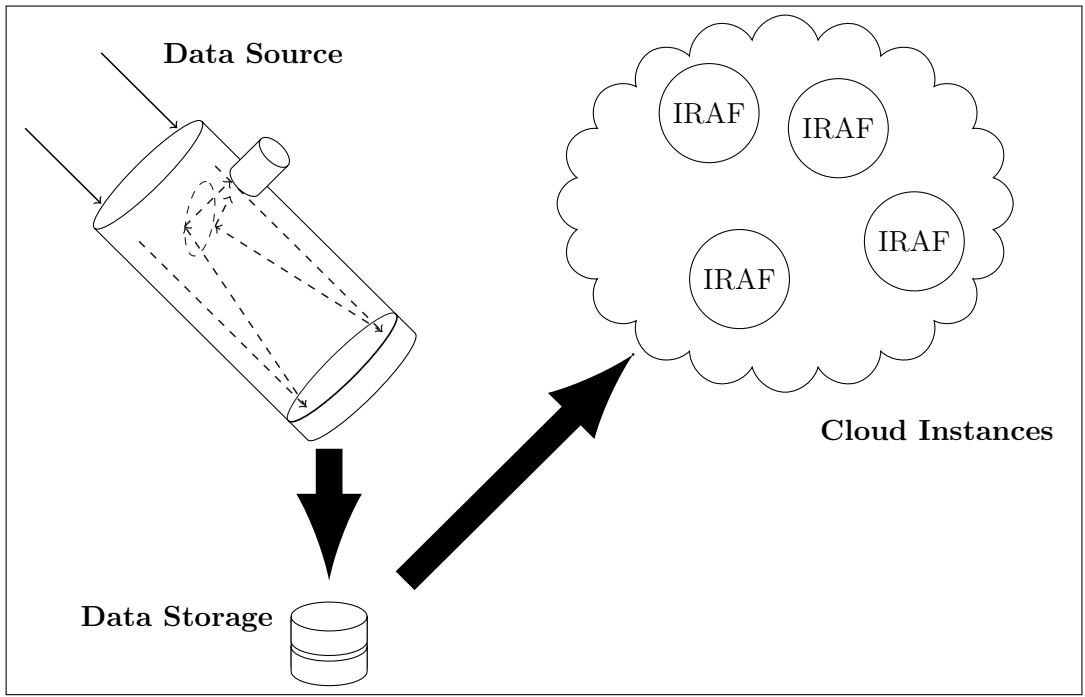

\begin{figure} [htbp]
 \begin{center}
\fbox { 

\begin{tikzpicture}
  \node [shape=cloud,name path=cloud,cloud puffs=20,aspect=2,draw,fill=white,minimum height=5cm,minimum width=6cm](cloud) at (2,2) {\strut};

  \draw node [fill,circle,inner sep=0pt,minimum size=1pt] {};
    \node[iraf] (a)    at (2.7,3)  {IRAF};
        \node[iraf] (a)    at (3.7,1.5)  {IRAF};

    \node[iraf] (a)    at (1.4,1)  {IRAF};

    \node[iraf] (a)    at (1,3.2)  {IRAF};

  \node [shape=cloud,name path=cloud,cloud puffs=20,aspect=2,draw,fill=white,minimum height=5cm,minimum width=6cm](cloud) at (-3,-3) {\strut};

    \node[draw, cylinder, shape aspect=1.5, inner sep=0.3333em, fill=white,
    rotate=90, minimum width=1cm, minimum height=1cm] (cyl1) at (-1.5,-4) {};
    \node[draw, cylinder, shape aspect=1.5, inner sep=0.3333em, fill=white,
    rotate=90, minimum width=1cm, minimum height=0.2cm] (cyl) at (-1.5,-4.2) {};
    \node[draw, cylinder, shape aspect=1.5, inner sep=0.3333em, fill=white,
    rotate=90, minimum width=1cm, minimum height=0.2cm] (cyl) at (-1.5,-3.8) {};
    
    \node[draw, cylinder, shape aspect=1.5, inner sep=0.3333em, fill=white,
    rotate=90, minimum width=1cm, minimum height=1cm] (cyl2) at (-4.5,-4) {};
    \node[draw, cylinder, shape aspect=1.5, inner sep=0.3333em, fill=white,
    rotate=90, minimum width=1cm, minimum height=0.2cm] (cyl) at (-4.5,-4.2) {};
    \node[draw, cylinder, shape aspect=1.5, inner sep=0.3333em, fill=white,
    rotate=90, minimum width=1cm, minimum height=0.2cm] (cyl) at (-4.5,-3.8) {};
    
    \node[draw, cylinder, shape aspect=1.5, inner sep=0.3333em, fill=white,
    rotate=90, minimum width=1cm, minimum height=1cm] (cyl3) at (-3,-2) {};
    \node[draw, cylinder, shape aspect=1.5, inner sep=0.3333em, fill=white,
    rotate=90, minimum width=1cm, minimum height=0.2cm] (cyl) at (-3,-2.2) {};
    \node[draw, cylinder, shape aspect=1.5, inner sep=0.3333em, fill=white,
    rotate=90, minimum width=1cm, minimum height=0.2cm] (cyl) at (-3,-1.8) {}; 
    
   
\draw (-6,2) -- ++(2cm,0) -- ++(0,-1.5cm) -- ++(-2cm,0);
\foreach \i in {1,...,4}
  \draw (-4cm-\i*10pt,2) -- +(0,-1.5cm);

\draw (-3.25,1.25cm) circle [radius=0.75cm];

\draw[->,>=latex,line width=4pt] (-2.5,1.25) -- +(40pt,0);
\node at (-3.25,1.25cm) {$\mu$};
\node at (-5.8,1.25cm) {$\lambda$};

\node[align=center] at (-3.25cm,0.2cm) {Service};
\node[align=center] at (-3.25cm,-0.2cm) {Node};
\node[align=center] at (-5.25cm,0.2cm) {Queue};

\node[align=center] at (-3cm,-3.25cm) {Torrents};

     \draw[<->, >=latex, line width=2pt]  (-4, -3.25) to (-3.5,-2.5);
     \draw[<->, >=latex, line width=2pt]  (-2, -3.25) to (-2.5,-2.5);
      \draw[<->, >=latex, line width=2pt]  (-3.75, -3.75) to (-2.25,-3.75);

     \draw (-5,3 ) node {{\textbf{Data\ Queue}}};
     \draw (-6,-5.5 ) node {{\textbf{Data\ Storage}}};
     \draw (4,-1 ) node {{\textbf{Cloud\ Instances}}};
     
     \draw[-latex,line width=2pt] (cloud) {}  to[out=180,in=180] (-6,1.25);

\end{tikzpicture} }

  \caption{Torrents for data distribution using a central queue. A messages $\mu$ is downloaded by an IRAF instance containing a torrent file $\lambda$ which is incomplete. The entire file is downloaded using a torrent client within the IRAF instance which connects to multiple torrent servers.}
  \label{fig:torrents}
 \end{center}
\end{figure}
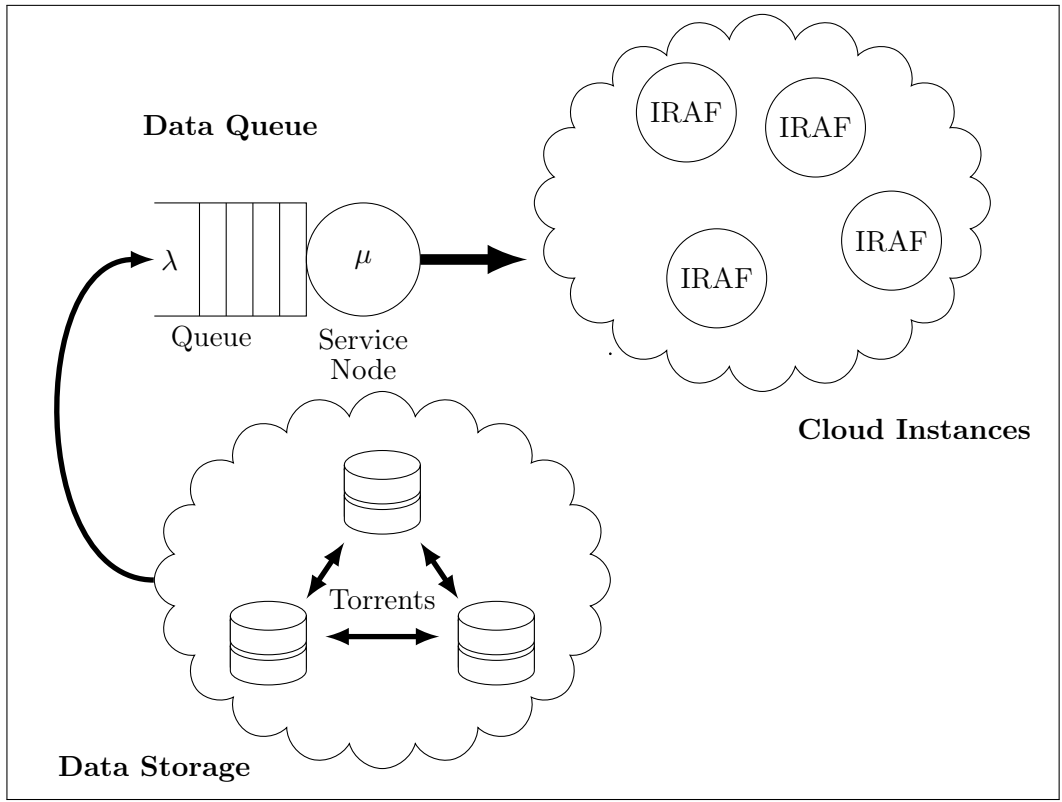 

\par Figure \ref{fig:torrents}, shows a torrent \cite{cohen2008bittorrent} based solution for data movement and replication where there were multiple copies of data distributed to various storage nodes. This design considered a message swapping system similar to the blackboard model within the OPUS pipeline.  When data is uploaded to a remote storage location it is first replicated to other data centres using a torrent  based infrastructure. All data uploaded to a data store would generate a message containing information about the data such as where it was, and what processing was required. Worker nodes running in virtual machines would read the message queue to consume a job which it would then download, process and upload to a location also specified in the message.

\par Issues arose with this design quite quickly however. The complexity of the IRAF solution and the size of a full virtual machine image imposed a large data copying, installation and management overhead for creating computing nodes. Instead of virtualizing an IRAF instance, a smaller appliance could be virtualized or installed natively on a number of platforms. A smaller appliance could be distributed more easily allowing for a greater number of computing nodes to participate using fewer resources on the computing node. Other reasons were also identified which ultimately led to a rejection of this design. 
\begin {itemize}
\item IRAF contains components which are not relevant to the cleaning process.
\item	IRAF performance is reasonably slow \cite{tody1993iraf}. 
\item	The installation and configuration of IRAF was non-trivial. 
\item	There are many manual steps in using IRAF although batch processing is possible.
\item	IRAF would reduce the requirement to fully appreciate the details of CCD image processing which was considered an essential skill to develop in order to potentially understand future optimisations. 
\end {itemize}

\par The BCO group are no longer using IRAF for cleaning and analysis of CCD images instead using MFITSIO with MATLAB. MATLAB instead of IRAF was also considered and rejected due to the license requirements for MATLAB, which would limit the number of deployable instances when processing a large dataset in a distributed environment. OCTAVE, a free alternative to MATLAB was also considered but this is based on the libcfitsio library and it was considered more appropriate to implement CCD reduction with the CFITSIO library using standard processing techniques. A lightweight processing utility  was considered preferable to facilitate a faster deployment of software to distributed computing nodes. 

\par Another issue with this design was the use of torrents for data distribution. Many firewalls are configured to block traffic of this type. A private torrent server was built internally for testing but this would limit data transfer to internal nodes. The data store was still considered a generic storage facility and not specific enough at this point. This design also required the creation of torrent files for all files. Unless the data was to be downloaded by multiple sources there appeared also to be a considerable overhead in storing the data multiple times to facilitate downloading. Unless the same data was to be downloaded by multiple users then the swarm based download mechanism would not be used which meant that the benefits of torrents were not being utilised. For large archive storage sites however this might be a reasonably interesting protocol to support.

\subsection{The ACN Pipeline - Design 2}

\par Following the initial IRAF design, a proposal was made to HEAnet, (Ireland's National Educational and Research Network), to facilitate a peer-to-peer network between DIT, and BCO  for the construction of an Astronomical Data Processing Cloud. This proposal was for a private network connection between the two institutes where data was moved to the DIT data centre and replicated to other storage facilities. Linux computing nodes are distributed to locations within this cloud to process image data. 
 
\par
\begin{figure}[htbp]
 \begin{center}
\fbox { 

\begin{tikzpicture}
  \node [shape=cloud,name path=cloud,cloud puffs=20,aspect=2,draw,fill=white,minimum height=10cm,minimum width=13cm](cloud) at (0,0) {\strut};

   \node [shape=circle,name=controller,radius=1.25] (controller) at (0,3) {};
   \draw[-latex,line width=2pt] (0,1.75) {}  to[out=270,in=180] (2,1.85);
   \draw[-latex,line width=1pt] (4.15,1) {}  to[out=0,in=90] (4.5,-0);
   \draw[-latex,line width=1pt] (4.15,1) {}  to[out=270,in=45] (0.25,-0.5);
   \draw[-latex,line width=1pt] (4.15,1) {}  to[out=225,in=45] (-2.5,-0.25);

   \draw (0,3cm) circle [radius=1.15cm];
   \draw (0,3cm) circle [radius=1.05cm];
  \draw (0,3cm) circle [radius=0.95cm];
  \draw (0,3cm ) node {{\textbf{Controller}}};

     \draw[->, >=latex, line width=2pt]  (-4, 2) to (-1.2,2.8);

  \draw node [fill,circle,inner sep=0pt,minimum size=1pt] {};


  \node [shape=cloud,name path=cloud,cloud puffs=20,aspect=2,draw,fill=white,minimum height=3cm,minimum width=4cm](cloud) at (4,-1.5) {\strut};
  \node [shape=cloud,name path=cloud,cloud puffs=20,aspect=2,draw,fill=white,minimum height=3cm,minimum width=4cm](cloud) at (0.25,-2) {\strut};
  \node [shape=cloud,name path=cloud,cloud puffs=20,aspect=2,draw,fill=white,minimum height=3cm,minimum width=4cm](cloud) at (-3.5,-1.5) {\strut};
    \node[iraf] (a)    at (-4.5,-2)  {ACN};
    \node[iraf] (a)    at (-3.5,-2)  {ACN};
    \node[iraf] (a)    at (-2.5,-2)  {ACN};
    
    \node[iraf] (a)    at (-0.75,-2.5)  {ACN};
    \node[iraf] (a)    at (0.25,-2.5)  {ACN};
    \node[iraf] (a)    at (1.25,-2.5)  {ACN};
    
    \node[iraf] (a)    at (5,-2)  {ACN};
    \node[iraf] (a)    at (4,-2)  {ACN};
    \node[iraf] (a)    at (3,-2)  {ACN};

    \node[draw, cylinder, shape aspect=1.5, inner sep=0.3333em, fill=white,
    rotate=90, minimum width=1cm, minimum height=1cm] (cyl1) at (-3.5,2) {};
    \node[draw, cylinder, shape aspect=1.5, inner sep=0.3333em, fill=white,
    rotate=90, minimum width=1cm, minimum height=0.2cm] (cyl) at (-3.5,1.8) {};
    \node[draw, cylinder, shape aspect=1.5, inner sep=0.3333em, fill=white,
    rotate=90, minimum width=1cm, minimum height=0.2cm] (cyl) at (-3.5,2.2) {};
    
    \node[draw, cylinder, shape aspect=1.5, inner sep=0.3333em, fill=white,
    rotate=90, minimum width=1cm, minimum height=1cm] (cyl2) at (-4.5,2) {};
    \node[draw, cylinder, shape aspect=1.5, inner sep=0.3333em, fill=white,
    rotate=90, minimum width=1cm, minimum height=0.2cm] (cyl) at (-4.5,1.8) {};
    \node[draw, cylinder, shape aspect=1.5, inner sep=0.3333em, fill=white,
    rotate=90, minimum width=1cm, minimum height=0.2cm] (cyl) at (-4.5,2.2) {};
    
    \node[draw, cylinder, shape aspect=1.5, inner sep=0.3333em, fill=white,
    rotate=90, minimum width=1cm, minimum height=1cm] (cyl2) at (-4,1.5) {};
    \node[draw, cylinder, shape aspect=1.5, inner sep=0.3333em, fill=white,
    rotate=90, minimum width=1cm, minimum height=0.2cm] (cyl) at (-4,1.3) {};
    \node[draw, cylinder, shape aspect=1.5, inner sep=0.3333em, fill=white,
    rotate=90, minimum width=1cm, minimum height=0.2cm] (cyl) at (-4,1.7) {};
         \draw (-4,0.8 ) node {{\textbf{S3\ Storage}}};
         \draw (-4,3 ) node {{\textbf{Amazon}}};

   
\draw (1.5,2.5) -- ++(2cm,0) -- ++(0,-1.5cm) -- ++(-2cm,0);
\foreach \i in {1,...,4}
  \draw (3.5cm-\i*10pt,2.5) -- +(0,-1.5cm);

\draw (4.25,1.75cm) circle [radius=0.75cm];

\node at (4.25,1.75cm) {$\mu$};



     \draw (3,3 ) node {{\textbf{Data\ Queue}}};
     \draw (4,-1 ) node {{\textbf{ITTD}}};
     \draw (0.25,-1 ) node {{\textbf{DIT}}};
     \draw (-3.5,-1 ) node {{\textbf{BCO}}};


\end{tikzpicture} }

  \caption{ACN Pipeline: Multi-institute private cloud using AWS S3 storage. Worker nodes download messages $\mu$ from an NFS based queue.}
  \label{ACNpipeline}
 \end{center}
\end{figure}
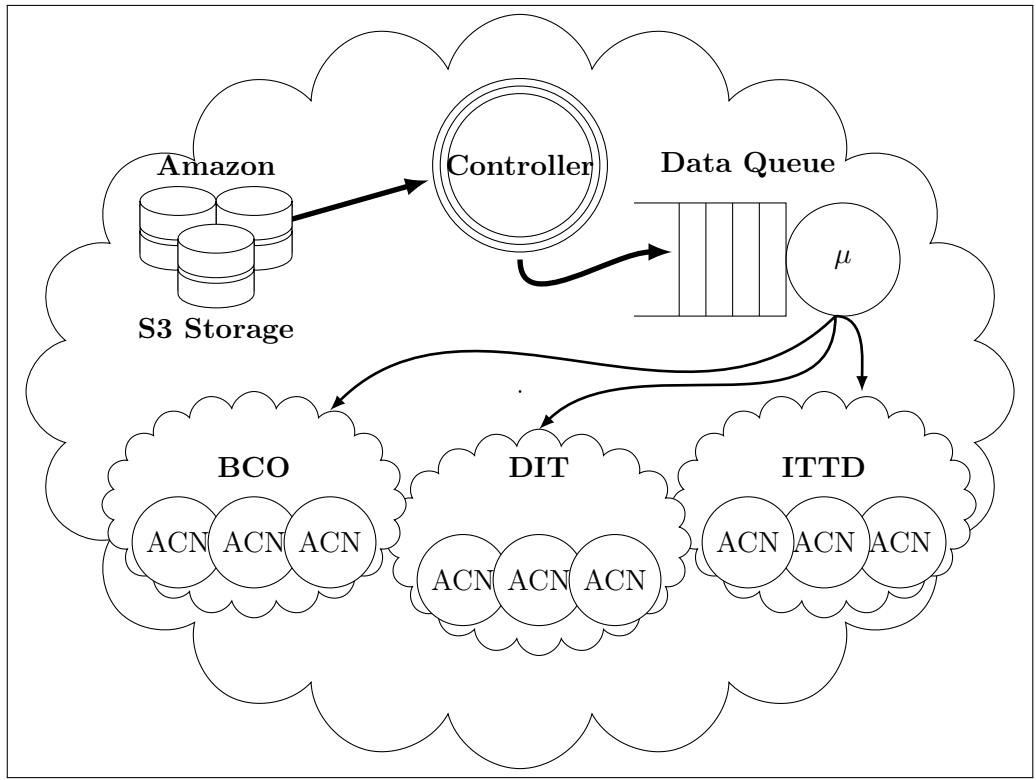

 The Astronomical Data Process Cloud was expanded to include the \gls{ittd} to allow a wider distribution of nodes as shown in Figure \ref{ACNpipeline}. The private layer-2, point-to-point network, was constructed in 2011 with the assistance of HEAnet. 

\par This pipeline reduction software was developed in two phases. The first phase integrated the previously verified calibration software used in the FEBRUUS pilot and the second phase  focused on increasing the functionality of the image processing by producing magnitude values for stars. This required the creation of a single appliance written in C which cleaned pixels and then performed aperture photometry, producing a range of magnitude values for each star using multiple aperture sizes. This program ran on a Linux ubuntu 12.04 server finding a filename using a shared NFS based queue before downloading the actual data file from an Amazon \gls{s3} bucket.  The S3 storage contained compressed versions of the BCO dataset, and the compressed and upload process was included as part of the running pipeline.  Table \ref{tab:acn-experiments} summarises the experiments used to test the ACN pipeline.

  \begin{table} [htbp]
\centering

\begin{tabular}{p{2cm} p{3.2cm} p{8cm}}

  \toprule
Reference & Measure & Experimental Objectives \\
  \midrule
\textbf{\small{Exp:ACN1}}& \small{ACN-APHOT Performance} & \small {Determining the performance of this program by running in multiple modes using various storage devices. Two step versus one step cleaning is examined. }\\
\textbf{\small{Exp:ACN2}}& \small{Storage Performance} & \small {Determining the Impact of the location of the storage devices and their ability to support multiple queries. }\\
\textbf{\small{Exp:ACN3}}& \small{Data Compression} & \small {Compression of data reduces the size of data for both storage and transfer. Data compression techniques and approaches are considered.   }\\
\textbf{\small{Exp:ACN4}}& \small{Data Transfer} & \small {Data stores are compared in terms of data transfer rates. }\\
\textbf{\small{Exp:ACN5}}& \small{Pipeline Limits} & \small {Determining how fast the pipeline can operate within the proposed architecture.}\\

\hline
  \bottomrule
\end{tabular}
  \caption{ACN performance experimental set }
\label{tab:acn-experiments}
\end{table}

\subsection{NIMBUS Pipeline - Design 3}

\par
\begin{figure} [htbp]
 \begin{center}
\fbox { 

  \begin{tikzpicture}[node distance=2.5cm, auto, >=stealth,cross line/.style={preaction={draw=white,->,line width=4pt}}]

%
%
    \node[draw, cylinder, shape aspect=1.5, inner sep=0.3333em, fill=white, minimum width=1cm, minimum height=1cm] (cyl1) at (2,-1.25) {\footnotesize{Q}};
    \node[draw, cylinder, shape aspect=1.5, inner sep=0.3333em, fill=white, minimum width=1cm, minimum height=1cm] (cyl1) at (1.5,-2) {\footnotesize{Q}};

  \node[draw, cylinder, shape aspect=1.5, inner sep=0.3333em, fill=white, minimum width=1cm, minimum height=1cm] (cyl1) at (2.5,-1.75) {\footnotesize{Q}};
  \node[shape=cloud, cloud puffs=15.7, cloud ignores aspect, minimum width=4cm, minimum height=3.5cm, align=center, draw] (wrkq) at (2.25, -1.5) {};
  \draw  ($(wrkq.north)+(-0mm,-5mm)$)                     node                {\footnotesize {Distributed}};
  \draw  ($(wrkq.north)+(-0mm,-8mm)$)                     node                {\footnotesize {Worker Queue}};
 
%
%
    \node[shape=cloud, cloud puffs=15.7, cloud ignores aspect, minimum width=6cm, minimum height=3.5cm, align=center, draw] (ARCHIVECloud) at (6.25, 2) {};
    \draw  ($(ARCHIVECloud.north)+(-0mm,-7mm)$)                     node                {\footnotesize {Data Archive Cloud}};
  
    \node[draw, cylinder, shape aspect=1.5, inner sep=0.3333em, fill=white,
    rotate=90, minimum width=1cm, minimum height=0.05cm] (cyl2a) at (5.2,1.8) {};
    \node[draw, cylinder, shape aspect=1.5, inner sep=0.3333em, fill=white,
    rotate=90, minimum width=1cm, minimum height=0.05cm] (cyl3) at (5.2,2.2) {};  
    \draw  ($(cyl3.south)+(-5.5mm,-8mm)$)                     node                {\footnotesize {Web }};
    \draw  ($(cyl3.south)+(-5.5mm,-11mm)$)                     node                {\footnotesize {Server }};

      \node[shape=circle, fill=white,minimum width=.7cm, minimum height=.7cm, align=center, draw] (web) at (5.6, 1.8) {};

    \node[draw, cylinder, shape aspect=1.5, inner sep=0.3333em, fill=white,
    rotate=90, minimum width=1cm, minimum height=0.05cm] (cyl2a) at (7.2,1.8) {};
    \node[draw, cylinder, shape aspect=1.5, inner sep=0.3333em, fill=white,
    rotate=90, minimum width=1cm, minimum height=0.05cm] (cyl3) at (7.2,2.2) {};  
    \draw  ($(cyl3.south)+(-5.5mm,-8mm)$)                     node                {\footnotesize {Web }};
    \draw  ($(cyl3.south)+(-5.5mm,-11mm)$)                     node                {\footnotesize {Server }};

      \node[shape=circle, fill=white,minimum width=.7cm, minimum height=.7cm, align=center, draw] (web) at (7.6, 1.8) {};    
      
%
%
    \node[shape=circle, fill=white,minimum width=2.4cm,  align=center, draw] (controller) at (8, -1) {};    
    \node[shape=circle, fill=white,minimum width=2.3cm,  align=center, draw] (controller) at (8,-1) {};    
    \node[shape=circle, fill=white,minimum width=2.2cm,  align=center, draw] (controller) at (8, -1) {};    
  \draw  ($(controller)+(-0mm,-0mm)$)                     node                {\footnotesize {Control System}};

%
%
    \node[shape=cloud, cloud puffs=15.7, cloud ignores aspect, minimum width=6cm, minimum height=3.5cm, align=center, draw] (DCCloud) at (-2, 2) {};
    \draw  ($(DCCloud.north)+(-0mm,-7mm)$)                     node                {\footnotesize {Data Capture Cloud}};
    
     \node[shape=trapezium, minimum width=.7cm, minimum height=.7cm, align=center, draw,trapezium right angle=120,trapezium left angle=60] (CCD) at (-3.75, 1.8) {};
    \draw  ($(CCD.north)+(-0mm,-4mm)$)                     node                {\footnotesize {CCD}};

    \node[draw, cylinder, shape aspect=1.5, inner sep=0.3333em, fill=white,
    rotate=90, minimum width=1cm, minimum height=0.05cm] (cyl1) at (-2,1.4) {};
    \node[draw, cylinder, shape aspect=1.5, inner sep=0.3333em, fill=white,
    rotate=90, minimum width=1cm, minimum height=0.05cm] (cyl2-1) at (-2,1.8) {};
    \node[draw, cylinder, shape aspect=1.5, inner sep=0.3333em, fill=white,
    rotate=90, minimum width=1cm, minimum height=0.05cm] (cyl3) at (-2,2.2) {};  
  
    \draw  ($(cyl1.south)+(-5mm,-5mm)$)                     node                {\footnotesize {Data }};  
    \node[draw, cylinder, shape aspect=1.5, inner sep=0.3333em, fill=white,
    rotate=90, minimum width=1cm, minimum height=0.05cm] (cyl2a) at (-0.2,1.8) {};
    \node[draw, cylinder, shape aspect=1.5, inner sep=0.3333em, fill=white,
    rotate=90, minimum width=1cm, minimum height=0.05cm] (cyl3) at (-0.2,2.2) {};  
    \draw  ($(cyl3.south)+(-5.5mm,-8mm)$)                     node                {\footnotesize {Web }};
    \draw  ($(cyl3.south)+(-5.5mm,-11mm)$)                     node                {\footnotesize {Server }};

      \node[shape=circle, fill=white,minimum width=.7cm, minimum height=.7cm, align=center, draw] (web) at (0.2,1.8) {};    
%
 %
%
  \node[draw, cylinder, shape aspect=1.5, inner sep=0.3333em, fill=white, minimum width=1cm, minimum height=1cm] (cyla) at (-3.75,-2.8) {\footnotesize{Q}};
  \node[draw, cylinder, shape aspect=1.5, inner sep=0.3333em, fill=white, minimum width=1cm, minimum height=1cm] (cylb) at (-4.25,-3.55) {\footnotesize{Q}};
  \node[draw, cylinder, shape aspect=1.5, inner sep=0.3333em, fill=white, minimum width=1cm, minimum height=1cm] (cylc) at (-3.25,-3.3) {\footnotesize{Q}};
  \node[shape=circle, minimum width=4cm, minimum height=3.5cm, align=center, draw] (moncloud) at (-3.5, -3) {};
    \node[shape=circle, minimum width=3.8cm, minimum height=3.5cm, align=center, draw] (moncloud) at (-3.5, -3) {};

  \draw  ($(moncloud.north)+(-0mm,-8mm)$)                     node                {\footnotesize {Monitoring System}};

%
  \node[shape=cloud, cloud puffs=20, cloud ignores aspect, minimum width=10cm, minimum height=5.5cm, align=center, draw] (heanetcloud) at (2.25, -7) {};
  \draw  ($(heanetcloud.north)+(-0mm,-7mm)$)                     node                {\footnotesize {Global Processing Cloud}};

       \node[shape=rectangle, fill=white,minimum width=.7cm, minimum height=.7cm, align=center, draw] (cpu1) at (4.5, -8) {};
  \node[shape=rectangle, fill=white,minimum width=.7cm, minimum height=.7cm, align=center, draw] (cpu1) at (4.7, -8.2) {};
  \node[shape=rectangle, fill=white,minimum width=.7cm, minimum height=.7cm, align=center, draw] (cpu1) at (4.9, -8.4) {};
  
    \node[shape=rectangle, fill=white,minimum width=.7cm, minimum height=.7cm, align=center, draw] (cpu1) at (3.5, -8) {};
  \node[shape=rectangle, fill=white,minimum width=.7cm, minimum height=.7cm, align=center, draw] (cpu1) at (3.7, -8.2) {};
  \node[shape=rectangle, fill=white,minimum width=.7cm, minimum height=.7cm, align=center, draw] (cpu1) at (3.9, -8.4) {};

 \node[shape=rectangle, fill=white,minimum width=.7cm, minimum height=.7cm, align=center, draw] (cpu1) at (2.5, -8) {};
  \node[shape=rectangle, fill=white,minimum width=.7cm, minimum height=.7cm, align=center, draw] (cpu1) at (2.7, -8.2) {};
  \node[shape=rectangle, fill=white,minimum width=.7cm, minimum height=.7cm, align=center, draw] (cpu1) at (2.9, -8.4) {};

 \node[shape=rectangle, fill=white,minimum width=.7cm, minimum height=.7cm, align=center, draw] (cpu1) at (1.5, -8) {};
  \node[shape=rectangle, fill=white,minimum width=.7cm, minimum height=.7cm, align=center, draw] (cpu1) at (1.7, -8.2) {};
  \node[shape=rectangle, fill=white,minimum width=.7cm, minimum height=.7cm, align=center, draw] (cpu1) at (1.9, -8.4) {};

  \node[shape=rectangle, fill=white,minimum width=.7cm, minimum height=.7cm, align=center, draw] (cpu1) at (0.5, -8) {};
  \node[shape=rectangle, fill=white,minimum width=.7cm, minimum height=.7cm, align=center, draw] (cpu1) at (0.7, -8.2) {};
  \node[shape=rectangle, fill=white,minimum width=.7cm, minimum height=.7cm, align=center, draw] (cpu1) at (0.9, -8.4) {};
  
    \node[shape=rectangle, fill=white,minimum width=.7cm, minimum height=.7cm, align=center, draw] (cpu1) at (-0.5, -8) {};
  \node[shape=rectangle, fill=white,minimum width=.7cm, minimum height=.7cm, align=center, draw] (cpu1) at (-0.3, -8.2) {};
  \node[shape=rectangle, fill=white,minimum width=.7cm, minimum height=.7cm, align=center, draw] (cpu1) at (-0.1, -8.4) {};

        \node[shape=rectangle, fill=white,minimum width=.7cm, minimum height=.7cm, align=center, draw] (cpu1) at (4.5, -7) {};
  \node[shape=rectangle, fill=white,minimum width=.7cm, minimum height=.7cm, align=center, draw] (cpu1) at (4.7, -7.2) {};
  \node[shape=rectangle, fill=white,minimum width=.7cm, minimum height=.7cm, align=center, draw] (cpu1) at (4.9, -7.4) {};
    
        \node[shape=rectangle, fill=white,minimum width=.7cm, minimum height=.7cm, align=center, draw] (cpu1) at (3.5, -7) {};
  \node[shape=rectangle, fill=white,minimum width=.7cm, minimum height=.7cm, align=center, draw] (cpu1) at (3.7, -7.2) {};
  \node[shape=rectangle, fill=white,minimum width=.7cm, minimum height=.7cm, align=center, draw] (cpu1) at (3.9, -7.4) {};

 \node[shape=rectangle, fill=white,minimum width=.7cm, minimum height=.7cm, align=center, draw] (cpu1) at (2.5, -7) {};
  \node[shape=rectangle, fill=white,minimum width=.7cm, minimum height=.7cm, align=center, draw] (cpu1) at (2.7, -7.2) {};
  \node[shape=rectangle, fill=white,minimum width=.7cm, minimum height=.7cm, align=center, draw] (cpu1) at (2.9, -7.4) {};

 \node[shape=rectangle, fill=white,minimum width=.7cm, minimum height=.7cm, align=center, draw] (cpu1) at (1.5, -7) {};
  \node[shape=rectangle, fill=white,minimum width=.7cm, minimum height=.7cm, align=center, draw] (cpu1) at (1.7, -7.2) {};
  \node[shape=rectangle, fill=white,minimum width=.7cm, minimum height=.7cm, align=center, draw] (cpu1) at (1.9, -7.4) {};

  \node[shape=rectangle, fill=white,minimum width=.7cm, minimum height=.7cm, align=center, draw] (cpu1) at (0.5, -7) {};
  \node[shape=rectangle, fill=white,minimum width=.7cm, minimum height=.7cm, align=center, draw] (cpu1) at (0.7, -7.2) {};
  \node[shape=rectangle, fill=white,minimum width=.7cm, minimum height=.7cm, align=center, draw] (cpu1) at (0.9, -7.4) {};
  
  \node[shape=rectangle, fill=white,minimum width=.7cm, minimum height=.7cm, align=center, draw] (cpu1) at (-0.5, -7) {};
  \node[shape=rectangle, fill=white,minimum width=.7cm, minimum height=.7cm, align=center, draw] (cpu1) at (-0.3, -7.2) {};
  \node[shape=rectangle, fill=white,minimum width=.7cm, minimum height=.7cm, align=center, draw] (cpu1) at (-0.1, -7.4) {};

      \node[shape=rectangle, fill=white,minimum width=.7cm, minimum height=.7cm, align=center, draw] (cpu1) at (4.5, -6) {};
  \node[shape=rectangle, fill=white,minimum width=.7cm, minimum height=.7cm, align=center, draw] (cpu1) at (4.7, -6.2) {};
  \node[shape=rectangle, fill=white,minimum width=.7cm, minimum height=.7cm, align=center, draw] (cpu1) at (4.9, -6.4) {};
    
      \node[shape=rectangle, fill=white,minimum width=.7cm, minimum height=.7cm, align=center, draw] (cpu1) at (3.5, -6) {};
  \node[shape=rectangle, fill=white,minimum width=.7cm, minimum height=.7cm, align=center, draw] (cpu1) at (3.7, -6.2) {};
  \node[shape=rectangle, fill=white,minimum width=.7cm, minimum height=.7cm, align=center, draw] (cpu1) at (3.9, -6.4) {};

 \node[shape=rectangle, fill=white,minimum width=.7cm, minimum height=.7cm, align=center, draw] (cpu1) at (2.5, -6) {};
  \node[shape=rectangle, fill=white,minimum width=.7cm, minimum height=.7cm, align=center, draw] (cpu1) at (2.7, -6.2) {};
  \node[shape=rectangle, fill=white,minimum width=.7cm, minimum height=.7cm, align=center, draw] (cpu1) at (2.9, -6.4) {};

 \node[shape=rectangle, fill=white,minimum width=.7cm, minimum height=.7cm, align=center, draw] (cpu1) at (1.5, -6) {};
  \node[shape=rectangle, fill=white,minimum width=.7cm, minimum height=.7cm, align=center, draw] (cpu1) at (1.7, -6.2) {};
  \node[shape=rectangle, fill=white,minimum width=.7cm, minimum height=.7cm, align=center, draw] (cpu1) at (1.9, -6.4) {};

  \node[shape=rectangle, fill=white,minimum width=.7cm, minimum height=.7cm, align=center, draw] (cpu1) at (0.5, -6) {};
  \node[shape=rectangle, fill=white,minimum width=.7cm, minimum height=.7cm, align=center, draw] (cpu1) at (0.7, -6.2) {};
  \node[shape=rectangle, fill=white,minimum width=.7cm, minimum height=.7cm, align=center, draw] (cpu1) at (0.9, -6.4) {};
 
   \node[shape=rectangle, fill=white,minimum width=.7cm, minimum height=.7cm, align=center, draw] (cpu1) at (-0.5, -6) {};
  \node[shape=rectangle, fill=white,minimum width=.7cm, minimum height=.7cm, align=center, draw] (cpu1) at (-0.3, -6.2) {};
  \node[shape=rectangle, fill=white,minimum width=.7cm, minimum height=.7cm, align=center, draw] (cpu1) at (-0.1, -6.4) {}; 
  
%
%
  \node[shape=cloud, cloud puffs=15.7, cloud ignores aspect, minimum width=4cm, minimum height=3.5cm, align=center, draw] (virginiacloud) at (2.25, -12) {};
  \draw  ($(virginiacloud.north)+(-0mm,-7mm)$)                     node                {\footnotesize {Results Cloud}};

    \node[draw, cylinder, shape aspect=1.5, inner sep=0.3333em, fill=white, rotate=90, minimum width=1cm, minimum height=0.05cm] (cyl2) at (3,-12.2) {};
    \node[draw, cylinder, shape aspect=1.5, inner sep=0.3333em, fill=white, rotate=90, minimum width=1cm, minimum height=0.05cm] (cyl3) at (3,-11.8) {};      
    
    \node[draw, cylinder, shape aspect=1.5, inner sep=0.3333em, fill=white,
    rotate=90, minimum width=1cm, minimum height=0.05cm] (cyl2) at (1.5,-12.2) {};
    \node[draw, cylinder, shape aspect=1.5, inner sep=0.3333em, fill=white,
    rotate=90, minimum width=1cm, minimum height=0.05cm] (cyl3) at (1.5,-11.8) {};    
    
    \node[draw, cylinder, shape aspect=1.5, inner sep=0.3333em, fill=white,    rotate=90, minimum width=1cm, minimum height=0.05cm] (cyl2) at (2.25,-13.2) {};
    \node[draw, cylinder, shape aspect=1.5, inner sep=0.3333em, fill=white,    rotate=90, minimum width=1cm, minimum height=0.05cm] (cyl3) at (2.25,-12.8) {};  

%
%

    \draw[->] (ARCHIVECloud) -- (wrkq) ;
    \draw[->] (ARCHIVECloud) -- (heanetcloud) ;
    \draw[->] (heanetcloud) -- (virginiacloud.north) ;   
    \draw[->] (DCCloud) -- (wrkq) ;
    \draw[->] (DCCloud) -- (heanetcloud) ;
    \draw[->] (wrkq) -- (heanetcloud) ;
    \draw[->] (heanetcloud) -- (moncloud) ;
    \draw[->] (CCD) -- (cyl2-1) ;
    \draw[->] (cyl2-1) -- (cyl2a) ;

    \draw[->] (controller) -- (heanetcloud) ;
    \draw[->] (controller) -- (wrkq) ;
    \draw[->] (controller) -- (ARCHIVECloud) ;
    \draw[->] (controller) -- (DCCloud) ;

 %
    \draw[->] (wrkq) -- (moncloud);

  \end{tikzpicture} }
  \caption{NIMBUS architecture. }
  \label{fig:NIMBUS}
 \end{center} 
\end{figure}
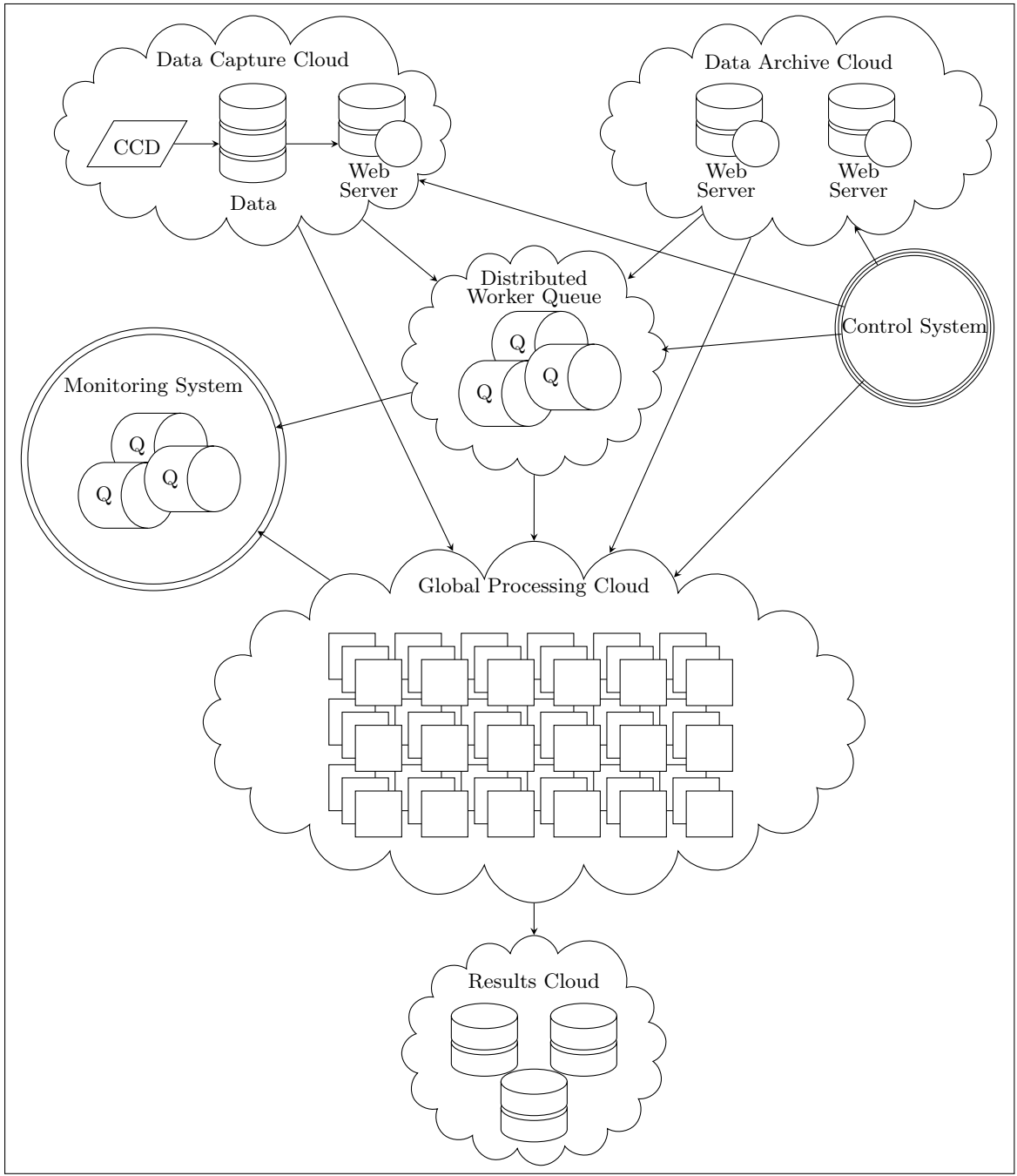

 This design moved the queuing system to the distributed Amazon AWS Simple Queue Service (SQS) \cite{cloud2011amazon} to allocate work to multiple computing nodes. This allowed global access to the queue potentially increasing the number of systems which could participate in image processing.  The data set was replicated and pushed to a number of different NGINX webservers within the DIT and within the AWS cloud to simulate distributed data sources. In this pipeline, the data was already compressed and in position before data processing began.  The AWS SQS message queue service was used to store the location of work, which a worker node could use to contribute to the overall experiment. The basic workflow is for a controller to instruct a storage node to publish the address of all files in its data store, and to then activate AWS EC2 nodes. Each node upgrades its software when activated, by downloading the latest version of the package software with instructions on how it should operate. The node then proceeds to take messages off the SQS system, download the file named within the message and processes the file. Once results are obtained they are written to an AWS S3 facility. Nodes can be added or removed at any time. Any work not completed is automatically reinserted onto the queue for another node to take. A node can run multiple threads, the number of files downloaded can be configured, the queue which is used can be updated and the software used for processing can be updated centrally. Multiple web servers containing data can all contribute to the worker queue, the instances can be of any size or configuration once they can run the software stack downloaded from the software distribution web server.  The NIMBUS architecture is shown in Figure \ref{fig:NIMBUS}.

\par A summary of the variables controlled for within the architecture is shown below with Table \ref{tab:nim-table} providing details of high level experiments run. 

\begin{itemize}
 \item Number of worker nodes
\item Size/type of the worker nodes
\item Number of webservers activated
\item Location of webservers activated
\item Number of worker instances per node
\item Batch size for a worker instance to process
\item Length of time for the experiment to run
\end{itemize}

\par The combinations of these variables causes a potentially large set of experiments to be required to develop a comprehensive view of the system behaviour. In total approximately 100 different experimental combinations were run to provide a reasonable view of the system capabilities.  

  \begin{table}[ht]
\centering
\begin{tabular}{p{2cm} p{3.2cm} p{8cm}}

  \toprule
Reference & Measure & Experimental Objectives \\
  \midrule
  
\textbf{\small{Exp:NIM1}}& \small{SQS Performance} & \small {Testing the read and writing times of the web message queues}\\
\textbf{\small{Exp:NIM2}}& \small{Single-Instance Node Performance.} & \small{ Determine the variables which affect the performance of the overall processing power of a single instance.} \\
\textbf{\small{Exp:NIM3}}& \small{Multi-Instances Node Performance.} & \small{ Focus on scaling the number of instances up to 100 looking for factors which could affect the scalability of the system.} \\
\textbf{\small{Exp:NIM4}}& \small{System Limits.} & \small{ Identify the full scalability of the pipeline and to identify strategies to continue improving the system performance} \\
\textbf{\small{Exp:NIM5}}& \small{System Scalability.} & \small{ The scalability and flexibility of the system is tested taking into account any limits observed in previous experiments. } \\

\hline
  \bottomrule
\end{tabular}
  \caption{NIMBUS: performance experimental set }
\label{tab:nim-table}
\end{table}

		\tikzstyle {webserver} = [draw,rectangle,node distance=2.5cm,rounded corners=0.5ex,minimum height=2em]
\tikzstyle {queue} = [draw,trapezium,fill=white,trapezium left angle=70,trapezium right angle=-70,shape border rotate=180,node distance=2.5cm,minimum height=2em]

\par There are six components central to this pipeline; (i) data capture and staging, (ii) serving archive data, (iii) distributed worker queues, (iv) distributed data processing, (v) results storage and (vi) monitoring. Each component is required to operate continuously and asynchronously, allowing for resource utilisation to be varied without interrupting the overall pipeline. While tested to a processing rate of 200 terabytes per day, the experiments were not at the limit of possible processing rates, with the primary restriction being a lack of additional resources available. Some of the larger experiments  utilised over 10,000 processing worker threads across 100 distributed servers.  

	\begin{description}
	\item [Data Capture Cloud]  \hfill \\
The data capture cloud consists of multiple distributed telescope sites containing CCD devices which record image data to a local storage device. Lossless data compression on  images is performed to reduce the bandwidth required for data transfer. Image clipping reduces images to only include the pixels used in the calculations of magnitude values. Compressed, clipped images are stored on fast storage disks attached to static web servers which serve http requests from the global processing cloud.  The Webservers inform the distributed worker queue that there is work to be performed.
	\item [Data Archive Cloud]  \hfill \\
The data archive cloud consists of multiple distributed websites containing image datasets. Images will already be compressed and possibly clipped. The images are stored on fast storage disks attached to static web servers which serve http requests from the global data processing cloud.  The Webservers advertise files to be processed via  the distributed worker queue.

	\item [Distributed Worker Queues]  \hfill \\
When the worker web queue is informed of a file  available for processing it stores the url of the file in a simple message which is available for worker nodes to read. The web queue ensures that only one copy of a message can be read from the queue at a time. When a worker completes its processing it permanently deletes the message. If a worker node fails to complete the processing of the image, the message will eventually reappear on the queue. This ensures that the overall system is resilient against compute node failures\footnote{A compute node failure is when a process terminates prior to the completion of the reduction process}.    
	\item [Global Processing Cloud]  \hfill \\
Worker nodes contain an initialisation boot script which installs worker sandboxes using tools downloaded from a predefined URL.  These tools ensure that the work performed is configurable both in terms of the work to be performed and web queues to listen or write to. Worker nodes within the processing cloud can be located anywhere in the world. Workers can join or leave the processing cloud at any point, without impacting the overall processing pipeline. 
	\item [Results Cloud]  \hfill \\
When a worker has completed its work, the resulting data file is uploaded to a distributed storage facility and a message is then written to the result queue that contains the URL for the location of the upload file. Using this queue, a processing cloud can be reconfigured to read the message queue to identify the URL of the result and to download results to a central location if required.   
	\item [Monitoring Cloud]  \hfill \\
For both experimental and control reasons a series of message queues is constructed to observe the performance of the pipeline. Worker nodes use queues to log their progress, and all queues are monitored to determine their read and write rates.    
	\end{description}

\subsection{Conclusion}

\par The architectures considered within this chapter have each contributed in an iterative way to the ultimate development of the NIMBUS pipeline. The methods used for evaluation are quantitative, focusing on the overall processing speed of the pipeline architecture, while considering the expandability and flexibility of the system. 

\par The pilot system established a baseline for performing accurate calibration ensuring that the tools required at the centre of the pipeline are representative of the work performed on CCD images by the existing BCO MATLAB pipeline.

\par The initial IRAF virtualisation design proposed a central queueing system designed to allow multiple virtual instances to obtain work. The method for data replication however was torrent based and practical limitations within secure network environments prohibited its use. The virtualisation image was large and would be an impediment to the dynamic provisioning of computing nodes within a pipeline. This design was not implemented. 

\par Design 2 focuses on the data compression and distribution using a central queue and experiments are focused on identifying the limits of the system, seeking bottlenecks and limitations within data compression and staging. Design 3 focuses on global scaling of the processing nodes using a web queuing model to ensure worker nodes can be easily added to the pipeline. Experiments for the ACN and NIMBUS pipelines were designed to identify the limits and capabilities of each design and more detail regarding their architectures and performance are presented in Chapter \ref{chapter4} and \ref{chapter5}. 

%% file: chapter5.tex

\chapter{The Astronomical Compute Node (ACN) Pipeline}
\label{chapter4}

{\centerline  {Abstract}}

\small{
The scientific community is in the midst of a data analysis crisis. The increasing capacity of scientific CCD instrumentation and their falling costs is contributing to an explosive generation of raw photometric data. This data must go through a process of cleaning and reduction before it can be used for high precision photometric analysis. Many existing data processing pipelines either assume a relatively small dataset or are batch processed by a High Performance Computing centre. A radical overhaul of these processing pipelines is required to allow reduction and cleaning rates to process terabyte sized datasets at near capture rates using an elastic processing architecture. The ability to access computing resources and to allow them to grow and shrink as demand fluctuates is essential, as is exploiting the parallel nature of the datasets. A distributed data processing pipeline is required. It should incorporate lossless data compression, allow for data segmentation and support processing of data segments in parallel. Academic institutes can collaborate and provide an elastic computing model without the requirement for large centralised high performance computing data centers. This paper demonstrates how a base 10 order of magnitude improvement in overall processing time has been achieved using the \emph{ACN pipeline}, a distributed pipeline spanning multiple academic institutes.
\\
\par \emph{Doyle, P; Mtenzi, F; Smith, N; Collins, A and O'Shea, B "Significantly reducing the processing times of high-speed photometry datasets using a distributed computing model", Proc. SPIE 8451, Software and Cyber infrastructure for Astronomy II, 84510C (September 24, 2012); doi: 10.1117/12.924863;}
}

\section{Overview}

\par The Astronomical Compute Node pipeline (ACN) \cite{doyle2012significantly}, was designed and implemented as part of this research to demonstrate the feasibility of performing distributed data processing using large data volumes hosted on a cloud based storage solution.  The ACN also validated the effective use of a centralised queue to manage multiple workers, facilitating a load balancing data processing solution for disparate server types. This pipeline architecture uses a distributed private cloud to demonstrate the scaling nature of the system across multiple Institutes of Technology. The ACN pipeline builds upon the \emph{rrf.c} program discussed in the previous chapter, which was a core component of the FEBRUUS pilot system. The \emph{acn-aphot.c} program  uses the NASA CFITSIO library,  and extends the cleaning features of the \emph{rrf.c} program to include magnitude calculations for each reference star within an image. This pipeline incorporates hardware capable of running a compiled instance of the \emph{acn-aphot.c} program to contribute to the processing of FITS images. The pipeline distributed design is shown in Figure \ref{flowchartdistrib} and the BCO dataset was used. The following components are central to this pipeline  and were designed and developed specifically for this experiment. These components are described in more detail within the experimental architecture section of this chapter.  The code used for this pipeline is available on github in the following repository. \url{https://github.com/paulfdoyle/acn.git} 

\begin {itemize}
\item Storage Control. Storage services including data compression, transport to central storage and downloading data to each of the ACN nodes. 
\item	Queue Control. Management of the NFS locking system including queue creation and management. 
\item	ACN Nodes. Individual workers and the infrastructure required to perform work. 
\item	Node Control. Activation and control of connected node systems including defining work to be performed. 

\end{itemize}

\tikzstyle{worker} = [rectangle, draw,fill=blue!10, text width=8em, rounded corners=3ex, minimum height=16em]
\tikzstyle{block} = [rectangle, draw,text centered, text width=10em, minimum height=2em]
\tikzstyle{block1} = [rectangle, draw,fill=white,text centered, text width=6em, minimum height=3em]

\tikzstyle{disk} =   [cylinder, draw,fill=white, text width=3em, text centered,shape border rotate=90, shape aspect=0.5, inner sep=0.3333em,  minimum width=2cm, minimum height=3em]
\tikzstyle {data} = [draw,trapezium,trapezium left angle=70,trapezium right angle=-70,node distance=2.5cm,minimum height=2em]
\tikzstyle{acn} = [draw, circle, fill=white,node distance=2.5cm, minimum height=2em, minimum width=2cm]

\begin{figure}
 \begin{center}
\fbox { 

  \begin{tikzpicture}[node distance=2.5cm, auto, >=stealth]
   \node[block] (a)                                     {\small {ACN Pipeline Controller}};
   \node[worker] (b)  [below of=a,node distance=4.5cm]                       {};
   \node[worker] (c)  [right of=b,node distance=4.5cm]                       {};
   \node[worker] (d)  [left of=b,node distance=4.5cm]                       {};
   \node[data] (f)  [above of=d,node distance=4.5cm]                       {Node File};

   \node[shape=cloud, cloud puffs=15.7, cloud ignores aspect, minimum width=7cm, minimum height=5cm, align=center, draw] (cloud) at (-4cm, -12cm) {};

  \node[below] at (b.north) {Queue Control};
  \node[below] at (c.north) {Storage Control};
  \node[below] at (d.north) {Node Control};
  \node[below] at ($(cloud.south)+(10mm,0mm)$) {Distributed ACN Node Cloud with 58 Nodes across 3 institutes};
  
  \node[iraf] (acn1) at ($(cloud.south)+(0mm,28mm)$) {ACN};
  \node[iraf] (acn1q) at ($(cloud.south)+(0mm,26mm)$) {ACN};
  \node[iraf] (acn1b) at ($(cloud.south)+(0mm,24mm)$) {ACN};
  \node[below] at (acn1b.south) {DIT};
  
 \node[iraf] (acn2) at ($(cloud.south)+(18mm,28mm)$) {ACN};
 \node[iraf] (acn2a) at ($(cloud.south)+(18mm,26mm)$) {ACN};
 \node[iraf] (acn2b) at ($(cloud.south)+(18mm,24mm)$) {ACN};
   \node[below] at (acn2b.south) {ITTD};

  \node[iraf] (acn3) at ($(cloud.south)+(-18mm,28mm)$) {ACN};
  \node[iraf] (acn3a) at ($(cloud.south)+(-18mm,26mm)$) {ACN};
  \node[iraf] (acn3b) at ($(cloud.south)+(-18mm,24mm)$) {ACN};
  \node[below] at (acn3b.south) {BCO};

   \node[block1] (NC1) at ($(d.north)-(0mm,12mm)$)                                     {\small {Reboot ACN Nodes}};
   \node[block1] (NC2) at ($(d.north)-(0mm,26mm)$)                                     {\small {Ping ACN Nodes}};
   \node[block1] (NC3) at ($(d.north)-(0mm,40mm)$)                                     {\small {Clean ACN Nodes}};
   \node[block1] (NC4) at ($(d.north)-(0mm,54mm)$)                                     {\small {Start ACN Nodes}};

   \node[block1] (QC1) at ($(b.north)-(0mm,12mm)$)                                     {\small {Compressed files}};
   \node[block1] (QC2) at ($(b.north)-(0mm,30mm)$)                                     {\small {Uncompressed files}};
   \node[block1] (QC3) at ($(b.north)-(0mm,48mm)$)                                     {\small {Clipped Compressed files}};
   
   \node[block1] (SC1) at ($(c.north)-(0mm,20mm)$)                                     {\small {FITS Compressed }};
   \node[block1] (SC2) at ($(c.north)-(0mm,40mm)$)                                     {\small {S3 Upload}};

   \node[disk] (e)  at (4.5cm, -12cm)                     {Amazon S3};

   \node[block] (g)  [below of=b,node distance=4cm]                       {NFS Queue};

\draw[->] (a.south) -- (b);
\draw[->] (b.south) -- (g);

\draw[->] (g.south) -- ($(cloud.east) -(6mm,-8mm) $);

\draw[->] (SC1) -- (SC2);
\draw[->] (SC2) -- (e);

\draw[->] ($(f.south)   -(4mm,0mm) $) -- ($(d.north)   -(4mm,0mm) $);

\draw[->] (a.south)  -- +(0mm,-5mm)  -| (d.north)  ; 
\draw[->] (a.south)  -- +(0mm,-5mm)  -| (c.north)  ; 

\draw[->] ($(cloud.east) -(0mm,3mm) $)  -- ($(e.west)-(0mm,3mm) $)  ; 
\draw[<-] ($(cloud.east) +(0mm,3mm) $)  -- ($(e.west)+(0mm,3mm) $)  ; 

\draw[->] (d.south) -- (acn1.north);
\draw[->] (d.south) -- (acn2.north);
\draw[->] (d.south) -- (acn3.north);

  \end{tikzpicture} }
  \caption{ACN distributed design.}
  \label{flowchartdistrib}
 \end{center} 
\end{figure}
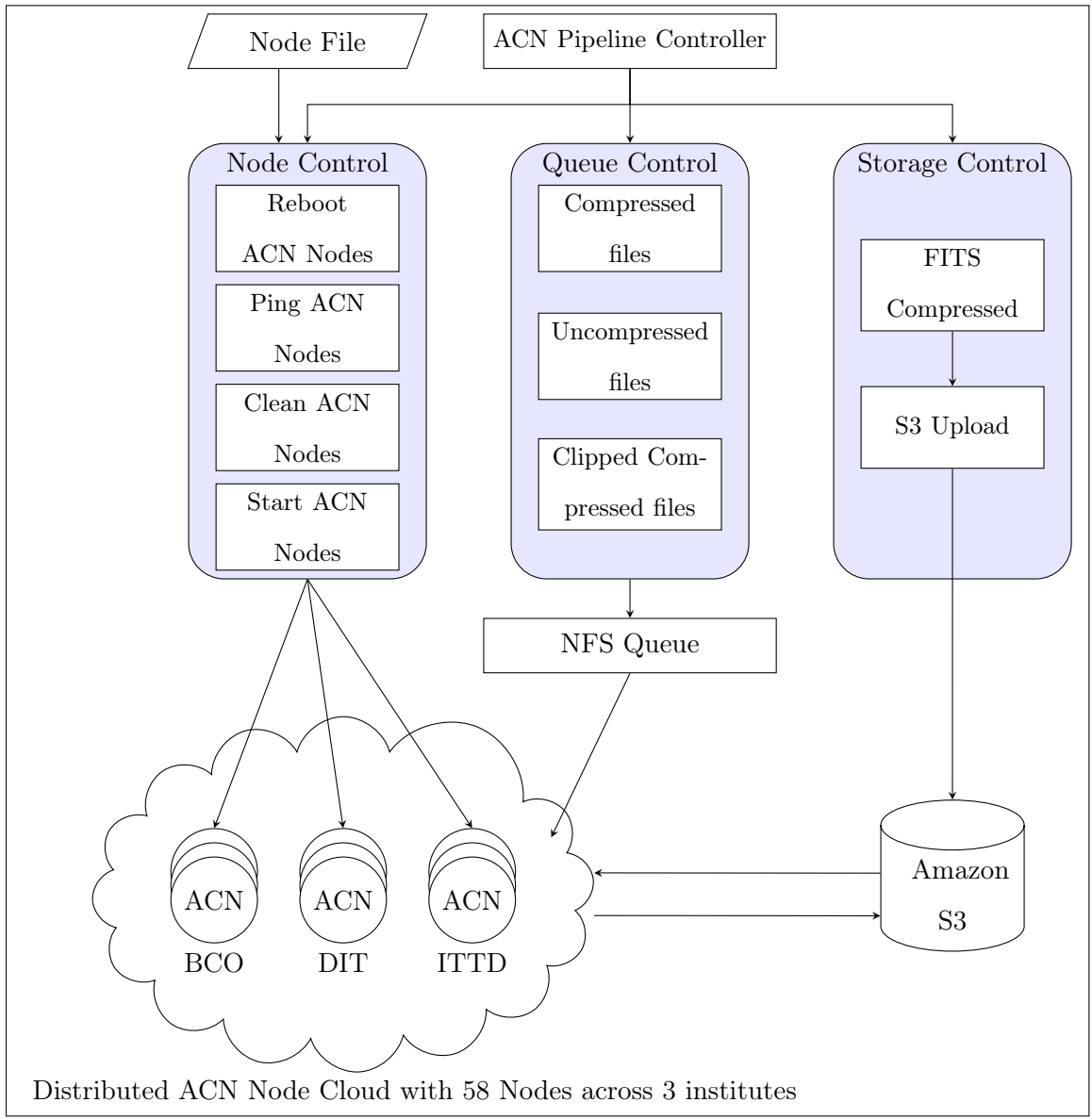 

\par
\begin{figure} 
\centering
\fbox{ \includegraphics[width=0.9\textwidth] {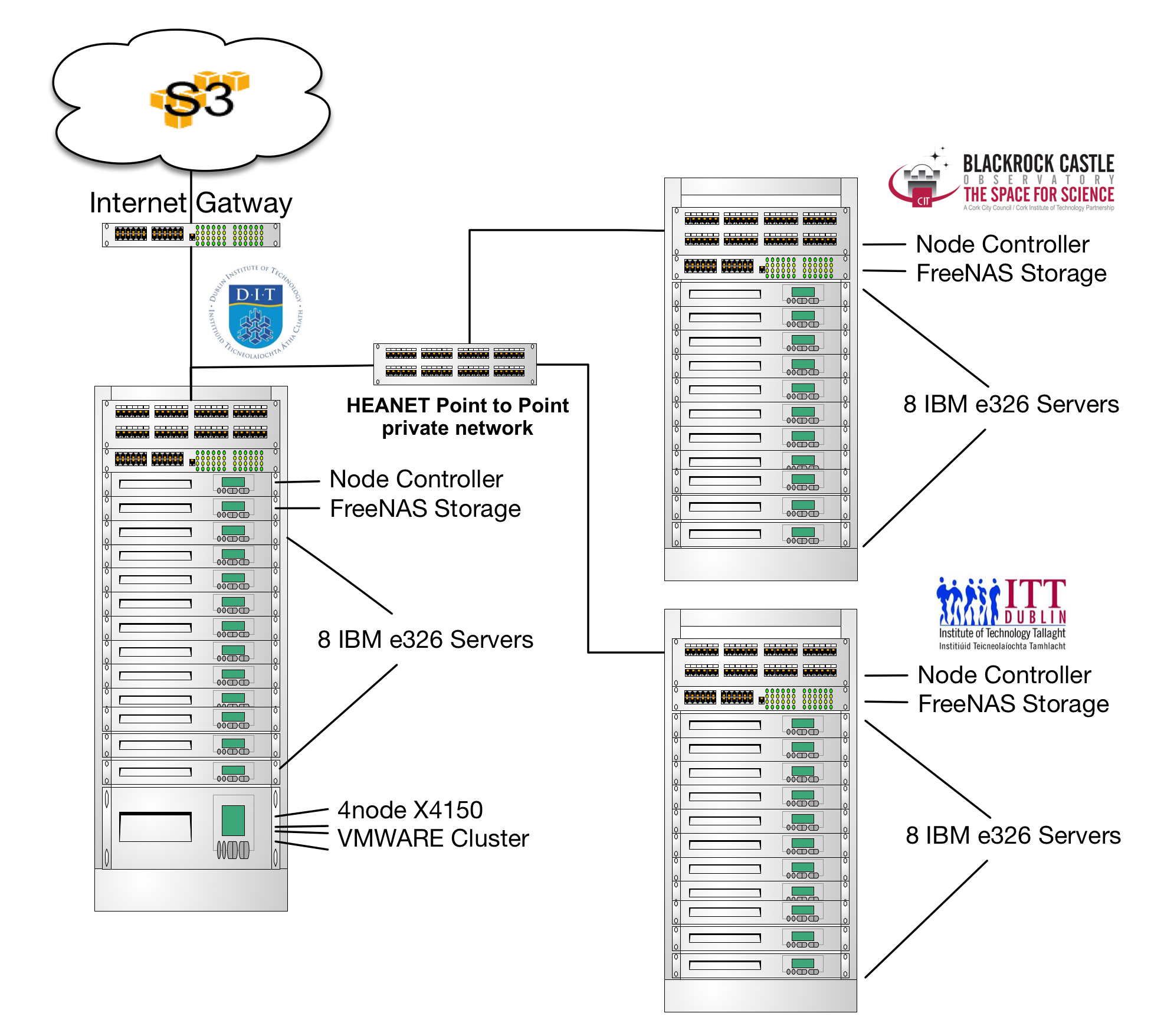} }
  \vskip -0.9em
    \caption{ACN network diagram connecting multiple institutes.} 
  \label{fig:ACN-Network2}
\end{figure}

 A private point to point IP network was constructed between three institutes\footnote{Dublin Institute of Technology, Blackrock Castle Observatory (Cork Institute of Technology), and the Institute of Technology Tallaght, Dublin} with 8 IBM eServer 326 machines connected to the network at each location with each server operating as an ACN node as shown in Figure \ref{fig:ACN-Network2}. Multiple FreeNAS storage devices were added to provide the ACN nodes with common utilities,  and a central NFS queue was implemented to allow nodes to obtain work. In addition to the physical IBM servers,  VMware based ACN instances were added to the network running on 4 Sun Microsystems x4150 servers. The internet was accessible through a gateway router which provided network address translation of addresses thereby allowing all nodes to access the S3 storage service. 

\section{System Architecture}

\par This pipeline has a number of distinct components, each of which collaborate to allow for a high processing rate of images. All of the systems are implemented within a Linux shell environment or are written using the $C$ programming language using the $CFITSIO$ library. The system spanned three institutes, each of which contained processing nodes for data processing but only one of which, the central control within the DIT,  managed the queue of work to be performed. Data was stored in a publicly accessible location which all nodes had access to, and a private queue only accessible from within the Hybrid Cloud system. The four primary components of the design as described in this section.

\subsection {Storage Control}
\par A high performance data storage system is essential to this design, as all nodes must have fast access to raw data files to allow for scaling of nodes without blocking on the I/O performance of the disks. Initially a single central FreeNAS \cite{cochard2007freenas} storage node was used which contained a full copy of the BCO dataset, but bottlenecks in processing file requests were quickly evident due to the limited number of disks physically used. Data was then replicated across a number of FreeNAS nodes, however the equipment used clearly didn't offer a scaling solution and bottlenecks continued to be observed. The Amazon's Simple Storage Service (S3) was selected instead given that it is a scalable storage solution that is available to all processing nodes, simplifying any assumptions regarding file location. The S3 storage also provides access to files using the HTTP protocol on port 80, which is typically not blocked by institutes. The Linux utility s3cmd was used to create storage "buckets" on S3 and transfer data to the S3 system. Each file is downloadable using a command line browser \emph{wget}, a Linux utility. The S3 storage supports concurrent file requests and while it has a significant delay in reading a single file, that delay is not cumulative and remains constant over multiple concurrent file requests \cite{vogels2009eventually}. The S3 storage provided publicly accessible files to all computing nodes and buckets were populated with compressed, clipped and uncompressed versions of the dataset for use in performance experiments. 

\par File compression was used to determine the difference in transfer costs for different file sizes.  Lossless compression was achieved on FITS files using \emph{fpack} \cite{pence2009lossless}, a utility available within the CFITSIO library. 

\par Compression was performed by a multi-core Dell 410 PowerEdge Server with 64 GBs of RAM. The assumption to test was that the cost of compressing a file and the transfer time for the compressed file would be less than the transfer time of an uncompressed file. Compression was seen as a potentially essential step to reduce  transfer times across networks. A third version of the dataset was prepared using a clipped  region \footnote{A clipped region refers to a portion of an image which is identified for a boundary which is used to identify a subset of the image}  of each image. For the calculation of a magnitude value for an object, the pixels around the star which contribute to that calculation are selected to create a subset of the file. Each of these clipped regions is stored as a FITS file and can be compressed independently of each other.  The more reference stars, the higher the number of files created. To simplify the process of image calibration, clipped master Bias, Dark, and Flat Field images can be created with identical dimensions. Figure \ref{fig:clip2} shows the actual pixels used in magnitude calculations, and in this dataset there are 5 clip regions. The processing of clipping this dataset involves an increase in the number of files in the dataset but a reduction in the overall number of pixels.

\subsection {Queue Control}
\par The process by which the ACN nodes determine what file is available for processing is the NFS file locking system which only allows one process to change a filename. The queue is a directory of empty filenames on an NFS mount point shared by all ACN nodes. Each filename corresponds to a data file to be processed on the S3 storage system. The queue is created using a utility which traverses the list of all available files and creates the set of empty files. The files conform to the following naming convention with the files being initialised with the \emph{Queued} name state tag. 

\begin{enumerate}
\item Queued\hbox{-}<filename.fits>
\item Locked\hbox{-}<filename.fits>
\end{enumerate}

\par Each ACN worker will traverse the queue as a directory listing looking at all filenames in turn. When it finds a file with the \emph{Queued} tag in the filename it will attempt to rename the file to add the \emph{Locked} tag. If the file is renamed successfully then an NFS lock was successfully obtained and the ACN node will then download the file from S3, process it and upload the results file to an S3 storage location. If the NFS lock is not secured then the next file in the directory is checked. 

\par This NFS locking system was both simple and reliable although there is a delay in obtaining a lock as new nodes are brought online and have to traverse the entire list of files from the start until it finds a file not yet processed (not marked as locked).  The greater the number of files, the longer it will be before a newly active worker node will be in finding a file to process.  Similar modifications to the process are possible, such as file removal once the uploading has occurred to S3. The principle algorithm used to secure a lock is given in Algorithm \ref{alg:locking}. 
\begin{algorithm}

  \caption{File Locking using NFS}\label{mba}
  \label{alg:locking}
  \begin{algorithmic}[1]
     \Procedure{obtainLock}{$QUEUEDIR$}\Comment{Directory containing queue}
     \State {$nameparts[]$}

        \For{$filename \ in\   \$ (  ls\  $QUEUEDIR ) }
        
                   \State {$nameparts  \gets \$splitfilename(\hbox{-})$} \Comment{Split filename into strings separated by \hbox{-}}

                 \If{ $nameparts[0] \neq "Queued" $}                         \Comment{Request NFS lock}

                       \State {$mv\ \$QUEUEDIR/\$filename\  to\ \$QUEUEDIR/Locked-\$nameparts[1]$} 
                 \If{ $\$?  = "0" $} \Comment{NFS lock obtained }
                        \State {$ProcessFile(nameparts[1])$} 
                        		\EndIf
				\EndIf

      \EndFor  \Comment{go to the next file in the queue}

    \EndProcedure
  \end{algorithmic}
\end{algorithm}

\subsection {Worker Nodes}

\par The core function of the ACN node is to find work in the queue, download image data, process it and upload the result, and to continue this process until there is no more work available.  The \emph{acn-aphot.c} program, which runs at the heart of each ACN node, is a C program compiled for multiple Linux systems including the Mac OSX. Its primary function is to generate a valid magnitude estimate for stars in an image, ensuring that the pixels used in the calculation are correctly calibrated. To support this process there are a number of supporting scripts within the node which are activated when the node first starts. The basic process for each ACN node is given in Figure \ref{flowchartworker}.   The queue of work is made available on an NFS shared drive. Once the worker node is activated it cycles through the available files until the queue is empty.

\tikzstyle{block1} = [rectangle, draw,fill=white,text centered, text width=7em, minimum height=2em]
\tikzstyle{disk} =   [cylinder, draw,fill=white, text width=3em, text centered,shape border rotate=90, shape aspect=0.5, inner sep=0.3333em,  minimum width=2cm, minimum height=3em]
\tikzstyle {data} = [draw,trapezium,trapezium left angle=70,trapezium right angle=-70,node distance=2.5cm,minimum height=2em]
\tikzstyle{decision} = [diamond, draw,fill=white,text width=4em, text badly centered, minimum height=2em, inner sep=0pt]

\begin{figure} [htbp]
 \begin{center}
\fbox { 

    \begin{tikzpicture}[node distance=1cm]

   \node[cloud] (a)                                              {\footnotesize Start};
   \node[block1] (b)  [below of=a]                       {\footnotesize Mount NFS Share};
   \node[block1] (c)  [below of=b]                       {\footnotesize Fetch Utilities};
   \node[block1] (d)  [below of=c]                       {\footnotesize Read Queue};
   \node[decision] (e)  [below of=d,node distance=2cm]                       {\footnotesize Secured Lock?};
    \node[block1] (f)  [below of=e,node distance=2cm]                       {\footnotesize Download File};
    \node[block1] (g)  [below of=f]                       {\footnotesize Process File};
    \node[decision] (h)  [below of=g,node distance=2cm]                       {\footnotesize Queue Empty?};
    \node[cloud] (i)  [below of=h,node distance=2cm]                       {\footnotesize Stop};

   \node[disk] (j)  [right of=b, node distance=4cm]                       {\footnotesize Disk};
    \node[data] (k)  [right of=f, node distance=4cm]                       {\footnotesize S3 Data};
      \node[data] (l)  [right of=g, node distance=4cm]                       {\footnotesize Result};

   \draw[->, thick] (a) -- (b);
   \draw[->, thick] (b) -- (c);
   \draw[->, thick] (c) -- (d);
   \draw[->, thick] (d) -- (e);
   \draw[->, thick] (e) -- (f);
   \draw[->, thick] (f) -- (g);
   \draw[->, thick] (g) -- (h);
   \draw[->, thick] (h) -- (i);

   \draw[->, thick] (j) -- (b);
   \draw[->, thick] (k) -- (f);
   \draw[->, thick] (g) -- (l);
   
   \draw[->, thick]  (e.west)    -- + (-10mm,0mm) -| ($ (d.west) - (10mm,1mm) $)  -- ($(d.west) -(0mm,1mm)$) ; 
   \draw[->, thick]  (h.west)    -- + (-13mm,0mm) -| ($ (d.west) - (13mm,-1mm) $)  -- ($(d.west) -(0mm,-1mm)$) ;

\node at (0.3,-6.5) {\small Yes};
\node at (-1.5,-4.75) {\small No};

\node at (0.3,-11.4) {\small Yes};
\node at (-1.5,-9.75) {\small No};

  \end{tikzpicture} }
  \caption{The ACN worker node processing flowchart.}
  \label{flowchartworker}
 \end{center}
\end{figure}
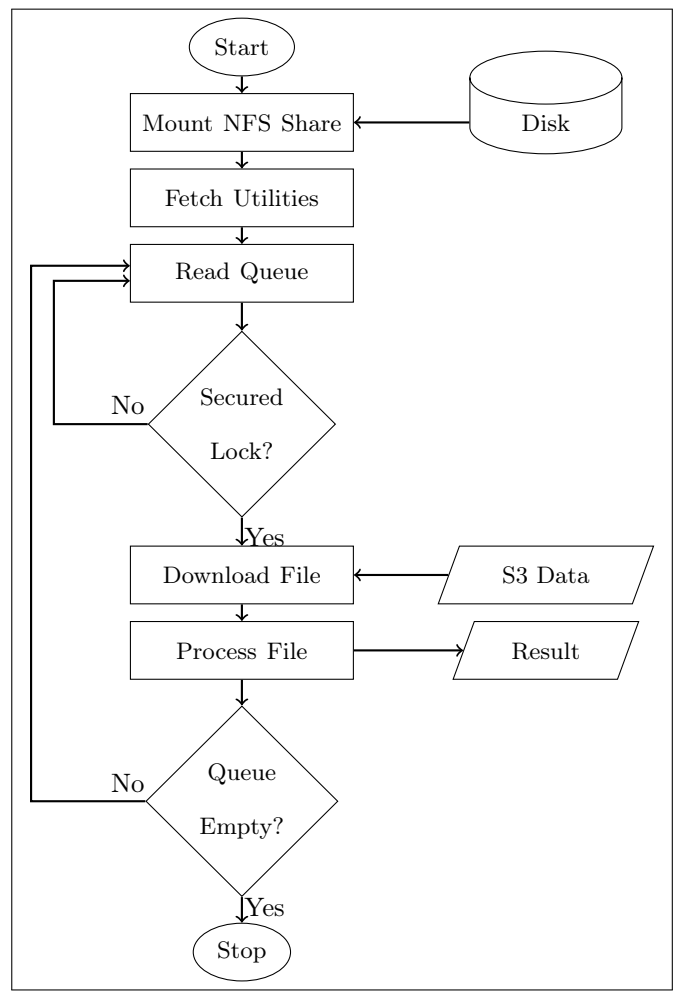 

\par The init scripts within the node will mount the central NFS directory containing the queue and worker utilities. Once the directory is mounted, all cleaning data and utilities within the node are removed. The latest version of the cleaning tools, including updated master frames, and any node control scripts will be downloaded and installed onto the node. This is required to ensure that with minimal effort, each node is always running the latest software versions. A script will traverse the NFS queue and when it succeeds in obtaining a lock on a file, it downloads that file from S3 to a working directory. The \emph{acn-aphot.c} program is then called to perform the image cleaning. If the program completes without any errors, then the results file is uploaded to the S3 storage. 

\par The ACN node can process compressed, uncompressed or clipped files. All that is required is that the latest cleaning scripts downloaded are configured to uncompress data as required. This process continues until there are no files left in the queue.

\subsubsection {Aperture Photometry in ACN-APHOT}

\par The \emph{acn-aphot.c} program is built upon the \emph{rrf.c} program from the initial pilot system. These changes extend the functionality to include aperture photometry and the following new features were developed. 

\begin{itemize}
\item Use an initial set of coordinates find the centre of an object on the image file.
\item Calculate the sky background.
\item Calculate the flux intensity of the object using a partial pixel algorithm for a given aperture size.
\item Generate multiple magnitude values for an object using a range of aperture sizes.
\item Calibrate pixels only as they are used.
\item Calibrate all pixels without performing aperture photometry.
\end{itemize}

\par
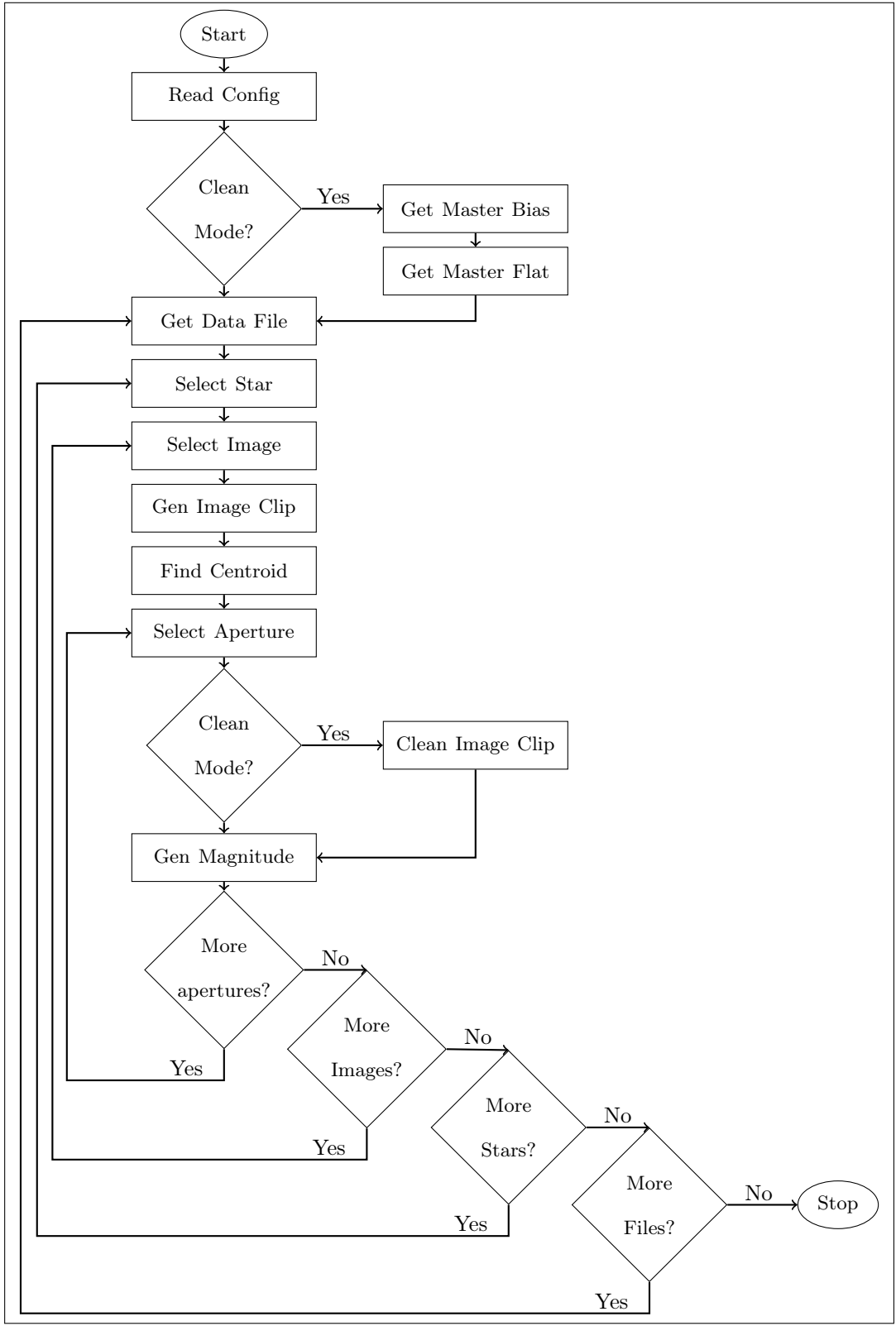
\begin{figure} [htbp]
 \begin{center}
\fbox { 

    \begin{tikzpicture}[node distance=1.5cm]

   \node[cloud] (a)                                              {\footnotesize Start};
   \node[block1] (b)     [below of=a, node distance=1cm]                       {\footnotesize Read Config};
   \node[decision] (c)  [below of=b, node distance=1.8cm]                       {\footnotesize Clean Mode?};
   \node[block1] (c1)  [right of=c,node distance=4cm]                       {\footnotesize Get Master Bias};
   \node[block1] (c2)  [below of=c1,node distance=1cm]               {\footnotesize Get Master Flat};
   
   \node[block1] (d)  [below of=c, node distance=1.8cm]                       {\footnotesize Get Data File};
   \node[block1] (e)  [below of=d, node distance=1cm]                       {\footnotesize Select Star};
   \node[block1] (f)  [below of=e, node distance=1cm]                       {\footnotesize Select Image};
   \node[block1] (g)  [below of=f, node distance=1cm]                       {\footnotesize Gen Image Clip};

   \node[block1] (h)  [below of=g, node distance=1cm]                       {\footnotesize Find Centroid};
   \node[block1] (i)  [below of=h, node distance=1cm]                       {\footnotesize Select Aperture};

   \node[decision] (j)  [below of=i, node distance=1.8cm]                       {\footnotesize Clean Mode?};
   \node[block1] (j1)  [right of=j,node distance=4cm]                       {\footnotesize Clean Image Clip};
 
   \node[block1] (k)  [below of=j, node distance=1.8cm]                       {\footnotesize Gen Magnitude};
   \node[decision] (l)  [below of=k, node distance=1.8cm]                       {\footnotesize More apertures?};
   \node[decision] (m)  [below right=0cm and 1cm of l]                       {\footnotesize More Images?};
   \node[decision] (n)  [below right=0cm and 1cm of m]                       {\footnotesize More Stars?};
   \node[decision] (o)  [below right=0.cm and 1cm of n]                       {\footnotesize More Files?};

   \node[cloud] (p)       [right of=o, node distance=3cm]                                        {\footnotesize Stop};

   \draw[->, thick] (a) -- (b);
   \draw[->, thick] (b) -- (c);
   \draw[->, thick] (c) -- (d);
   \draw[->, thick] (c.east) -- (c1.west);
   \draw[->, thick] (c1) -- (c2);
   \draw[->, thick] (c2) |- (d.east);

   \draw[->, thick] (d) -- (e);
   \draw[->, thick] (e) -- (f);
   \draw[->, thick] (f) -- (g);
   \draw[->, thick] (g) -- (h);
   \draw[->, thick] (h) -- (i);
   \draw[->, thick] (i) -- (j);
   \draw[->, thick] (j.east) -- (j1.west);
   \draw[->, thick] (j1.south) |- (k.east);
   \draw[->, thick] (j) -- (k);
   \draw[->, thick] (k) -- (l);
   \draw[->, thick] (l.east) -- (m.north);
   \draw[->, thick] (m.east) -- (n.north);
     \draw[->, thick] (n.east) -- (o.north);
     \draw[->, thick] (o.east) -- (p);

      \draw[->, thick] (o.south) -- +(-0mm,-5mm) -- +(-100mm,-5mm) |- ($ (d.west) - (-0mm,0mm) $) ; 
     \draw[->, thick] (n.south) -- +(-0mm,-5mm) -- +(-75mm,-5mm) |- ($ (e.west) - (-0mm,0mm) $) ; 
     \draw[->, thick] (m.south) -- +(-0mm,-5mm) -- +(-50mm,-5mm) |- ($ (f.west) - (-0mm,0mm) $) ; 
     \draw[->, thick] (l.south) --  +(-0mm,-5mm) -- +(-25mm,-5mm) |- ($ (i.west) - (-0mm,0mm) $) ;    
     
\node at ($(c.east) +(5mm,2mm)$)  {\small Yes};
\node at ($(j.east) +(5mm,2mm)$)  {\small Yes};
\node at ($(l.east) +(5mm,2mm)$)  {\small No};
\node at ($(m.east) +(5mm,2mm)$)  {\small No};
\node at ($(n.east) +(5mm,2mm)$)  {\small No};
\node at ($(o.east) +(5mm,2mm)$)  {\small No};

\node at ($(l.south) +(-6mm,-3mm)$)  {\small Yes};
\node at ($(m.south) +(-6mm,-3mm)$)  {\small Yes};
\node at ($(n.south) +(-6mm,-3mm)$)  {\small Yes};
\node at ($(o.south) +(-6mm,-3mm)$)  {\small Yes};

  \end{tikzpicture} }
  \caption{ACN-APHOT Image cleaning and reduction process work flow.}
  \label{flowchartacnaphot}
 \end{center}
\end{figure}

 These features extended the core software within the pipeline to reduce the data to a range of magnitude values for each star on each image.  The result files are measured in kilobytes and not megabytes allowing for faster uploading of results to improve the overall performance of the system. The option to have pixels cleaned before they were used allows for a comparison of the performance of a one phase reduction compared to a two phase reduction. In one phase reduction, the program loads the master frames into memory and only when a pixel value is used in a calculation such as  sky background calculation or flux intensity is the pixel calibrated. If a pixel in the image is not used, then it will never be calibrated. In two phase reduction the images are first calibrated using the \emph{rrf.c} program from the pilot system and these calibrated image files are passed to the \emph{acn-aphot.c} program which this time does not clean the pixels. Figure \ref{flowchartacnaphot} provides an overview of the \emph{acn-aphot.c} work flow. The algorithms used to extend the program are briefly described below.  

\subsubsection {Centroid Algorithm}
\par The gradient centroid approach described in Chapter \ref{chapter2} was used and Algorithm \ref{alg:centroid} was implemented. The initial x,y co-ordinate of the object is provided in this case using a preprepared co-ordinate file containing references to the objects in the image as shown in Table \ref{tab:acn-config}. Additional information is provided within this configuration file which will be discussed later. For the BCO dataset there are 5 objects of interest. For each of these objects the centroid algorithm was run to determine the centre of the object. 

  \begin{table}
\centering
\begin{tabular}{p{1.2cm} p{1.2cm} p{1.2cm} p{1.2cm}p{1.2cm}p{1.2cm}p{1.2cm}}
  \toprule
Xval & Yval & Radius & Annulus & Dannulus & Box & Threshold \\
  \midrule
\small{112}& \small{320} & \small {15}   & \small {10} & \small {15} & \small {80} & \small {660}    \\
\small{127}& \small{221} & \small {15}   & \small {10} & \small {15} & \small {80} & \small {660}    \\
\small{119}& \small{104} & \small {15}   & \small {10} & \small {15} & \small {80} & \small {660}    \\
\small{260}& \small{43} & \small {15}   & \small {10} & \small {15} & \small {80} & \small {660}    \\
\small{380}& \small{378} & \small {15}   & \small {10} & \small {15} & \small {80} & \small {660}    \\

\hline
  \bottomrule
\end{tabular}
  \caption{ACN Configuration File for the ACN-APHOT program }
\label{tab:acn-config}
\end{table}

\begin{algorithm} [htbp]
  \caption{Gradient Centroid Algorithm}
  \label{alg:centroid}
  \begin{algorithmic}[1]
     \Procedure{centroid}{$col,row,threshold,img$}

      \For{$i = 1$ \textbf{to} $row$}
                 \For {$k = 1$ \textbf{to}  $col$} 
                 \If{ $img[i,k] \leq\ threshold$} \Comment{test threshold to create a binary mask}
                       \State {$img[i,k] \gets 0$} 
		\Else 
                       \State {$img[i,k] \gets 1$} 
			\State{$TotalMasks \gets  TotalMasks + 1$} 
		\EndIf

            \EndFor \Comment{go to the next column}
            
            \EndFor \Comment{go to the next row}
            
      \For{$i = 1$ \textbf{to} $row$}
                 \For {$k = 1$ \textbf{to}  $col$} 
                 \If{ $img[i,k] = 1$} \Comment{test threshold to create a binary mask}
                       \State {$Tx \gets Tx + i$} \Comment{Sum all row gradients}
                       \State {$Ty \gets Ty + k$} \Comment{Sum all col gradients}
		\EndIf

            \EndFor 
            
            \EndFor \Comment{go to the next row}
            \State \texttt{$Cx \gets Tx / TotalMasks$} \Comment{Find X center point}
            \State $Cy \gets Ty / TotalMasks$ \Comment{Find Y center point}

   \Return $Cx, Cy$ 
    \EndProcedure
  \end{algorithmic}
\end{algorithm}

\subsubsection {Sky Background Algorithm}

The sky background is a measure of the average amount of light contained within each pixel which should be removed before calculating the intensity flux value. For a pixel value to be included in the sky background calculation it must be fully within the sky annulus.  Algorithm \ref{alg:sky} shows how to determine if a pixel is within the sky annulus. The sky background value per pixel is calculated as the average value of all of the pixels fully within the sky annulus. The inner radius of the annulus is the radius of the aperture used plus the annulus value given in the configuration file in Table \ref{tab:acn-config}. The dannulus is the width of the sky annulus, also given within Table \ref{tab:acn-config}. 

\begin{algorithm} [htbp]
  \caption{Sky Background}
  \label{alg:sky}
  \begin{algorithmic}[1]
     \Procedure{skybackground}{$x,y,centX,centY,radius,annulusval,dannulusval$}
 
                \Comment {The x,y values are the pixel coordinates}
                
                \Comment {CentX,centY are the centroid values}

		\State {$annulus = radius + annulusval$}
		\State {$dannulus = annulusval + dannulusval$}
		\State {$distance = euclid_distance(x,y,centX,centY)$}
		
                 \If{ $distance < dannulus - 0.5 \ \&\& \ distance > annulus + 0.5$ } 

                            \Return {$TRUE$}
		 \Else

                            \Return {$FALSE$}
        		\EndIf
    \EndProcedure
  \end{algorithmic}
\end{algorithm}

\subsubsection {Partial Pixel Algorithm}

\par Unlike the sky background, the calculation of the flux intensity required for the magnitude calculation uses a partial pixel algorithm. This is an extension of the sky background algorithm where pixels can be partially within the radius value used for the aperture. Algorithm \ref{alg:pp} was implemented within the \emph{acn-aphot.c} program.

\begin{algorithm} [htbp]
  \caption{Partial Pixel}
  \label{alg:pp}
  \begin{algorithmic}[1]
     \Procedure{intensitymeasure}{$pixval,radius,distance$}

                 \If{ $distance = radius$} \Comment{test if pixel is on the line of the aperture}
                       \State {$val \gets (0.5 * pixval) $} 		
		 \Else
		 	\If  { $distance <( radius - 0.5)$}
                       		\State {$val \gets pixval $} \Comment{ pixel is fully within the aperture}
                       
			\Else
				\If  { $distance >( radius + 0.5)$}
                       			\State {$val \gets 0 $} \Comment{ pixel is fully outside of the aperture}
              
				\Else
                       				\State {$val \gets radius + 0.5 - distance $} \Comment{ pixel is partially within the aperture}

				\EndIf
			\EndIf
		\EndIf

   \Return $val$ 
    \EndProcedure
  \end{algorithmic}
\end{algorithm}

\subsection {Node Control}

\par The pipeline is controlled via the ACN-Control script run on the Remote Control command console system which must be on the same logical network as the DIT. From this command console a number of functions are supported which control the running of the pipeline. These controls are primarily for experimental execution.

\par All ACN nodes be registered in a nodes file allowing experiments to be run with different numbers of nodes. A nodes file contains the name, IP address and a storage location for data retrieval is specified.  The following commands are supported and executable from the ACN Pipeline Controller via an SSH script which remotely connects to each node. A brief description of each command is provided under each component of the pipeline. 

\begin {itemize}
\item \emph{Activate-ACN -r NODEFILENAME} This command will activate all ACN Nodes in the nodes file so they start looking to a queue for image files to process. The ACN nodes will wait for a queue to be created, but this option provides better control for experimentation so it is clear which nodes are active. Once a node is running, it checks the queue for unlocked files and locks them as already described. Once locked, the full image file is downloaded from an S3 bucket and cleaned using the \emph{acn-aphot.c} program. This program runs once for each file downloaded.
\item \emph{Activate-ACN -c NODEFILENAME}   This command will reset all ACN Nodes in the nodes file so they stop running, remove all temporary data and download the latest utility files, configuration files and Master Bias/Master Flat images ready for the next round of processing.  
\item \emph{Activate-ACN -p NODEFILENAME}  Ping all of the nodes to ensure that they are accessible to the pipeline. 
\item \emph{Activate-ACN -x NODEFILENAME}  Reboot all of the nodes to ensure that they have flushed all caches. If a node processes the same data multiple times, it may operate faster on the subsequent executions due to caching of data in memory.  
\item \emph{ACN-Control -q compressed | standard | clipped} This command will create a list of empty files in a directory, which can be used as a simple queue. Files are named with a prefix source and traverse a dataset creating an entry in the queue for all files found. A compressed source file of 0000001.fits.fz has a corresponding queue entry of Queued-0000001.fits.fz. When successfully locked for processing by an ACN-Node this is changed to LOCKED-0000001.fits.fz. The NFS file system ensures only one lock can be obtained.
\item \emph{fits-compress -p DATADIRECTORY} This script has the option of performing compression in parallel or in sequence. The parallel execution spawns off processes and requires a machine with good RAM and processing capabilities. A comparison of performance for this script running in both modes is given in this chapter. 

\item \emph{s3-upload -p DATADIRECTORY} This script has the option of performing upload in parallel or in sequence. The parallel execution spawns off processes and requires a machine with sufficient RAM and processing capabilities. A comparison of performance for this script running in both modes is given in this chapter. Compressed or uncompressed data (depending on what is in the data directory) is uploaded to an S3 bucket.  A comparison of performance for this script running in both modes is given in this chapter. 
\end {itemize}
\definecolor{dkgreen}{rgb}{0,0.6,0}
\definecolor{gray}{rgb}{0.5,0.5,0.5}
\definecolor{mauve}{rgb}{0.58,0,0.82}

\lstset{
language=bash,
basicstyle=\small\sffamily,
  breaklines=true,
  breakatwhitespace=true,
numbers=left,
  commentstyle=\color{dkgreen},
numberstyle=\tiny\color{gray},
frame=tb,
 keywordstyle=\color{blue},
  identifierstyle=\bfseries\color{black},
columns=fullflexible,xleftmargin=5.0ex,
showstringspaces=false
}

\section{Experimental Methodology}

\par To ensure that all experiments are executed in a consistent manner, and that experiments were repeatable, a strict process of experimental setup and execution was followed. Where possible issues of caching were eliminated along with network contention issues. For example all experiments were run late at night because of the requirement to use the Institute network to route all traffic to the internet.  

\par The network setup remained the same for all experiments. The storage servers used the same disk types, ran on the same hardware and used a central FreeNAS server for NFS locking.  Each of the worker instances ran the same version of Ubuntu 12.04, and were fully cleaned before each experiment ran. The central control script logged into each of the workers before an experiment via SSH and removed all of the files in the user directory, then installed a new version of the software taken from the central NFS server. This ensured that all worker instances ran the same version of software. After each server was updated they were rebooted. 

\par The queue was also constructed using a central script which created a list of empty files after first archiving previous versions if they exist. The order of files created was consistent in each case, as was the length of the queue. 

\par The queue refers to files which have been uploaded to the S3 website prior to the experiment and the scripts which take the files from the NFS queue will then construct a URL which be used to access the file on the AWS service. 

\par Once the queue has been created, the workers are reinstalled and rebooted and the files uploaded to S3, so the experiment is initiated. A central script forks off a process to remotely connect to each of the worker instances and starts the cleaning script which obtains NFS locks, downloads files, cleans them and uploads results to S3. All workers however remain in standby mode until a specific file is detected in the central NFS server. When that file is detected by the worker it begins the processing cycle. This process ensures that all workers are running before the experiment begins.  Each worker contains a  timer which is written to a central directory on the NFS server when they have completed all available work. Each worker records their name, the time spent cleaning, the number of files cleaned and the average cleaning rate achieved. 

\par Each worker instance only runs a single process for downloading, processing and uploading results. 

\par Multiple experiments were run to determine that the procedures were functional, then formal experiment ran using the same experimental software to ensure they could be compared. Each experiment focused on an increase in the number of servers processing data to record the incremental performance of each worker. 

\par The experimental run script used for the ACN experiments is shown in Figure \ref{flowchartrunexpacn}.

\tikzstyle{block1} = [rectangle, draw,fill=white,text centered, text width=7em, minimum height=2em]
\tikzstyle{disk} =   [cylinder, draw,fill=white, text width=3em, text centered,shape border rotate=90, shape aspect=0.5, inner sep=0.3333em,  minimum width=2cm, minimum height=3em]
\tikzstyle {data} = [draw,trapezium,trapezium left angle=70,trapezium right angle=-70,node distance=2.5cm,minimum height=2em]

\begin{figure} [htbp]
 \begin{center}
\fbox { 

    \begin{tikzpicture}

   \node[cloud] (a)                                              {\footnotesize Start};
   \node[block1] (b)  [below of=a,node distance=1.2cm]                       {\footnotesize Create Queue};
   \node[block1] (c)  [below of=b,node distance=1.2cm]                       {\footnotesize Read Node File};
   \node[block1] (d)  [below of=c,node distance=1.2cm]                       {\footnotesize Ping all Nodes  };
   
   \node[block1] (h)  [below of=d,node distance=1.5cm]                       {\footnotesize Clean Worker 2} ;
   \node[block1] (f)  [left of=h, node distance =4cm ]                         {\footnotesize Clean Worker 1 };
   \node[block1] (g)  [right of=h, node distance=4cm]                       {\footnotesize Clean Worker N};

   \node[block1] (i)  [below of=h,node distance=1.5cm]                       {\footnotesize Reboot Worker 2} ;
   \node[block1] (j)  [left of=i, node distance =4cm ]                         {\footnotesize Reboot Worker 1 };
   \node[block1] (k)  [right of=i, node distance=4cm]                       {\footnotesize Reboot Worker N};

   \node[block1] (l)  [below of=i,node distance=1.5cm]                       {\footnotesize Start Worker 2} ;
   \node[block1] (m)  [left of=l, node distance =4cm ]                         {\footnotesize Start Worker 1 };
   \node[block1] (n)  [right of=l, node distance=4cm]                       {\footnotesize Start Worker N};

   \node[block1] (o)  [below of=l,node distance=1.5cm]                       {\footnotesize Check for GO Cmd} ;
   \node[block1] (p)  [left of=o, node distance =4cm ]                         {\footnotesize Check for GO Cmd };
   \node[block1] (q)  [right of=o, node distance=4cm]                       {\footnotesize Check for GO Cmd};

   \node[block1] (r)  [below of=o,node distance=1.5cm]                       {\footnotesize Clean Files} ;
   \node[block1] (s)  [left of=r, node distance =4cm ]                         {\footnotesize Clean Files };
   \node[block1] (t)  [right of=r, node distance=4cm]                       {\footnotesize Clean Files};

 \node[cloud] (u) [below of=r,node distance=1.5cm]                                             {\footnotesize Stop};
 \node[cloud] (v) [below of=s,node distance=1.5cm]                                             {\footnotesize Stop};
 \node[cloud] (w) [below of=t,node distance=1.5cm]                                             {\footnotesize Stop};


   \draw[->, thick] (a) -- (b);
   \draw[->, thick] (b) -- (c);
   \draw[->, thick] (c) -- (d);
   \draw[->, thick] (d) -- (h);
   \draw[->, thick] (h) -- (i);
   \draw[->, thick] (i) -- (l);
   \draw[->, thick] (l) -- (o);
   \draw[->, thick] (o) -- (r);
   \draw[->, thick] (m) -- (p);
   \draw[->, thick] (p) -- (s);
   \draw[->, thick] (q) -- (t);
   \draw[->, thick] (s) -- (v);
   \draw[->, thick] (t) -- (w);

   \draw[->, thick] (n) -- (q);
   
      \draw[->, thick] (r) -- (u);

   \draw[->] (d.south)  -- +(0mm,-2mm)  -| (f.north)  ; 
   \draw[->] (d.south)  -- +(0mm, -2mm)  -| (g.north)  ; 
 
   \draw[->] (h.south)  -- +(0mm,-2mm)  -| (j.north)  ; 
   \draw[->] (h.south)  -- +(0mm, -2mm)  -| (k.north)  ; 

   \draw[->] (i.south)  -- +(0mm,-2mm)  -| (m.north)  ; 
   \draw[->] (i.south)  -- +(0mm,-2mm)  -| (n.north)  ;

  \end{tikzpicture} }
  \caption{ACN experimental run script flow chart.}
  \label{flowchartrunexpacn}
 \end{center}
\end{figure}
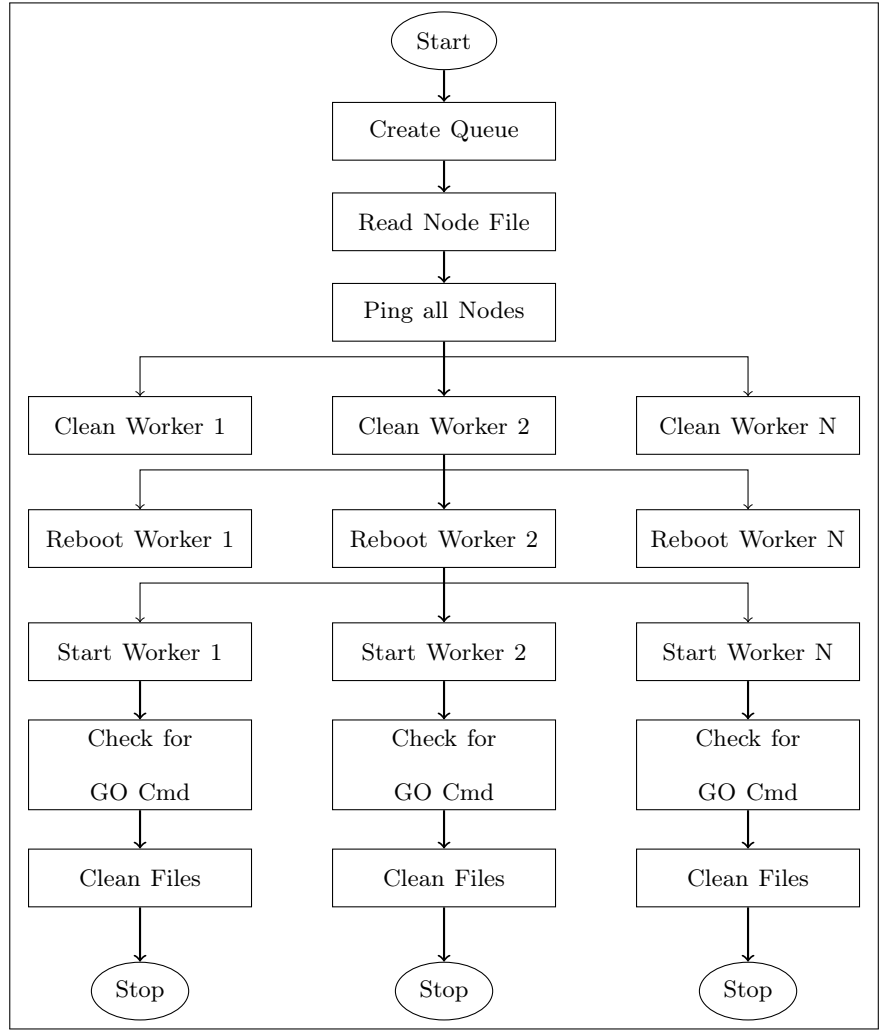

\section{Results and Discussion}

\par In this section, the performance of the ACN pipeline is reviewed seeking insights into the system design which can be later used within the NIMBUS pipeline. Each of the nodes central to the pipeline are performance tested.  Table \ref {tab:acn-experiements-summary} provide a list of the primary experiments run.  Each experiment is discussed and the results presented. A summary and conclusion for these experiments is provided at the end of the chapter.  All data sources for all experiments and graphs are identified in Appendix Table \ref{tab:datasets} which references an accompanying supplementary USB disk which contains raw and processed data relevant to these experiments. 

  \begin{table}[H]
\centering
\begin{tabular}{p{2cm} p{3.2cm} p{8.5cm}}
  \toprule
Reference & Measure & Description \\
  \midrule
\textbf{\small{Exp:ACN1}}& \small{ACN-APHOT Performance} & \small {Determining the performance of this program by running in multiple modes using various storage devices. Two step versus one step cleaning is examined. }\\
\textbf{\small{Exp:ACN2}}& \small{Storage Performance} & \small {Determining the Impact of the location of the storage devices and their ability to support multiple queries. }\\
\textbf{\small{Exp:ACN3}}& \small{Data Compression} & \small {Compression of data reduces the size of data for both storage and transfer. Data compression techniques and approaches are considered.   }\\
\textbf{\small{Exp:ACN4}}& \small{Data Transfer} & \small {Data stores are compared in terms of data transfer rates. }\\
\textbf{\small{Exp:ACN5}}& \small{Pipeline Limits} & \small {Determining how fast the pipeline can operate within the propose architecture.}\\

\hline
  \bottomrule
\end{tabular}
  \caption{ACN performance experimental set.}
\label{tab:acn-experiements-summary}
\end{table}

\subsection{ACN1: ACN-APHOT Performance}

\par The purpose of this experiment is to look at how long it take for the BCO dataset to be fully calibrated with magnitude values generated for all objects within all images using different hardware, different approaches to cleaning and different file storage systems. The two programs used are  \emph{rrf.c} and \emph{acn-aphot.c}, and the IBMe326 and a Macbook Pro\footnote{MacBook Pro 2.4 Ghz Intel Core i5 and 8 GB RAM,  IBM eServer 326 Operon 2.8Ghz and 4 GB RAM} were the hardware platforms. The dataset was either stored on the local disk of the hardware, or mounted as an NFS drive over the network. Table \ref{tab:acn-experiements1} shows the results of this experiment with the processing rate given as file per second. To compare with later experiments we convert the processing rate to gigabytes per second\footnote{Data was not compressed so with each file = 7.297920 MB,  GB/s =$ rate * 7.29792/1024$}. The \emph{Step 1} time is the time it took for all raw image files to be pixel calibrated. In each case, all of the pixels in all of the images were calibrated. When \emph{Step 1} has a time, the \emph{Step 2} time refers to the time taken to read in calibrated images and to calculate the magnitude values. When \emph{Step 1} has an N/A then pixels in \emph{Step 2} were calibrated using the previously discussed just in time process.

\begin{table}[htbp]
\centering
\begin{tabular}{p{1cm}  p{1cm}p{0.8cm}p{0.8cm}p{1.5cm}p{1.2cm}p{1.2cm}p{1cm}p{1.2cm}p{1cm}  }
  \toprule
Ref & Data Store & ACNs & Steps & Hardware & Step 1 & Step 2  & Rate & Total &GB/s \\
  \midrule
\small{P1-1}& \small{Local} & \small {1} & \small {2} & \small {Macbook} & \small {11:21:45}& \small {00:54:48}& \small {0.083}& \small {12:16:33} & \small {0.0006 } \\
\small{P1-2}& \small{Local} & \small {1} & \small {1} & \small {Macbook} & \small {N/A}& \small {00:55:53}& \small {1.098}& \small {00:55:53} & \small {0.0078 } \\
\small{P1-3}& \small{Local} & \small {1} & \small {2} & \small {IBMe326} & \small {00:49:00}& \small {01:15:23}& \small {0.493}& \small {02:04:23} & \small {0.0035 } \\
\small{P1-4}& \small{Local} & \small {1} & \small {1} & \small {IBMe326} & \small {N/A}& \small {01:22:30}& \small {0.744}& \small {01:22:30} & \small {0.0053 } \\
\small{P1-5}& \small{NFS}  & \small {1} & \small {2} & \small {IBMe326} & \small {00:49:53}& \small {01:15:08}& \small {0.490}& \small {02:05:01}& \small {0.0035 }  \\
\small{P1-6}& \small{NFS} & \small {1} & \small {1} & \small {IBMe326} & \small {N/A}& \small {01:17:25}& \small {0.793}& \small {01:17:25} & \small {0.0057 } \\
\small{P1-7}& \small{NFS} & \small {2} & \small {1} & \small {IBMe326} & \small {N/A}& \small {01:07:43}& \small {0.906}& \small {01:07:43} & \small {0.0065 } \\
\small{P1-8}& \small{NFS} & \small {4} & \small {1} & \small {IBMe326} & \small {N/A}& \small {00:43:15}& \small {1.413}& \small {00:43:26} & \small {0.0101 } \\

\hline
  \bottomrule
\end{tabular}
  \caption{P1: Calibrating the processing time for full BCO Dataset using one or two pass cleaning.}
\label{tab:acn-experiements1}
\end{table}

\par
\begin{figure} [htbp]
\centering
\fbox{     

\begin{tikzpicture}
\begin{axis}[
ylabel=Seconds,
enlarge x limits=0.2,
legend style={
    at={(0.5,-0.15)},
    anchor=north,legend columns=-1 },
ymin=0,
ybar,
xtick=data,
symbolic x coords={eServer326 SCSI, eServer326 NFS, MacbookPro SATA},
grid=major, 
bar width=25pt,
width=12cm,
xmajorgrids=false,nodes near coords,
every node near coords/.append style={anchor=mid west,rotate=70}
]

\addplot coordinates {(eServer326 SCSI,4950) (eServer326 NFS,4645)  (MacbookPro SATA,3353) };
\addplot coordinates {(eServer326 SCSI,7463) (eServer326 NFS,7501)  (MacbookPro SATA,) };

\legend{1 Pass,2 Pass}
\end{axis}
\end{tikzpicture}
    
    }
    \caption{ACN1: One step versus two step cleaning using different storage media on full dataset. }
  \label{fig:passes}

\end{figure}
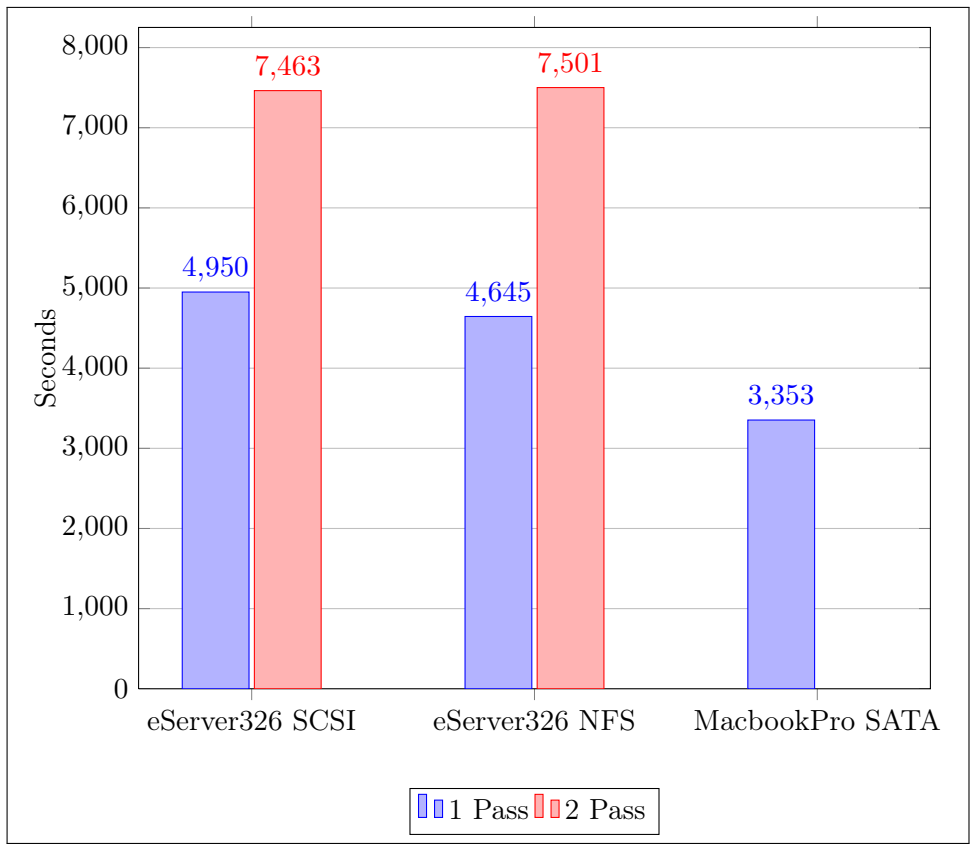

 With the exception of P1-1, the experiment results are reasonably consistent and clear. The time required to process images using two steps is almost double the time taken to use a single step as shown in Figure \ref{fig:passes}. This can be attributed to the number of file input and output events and using the CFITSIO library and the disk seek time. The amount of data read is only half of a 2 step approach, and the write operations are also dramatically reduced given the fact that result files are now measured in kilobytes rather than in megabytes. Raw data for this graph can be found in the Appendix \ref{app:chapter4},  Table \ref{tab:singlenode}.

\par When \emph{Step 1} is not used, the processing performed in \emph{Step 2} is slightly higher than in \emph{Step 1} but the processing times remain very similar. \emph{Step 2}  holds the Master Bias and the Master Flat file in memory and as each pixel value is used, it is first calibrated.  The output result files from these two approaches were compared and the results were identical.  The disk storage had some effect with the NFS disks operating slightly faster. This may have been due to the fact that they were faster disks, but overall the storage location for these experiments was not significant. 

\par In addition to testing the two step versus one step cleaning, the impact of adding additional nodes was introduced. In P1-7 the two nodes accessed NFS storage, with each having access to exactly half of the dataset. In P1-8 there were four nodes running each having access to a quarter of the data. No queuing system was used, just a copy of the data made available in a private location for each node. The processing times were reduced but not linearly with the number of nodes. The NFS storage used contains a single disk and is most likely the bottleneck in processing.  The data transfer rates as shown are small and the network was unlikely to be the issue. The next experiment, \emph{ACN2: Disk Testing} explored this further.

\subsection{ACN2: Disk Storage Testing} 

\par This experiment builds on the observations from experiment P1-4, where the file processing rate was 0.744 files for a single node.  The aim was to see how additional nodes could run in parallel and still retain this processing rate by specifically looking at the disks used for serving data. In these experiments the primary difference is the location of the data during processing. The data was split into eight equal parts and copied to the local storage of each of the 8 nodes. Three experiments were run.

\begin{itemize}
\item P2-1: Eight nodes processing $\frac{1}{8}th$ of the data from a local disk 
\item P2-2: Eight nodes processing $\frac{1}{8}th$ of the data from 3 NFS mounted disks 
\item P2-3: Comparing the NFS Storage Systems 
\end{itemize}

The results of P2-1 are given in Table \ref{tab:acn-experiements2} showing a consistent processing rate for each of the nodes, although the rate is less than seen in P1-4. In these experiment the overall data processing time for the full dataset is the time at which the last node stopped processing. The overall processing rate of the system is therefore $3682 files/928seconds = 3.968$ files per second, or $0.028 GB/s$. 

\begin{table}[htbp]
\centering
\begin{tabular}{p{1.5cm}  p{1.5cm}p{1.5cm}p{1.5cm}p{1.5cm}  p{1.5cm}}
  \toprule
Ref & Data Store & Node Number & Hardware & Rate & Time \\
  \midrule
\small{P2-1-1}& \small{Local} & \small {1}  & \small {IBMe326} & \small {0.522} & \small {00:14:41} \\
\small{P2-1-2}& \small{Local} & \small {2}  & \small {IBMe326} & \small {0.512} & \small {00:14:59} \\
\small{P2-1-3}& \small{Local} & \small {3}  & \small {IBMe326} & \small {0.514} & \small {00:14:56} \\
\small{P2-1-4}& \small{Local} & \small {4}  & \small {IBMe326} & \small {0.500} & \small {00:15:28} \\
\small{P2-1-5}& \small{Local} & \small {5}  & \small {IBMe326} & \small {0.499} & \small {00:15:23} \\
\small{P2-1-6}& \small{Local} & \small {6}  & \small {IBMe326} & \small {0.496} & \small {00:15:28} \\
\small{P2-1-7}& \small{Local} & \small {7}  & \small {IBMe326} & \small {0.497} & \small {00:15:27} \\
\small{P2-1-8}& \small{Local} & \small {8}  & \small {IBMe326} & \small {0.504} & \small {00:15:14} \\

\hline
  \bottomrule
\end{tabular}
  \caption{P2-1: File processing rates on local disks using 8 ACN Nodes.}
\label{tab:acn-experiements2}
\end{table}

\par Because the NFS processing times were faster than the local disks, P2-2 was run using 3 different NFS storage servers on the local network to see if further improvements were possible. Since each of the nodes are identical, the differences observed in Table \ref{tab:acn-experiement3} are either related to the NFS system or the network. Additional testing was then performed using P2-3 which repeated eight  single node tests against each of the NFS stores with no two tests running at the same time.  Using a smaller dataset of only $\frac{1}{24}$ of the total, each of these experiments ran against each NFS storage device. The results for P2-3 are given in Table \ref{tab:acn-experiement4}  and clearly show a consistent processing rate depending on the storage node used, with NFS1 providing the best rate. The reduction in processing rates as the number of nodes increased, and the variability of the nodes resulted in a move to the S3 storage system which offered more consistent download times  

\begin{table}[htbp]
\centering
\begin{tabular}{p{1.5cm}  p{1.5cm}p{1.5cm}p{1.5cm}p{1.5cm}  p{1.5cm}}
  \toprule
Ref & Data Store & Node Number & Hardware & Rate & Time \\
  \midrule
\small{P2-2-1}& \small{NFS1} & \small {1}  & \small {IBMe326} & \small {0.522} & \small {00:08:03} \\
\small{P2-2-2}& \small{NFS1} & \small {2}  & \small {IBMe326} & \small {0.512} & \small {00:07:54} \\
\small{P2-2-3}& \small{NFS1} & \small {3}  & \small {IBMe326} & \small {0.514} & \small {00:07:30} \\
\small{P2-2-4}& \small{NFS2} & \small {4}  & \small {IBMe326} & \small {0.500} & \small {00:12:43} \\
\small{P2-2-5}& \small{NFS2} & \small {5}  & \small {IBMe326} & \small {0.499} & \small {00:12:39} \\
\small{P2-2-6}& \small{NFS3} & \small {6}  & \small {IBMe326} & \small {0.496} & \small {00:11:14} \\
\small{P2-2-7}& \small{NFS3} & \small {7}  & \small {IBMe326} & \small {0.497} & \small {00:11:07} \\
\small{P2-2-8}& \small{NFS3} & \small {8}  & \small {IBMe326} & \small {0.504} & \small {00:11:00} \\

\hline
  \bottomrule
\end{tabular}
  \caption{P2-2: File processing rates using 3 different NFS servers and 8 ACN Nodes}
\label{tab:acn-experiement3}
\end{table}

\begin{table}[htbp]
\centering
\begin{tabular}{p{1.5cm}  p{1.5cm}p{1.5cm}p{1.5cm}p{1.5cm}  p{1.5cm}}
  \toprule
Ref & Data Store & Node Number & Hardware & Rate & Time \\
  \midrule
\small{P2-3-1}& \small{NFS1} & \small {1}  & \small {IBMe326} & \small {0.546} & \small {00:04:41} \\
\small{P2-3-2}& \small{NFS1} & \small {1}  & \small {IBMe326} & \small {0.562} & \small {00:04:33} \\
\small{P2-3-3}& \small{NFS1} & \small {1}  & \small {IBMe326} & \small {0.542} & \small {00:04:43} \\
\small{P2-3-4}& \small{NFS1} & \small {1}  & \small {IBMe326} & \small {0.544} & \small {00:04:42} \\
\small{P2-3-5}& \small{NFS1} & \small {1}  & \small {IBMe326} & \small {0.540} & \small {00:04:44} \\
\small{P2-3-6}& \small{NFS1} & \small {1}  & \small {IBMe326} & \small {0.542} & \small {00:04:43} \\
\small{P2-3-7}& \small{NFS1} & \small {1}  & \small {IBMe326} & \small {0.540} & \small {00:04:44} \\
\small{P2-3-8}& \small{NFS1} & \small {1}  & \small {IBMe326} & \small {0.546} & \small {00:04:41} \\
\hline
\small{P2-3-9}& \small{NFS2} & \small {1}  & \small {IBMe326} & \small {0.248} & \small {00:10:20} \\
\small{P2-3-10}& \small{NFS2} & \small {1}  & \small {IBMe326} & \small {0.195} & \small {00:13:07} \\
\small{P2-3-11}& \small{NFS2} & \small {1}  & \small {IBMe326} & \small {0.214} & \small {00:11:56} \\
\small{P2-3-12}& \small{NFS2} & \small {1}  & \small {IBMe326} & \small {0.222} & \small {00:10:30} \\
\small{P2-3-13}& \small{NFS2} & \small {1}  & \small {IBMe326} & \small {0.223} & \small {00:10:29} \\
\small{P2-3-14}& \small{NFS2} & \small {1}  & \small {IBMe326} & \small {0.194} & \small {00:13:11} \\
\small{P2-3-15}& \small{NFS2} & \small {1}  & \small {IBMe326} & \small {0.243} & \small {00:10:31} \\
\small{P2-3-16}& \small{NFS2} & \small {1}  & \small {IBMe326} & \small {0.245} & \small {00:10:27} \\
\hline
\small{P2-3-17}& \small{NFS3} & \small {1}  & \small {IBMe326} & \small {0.376} & \small {00:06:48} \\
\small{P2-3-18}& \small{NFS3} & \small {1}  & \small {IBMe326} & \small {0.384} & \small {00:06:48} \\
\small{P2-3-19}& \small{NFS3} & \small {1}  & \small {IBMe326} & \small {0.372} & \small {00:06:48} \\
\small{P2-3-20}& \small{NFS3} & \small {1}  & \small {IBMe326} & \small {0.369} & \small {00:06:48} \\
\small{P2-3-21}& \small{NFS3} & \small {1}  & \small {IBMe326} & \small {0.369} & \small {00:06:48} \\
\small{P2-3-22}& \small{NFS3} & \small {1}  & \small {IBMe326} & \small {0.367} & \small {00:06:48} \\
\small{P2-3-23}& \small{NFS3} & \small {1}  & \small {IBMe326} & \small {0.369} & \small {00:06:48} \\
\small{P2-3-24}& \small{NFS3} & \small {1}  & \small {IBMe326} & \small {0.373} & \small {00:06:48} \\

\hline
  \bottomrule
\end{tabular}
  \caption{P2-3: Comparing File Processing rates against each of the NFS storage systems}
\label{tab:acn-experiement4}
\end{table}

\subsection{ACN3: Data Compression}

\par
\begin{figure} [htbp]
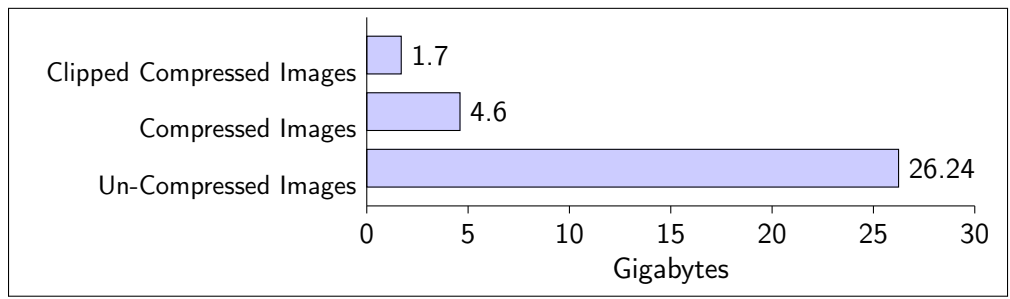

\centering
\fbox{      \begin{bchart}[step=5,max=30]
        \bcbar{1.7}
        \bclabel{\small{Clipped Compressed Images }}
         \smallskip
        \bcbar{4.6}
         \bclabel{\small{Compressed Images }}
         \smallskip
        \bcbar{26.24}
        \bclabel{\small{Un-Compressed Images }}
	\bcxlabel{Gigabytes}
    \end{bchart}}
    \caption{ACN3: BCO Dataset sizes in Gigabytes}
  \label{fig:BCOclipped}

\end{figure}

 The compression used on the BCO data was the CFITSIO  \emph{fpack} utility. The raw dataset is 26.4GB, but when compressed this is reduced to 4.6GB. If images are first clipped to only include the regions around the objects for which magnitude values are required, then the dataset is further reduced to 1.7 GB as shown in Figure \ref{fig:BCOclipped}. By compressing the data using a powerful server, transfer times should be shorter. The \emph{fpack} utility in this case reduced each image by over 80\% making a significant difference to the size of the total dataset. 

\par
\begin{figure} [htbp]
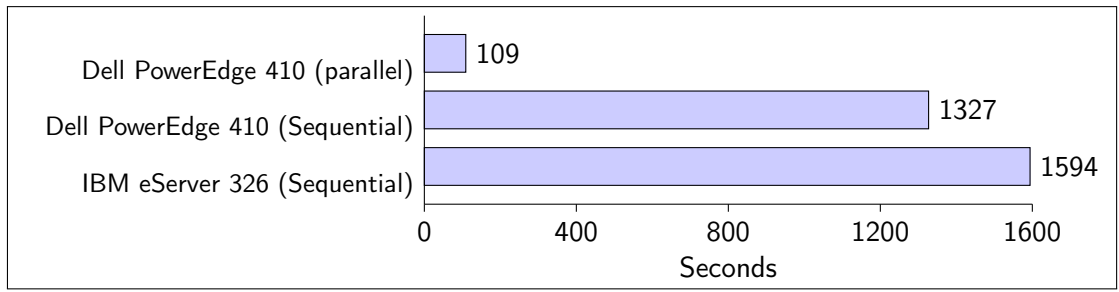

\centering
\fbox{      \begin{bchart}[step=400,max=1600]
        \bcbar{109}
        \bclabel{\small{Dell PowerEdge 410 (parallel)}}

            \smallskip
        \bcbar{1327}
                \bclabel{\small{Dell PowerEdge 410 (Sequential) }}

         \smallskip

        \bcbar{1594}
                        \bclabel{\small{IBM eServer 326 (Sequential) }}

\bcxlabel{Seconds}

    \end{bchart}}
    \caption{ACN3: Comparison of Data Compression times using different modes and hardware }
  \label{fig:compressiontime}

\end{figure}

 To determine the optimal performance of the Dell 410 server, FITS lossless compression was run using two different methods. The first compressed files in sequential order where a file-list was given to a single instance of the utility. The second approach started thousands of processes at staggered intervals where each process was running concurrently utilising large portions of the available memory on the server. This approach is valid for a server with sufficiently large quantities of memory and  CPU cores. If there are not enough resources then the system begins to thrash. The compression time reading and writing files on the NFS server mount point are shown in Figure \ref{fig:compressiontime}. The parallel performance of the Dell 410 was instrumental in the pipeline providing a compression time considerably faster than the sequential method. It was interesting to note that the older IBM systems provided comparable compression times to the Dell when run in sequential mode.

\subsection{ACN4: Data Transfer}

\par
\begin{figure} [htbp]
\centering
\fbox{     
\begin{tikzpicture}
\begin{axis}[
ylabel=Seconds,
enlarge x limits=0.2,
legend style={
    at={(0.5,-0.15)},
    anchor=north,legend columns=-1 },
ymin=0,
ybar,
xtick=data, bar width=20pt,
symbolic x coords={{Uncompressed},{Compressed},{Clipped}}, 
grid=major,  width=10cm,  
xmajorgrids=false,nodes near coords, every node near coords/.append style={anchor=mid west,rotate=70}
]

\addplot coordinates {(Uncompressed,246)   (Compressed,138) (Clipped,186)};
\addplot coordinates {({Uncompressed},633)({Compressed},114)({Clipped},42)};
\addplot coordinates {(Uncompressed,1050) (Compressed, 200) (Clipped,148)};

\legend{Local to NFS,NFS to Local, Local to S3}
\end{axis}
\end{tikzpicture}
}
    \caption{ACN4: Comparison of data transfer times between storage types. }
  \label{fig:copyingtime}

\end{figure}
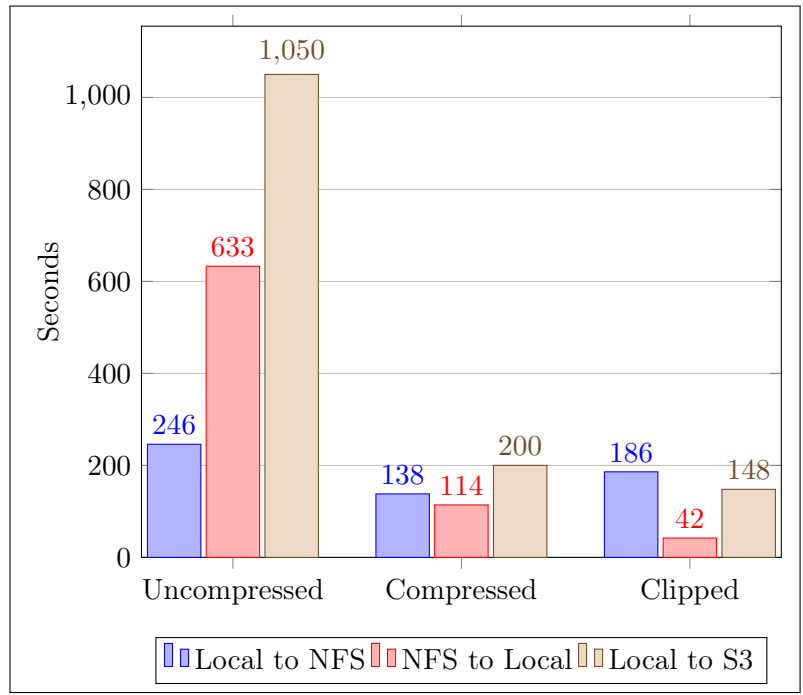

 The transfer times of the datasets between different types of storage is shown in Figure \ref{fig:copyingtime}. The private network was connected via a gigabit switch to the DIT network, which has an external institute-wide Internet connection of 1 Gigabits per second. Approximately 20\% of the bandwidth was used in this transfer although the use decreased to approximately 8\% as the number of files was increased. The nature of the network is that it is variable, however the experiments were run late at night when the Institute's network was lightly loaded.   The clipped and compressed data files were files where the FITS files were clipped into the smallest possible region around the star so pixels, which were not going to be cleaned, were not transported. The time saved in data transfer however was negligible. 

\subsection {ACN5: Pipeline Limits}

\par
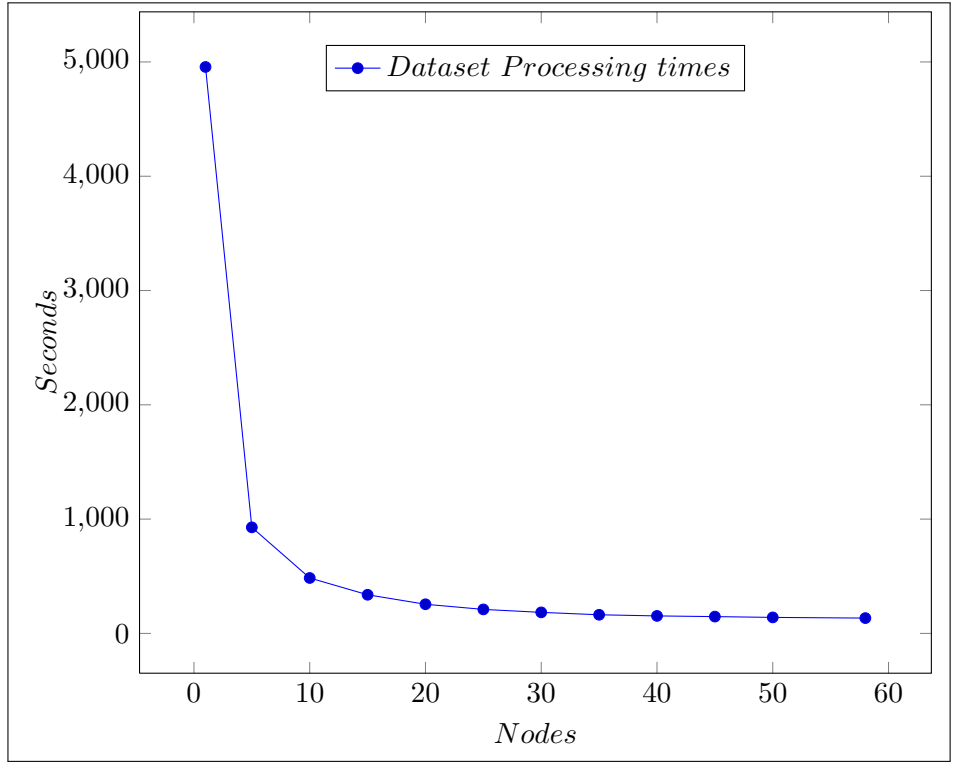
\begin{figure} [htbp]
\centering
\fbox {
\begin{tikzpicture}
    \begin{axis}[width=0.8\textwidth,
    legend style={at={(0.5,0.95)},anchor=north,legend cell align=left},
        xlabel=$Nodes$,
        ylabel=$Seconds$]
    \addplot plot coordinates {
    (1,4956) (5,928)  (10,485) (15,338)  (20,255)    (25,210)  (30,184) (35,163) (40,153) (45,147) (50,140) (58,134) };
       \legend{$Dataset\ Processing\ times$\\}   
    \end{axis}

\end{tikzpicture} }
  \vskip -0.8em
    \caption{ACN5: Data cleaning for increasing numbers of ACN nodes in seconds. Includes compression time of 109 seconds and upload time of 200 seconds }
  \label{fig:sizes}
\end{figure}

 Once all data was compressed and uploaded to the S3 storage system, the central queue was activated to allow for as many nodes as possible to come online. It took approximately 20 seconds to create the 3682 empty files in the NFS queue. Experiments were run using the compressed dataset on S3 and using a one pass approach where data was cleaned only as pixels were used. In each experiment the full dataset needed to be fully processed for the experiment to finish. Experiments were run for increments of 5  ACN Nodes up to a maximum of 58 Nodes.  The overall processing time in seconds is given, and includes the data compression and data upload time to S3. The single node processing time is taken from experiment P1-4 which had a time of 1 hour, 22 minutes and 30 seconds. Figure \ref{fig:sizes} shows a graph of the time in seconds for the full dataset to be processed for varying numbers of ACN Nodes. Because of the way the compression and upload process was performed, it is the limiting factor on the pipeline. As more nodes are introduced the processing time moves closer to the compression and upload times. Table \ref{tab:acn-experiement5} provides full details of the times and file processing rates for each experiment run. While these experimental results have the compression and upload time costs added there is no specific dependancy on these steps completing before the next step can start. A file could be compressed, uploaded and processed while other files are going through the same process.

\begin{table}
\centering
\begin{tabular}{p{1.8cm} p{1cm}   p{2cm}p{1.5cm}p{1.5cm}p{1.5cm} p{1.3cm} p{1cm}}
  \toprule
Ref & Nodes & Compression & Upload & Processing & Total & File/s& GB/s \\
  \midrule
  
\textbf{   \small{P1-4}  }& \small{1}      & \small {00:01:49}  &  \small{00:03:20} & \small{01:22:36}  & \small{01:27:45} & \small{0.699} & \small{0.005}  \\
\textbf{    \small{ACN5-1} } &       \small{5}      & \small {00:01:49}  &  \small{00:03:20} & \small{00:15:28}  & \small{00:20:37} & \small{2.977} &  \small{0.021}  \\
\textbf{    \small{ACN5-2}  }&        \small{10}      & \small {00:01:49}  &  \small{00:03:20} & \small00:08:05  & \small{00:13:14} & \small{4.637} &  \small{0.033}  \\
\textbf{     \small{ACN5-3}  }&      \small{15}      & \small {00:01:49}  &  \small{00:03:20} & \small{00:05:38}  & \small{00:10:47} & \small{5.691} &  \small{0.041}  \\
\textbf{     \small{ACN5-4}  }&      \small{20}      & \small {00:01:49}  &  \small{00:03:20} & \small{00:04:15}  & \small{00:09:24} & \small{6.528} &  \small{0.047}  \\
\textbf{     \small{ACN5-5}  }&      \small{25}      & \small {00:01:49}  &  \small{00:03:20} & \small{00:03:30}  & \small{00:08:39} & \small{7.094} &  \small{0.051}  \\
\textbf{     \small{ACN5-6}  }&      \small{30}      & \small {00:01:49}  &  \small{00:03:20} & \small{00:03:04}  & \small{00:08:13} & \small{7.469} &  \small{0.053}  \\
\textbf{     \small{ACN5-7}  }&      \small{35}      & \small {00:01:49}  &  \small{00:03:20} & \small{00:02:43}  & \small{00:07:52} & \small{7.801} &  \small{0.056}  \\
\textbf{     \small{ACN5-8}  }&      \small{40}      & \small {00:01:49}  &  \small{00:03:20} & \small{00:02:33}  & \small{00:07:42} & \small{7.970} &  \small{0.057}  \\
\textbf{     \small{ACN5-9}  }&      \small{45}      & \small {00:01:49}  &  \small{00:03:20} & \small{00:02:27}  & \small{00:07:36} & \small{8.075} &  \small{0.058}  \\
\textbf{     \small{ACN5-10}}  &      \small{50}      & \small {00:01:49}  &  \small{00:03:20} & \small{00:02:20}  & \small{00:07:29} & \small{8.200} &  \small{0.058}  \\
\textbf{     \small{ACN5-11} } &      \small{58}      & \small {00:01:49}  &  \small{00:03:20} & \small{00:02:14}  & \small{00:07:23} & \small{8.312} & \small{0.059}  \\

\hline
  \bottomrule
\end{tabular}
  \caption{ACN5: Clean rates per node and GB/s processing rate}
\label{tab:acn-experiement5}
\end{table}

\par
\begin{figure} [htbp]
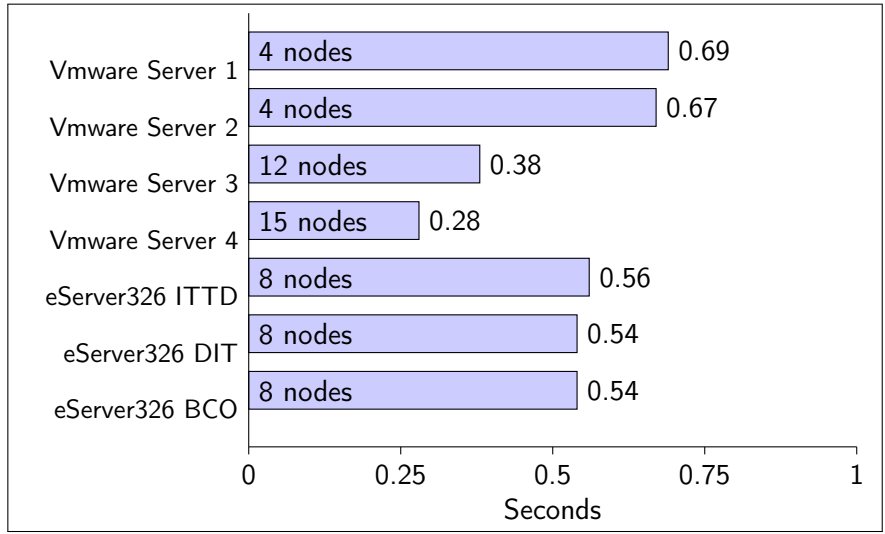

\centering
\fbox{      \begin{bchart}[step=0.25,max=1]
        \bcbar[text=4 nodes]{0.69}
        \bclabel{\small{Vmware Server 1}}
        \smallskip
 
         \bcbar[text=4 nodes]{0.67}
        \bclabel{\small{Vmware Server 2}}
        \smallskip

         \bcbar[text=12 nodes]{0.38}
        \bclabel{\small{Vmware Server 3}}
        \smallskip

          \bcbar[text=15 nodes]{0.28}
        \bclabel{\small{Vmware Server 4}}
        \smallskip

          \bcbar[text=8 nodes]{0.56}
        \bclabel{\small{eServer326 ITTD}}
        \smallskip

          \bcbar[text=8 nodes]{0.54}
        \bclabel{\small{eServer326 DIT}}
        \smallskip

          \bcbar[text=8 nodes]{0.54}
        \bclabel{\small{eServer326 BCO}}
        \smallskip

 \bcxlabel{Seconds}

    \end{bchart}}
    \caption{ACN5: Average file processing rate (files per second) per server type}
  \label{fig:cleaningtime}

\end{figure}

 Due to the mixture of server types used within the last experimental set it is worth taking a closer look at the breakdown of the nodes and their overall processing contribution to the pipeline. Specifically we will look at ACN5-11 which had 58 nodes running. In Figure \ref{fig:cleaningtime} the number of nodes for each server type and server location is shown. Each of the institutes had 8 IBM e326 servers which were all configured exactly the same and had the same hardware specification. The VMWare servers were all within the DIT and contained varying numbers of virtualised instances of the ACN Nodes. All nodes were started within a few seconds of each other by the ACN-Control system. The average file processing rate is also given for each of the servers.  The IBM servers are operating in line with previous experiments when running against an NFS server, but there is no degradation of the processing rate as the number of ACN nodes increased. Two other interesting observations can made by looking at Figures \ref{fig:finishtimes} and \ref{fig:finishtimes2}. Because of the introduction of the queue to the system, each of the nodes will clean as fast as it can then move on to the next file. Even when some of the nodes are considerably slower and cleaning less files as shown in Figure \ref{fig:finishtimes2}, all processing finished roughly at the same time.

\pgfplotstableread{Data/runningtime.dat}
\datatable

\begin{figure} [htbp]
\centering
\fbox {
\begin{tikzpicture}
    \begin{axis}[xmin=0,
    ymin=100,
    width=0.8\textwidth,height=6cm,
    legend style={at={(0.5,0.8)},anchor=north,legend cell align=left},
        xlabel=$Node\ Number$,
        ylabel=$Seconds$]

\addplot table[y = Seconds] from \datatable ;
\legend{$Node\ Running\ Time$\\}   

    \end{axis}

\end{tikzpicture} }
  \vskip -0.8em
    \caption{ACN5: Running time for each individual node.}
  \label{fig:finishtimes}
\end{figure}
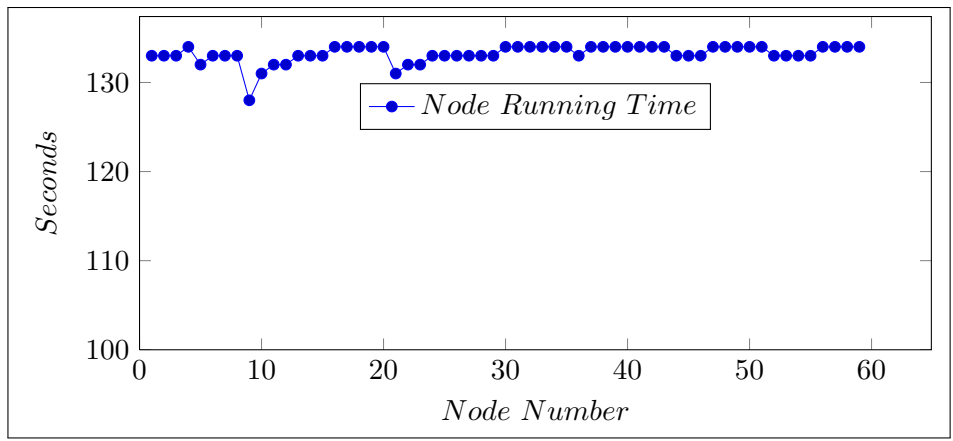

\pgfplotstableread{Data/filescleaned.dat}
\datatable

\begin{figure} [htbp]
\centering
\fbox {
\begin{tikzpicture}
    \begin{axis}[xmin=0,
    ymin=30,
    width=0.8\textwidth,height=6cm,
    legend style={at={(0.5,0.95)},anchor=north,legend cell align=left},
        xlabel=$Node\ Number$,
        ylabel=$Seconds$]

\addplot table[y = Cleaned] from \datatable ;
\legend{$Node\ Files\ Cleaned$\\}   

    \end{axis}

\end{tikzpicture} }
  \vskip -0.8em
    \caption{ACN5: Number of files cleaned by each individual node.}
  \label{fig:finishtimes2}
\end{figure}
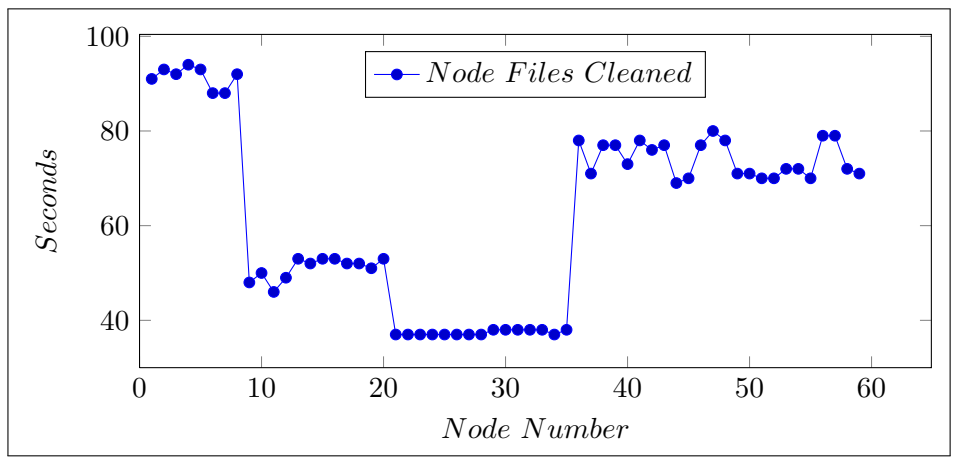

\section{Conclusion}

\par The ACN pipeline provides a strong argument for a storage device which can service requests at a consistent rate as the rate of requests increases. Experiments with local NFS devices proved to be inconsistent and required that multiple copies of the dataset had to be available in multiple locations to avoid network or read bottlenecks. The cost of compression and subsequent upload were not significant compared to the initial cleaning time of the data when a single node was used. By moving the dataset to a disturbed storage system, data replication was no longer required. 

\par The NFS based queue, while simple and effective, limited access to worker nodes requiring that they remain within the private network. A more public queue would allow ACN nodes to join in the pipeline from other locations. The requirement to use SSH to access these nodes for the purpose of control is also a limiting factor for further scaling. SSH ports are often restricted. There is also a cost for Amazon S3 storage and an alternative \gls{hdfs} implementation or high performance static web server could provide similar functionality. 

\par The pipeline does provide clear evidence that the parallelised processing approach does significantly reduce the processing time by allowing additional ACN Nodes to contributed. It has also been verified that this process does not impact the calculation of the instrument magnitude values and that the reduction in file read and write operations has also contributed to some of the performance enhancements observed. 

\par The systems used to control the experiments required that operations could be run in parallel and in sequence for the purpose of comparison. Data compression and uploading should be incorporated into the data production cycle so that there is a constant flow of data into the pipeline. The worker nodes should also be able to play a more proactive role in monitoring the queue and ensuring they contain the latest software. This pipeline used a push model, but a pull model would suit a distributed system. 

\par  Using the ACN pipeline it has been demonstrated that the processing rate for this dataset can be reduced from the initial MATLAB time of 0.1 files per second to 8.3 files per second when using 58 nodes. If the overhead cost of the compression and uploading are removed that cleaning rate could increase to 27.48 files per second however the rate of performance improvements as nodes were added fell dramatically for larger numbers of nodes. 

\par The lessons taken from this pipeline which are to be used within the NIMBUS system are as follows. 

\begin{itemize}
\item The queue needs to be globally accessible
\item The workers need to be more self configurable
\item Data stores need to be more distributed and globally available
\item The linear increase in computing resources did not produce a linear increase in performance, most likely due to resource sharing. 
\item One pass pixel cleaning and magnitude generation speeds up individual ACN Nodes by reducing the number of I/O operations. 
\item Compression is a net contributor to the system and should be continued. 
\end {itemize}

%% file: chapter6.tex
\chapter{The \emph{NIMBUS} Pipeline}
\label{chapter5}
\section{Overview}
\par This NIMBUS pipeline, uses a public web queue to publish work to distributed computing nodes build explicitly for this pipeline which are referred to as \emph{workers}. Workers are computing instances that can reside anywhere on the internet but are required to have internet access using port 80, as all services accessed are HTTP based. Each worker uses at its core the \emph{acn-aphot.c} program used in the ACN Pipeline which runs in single step mode. For this pipeline the BCO dataset is also used but it has been renamed and replicated so that there are multiple terabytes of data available for processing, not just 26GB. The assumption with this pipeline is that the data is already staged for processing, and the emphasis is on processing image data rather than compression and uploading image files. From the previous experiments, the compression and upload costs were already established. The architecture is shown in Figure \ref{fig:NIMBUSarch} and breaks into the following fundamental components which are described in more detail within this chapter. The code used for this pipeline is available on github in the following repository. \url{https://github.com/paulfdoyle/NIMBUS.git} 

Figure \ref{flowchartnimbus} gives the basic work flow of the pipeline.

\begin {itemize}
\item Data Archive Cloud. A distributed collection of web servers which provide data for the pipeline to process. Data has already been compressed and servers within this cloud can be both virtual and physical. The archive is required to advertise the files it contains via a distributed web message queue, and to service those files via HTTP requests.
\item	Distributed Worker Queue Cloud. The entire pipeline is centred  around a series of web message queues, and this is the central queue which contains work to be performed. The queue cloud is a series of communication protocols which allow the overall pipeline to orchestrate work in such a way that all workers operate at their peak performance by working as independent consumers of work to produce a series of result files. Data is advertised as a web message, and workers are controlled using command queues. The worker queue specifically contains the messages for workers to process. 
\item	Monitoring System. Due to the highly distributed nature of the pipeline, monitoring also needs to be distributed. Used primarily for experimental measurement monitoring is performed on queues and workers.  All workers record their progress in a distributed log queue and queue sizes and rates of change are monitored. The web servers and worker nodes contained monitoring software to review CPU and Network performance.  

\item Global Processing Cloud. All workers which contribute to the processing are required to have an initialisation script within their boot sequence, and are assumed to be a Linux based system. This script initiates worker processes on the server by downloading a customer package which  contains all instructions and tools for use by the workers. Workers can be globally located and are self contained processing units which are assumed to be transient resources.   
\item Results Cloud. When workers complete the processing for an image, the results are posted to a distributed storage service and a web message is constructed which provides a reference to the result. The reference can be used to reconstruct the sequence of the original images if required. 
\item	Control System. A central system was created to control experiments. Using an AWS API in Python, workers and queues were created and monitored. All experiments are started and shutdown using well defined procedures to ensure consistency across experiments. The control system initialises the pipeline and the queues,  instructs the storage nodes to advertise work and the processing nodes to start processing.  

\end{itemize}

		\tikzstyle {webserver} = [draw,rectangle,node distance=2.5cm,rounded corners=0.5ex,minimum height=2em]
\tikzstyle {queue} = [draw,trapezium,fill=white,trapezium left angle=70,trapezium right angle=-70,shape border rotate=180,node distance=2.5cm,minimum height=2em]

\begin{figure} [htbp]
 \begin{center}
\fbox { 

  \begin{tikzpicture}[node distance=2.5cm, auto, >=stealth,cross line/.style={preaction={draw=white,->,line width=4pt}}]

%
%
    \node[draw, cylinder, shape aspect=1.5, inner sep=0.3333em, fill=white, minimum width=1cm, minimum height=1cm] (cyl1) at (2,-1.25) {\footnotesize{Q}};
    \node[draw, cylinder, shape aspect=1.5, inner sep=0.3333em, fill=white, minimum width=1cm, minimum height=1cm] (cyl1) at (1.5,-2) {\footnotesize{Q}};

  \node[draw, cylinder, shape aspect=1.5, inner sep=0.3333em, fill=white, minimum width=1cm, minimum height=1cm] (cyl1) at (2.5,-1.75) {\footnotesize{Q}};
  \node[shape=cloud, cloud puffs=15.7, cloud ignores aspect, minimum width=4cm, minimum height=3.5cm, align=center, draw] (wrkq) at (2.25, -1.5) {};
  \draw  ($(wrkq.north)+(-0mm,-5mm)$)                     node                {\footnotesize {Distributed}};
  \draw  ($(wrkq.north)+(-0mm,-8mm)$)                     node                {\footnotesize {Worker Queue}};
 
%
%
    \node[shape=cloud, cloud puffs=15.7, cloud ignores aspect, minimum width=6cm, minimum height=3.5cm, align=center, draw] (ARCHIVECloud) at (6.25, 2) {};
    \draw  ($(ARCHIVECloud.north)+(-0mm,-7mm)$)                     node                {\footnotesize {Data Archive Cloud}};
  
    \node[draw, cylinder, shape aspect=1.5, inner sep=0.3333em, fill=white,
    rotate=90, minimum width=1cm, minimum height=0.05cm] (cyl2a) at (5.2,1.8) {};
    \node[draw, cylinder, shape aspect=1.5, inner sep=0.3333em, fill=white,
    rotate=90, minimum width=1cm, minimum height=0.05cm] (cyl3) at (5.2,2.2) {};  
    \draw  ($(cyl3.south)+(-5.5mm,-8mm)$)                     node                {\footnotesize {Web }};
    \draw  ($(cyl3.south)+(-5.5mm,-11mm)$)                     node                {\footnotesize {Server }};

      \node[shape=circle, fill=white,minimum width=.7cm, minimum height=.7cm, align=center, draw] (web) at (5.6, 1.8) {};

    \node[draw, cylinder, shape aspect=1.5, inner sep=0.3333em, fill=white,
    rotate=90, minimum width=1cm, minimum height=0.05cm] (cyl2a) at (7.2,1.8) {};
    \node[draw, cylinder, shape aspect=1.5, inner sep=0.3333em, fill=white,
    rotate=90, minimum width=1cm, minimum height=0.05cm] (cyl3) at (7.2,2.2) {};  
    \draw  ($(cyl3.south)+(-5.5mm,-8mm)$)                     node                {\footnotesize {Web }};
    \draw  ($(cyl3.south)+(-5.5mm,-11mm)$)                     node                {\footnotesize {Server }};

      \node[shape=circle, fill=white,minimum width=.7cm, minimum height=.7cm, align=center, draw] (web) at (7.6, 1.8) {};    
      
%
%
    \node[shape=circle, fill=white,minimum width=2.4cm,  align=center, draw] (controller) at (8, -1) {};    
    \node[shape=circle, fill=white,minimum width=2.3cm,  align=center, draw] (controller) at (8,-1) {};    
    \node[shape=circle, fill=white,minimum width=2.2cm,  align=center, draw] (controller) at (8, -1) {};    
  \draw  ($(controller)+(-0mm,-0mm)$)                     node                {\footnotesize {Control System}};

%
%
    \node[shape=cloud, cloud puffs=15.7, cloud ignores aspect, minimum width=6cm, minimum height=3.5cm, align=center, draw] (DCCloud) at (-2, 2) {};
    \draw  ($(DCCloud.north)+(-0mm,-7mm)$)                     node                {\footnotesize {Data Capture Cloud}};
    
     \node[shape=trapezium, minimum width=.7cm, minimum height=.7cm, align=center, draw,trapezium right angle=120,trapezium left angle=60] (CCD) at (-3.75, 1.8) {};
    \draw  ($(CCD.north)+(-0mm,-4mm)$)                     node                {\footnotesize {CCD}};

    \node[draw, cylinder, shape aspect=1.5, inner sep=0.3333em, fill=white,
    rotate=90, minimum width=1cm, minimum height=0.05cm] (cyl1) at (-2,1.4) {};
    \node[draw, cylinder, shape aspect=1.5, inner sep=0.3333em, fill=white,
    rotate=90, minimum width=1cm, minimum height=0.05cm] (cyl2-1) at (-2,1.8) {};
    \node[draw, cylinder, shape aspect=1.5, inner sep=0.3333em, fill=white,
    rotate=90, minimum width=1cm, minimum height=0.05cm] (cyl3) at (-2,2.2) {};  
  
    \draw  ($(cyl1.south)+(-5mm,-5mm)$)                     node                {\footnotesize {Data }};  
    \node[draw, cylinder, shape aspect=1.5, inner sep=0.3333em, fill=white,
    rotate=90, minimum width=1cm, minimum height=0.05cm] (cyl2a) at (-0.2,1.8) {};
    \node[draw, cylinder, shape aspect=1.5, inner sep=0.3333em, fill=white,
    rotate=90, minimum width=1cm, minimum height=0.05cm] (cyl3) at (-0.2,2.2) {};  
    \draw  ($(cyl3.south)+(-5.5mm,-8mm)$)                     node                {\footnotesize {Web }};
    \draw  ($(cyl3.south)+(-5.5mm,-11mm)$)                     node                {\footnotesize {Server }};

      \node[shape=circle, fill=white,minimum width=.7cm, minimum height=.7cm, align=center, draw] (web) at (0.2,1.8) {};    
%
 %
%
  \node[draw, cylinder, shape aspect=1.5, inner sep=0.3333em, fill=white, minimum width=1cm, minimum height=1cm] (cyla) at (-3.75,-2.8) {\footnotesize{Q}};
  \node[draw, cylinder, shape aspect=1.5, inner sep=0.3333em, fill=white, minimum width=1cm, minimum height=1cm] (cylb) at (-4.25,-3.55) {\footnotesize{Q}};
  \node[draw, cylinder, shape aspect=1.5, inner sep=0.3333em, fill=white, minimum width=1cm, minimum height=1cm] (cylc) at (-3.25,-3.3) {\footnotesize{Q}};
  \node[shape=circle, minimum width=4cm, minimum height=3.5cm, align=center, draw] (moncloud) at (-3.5, -3) {};
    \node[shape=circle, minimum width=3.8cm, minimum height=3.5cm, align=center, draw] (moncloud) at (-3.5, -3) {};

  \draw  ($(moncloud.north)+(-0mm,-8mm)$)                     node                {\footnotesize {Monitoring System}};

%
  \node[shape=cloud, cloud puffs=20, cloud ignores aspect, minimum width=10cm, minimum height=5.5cm, align=center, draw] (heanetcloud) at (2.25, -7) {};
  \draw  ($(heanetcloud.north)+(-0mm,-7mm)$)                     node                {\footnotesize {Global Processing Cloud}};

       \node[shape=rectangle, fill=white,minimum width=.7cm, minimum height=.7cm, align=center, draw] (cpu1) at (4.5, -8) {};
  \node[shape=rectangle, fill=white,minimum width=.7cm, minimum height=.7cm, align=center, draw] (cpu1) at (4.7, -8.2) {};
  \node[shape=rectangle, fill=white,minimum width=.7cm, minimum height=.7cm, align=center, draw] (cpu1) at (4.9, -8.4) {};
  
    \node[shape=rectangle, fill=white,minimum width=.7cm, minimum height=.7cm, align=center, draw] (cpu1) at (3.5, -8) {};
  \node[shape=rectangle, fill=white,minimum width=.7cm, minimum height=.7cm, align=center, draw] (cpu1) at (3.7, -8.2) {};
  \node[shape=rectangle, fill=white,minimum width=.7cm, minimum height=.7cm, align=center, draw] (cpu1) at (3.9, -8.4) {};

 \node[shape=rectangle, fill=white,minimum width=.7cm, minimum height=.7cm, align=center, draw] (cpu1) at (2.5, -8) {};
  \node[shape=rectangle, fill=white,minimum width=.7cm, minimum height=.7cm, align=center, draw] (cpu1) at (2.7, -8.2) {};
  \node[shape=rectangle, fill=white,minimum width=.7cm, minimum height=.7cm, align=center, draw] (cpu1) at (2.9, -8.4) {};

 \node[shape=rectangle, fill=white,minimum width=.7cm, minimum height=.7cm, align=center, draw] (cpu1) at (1.5, -8) {};
  \node[shape=rectangle, fill=white,minimum width=.7cm, minimum height=.7cm, align=center, draw] (cpu1) at (1.7, -8.2) {};
  \node[shape=rectangle, fill=white,minimum width=.7cm, minimum height=.7cm, align=center, draw] (cpu1) at (1.9, -8.4) {};

  \node[shape=rectangle, fill=white,minimum width=.7cm, minimum height=.7cm, align=center, draw] (cpu1) at (0.5, -8) {};
  \node[shape=rectangle, fill=white,minimum width=.7cm, minimum height=.7cm, align=center, draw] (cpu1) at (0.7, -8.2) {};
  \node[shape=rectangle, fill=white,minimum width=.7cm, minimum height=.7cm, align=center, draw] (cpu1) at (0.9, -8.4) {};
  
    \node[shape=rectangle, fill=white,minimum width=.7cm, minimum height=.7cm, align=center, draw] (cpu1) at (-0.5, -8) {};
  \node[shape=rectangle, fill=white,minimum width=.7cm, minimum height=.7cm, align=center, draw] (cpu1) at (-0.3, -8.2) {};
  \node[shape=rectangle, fill=white,minimum width=.7cm, minimum height=.7cm, align=center, draw] (cpu1) at (-0.1, -8.4) {};

        \node[shape=rectangle, fill=white,minimum width=.7cm, minimum height=.7cm, align=center, draw] (cpu1) at (4.5, -7) {};
  \node[shape=rectangle, fill=white,minimum width=.7cm, minimum height=.7cm, align=center, draw] (cpu1) at (4.7, -7.2) {};
  \node[shape=rectangle, fill=white,minimum width=.7cm, minimum height=.7cm, align=center, draw] (cpu1) at (4.9, -7.4) {};
    
        \node[shape=rectangle, fill=white,minimum width=.7cm, minimum height=.7cm, align=center, draw] (cpu1) at (3.5, -7) {};
  \node[shape=rectangle, fill=white,minimum width=.7cm, minimum height=.7cm, align=center, draw] (cpu1) at (3.7, -7.2) {};
  \node[shape=rectangle, fill=white,minimum width=.7cm, minimum height=.7cm, align=center, draw] (cpu1) at (3.9, -7.4) {};

 \node[shape=rectangle, fill=white,minimum width=.7cm, minimum height=.7cm, align=center, draw] (cpu1) at (2.5, -7) {};
  \node[shape=rectangle, fill=white,minimum width=.7cm, minimum height=.7cm, align=center, draw] (cpu1) at (2.7, -7.2) {};
  \node[shape=rectangle, fill=white,minimum width=.7cm, minimum height=.7cm, align=center, draw] (cpu1) at (2.9, -7.4) {};

 \node[shape=rectangle, fill=white,minimum width=.7cm, minimum height=.7cm, align=center, draw] (cpu1) at (1.5, -7) {};
  \node[shape=rectangle, fill=white,minimum width=.7cm, minimum height=.7cm, align=center, draw] (cpu1) at (1.7, -7.2) {};
  \node[shape=rectangle, fill=white,minimum width=.7cm, minimum height=.7cm, align=center, draw] (cpu1) at (1.9, -7.4) {};

  \node[shape=rectangle, fill=white,minimum width=.7cm, minimum height=.7cm, align=center, draw] (cpu1) at (0.5, -7) {};
  \node[shape=rectangle, fill=white,minimum width=.7cm, minimum height=.7cm, align=center, draw] (cpu1) at (0.7, -7.2) {};
  \node[shape=rectangle, fill=white,minimum width=.7cm, minimum height=.7cm, align=center, draw] (cpu1) at (0.9, -7.4) {};
  
  \node[shape=rectangle, fill=white,minimum width=.7cm, minimum height=.7cm, align=center, draw] (cpu1) at (-0.5, -7) {};
  \node[shape=rectangle, fill=white,minimum width=.7cm, minimum height=.7cm, align=center, draw] (cpu1) at (-0.3, -7.2) {};
  \node[shape=rectangle, fill=white,minimum width=.7cm, minimum height=.7cm, align=center, draw] (cpu1) at (-0.1, -7.4) {};

      \node[shape=rectangle, fill=white,minimum width=.7cm, minimum height=.7cm, align=center, draw] (cpu1) at (4.5, -6) {};
  \node[shape=rectangle, fill=white,minimum width=.7cm, minimum height=.7cm, align=center, draw] (cpu1) at (4.7, -6.2) {};
  \node[shape=rectangle, fill=white,minimum width=.7cm, minimum height=.7cm, align=center, draw] (cpu1) at (4.9, -6.4) {};
    
      \node[shape=rectangle, fill=white,minimum width=.7cm, minimum height=.7cm, align=center, draw] (cpu1) at (3.5, -6) {};
  \node[shape=rectangle, fill=white,minimum width=.7cm, minimum height=.7cm, align=center, draw] (cpu1) at (3.7, -6.2) {};
  \node[shape=rectangle, fill=white,minimum width=.7cm, minimum height=.7cm, align=center, draw] (cpu1) at (3.9, -6.4) {};

 \node[shape=rectangle, fill=white,minimum width=.7cm, minimum height=.7cm, align=center, draw] (cpu1) at (2.5, -6) {};
  \node[shape=rectangle, fill=white,minimum width=.7cm, minimum height=.7cm, align=center, draw] (cpu1) at (2.7, -6.2) {};
  \node[shape=rectangle, fill=white,minimum width=.7cm, minimum height=.7cm, align=center, draw] (cpu1) at (2.9, -6.4) {};

 \node[shape=rectangle, fill=white,minimum width=.7cm, minimum height=.7cm, align=center, draw] (cpu1) at (1.5, -6) {};
  \node[shape=rectangle, fill=white,minimum width=.7cm, minimum height=.7cm, align=center, draw] (cpu1) at (1.7, -6.2) {};
  \node[shape=rectangle, fill=white,minimum width=.7cm, minimum height=.7cm, align=center, draw] (cpu1) at (1.9, -6.4) {};

  \node[shape=rectangle, fill=white,minimum width=.7cm, minimum height=.7cm, align=center, draw] (cpu1) at (0.5, -6) {};
  \node[shape=rectangle, fill=white,minimum width=.7cm, minimum height=.7cm, align=center, draw] (cpu1) at (0.7, -6.2) {};
  \node[shape=rectangle, fill=white,minimum width=.7cm, minimum height=.7cm, align=center, draw] (cpu1) at (0.9, -6.4) {};
 
   \node[shape=rectangle, fill=white,minimum width=.7cm, minimum height=.7cm, align=center, draw] (cpu1) at (-0.5, -6) {};
  \node[shape=rectangle, fill=white,minimum width=.7cm, minimum height=.7cm, align=center, draw] (cpu1) at (-0.3, -6.2) {};
  \node[shape=rectangle, fill=white,minimum width=.7cm, minimum height=.7cm, align=center, draw] (cpu1) at (-0.1, -6.4) {}; 
  
%
%
  \node[shape=cloud, cloud puffs=15.7, cloud ignores aspect, minimum width=4cm, minimum height=3.5cm, align=center, draw] (virginiacloud) at (2.25, -12) {};
  \draw  ($(virginiacloud.north)+(-0mm,-7mm)$)                     node                {\footnotesize {Results Cloud}};

    \node[draw, cylinder, shape aspect=1.5, inner sep=0.3333em, fill=white, rotate=90, minimum width=1cm, minimum height=0.05cm] (cyl2) at (3,-12.2) {};
    \node[draw, cylinder, shape aspect=1.5, inner sep=0.3333em, fill=white, rotate=90, minimum width=1cm, minimum height=0.05cm] (cyl3) at (3,-11.8) {};      
    
    \node[draw, cylinder, shape aspect=1.5, inner sep=0.3333em, fill=white,
    rotate=90, minimum width=1cm, minimum height=0.05cm] (cyl2) at (1.5,-12.2) {};
    \node[draw, cylinder, shape aspect=1.5, inner sep=0.3333em, fill=white,
    rotate=90, minimum width=1cm, minimum height=0.05cm] (cyl3) at (1.5,-11.8) {};    
    
    \node[draw, cylinder, shape aspect=1.5, inner sep=0.3333em, fill=white,    rotate=90, minimum width=1cm, minimum height=0.05cm] (cyl2) at (2.25,-13.2) {};
    \node[draw, cylinder, shape aspect=1.5, inner sep=0.3333em, fill=white,    rotate=90, minimum width=1cm, minimum height=0.05cm] (cyl3) at (2.25,-12.8) {};  

%
%

    \draw[->] (ARCHIVECloud) -- (wrkq) ;
    \draw[->] (ARCHIVECloud) -- (heanetcloud) ;
    \draw[->] (heanetcloud) -- (virginiacloud.north) ;   
    \draw[->] (DCCloud) -- (wrkq) ;
    \draw[->] (DCCloud) -- (heanetcloud) ;
    \draw[->] (wrkq) -- (heanetcloud) ;
    \draw[->] (heanetcloud) -- (moncloud) ;
    \draw[->] (CCD) -- (cyl2-1) ;
    \draw[->] (cyl2-1) -- (cyl2a) ;

    \draw[->] (controller) -- (heanetcloud) ;
    \draw[->] (controller) -- (wrkq) ;
    \draw[->] (controller) -- (ARCHIVECloud) ;
    \draw[->] (controller) -- (DCCloud) ;

 %
    \draw[->] (wrkq) -- (moncloud);

  \end{tikzpicture} }
  \caption{NIMBUS Architecture }
  \label{fig:NIMBUSarch}
 \end{center} 
\end{figure}

\tikzstyle{block1} = [rectangle, draw,fill=white,text centered, text width=7em, minimum height=2em]
\tikzstyle{disk} =   [cylinder, draw,fill=white, text width=3em, text centered,shape border rotate=90, shape aspect=0.5, inner sep=0.3333em,  minimum width=2cm, minimum height=3em]
\tikzstyle {data} = [draw,trapezium,trapezium left angle=70,trapezium right angle=-70,node distance=2.5cm,minimum height=2em]

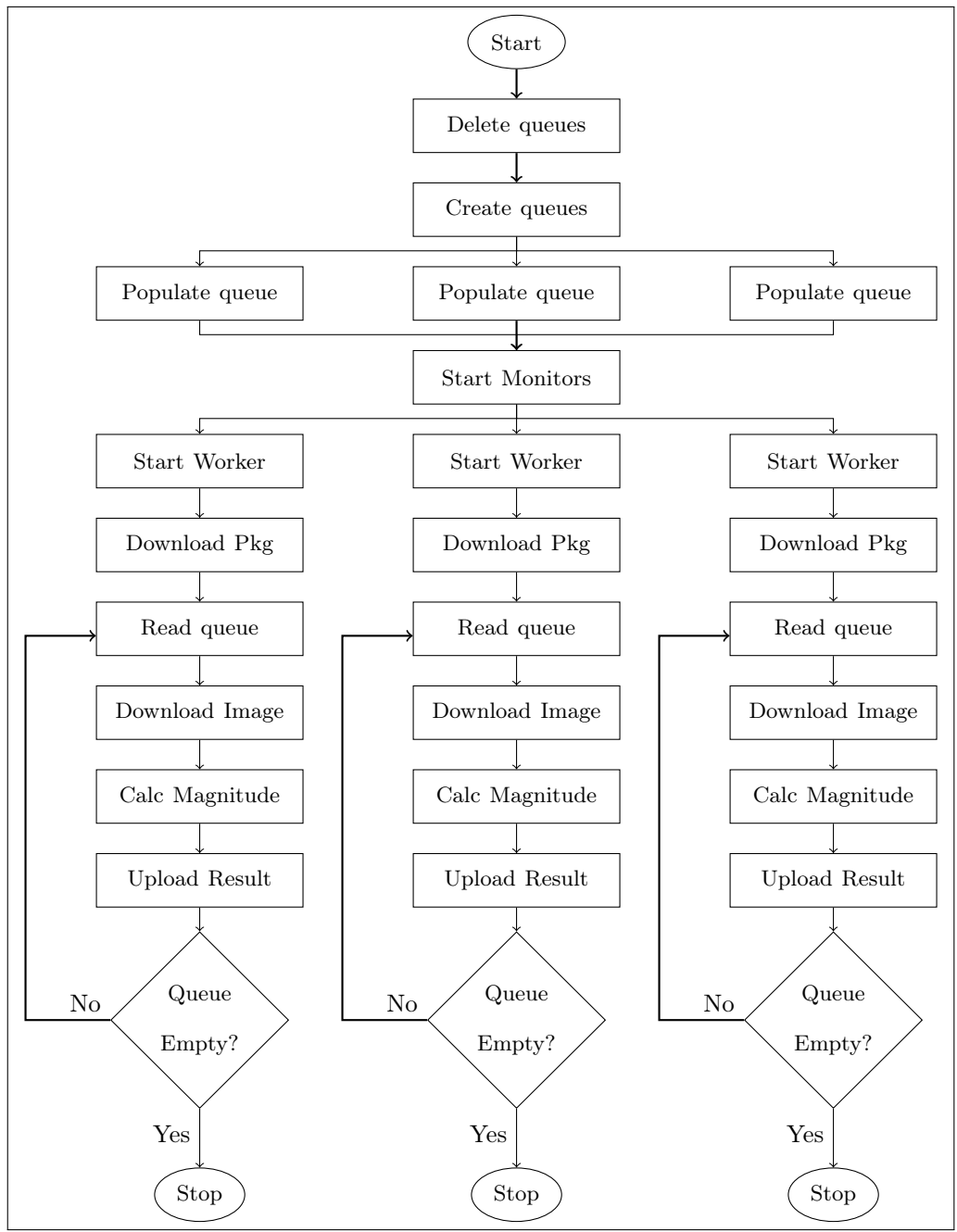
\begin{figure}[htbp]
 \begin{center}
\fbox { 

    \begin{tikzpicture}[node distance=1.2cm]

   \node[cloud] (a)                                              {\footnotesize Start};
   \node[block1] (b)  [below of=a]                       {\footnotesize Delete queues};
   \node[block1] (c)  [below of=b]                       {\footnotesize Create queues};
   
   \node[block1] (d)  [below of=c]                       {\footnotesize Populate queue };
   \node[block1] (e)  [right of=d,node distance=4.5cm]                        {\footnotesize Populate queue };
   \node[block1] (f)  [left of=d,node distance=4.5cm]                         {\footnotesize Populate queue };
 
   \node[block1] (g)  [below of=d]                       {\footnotesize Start Monitors};

   \node[block1] (h)  [below of=g]                       {\footnotesize Start Worker };
   \node[block1] (i)  [right of=h,node distance=4.5cm]                        {\footnotesize Start Worker };
   \node[block1] (j)  [left of=h,node distance=4.5cm]                          {\footnotesize Start Worker };

   \draw[->] (c.south) -- (d);
   \draw[->] (c.south)  -- +(0mm,-2mm)  -| (e.north)  ; 
   \draw[->] (c.south)  -- +(0mm,-2mm)  -| (f.north)  ;   
   
      \draw[->] (e.south)  -- +(0mm,-2mm) -| (g.north)  ; 
      \draw[->] (f.south)  -- +(0mm,-2mm)  -| (g.north)  ;

   \draw[->] (g.south) -- (h);
   \draw[->] (g.south)  -- +(0mm,-2mm)  -| (i.north)  ; 
   \draw[->] (g.south)  -- +(0mm,-2mm)  -| (j.north)  ;   
   
      \node[block1] (k)  [below of=h]                       {\footnotesize Download Pkg  };
      \node[block1] (l)  [below of=i]                       {\footnotesize Download Pkg };
      \node[block1] (m)  [below of=j]                       {\footnotesize Download Pkg };
   \draw[->] (h.south) -- (k);
   \draw[->] (i.south) -- (l);
   \draw[->] (j.south) -- (m);

        \node[block1] (n)  [below of=k]                       {\footnotesize Read queue  };
      \node[block1] (o)  [below of=l]                       {\footnotesize Read queue };
      \node[block1] (p)  [below of=m]                       {\footnotesize Read queue };
   \draw[->] (k.south) -- (n);
   \draw[->] (l.south) -- (o);
   \draw[->] (m.south) -- (p);
   
         \node[block1] (q)  [below of=n]                       {\footnotesize Download Image  };
      \node[block1] (r)  [below of=o]                       {\footnotesize Download Image };
      \node[block1] (s)  [below of=p]                       {\footnotesize Download Image };
   \draw[->] (n.south) -- (q);
   \draw[->] (o.south) -- (r);
   \draw[->] (p.south) -- (s);  
   
            \node[block1] (t)  [below of=q]                       {\footnotesize Calc Magnitude  };
      \node[block1] (u)  [below of=r]                       {\footnotesize Calc Magnitude };
      \node[block1] (v)  [below of=s]                       {\footnotesize Calc Magnitude };
   \draw[->] (q.south) -- (t);
   \draw[->] (r.south) -- (u);
   \draw[->] (s.south) -- (v);  
   
   \node[block1] (w)  [below of=t]                       {\footnotesize Upload Result  };
      \node[block1] (x)  [below of=u]                       {\footnotesize Upload Result };
      \node[block1] (y)  [below of=v]                       {\footnotesize Upload Result };
   \draw[->] (t.south) -- (w);
   \draw[->] (u.south) -- (x);
   \draw[->] (v.south) -- (y);

   \node[decision] (d1)  [below of=w,node distance=2cm]                       {\footnotesize Queue Empty?};
   \node[decision] (d2)  [below of=x,node distance=2cm]                       {\footnotesize Queue Empty?};
   \node[decision] (d3)  [below of=y,node distance=2cm]                       {\footnotesize Queue Empty?};
     \draw[->] (w.south) -- (d1);
   \draw[->] (x.south) -- (d2);
   \draw[->] (y.south) -- (d3);  
   
   \node[cloud] (s1)  [below of=d1]                                            {\footnotesize Stop};
   \node[cloud] (s2)  [below of=d2]                                             {\footnotesize Stop};
   \node[cloud] (s3)   [below of=d3]                                            {\footnotesize Stop};
   \draw[->] (d1.south) -- (s1);
   \draw[->] (d2.south) -- (s2);
   \draw[->] (d3.south) -- (s3);  
 
   \draw[->, thick]  (d1.west)    -- + (-10mm,0mm) -| ($ (n.west) - (10mm,1mm) $)  -- ($(n.west) -(0mm,1mm)$) ; 
   \draw[->, thick]  (d2.west)    -- + (-10mm,0mm) -| ($ (o.west) - (10mm,1mm) $)  -- ($(o.west) -(0mm,1mm)$) ; 
   \draw[->, thick]  (d3.west)    -- + (-10mm,0mm) -| ($ (p.west) - (10mm,1mm) $)  -- ($(p.west) -(0mm,1mm)$) ; 

\node (y1) [below left=0.1cm and 0cm of d1.south] {\small Yes};
\node (y2) [below left=0.1cm and 0cm of d2.south] {\small Yes};
\node (y3) [below left=0.1cm and 0cm of d3.south] {\small Yes};

\node (n1) [below left=-0.5cm and 0cm of d1.west] {\small No};
\node (n2) [below left=-0.5cm and 0cm of d2.west] {\small No};
\node (n3) [below left=-0.5cm and 0cm of d3.west] {\small No};

   \draw[->, thick] (a) -- (b);
   \draw[->, thick] (b) -- (c);
   \draw[->, thick] (d) -- (g);

  \end{tikzpicture} }
  \caption{Distributed processing pipeline where worker nodes use the queue to work in parallel.  }
  \label{flowchartnimbus}
 \end{center}
\end{figure}

\newpage
\section{System Architecture}

\par The NIMBUS pipeline is based on the requirement that all components operate in parallel with the ability to scale up or down as required. This is a distributed system with no requirements for locality built into the system, and where possible, sequential workflows are eliminated from the system. While this pipeline uses an archive data cloud, the only requirement for a facility producing new data to be included in the cloud is for  data to be made available via a web server, and be advertised on one of the worker queues. 

\subsection{Data Archive Cloud}
\par The data archive cloud is comprised of storage nodes which are distributed web servers with mounted file systems containing files to be processed by worker nodes. Each of the storage nodes within the cloud provides a set of services to the pipeline as shown in the Table \ref{tab:storagecloudservices}. Each service is accessed either via the control system or as HTTP requests. The function of servicing web based image file requests separated from the storage of image data allows for a flexible method of data storage and upload. The storage node architecture is shown in Figure \ref{fig:storagesrch}. 

\tikzstyle {service} = [draw,rectangle,fill=white,node distance=2.5cm,rounded corners=0.5ex,minimum height=1cm,minimum width=2cm]

\tikzstyle {nginxserver} = [draw,rectangle,fill=white,node distance=2.5cm,rounded corners=0.5ex,minimum height=1cm,minimum width=4.5cm]

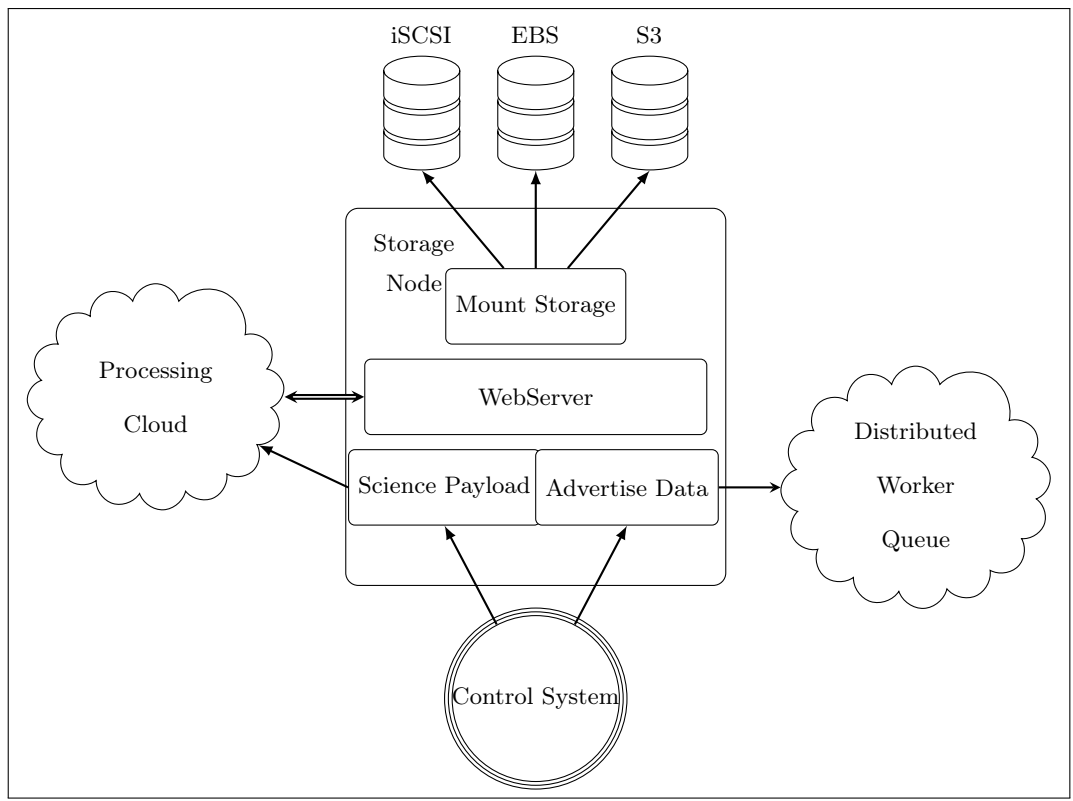
\begin{figure} [htbp]
 \begin{center}
\fbox { 

  \begin{tikzpicture}[node distance=2.5cm, auto, >=stealth]

\node [draw,rectangle,node distance=2.5cm,rounded corners=1ex,minimum height=5cm, minimum width=5cm] (webnode1) at (0,0) {};
  \draw  ($(webnode1.north)+(-16mm,-5mm)$)                     node                {\footnotesize {Storage}};
  \draw  ($(webnode1.north)+(-16mm,-10mm)$)                     node                {\footnotesize {Node}};

%
%
    \node[shape=circle, fill=white,minimum width=2.4cm,  align=center, draw] (ctrlcloud) at (0, -4) {};    
    \node[shape=circle, fill=white,minimum width=2.3cm,  align=center, draw] (ctrlcloud) at (0,-4) {};    
    \node[shape=circle, fill=white,minimum width=2.2cm,  align=center, draw] (ctrlcloud) at (0, -4) {};    
  \draw  ($(ctrlcloud)+(-0mm,-0mm)$)                     node                {\footnotesize {Control System}};

 \node[shape=cloud, cloud puffs=15.7, cloud ignores aspect, minimum width=3cm, minimum height=3cm, text width=2cm,align=center, draw] (queuecloud) at (5, -1.2) {\footnotesize{Distributed Worker Queue}};

    \node[shape=cloud, cloud puffs=15.7, cloud ignores aspect, minimum width=3cm, minimum height=3cm, text width=2cm, align=center, draw] (processingcloud) at (-5, 0) {\footnotesize{Processing Cloud}};


  \node[service] (pkg) at ($(webnode1)+(-12mm,-12mm)$) {\footnotesize {Science\ Payload }};
  \node[nginxserver] (webserver) at ($(webnode1)+(0mm,0mm)$) {\footnotesize  {WebServer}};
  \node[service] (mount) at ($(webnode1)+(0mm,12mm)$) {\footnotesize      {Mount\ Storage}};
  \node[service] (sqs) at ($(webnode1)+(+12mm,-12mm)$) {\footnotesize{Advertise\ Data}};

    
      \node[draw, cylinder, shape aspect=1.5, inner sep=0.3333em, fill=white,
    rotate=90, minimum width=1cm, minimum height=0.05cm] (cyl1) at (-1.5,3.2) {};
    \node[draw, cylinder, shape aspect=1.5, inner sep=0.3333em, fill=white,
    rotate=90, minimum width=1cm, minimum height=0.05cm] (cyl2) at (-1.5,3.6) {};
    \node[draw, cylinder, shape aspect=1.5, inner sep=0.3333em, fill=white,
    rotate=90, minimum width=1cm, minimum height=0.05cm] (cyl3) at (-1.5,4) {};  
    \draw  ($(cyl3.north)+(+5mm,8mm)$)                     node                {\footnotesize {iSCSI }};
 
          \node[draw, cylinder, shape aspect=1.5, inner sep=0.3333em, fill=white,
    rotate=90, minimum width=1cm, minimum height=0.05cm] (cyl4) at (1.5,3.2) {};
    \node[draw, cylinder, shape aspect=1.5, inner sep=0.3333em, fill=white,
    rotate=90, minimum width=1cm, minimum height=0.05cm] (cyl5) at (1.5,3.6) {};
    \node[draw, cylinder, shape aspect=1.5, inner sep=0.3333em, fill=white,
    rotate=90, minimum width=1cm, minimum height=0.05cm] (cyl6) at (1.5,4) {};  
    \draw  ($(cyl6.north)+(+5mm,8mm)$)                     node                {\footnotesize {S3 }};
    
           \node[draw, cylinder, shape aspect=1.5, inner sep=0.3333em, fill=white,
    rotate=90, minimum width=1cm, minimum height=0.05cm] (cyl7) at (0,3.2) {};
    \node[draw, cylinder, shape aspect=1.5, inner sep=0.3333em, fill=white,
    rotate=90, minimum width=1cm, minimum height=0.05cm] (cyl8) at (0,3.6) {};
    \node[draw, cylinder, shape aspect=1.5, inner sep=0.3333em, fill=white,
    rotate=90, minimum width=1cm, minimum height=0.05cm] (cyl9) at (0,4) {};  
    \draw  ($(cyl9.north)+(+5mm,8mm)$)                     node                {\footnotesize {EBS }};

          \draw[<->,double,thick] (processingcloud) -- (webserver) ;
          \draw[->,thick,-latex] (pkg.west) -- (processingcloud)  ;
          
          \draw[->,thick] (sqs.east) --(queuecloud) ;
          \draw[->,thick,-latex] (ctrlcloud) -- (sqs.south) ;
          \draw[->,thick,-latex] (ctrlcloud) -- (pkg.south) ;

          \draw[<-,thick,-latex] (mount) -- (cyl4.west) ;
          \draw[<-,thick,-latex] (mount) -- (cyl1.west) ;
          \draw[<-,thick,-latex] (mount) -- (cyl7.west) ;

  \end{tikzpicture} }
  \caption{Storage Node Architecture }
  \label{fig:storagesrch}
 \end{center} 
\end{figure}

%
%
%
%
%
  \begin{table}
\centering
\begin{tabular}{p{1.6cm} p{12cm}}
  \toprule
Service & Description \\
  \midrule
\small{Upload} & \small{There are three primary methods of upload used, the first is for a storage device containing images to be mounted by the web server, the second is via HTTP POST requests to send data to the web server, the third is to instruct the web server to initiate a download via HTTP from another data source. }\\

\small{Storage} & \small{ Image data can be stored on remote network storage system such as ISCSI devices, or it can be locally attached storage. For the purposes of the web server the method of mounting or attaching storage is irrelevant, however it may cause some delays in servicing files depending on the read time of the storage. Where possible files should be stored in a compressed state to reduce the file transfer times.}\\

\small{Advertise}& \small{A storage node will be required to populate an AWS \gls{sqs} message queue with the URL of all of the files that it currently has stored. The storage node will do this via direct commands from the control system, but could also perform this via a HTTP command. Work is advertised and the contents of the message describe the location of the stored files and how they may be accessed.}\\
\small{Download} & \small{A web server is used to service HTTP requests providing access to files. This NginX webserver is highly optimised for servicing static pages and can be tuned to allow large number of simultaneous connections. Any web server however is permissible, including the Apache server and S3. }\\
  \bottomrule
\small{Science Payload} & \small{This service provides a static location for the downloading of the worker package to each worker. This is a small package containing all of the instructions a worker node will need to operate, including where to obtain work, where to put results,  what work to perform and how to perform the work. The function of the payload and its operations are discussed within the Control system }\\
\hline
  \bottomrule

\end{tabular}
\caption{Storage Node Services.}
\label{tab:storagecloudservices}
\end{table}

\begin {itemize}
\item \textbf{Upload} is the process of placing image files into a disk storage devices which is accessible to the web server of the storage node. No specific method is presented within the NIMBUS pipeline, but data transfer times have been reviewed in the ACN Pipeline in Figure \ref{fig:copyingtime}.  Within the NIMBUS pipeline the data archive cloud was populated with images downloaded from the ACN AWS S3 storage buckets using the s3fs FUSE system which mounts S3 storage buckets allowing files to be copied to mounted storage blocks.  

\item \textbf{Storage} within the pipeline is flexible to the point that any mounted storage device which can be accessed by a webserver  can be accommodated. The speed at which a storage node serves file requests will be determined by the network, the number of concurrent request supported and the read speed of an image from the storage node. Three different configurations were constructed to demonstrate this flexibility within the implemented pipeline. The HEAnet iSCSI storage was mount on DIT based storage nodes while the AWS based storage systems used the AWS Elastic Block Storage service.  
\par An advantage of these devices is that through the use of the NFS file system, the storage device can be mounted for use by multiple servers. An NFS storage for example could be written to from a telescope location, but function as a read only mounted device running a web server. The performance of the read and write of the storage does depend on the configuration of the raw storage. If the storage uses multiple spindles then the write or read times may be quite reasonable. A disturbed file system such as S3 is likely to provide a more scaleable storage solution if there is a high rate of concurrent reading or writing. 
              
\item \textbf{Advertise} is the process of requesting a storage node to review the contents of its datastore and to create a message for each file found. That file is then written to an SQS queue as an advertisement of that file, indicating that it is accessible and ready for processing. The control system can instruct a storage node to review its storage and write the messages. In a production system it would be more realistic for the storage server to monitor for changes in storage and to create new messages when files were added to the data storage devices. The ability to reset and advertise everything rather than operate in an incremental fashion would also be a reasonable requirement. Message formats are explained further under the Distributed Queue Worker cloud section in this chapter. 
             
\item \textbf{Download} services the simple requirement of servicing static urls which provide access to files. Port 80 is a ubiquitously open port which allows all workers to access the image files. A worker node will read a message containing the URL of the image file and simply issue a request to download using a Python script. The NginX web server is a fast static web server which was used for most storage nodes. The entire pipeline works on the assumption that work is obtained through URL downloads.   

\item \textbf{Science Payload} is essential to the creation of workers which can be reconfigured easily. When creating $100's$ of workers it is a requirement that a worker can be dynamically reconfigured before it starts performing any work. The science payload solution is initiated from the control system, but uses the storage node as a central point of  advertising packages.   When requested, the central web node responsible for package management removes the existing package from a standard location on the web server and creates and publishes a new version of the package. New package details are obtained from a central GIT repository.        

\end {itemize}

\subsection{Distributed Worker Queue Cloud}
\par Central to the design of this architecture is the Amazon Simple SQS which is a distributed web based message delivery system. The service defines itself as \emph{Reliable, Scalable, Simple, Secure} and \emph{Inexpensive} and is one of many similar distributed web messages services such as RabbitMQ. A public message queue such as the AWS SQS system provides a transactional message system allowing for distributed processes to communicate. A transactional system ensures that messages can be read, held exclusively by one process and removed from the queue as required. The basic lifecycle of a message is shown in Figure \ref{fig:messagevisability}. Messages are sent to a distributed queue where they are replicated and stored. When the message is read, the message visibility timer starts ensuring that nobody else can receive the message. During the visibility timeout the process which received the message can process the data and remove the message from the queue by deleting it. If the process fails to compete and does not delete the message, then the message will become visible on the queue once the visibility timeout has elapsed.  The default timeout for messages within the system is 120 seconds.

\tikzstyle {msg} = [draw,rectangle,node distance=9cm,,minimum height=1cm]

\begin{figure}[htbp]  
 \begin{center}
 
 \fbox{ 
\begin{tikzpicture}

  	\node[msg] (msg1) at (0,0) {\small{\ Visibility Timeout (in seconds)\ }};
 	 \draw  ($(msg1.north)+(2mm,+8mm)$)                     node                {\footnotesize {Receive\ Message}};
 	 \draw  ($(msg1.north)+(2mm,+5mm)$)                     node                {\footnotesize {Requests}};

	 \draw  ($(msg1.north)+(-27mm,+15mm)$)                     node                {\footnotesize {Receive\ Message}};
 	 \draw  ($(msg1.north)+(-27mm,+12mm)$)                     node                {\footnotesize {Requests}};

	 \draw  ($(msg1.north)+(35mm,+15mm)$)                     node                {\footnotesize {Receive\ Message}};
 	 \draw  ($(msg1.north)+(35mm,+12mm)$)                     node                {\footnotesize {Requests}};

	\draw[->,-latex,line width=1mm] ($(msg1.south)-(4,0)$) -- ($(msg1.south)+(4.5,0)$);		
	\draw[->,-latex,line width=0.5mm] ($(msg1.west)+(0,1.5)$) -- ($(msg1.west)-(0,1.5)$);		
	\draw[->,-latex,line width=0.5mm] ($(msg1.east)+(1,1.5)$) -- ($(msg1.east)-(-1,1.5)$);		

	\draw[->,-latex,line width=0.5mm] ($(msg1.west)+(1.5,1)$) -- ($(msg1.west)-(-1.5,-0.5)$);		
	\draw[->,-latex,line width=0.5mm] ($(msg1.west)+(4,1)$) -- ($(msg1.west)-(-4,-0.5)$);		

	\draw  ($(msg1.south)+(-0mm,-4mm)$)                     node                {\footnotesize {Message\ Not\ Returned}};
	\draw  ($(msg1.south)+(-27mm,-14mm)$)                     node                {\footnotesize {Message\ returned}};
	\draw  ($(msg1.south)+(35mm,-14mm)$)                     node                {\footnotesize {Message\ returned}};

\end{tikzpicture}
}
  \vskip -0.8em
  \caption{SQS message visibility.}
   \label{fig:messagevisability}	
\end{center}
\end{figure}
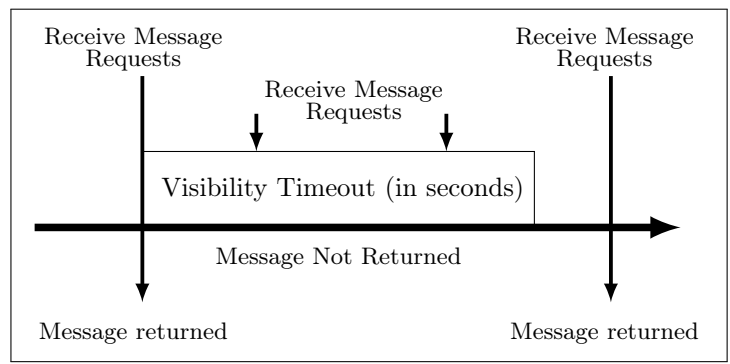

\par A 2007 performance review of the SQS service supports the reliability of the service \cite{Garfinkel07anevaluation} and although these tests were run using EC2 instances, the service is available outside of the Amazon Cloud. It should be noted that these performance tests indicated a bottleneck reading messages at high speed, approximately 5 messages per second, however this was done using a single threaded system rather than using a distributed or multi-threaded system. As with many of the amazon services SQS offers distributed computing opportunities. A more recent study of the lag time for messages being available from time of submission shows variability between 1 and 7 seconds \cite{Iosup:2011if}, however for a queue that grows faster than consumption this is not necessarily an issue. Writing messages in parallel to the queue is also possible when there is little worry about the message order. 

\par A key feature of a distributed queue is that it relies on the principle of eventual consistency, which means that given multiple SQS servers, each containing copies of the queue, when messages are written to one queue, there is a delay in syncing the message queues. It is also for this reason that message order is not preserved or guaranteed during delivery as shown in Figure \ref{sqsservers}. The advantage offered by the distributed system is that writing to a single queue is a distributed write operation allowing for the write operations to potentially occur in parallel. A series of tests were performed on message writing as shown later in the chapter. 
\tikzstyle {msg} = [draw,rectangle,node distance=2.5cm,rounded corners=1ex,minimum height=2em]

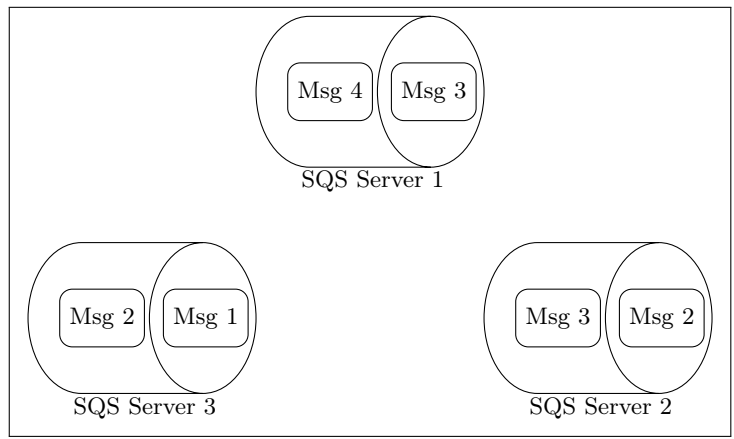
\begin{figure}[htbp]
 \begin{center}
\fbox { 

  \begin{tikzpicture}[node distance=2.5cm, auto, >=stealth]

  \node[draw, cylinder, shape aspect=5.5, inner sep=0.3333em, fill=white, minimum width=2cm, minimum height=3cm] (cyl1) at (0,0) {};
  \node[msg] (sqs1) at (0,0) {\footnotesize{Msg 4}};
  \node[msg] (sqs2) at ($(sqs1.east)+  (8mm,0 )$) {\footnotesize{Msg 3}};
  \draw  ($(sqs2.south)+(-8mm,-8mm)$)                     node                {\footnotesize {SQS Server 1}};

  \node[draw, cylinder, shape aspect=5.5, inner sep=0.3333em, fill=white, minimum width=2cm, minimum height=3cm] (cyl1) at (3,-3) {};
  \node[msg] (sqs3) at (3,-3) {\footnotesize{Msg 3}};
  \node[msg] (sqs4) at ($(sqs3.east)+  (8mm,0 )$) {\footnotesize{Msg 2}};
  \draw  ($(sqs4.south)+(-8mm,-8mm)$)                     node                {\footnotesize {SQS Server 2}};

  \node[draw, cylinder, shape aspect=5.5, inner sep=0.3333em, fill=white, minimum width=2cm, minimum height=3cm] (cyl1) at (-3,-3) {};
  \node[msg] (sqs5) at (-3,-3) {\footnotesize{Msg 2}};
  \node[msg] (sqs6) at ($(sqs5.east)+  (8mm,0 )$) {\footnotesize{Msg 1}};
  \draw  ($(sqs6.south)+(-8mm,-8mm)$)                     node                {\footnotesize {SQS Server 3}};

  \end{tikzpicture} }
  \caption{Distributed SQS Servers}
  \label{sqsservers}
 \end{center} 
\end{figure}

\par Webservers were configured with data mounted from various data storage devices and used to advertise the raw images as URLs for download. An SQS message was constructed to represent each file in the data store which advertises a file available for processing, using the following form \emph{http://webnode1.dit.ie/data/compressed/00-0001487.fits.fz}

A worker node will read the queue, download the file, and process it. There is an implicit relationship between the work performed and the work advertised. This is controlled by the work initialisation processes where workers download the worker scripts which include pointers to specific queues relevant to the worker script capabilities.

\par Within the NIMBUS architecture seven queues are used. Each queue allows for a worker process to work asynchronously while allowing a form of central control throughout the system.  For each experiment each queue is deleted and then recreated in an empty state to ensure that it only contains data relevant to a specific experiment. Queues are deleted rather than emptied, for performance reasons, as it is possible for  queues to contain hundreds of thousands of messages at the end of an experiment. Each of the queues and their function are explained below.

\begin {itemize}
\item \textbf{workerq} is the primary SQS queue containing the location of data files to be processed.  Items are added to the queue by storage nodes which advertise their files by writing them to the queue as shown in Figure \ref{fig:queuecontrols}.  Items from the queue are consumed by worker nodes which process the image and post the result prior to deleting the message from the queue. It is essential that the visibility timeout for the queue is set long enough for the worker node to complete the image processing before the message is visible. If a node fails to complete then the message reappears on the queue for another node to download and process. Storage nodes are instructed to access their storage directories and write a single message for each file found. The URL provided must be supported by the storage node through a web server running on the storage node. In all cases the NginX web server runs on each storage node and servers up files to requesting worker nodes.  The message written to this queue is in the following form. \emph{<NODEURL><filename>}. For example \emph{http://webnode1.dit.ie/data/compressed/00-0001487.fits.fz}.
\tikzstyle {webserver} = [draw,rectangle,node distance=2.5cm,rounded corners=0.5ex,minimum height=2em]
\tikzstyle {queue} = [draw,trapezium,fill=white,trapezium left angle=70,trapezium right angle=-70,shape border rotate=180,node distance=2.5cm,minimum height=2em]

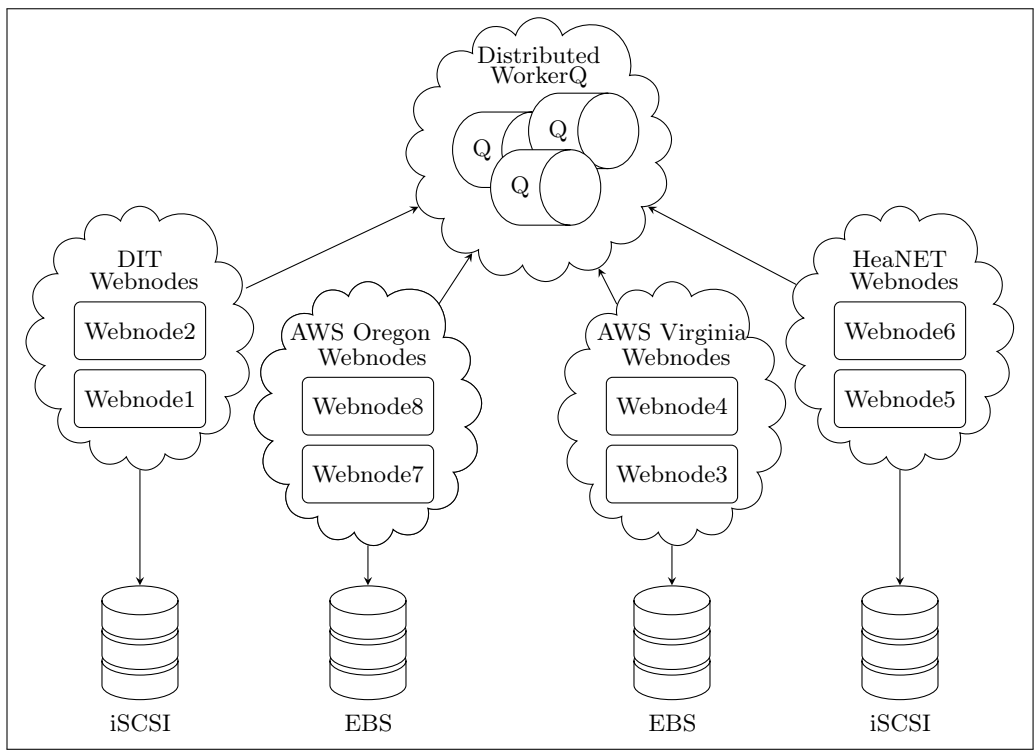
\begin{figure}[htbp]
 \begin{center}
\fbox { 

  \begin{tikzpicture}[node distance=2.5cm, auto, >=stealth]

  \node[draw, cylinder, shape aspect=1.5, inner sep=0.3333em, fill=white, minimum width=1cm, minimum height=1cm] (cyl1) at (-0.5,0.5) {\footnotesize{Q}};
  \node[draw, cylinder, shape aspect=1.5, inner sep=0.3333em, fill=white, minimum width=1cm, minimum height=1cm] (cyl1) at (0.5,0.75) {\footnotesize{Q}};
  \node[draw, cylinder, shape aspect=1.5, inner sep=0.3333em, fill=white, minimum width=1cm, minimum height=1cm] (cyl1) at (0,0) {\footnotesize{Q}};
  \node[shape=cloud, cloud puffs=15.7, cloud ignores aspect, minimum width=3.5cm, minimum height=3.5cm, align=center, draw] (sqscloud) at (0.25, 0.5) {};
  \draw  ($(sqscloud.north)+(-0mm,-5mm)$)                     node                {\footnotesize {Distributed}};
  \draw  ($(sqscloud.north)+(-0mm,-8mm)$)                     node                {\footnotesize {WorkerQ}};

  \node[shape=cloud, cloud puffs=15.7, cloud ignores aspect, minimum width=3cm, minimum height=3.5cm, align=center, draw] (ditcloud) at (-5, -2) {};
  \draw  ($(ditcloud.north)+(-0mm,-7mm)$)                     node                {\footnotesize {DIT}};
  \draw  ($(ditcloud.north)+(-0mm,-10mm)$)                     node                {\footnotesize { Webnodes}};

  \node[shape=cloud, cloud puffs=15.7, cloud ignores aspect, minimum width=3cm, minimum height=3.5cm, align=center, draw] (heanetcloud) at (5, -2) {};
  \draw  ($(heanetcloud.north)+(-0mm,-7mm)$)                     node                {\footnotesize {HeaNET}};
  \draw  ($(heanetcloud.north)+(-0mm,-10mm)$)                     node                {\footnotesize { Webnodes}};

  \node[shape=cloud, cloud puffs=15.7, cloud ignores aspect, minimum width=3cm, minimum height=3.5cm, align=center, draw] (oregoncloud) at (-2, -3) {};
  \draw[->] (oregoncloud) -- (sqscloud) ;

  \node[shape=cloud, cloud puffs=15.7, cloud ignores aspect, fill=white,minimum width=3cm, minimum height=3.5cm, align=center, draw] (oregoncloud) at (-2, -3) {};
  \draw  ($(oregoncloud.north)+(-1mm,-7mm)$)                     node                {\footnotesize {AWS Oregon}};
  \draw  ($(oregoncloud.north)+(-0mm,-10mm)$)                     node                {\footnotesize { Webnodes}};

  \node[shape=cloud, cloud puffs=15.7, cloud ignores aspect, minimum width=3cm, minimum height=3.5cm, align=center, draw] (virginiacloud) at (2, -3) {};
  \draw  ($(virginiacloud.north)+(-0mm,-7mm)$)                     node                {\footnotesize {AWS Virginia }};
  \draw  ($(virginiacloud.north)+(-0mm,-10mm)$)                     node                {\footnotesize { Webnodes}};



  \node[webserver] (webnode1) at ($(ditcloud)+(0mm,-8mm)$) {\footnotesize{Webnode1}};
  \node[webserver] (webnode2) at ($(ditcloud)+(0mm,+1mm)$) {\footnotesize{Webnode2}};

  \node[webserver] (webnode7) at ($(oregoncloud)+(0mm,-8mm)$) {\footnotesize{Webnode7}};
  \node[webserver] (webnode8) at ($(oregoncloud)+(0mm,+1mm)$) {\footnotesize{Webnode8}};

  \node[webserver] (webnode3) at ($(virginiacloud)+(0mm,-8mm)$) {\footnotesize{Webnode3}};
  \node[webserver] (webnode4) at ($(virginiacloud)+(0mm,+1mm)$) {\footnotesize{Webnode4}};

  \node[webserver] (webnode5) at ($(heanetcloud)+(0mm,-8mm)$) {\footnotesize{Webnode5}};
  \node[webserver] (webnode6) at ($(heanetcloud)+(0mm,+1mm)$) {\footnotesize{Webnode6}};

    
      \node[draw, cylinder, shape aspect=1.5, inner sep=0.3333em, fill=white,
    rotate=90, minimum width=1cm, minimum height=0.05cm] (cyl1) at (5,-6.6) {};
    \node[draw, cylinder, shape aspect=1.5, inner sep=0.3333em, fill=white,
    rotate=90, minimum width=1cm, minimum height=0.05cm] (cyl2) at (5,-6.2) {};
    \node[draw, cylinder, shape aspect=1.5, inner sep=0.3333em, fill=white,
    rotate=90, minimum width=1cm, minimum height=0.05cm] (cyl3) at (5,-5.8) {};  
  
    \draw  ($(cyl1.south)+(-5mm,-5mm)$)                     node                {\footnotesize {iSCSI }};

      \node[draw, cylinder, shape aspect=1.5, inner sep=0.3333em, fill=white,
    rotate=90, minimum width=1cm, minimum height=0.05cm] (cyl4) at (-5,-6.6) {};
    \node[draw, cylinder, shape aspect=1.5, inner sep=0.3333em, fill=white,
    rotate=90, minimum width=1cm, minimum height=0.05cm] (cyl5) at (-5,-6.2) {};
    \node[draw, cylinder, shape aspect=1.5, inner sep=0.3333em, fill=white,
    rotate=90, minimum width=1cm, minimum height=0.05cm] (cyl6) at (-5,-5.8) {};  
    \draw  ($(cyl4.south)+(-5mm,-5mm)$)                     node                {\footnotesize {iSCSI }};

        \node[draw, cylinder, shape aspect=1.5, inner sep=0.3333em, fill=white,
    rotate=90, minimum width=1cm, minimum height=0.05cm] (cyl7) at (2,-6.6) {};
    \node[draw, cylinder, shape aspect=1.5, inner sep=0.3333em, fill=white,
    rotate=90, minimum width=1cm, minimum height=0.05cm] (cyl8) at (2,-6.2) {};
    \node[draw, cylinder, shape aspect=1.5, inner sep=0.3333em, fill=white,
    rotate=90, minimum width=1cm, minimum height=0.05cm] (cyl9) at (2,-5.8) {};  
      \draw  ($(cyl7.south)+(-5mm,-5mm)$)                     node                {\footnotesize {EBS }};

          \node[draw, cylinder, shape aspect=1.5, inner sep=0.3333em, fill=white,
    rotate=90, minimum width=1cm, minimum height=0.05cm] (cyl10) at (-2,-6.6) {};
    \node[draw, cylinder, shape aspect=1.5, inner sep=0.3333em, fill=white,
    rotate=90, minimum width=1cm, minimum height=0.05cm] (cyl11) at (-2,-6.2) {};
    \node[draw, cylinder, shape aspect=1.5, inner sep=0.3333em, fill=white,
    rotate=90, minimum width=1cm, minimum height=0.05cm] (cyl12) at (-2,-5.8) {};    
    \draw  ($(cyl10.south)+(-5mm,-5mm)$)                     node                {\footnotesize {EBS }};

     \draw[->] (ditcloud) -- (sqscloud) ;
          \draw[->] (ditcloud) -- (cyl6) ;

     \draw[->] (heanetcloud) -- (sqscloud) ;
          \draw[->] (heanetcloud) -- (cyl3) ;

     \draw[->] (virginiacloud) -- (sqscloud) ;
          \draw[->] (virginiacloud) -- (cyl9) ;
          
       \draw[->] (oregoncloud) -- (cyl12) ;

  \end{tikzpicture} }
  \caption{SQS Worker Queue constructed from 8 different storage nodes }
  \label{fig:queuecontrols}
 \end{center} 
\end{figure} 
 
\item	\textbf{supervisor} is an SQS queue read by each worker before it looks for work, and is tested before a worker checks the workerq.  The supervisor queue contains approximately 100 messages all of which contain identical messages. The message visibility timeout for this queue is  set to one second so that the message is put back onto the queue as soon as possible. At the start of an experiment the queue is deleted and reconstructed to ensure that all messages are identical. The reason that the queue is used is to control the  behavior of a worker node, usually on startup and messages need to be all the same so that all workers behave in the same manner.  This queue contains one of four  messages shown in Table \ref {tab:supervisor} "REGISTER", "REBOOT","LISTEN", and "UPGRADE". The supervisor queue is checked before the worker queue is read, and after each worker node cycle. Once a command is performed the worker stops listening for REBOOT, UPGRADE or REGISTER commands to ensure that endless cycles are not initiated by the worker. To reset the  worker node to listen for these commands the LISTEN command is used. 

 \begin{table}
\centering
\begin{tabular}{p{2cm} p{11cm}}
  \toprule
Command & Description \\
  \midrule
\small{REGISTER} & \small{This command requires that the worker writes a message to the register queue before performing any other actions. The register message format is as follows. <threadname><ipaddress><timestamp>}\\
\small{REBOOT}& \small{This command requires that the worker performs a hardware reset and reboots the machine}\\
\small{UPGRADE} & \small{This command is an essential service where a worker node is required to go to the service package location and download install an update to the software being used to perform work on the worker node. It is through this mechanism that a worker can be provisioned to service alternative jobs. }\\
\small{LISTEN}& \small{After a command has been performed the worker stops listening to avoid endless reboots, upgrades or registrations. Because the workers may come online at different times, the supervisor queue must continue to advertise its instructions, but workers should only perform the action once. This behavior can be over rotten by issuing the LISTEN command which allows workers to once again listen for a supervisor command and act upon it.}\\
\hline
  \bottomrule
\end{tabular}
\caption{Supervisor Queue control commands.}
\label{tab:supervisor}
\end{table}
\newpage
\item	\textbf{cmdq} is an SQS queue read by the worker after the supervisor queue has been read. Assuming that the worker node is now registered and running the correct software, this queue gives explicit instructions to the worker on how to proceed. The supported queue commands are described in Table \ref{tab:cmdq}. 
 \begin{table}
\centering
\begin{tabular}{p{2cm} p{11cm}}
  \toprule
Command & Description \\
  \midrule
\small{START} & \small{This command informs the worker node to start processing messages from the worker queue. A timestamp for the start of work is taken and sent to the log queue. Messages are downloaded and a "do\_work" script is used to process the files.}\\
\small{STOP}& \small{This command suspends message processing. A timestamp for the stop event is taken and sent to the log queue.}\\
\small{SLEEP} & \small{This command is pauses the woker node for $60$ seconds after which time it will continue to look for commands on the cmdq queue. }\\
\small{QUIT}& \small{Nodes can be requested to terminate although this is an unlikely use case as the node will require a reboot to automatically join the worker node cloud. A QUIT command will stop the primary worker node software from running and it will not check any queues, effectively leaving the cloud. }\\
\hline
  \bottomrule
\end{tabular}

\caption{Command Queue control commands.}
\label{tab:cmdq}
\end{table}
  
\item \textbf{workerregister} is an SQS queue which is used to identify the workers as they become activated. For some of the larger experiments the number of workers dramatically increases as there are multiple workers per node. The message written to this queue is in the following form.\emph{<threadname><ipaddress>TIMESTAMP<data:time>.}

\item \textbf{logfile} is an SQS queue which is a source of data used in the analysis of the system performance. All workers write messages to the logfile queue to indicate progress and status. A worker writes to this queue for key events during processing. For all messages written to the logfile queue, core information about the worker is inserted with the message including the version of the software running within the worker node. The following events generate messages. \emph{Message Received} from the queue which is targeted for processing.   \emph{Message Processing Rate} which keeps a count of the total number of files processed since the worker has started and a private processing rate after each batch has been completed. \emph{Message Deleted} which is called just after a file has been processed, uploaded to the target storage node and the result has been written to the result queue.  This queue also contains some diagnostic information from the worker such as any supervisor or cmd queue events which are processed. This queue generates more than double the number of messages processed. Where $M$ is the number of messages processed by a single worker node, and $B$ is the number of files downloaded for processing as a batch\footnote{Workers can be configured to download a specific number of messages before attempting to perform any work on them. This is referred to as the worker Batch Number} a worker node will produce an estimate of log messages $\widehat{l}$ where 

\begin{equation}
\widehat{l} = M \times 2+\frac{M}{B} 
\end{equation}

\par This estimate ignores the comparatively few diagnostic messages generated by each worker.  To estimate $\widehat{L}$, the total number of log file messages created during an experiment, sum the values of all worker nodes threads where there are $n$ worker nodes\footnote{A worker node may run multiple threaded versions of the worker nodes which operate as distinct worker node processes}. 

\begin{equation}
\widehat{L} = \sum_{i=1}^{n} (M \times 2+\frac{M}{B} )
\end{equation}

\par If it is assumed that all messages in a worker queue are processed across all worker nodes, then the sum of all messages processed by all workers is equal to the total number of messages in the worker queue $T$. Hence the equation can be simplified as follows, with the assumption that batch sizes are equal across all worker nodes. 

\begin{equation}  \label{eq:lograte}
\widehat{L} = T \times 2+\frac{T}{B} 
\end{equation}

\par The number of log messages created for a worker message queue of $500,000$ is $2\times 500,000 + \frac{500,000}{10}$ which is equal to $1,050,000$.  The use of the log file queue creates additional work for the worker node, slowing it down slightly, and establishes a need to process the log file queue in order to review the results of the experiment. These log files are used as a monitor on the pipeline to provide an alternative measure of the processing rates and worker node behaviour. Similar to the writing of messages, a sequential approach to reading queues will result in poor performance. A distributed set of reading processes are required.  The message written to the log queue is in the following form. \\ \emph{<threadname><ipaddress>TIMESTAMP<data:time><Message> }

\item \textbf{resultq} is the SQS queue which is written to once a processed file has been successfully posted to its S3 destination.   There should be a single message for all files processed, so the number of messages in this queue should equate to the number of messages in the worker queue once all processing has been completed. The message written to this queue is in the following form.\\  \emph{<threadname><ipaddress>TIMESTAMP<data:time><resultfilename>.}

\item \textbf{canaryq} is used as an instrumentation queue to monitor the structure and behaviour of the system. When a storage node advertises its work to the worker it contains the URL and filename of an image. During the processing of the workerq it is a requirement that the message be deleted. While it is possible to extend the visibility of a message within the queue, only $120,000$ messages may be left in this condition. Messages in flight \footnote{Messages in flight are messages which have been read from the queue and are within a visibility timeout period} cannot exceed this value and thus the information within the workerq cannot be saved without the use of another queue. The canaryq provides the opportunity to create a mirror of the workerq which can be analyzed after an experiment has been run as it contains a record of the writing order of the queue by using a timestamp. Further analysis of the queue is provided later in this chapter. The message written to this queue is in the following form. \\ \emph{TIMESTAMP<data:time><URL><filename>}.

\end{itemize}

\subsection{System Monitoring}

\par The monitoring system is a collection of different subsystems each contributing to the overall monitoring and logging of the pipeline. This includes the logging queue which contains detailed information from each of the running worker processes, a Python monitor which calculates the rate that the workerq is being reduced over time, the canaryq which can be used to reconstruct message delivery sequence, web server statistics using a system called Munin, and processing node performance for AWS EC2 instances using their AWS monitoring service for special \emph{canary worker nodes}.  This node has additional monitoring configured for each experiment with the assumption that it is representative of other nodes performance. This is checked by seeing if the canary processes clean a similar number of files to the rest of the nodes. Given the number of instances potentially running in an experiment, this approach provides a snapshot of a single instance performance which can be used as a generalisation of the other instances performance. When the experiments are running, the primary monitor is the workerq monitor which estimates the cleaning rate over time.  The components within the  monitoring system summaries below. 

	\begin{itemize}
 \item\textbf{Web Server Monitoring.} Each web server is configured with the Munin monitoring software. This is used to track system performance helping identify bottlenecks and resource utilisation. The web server data is collected and shown as a series of graphs of system performance covering a wide range of system resources including Disk I/O, CPU, Networking and Memory utilisation.  Using these metrics an experiment can be reviewed to determine how the web servers were performing.   
 
  \item\textbf{Canary Instance Monitoring.} The AWS service provides monitoring facilities for EC2 instances and for SQS queues. Detailed monitoring is a paid service which is not enabled by default. When activated the time resolution is for 1 minute time slots and graphs are provided for Network In, Network Out and CPU utilisation. Assuming that the work performed by the canary instance is similar to that of the other others, then it's system performance metrics are a reasonable estimate of all instances performance. For most experiments the canary monitor is configured for AWS cloud monitoring.
  
 \item\textbf{Canary Queue Monitoring.} For some experiments, knowing the order of messages processed is required to provide an accurate view of web server contribution to the experiment over time. The canaryq queue is a full copy of the messages as they are processed including a central timestamp so all messages can be reordered. 
 
  \item\textbf{Logging Queue.} This has already been discussed, but it provides an alternative view to the processing rate by allowing to determine how many files the pipeline or individual worker processed. Differences in processing rates by web servers and workers can also be observed. 

  \item\textbf{Workerq monitor.} Use a call to the workerq every second, requesting  the size of the queue, the rate of processing since the experiment began can be estimated. Because messages which are consumed by workers nodes may not complete, this is not a cleaning rate, just a consumption of messages rate. If for any reason a worker node fails to delete a message then it will reappear on the queue. In a live pipeline with images potentially being added on a regular basis, this queue monitor would not necessarily be as useful. For these experiments however all data is in place and queues allowed to settle before monitoring begins. 

 	\end{itemize}

\subsection{Processing Cloud}

\par Worker nodes, or worker instances, are servers which run worker processes. Worker processes  are individual program threads running on a physical or virtual machine. The majority of instances in this pipeline, for the purpose of experimental control, are virtual machines hosted on the Amazon Web Service using the EC2 service. For an experiment all instances are usually set to be the same type as explained later in the Control System subsection in this chapter. Depending on the configuration of the instance it may be possible to run multiple worker threads  which could result in an overall improvement in the amount of files processed by the instance.  For simplicity, the worker process is not written to be multi-threaded, rather the operating system is instructed to run multiple copies of the same program which are kept in insolation from each other. Resource sharing is handled by the operating system. 

\par As with the previous pipelines the \emph{acn-aphot.c} program is run at the heart of the worker process. A control system implemented in Python wraps this program in a service which provides image data for cleaning. The core cleaning program was not modified for this pipeline. 

\par In order for a server to join the global processing cloud it must first install an initialisation script which is run when the server is first started. This script call be installed into an existing server, or as in the case of this pipleine a reference virtual machine was constructed with this script installed and an \gls{ami} produced from which large number of virtual machine instance can be created.  

\par This initialisation script will look for a science payload on a public web server then download and execute it. The payload is designed to clean the instance directory spaces, install scripts and utilities, and clone itself to create multiple programming processes if required and then begin whatever tasks are set by the worker utility script. 

\par To ensure that all experiments started in a consistent manner all experimental worker instances were deleted and new instances were created from the reference AMI.  The primary lifecycle of a worker instance is shown in Figure \ref{flowchartinitworker} and described below.  

\tikzstyle{block1} = [rectangle, draw,fill=white,text centered, text width=7em, minimum height=2em]
\tikzstyle{disk} =   [cylinder, draw,fill=white, text width=3em, text centered,shape border rotate=90, shape aspect=0.5, inner sep=0.3333em,  minimum width=2cm, minimum height=3em]
\tikzstyle {data} = [draw,trapezium,trapezium left angle=70,trapezium right angle=-70,node distance=2.5cm,minimum height=2em]

\begin{figure}[htbp]
\begin{center}
\fbox {
    \begin{tikzpicture}[node distance=1cm]

   \node[cloud] (a)                                              {\footnotesize Start};
   \node[block1] (b)  [below of=a,node distance=1.2cm]                       {\footnotesize Boot Instance};
   \node[block1] (c)  [below of=b,node distance=1.2cm]                       {\footnotesize Run init.d Script};
   \node[block1] (d)  [below of=c,node distance=1.2cm]                       {\footnotesize Download Science Payload};
   \node[block1] (e)  [below of=d,node distance=1.5cm]                       {\footnotesize Install Primary Worker};
   \node[block1] (f)  [right of=e, node distance=4cm]                       {\footnotesize Install Worker Thread N};
   \node[block1] (g)  [left of=e, node distance=4cm]                       {\footnotesize Install Worker Thread 2};

    \node[block1] (h)  [below of=e,node distance=1.5cm]                       {\footnotesize Start Worker};
   \node[block1] (i)  [right of=h, node distance=4cm]                       {\footnotesize Start Worker N};
   \node[block1] (j)  [left of=h, node distance=4cm]                       {\footnotesize Start Worker 2};

   \node[cloud] (k)  [below of=h,node distance=1.2cm]                                             {\footnotesize Stop};

   \draw[->, thick] (a) -- (b);
   \draw[->, thick] (b) -- (c);
   \draw[->, thick] (c) -- (d);
   \draw[->, thick] (d) -- (e);
   \draw[->, thick] (e) -- (h);
    \draw[->, thick] (h) -- (k);

   \draw[->] (d.south)  -- +(0mm,-2mm)  -| (f.north)  ; 
   \draw[->] (d.south)  -- +(0mm,-2mm)  -| (g.north)  ;   
     
   \draw[->] (e.south)  -- +(0mm,-2mm)  -| (i.north)  ; 
   \draw[->] (e.south)  -- +(0mm,-2mm)  -| (j.north)  ;   
   
  \end{tikzpicture} }
  \caption{Flow Chart showing worker node initialisation during the boot up process}
  \label{flowchartinitworker}
 \end{center}
\end{figure}
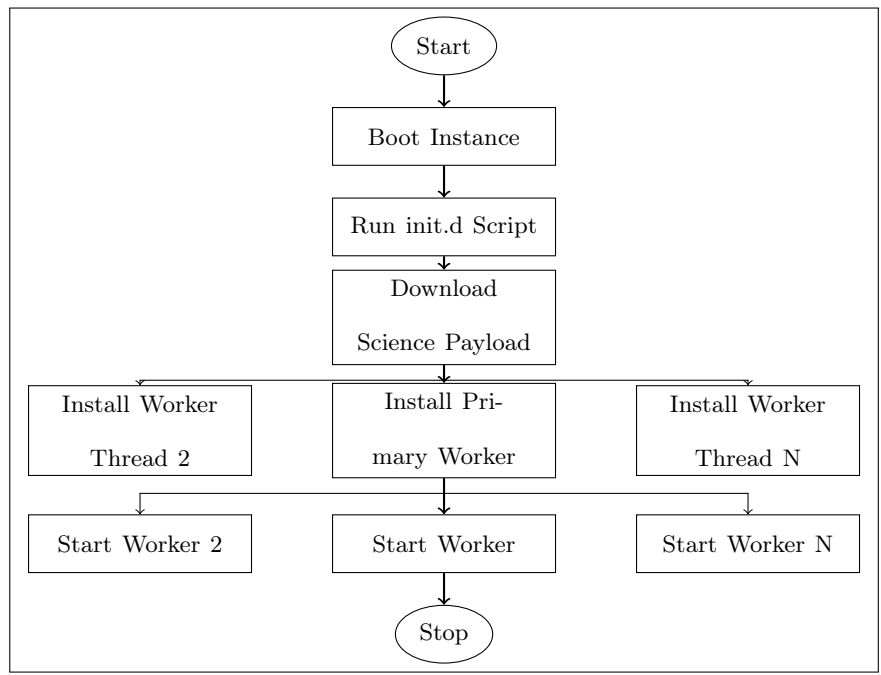 
 
	\begin{itemize}
\item\textbf{Start}  The control system is responsible for starting the worker instances. A standard preconfigured AMI is used. In some experiments physical instances were also used. 
\item\textbf{Boot Instance} The standard Linux boot sequence brings the instance onto the network and mounts the local storage. 
\item\textbf{Run init script} The nimbus-worker init script is run which ensure that the server is cleaned of any previous experimental data. 
\item\textbf{Download Science payload}  The science payload is downloaded, moved to the worker home directory on the server and installed. The work performed by a worker is detailed in the this package. This ensure that there is a flexibility to the system allowing for multiple types of processing to be performed depending on the payload. In most cases the worker will perform image cleaning and calibration of the FITS files, however any tasks are feasible. For the purpose of this pipeline the payload was altered to test network performance and web server performance by downloading files but not processing them, and by doubling the processing in an alternative payload to show the effects of a more CPU bound problem. 
\item\textbf{Install Worker thread 1-N}  The installed system will create the required number of independent threads which are then run. 
\item\textbf{Start worker thread}. The workers are ready to process data and listen to the supervisor and command queue for the instruction to start downloading data to process.  

\item\textbf{Stop/Delete} The worker threads will sleep when there are no more messages to process, periodically checking for work. When an experiment is finished the control system may stop or terminate the running instance. 
	\end {itemize}

\par
\begin{figure}[htbp]
 \begin{center}
\fbox { 

    \begin{tikzpicture}[node distance=1.5cm]

   \node[cloud] (a)                                              {\footnotesize Start};
   \node[decision] (b)  [below of=a,node distance=2.2cm]                       {\footnotesize Read Sup Queue};
   \node[block1] (c)  [right of=b, node distance=4cm]                       {\footnotesize Reg Worker};
   \node[block1] (d)  [left of=b, node distance=4cm]                       {\footnotesize Reboot Instance};
   \node[decision] (e)  [below of=b, node distance=2.8cm]                       {\footnotesize Read Cmd Queue};
    \node[block1] (f)  [right of=e, node distance=4cm]                       {\footnotesize Wait 60 Secs};
   \node[cloud] (g)  [left of=e, node distance=4cm]                       {\footnotesize Stop};

   \node[block1] (h)  [below of=e, node distance=1.8cm]                       {\footnotesize Start Timer};
   \node[block1] (i)  [below of=h, node distance=1cm]                       {\footnotesize Clear Directories};
   \node[block1] (j)  [below of=i, node distance=1.2cm]                       {\footnotesize Read Worker Queue};
   \node[decision] (j1)  [below of=j, node distance=2.1cm]                       {\footnotesize Msg Available};
       \node[block1] (j2)  [right of=j1, node distance=4cm]                       {\footnotesize Wait 60 Secs};

   \node[block1] (k)  [below of=j1, node distance=1.8cm]                       {\footnotesize Download Files};
   \node[block1] (l)  [below of=k, node distance=1.2cm]                       {\footnotesize Call do\_work Script};
   \node[block1] (m)  [below of=l, node distance=1.2cm]                       {\footnotesize Upload Results};
   \node[block1] (n)  [below of=m, node distance=1.2cm]                       {\footnotesize Delete Msg from Worker Queue};

   \draw[->, thick] (a) -- (b);
   \draw[->, thick] (b) -- (e);
   \draw[->, thick] (b) -- (c);
   \draw[->, thick] (b) -- (d);
   \draw[->, thick] (e) -- (f);
   \draw[->, thick] (d) -- (g);

   \draw[->, thick] (e) -- (g);
   \draw[->, thick] (e) -- (h);
   \draw[->, thick] (h) -- (i);
   \draw[->, thick] (i) -- (j);
   \draw[->, thick] (j) -- (j1);
      \draw[->, thick] (j1) -- (k);
      \draw[->, thick] (j1) -- (j2);
   \draw[->, thick] (k) -- (l);
   \draw[->, thick] (l) -- (m);   
   \draw[->, thick] (m) -- (n);

   \draw[->, thick]  (n.south)   -- +(0mm,-4mm)  -- + (-40mm,-4mm) -| ($ (a) - (60mm,7mm) $)  -- ($(a) -(1mm,7mm)$) ; 
   \draw[->, thick]  (c.east)    -- + (4mm,0mm) |- ($ (e.north) + (1mm,2mm) $) ; 
   \draw[->, thick]  (f.east)    -- + (6mm,0mm) |- ($ (a) - (-1mm,7mm) $) ; 
   \draw[->, thick]  (j2.east)    -- + (7mm,0mm) |- ($ (a) - (-1mm,5mm) $) ; 



\node at ($(j1.east) +(5mm,2mm)$)  {\small No};
\node at ($(e.east) +(5mm,2mm)$)  {\small Sleep};
\node at ($(e.west) +(-5mm,2mm)$)  {\small Stop};

\node at ($(b.east) +(6mm,2mm)$)  {\small Register};
\node at ($(b.west) +(-6mm,2mm)$)  {\small Reboot};


  \end{tikzpicture} }
  \caption{Worker control script managing the flow of work based on message reading status}
  \label{flowchartworkernimbus}
 \end{center}
\end{figure}
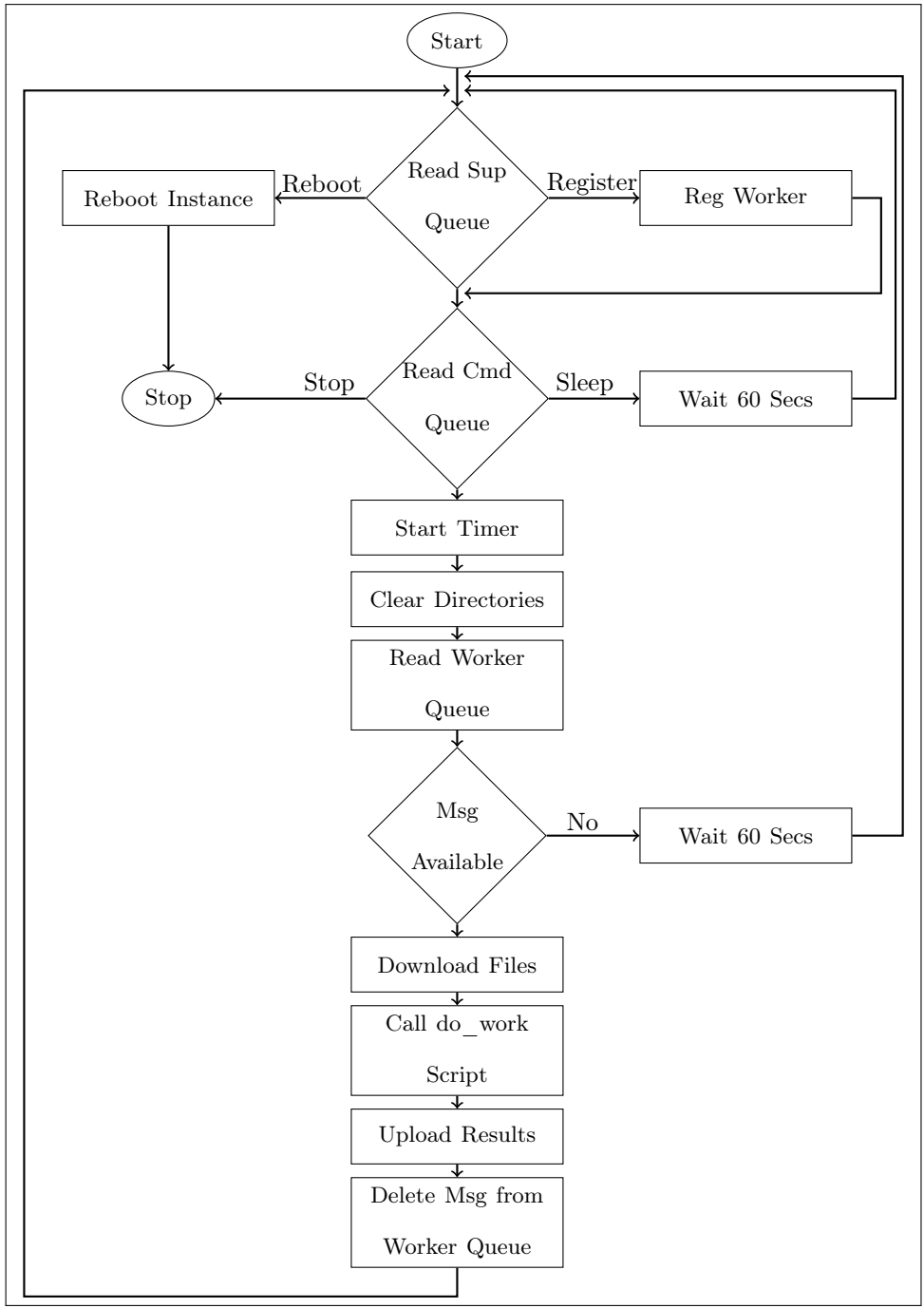

 The flowchart for a running worker thread is shown in Figure \ref{flowchartworkernimbus}. Each worker instance may have multiple worker threads all operating at the same time. All threads are isolated from each other however, with their own storage space and directory structure on the worker instance. The control system which operates this thread is a Python script inside which any science payload could potentially be inserted. The function of the worker thread is to listen to the sqs command queues for instructions on how to operated. Instructions such as start, stop, and sleep provide simple interfaces into the system. What to download and where to get it, is provided by the configurable message queue, and how to process it is determined by a central \emph{do\_work} script which then performs the required work. The output of the processing is then uploaded to a central web server and information on its location is pushed to a sqs based results queue. 

\section{Experimental Methodology}

\par The function of the control system is to initiate all experiments and ensure that all systems are available and functioning correctly. It is important that experiments can be compared, and to do this the starting state must be consistent in all cases. The control system runs a Python script which tears down the experimental infrastructure, then rebuilds it before the start of the experiment. All systems must be accessible from the control system which resides on a virtual machine within the AWS cloud, running an Ubuntu instance on the EC2 service. A batch script contains the series of experiments to run, which in turn calls a script to start and experiment. The batch script contains a series of calls to the \emph{run-experiment.sh} script, which takes a set of parameters, shown in Table \ref{tab:run-experiment}. The pipeline can also be left in a running state, which means that it will continue to process image files as they appear on queues one it has started, and new workers can be added or removed dynamically. The full workflow of the experimental run system \emph{run-experiment.sh} is given in Figure \ref{flowchartrunexp}.  The experimental run script was required to ensure  that the starting point of each experiment was consistent for each experimental iteration. A sample of the batch script used to run multiple experiments is shown in Appendix \ref{code:batchrunexperiments}.

  \begin{table} [H]
\centering
\begin{tabular}{p{2.3cm} p{11.5cm}}
  \toprule
Option & Description \\
  \midrule
\small{-a | -d | -n | -x} & \small{Identify the combination of web servers to use in the experiment. -d uses the DIT based web servers, -a indicates the use of the AWS based web servers, -n uses the heanet based web servers and -x indicates that all web servers should be used. For a web node to be used in an experiment it will advertise the files it has storage to the SQS worker queue.}\\
\small{Instances} & \small{The number of instances of workers to run for an experiment less the monitoring node which always runs. The maximum number of concurrent instances is set to 100 within these experiments. This limit required explicit permission from AWS Ireland to run instances within the Irish region. }\\
\small{Time}& \small{The maximum amount of time in seconds that the worker nodes should be allowed to run. In most cases experiments were set to 20 minutes. An experimental timer was set only when the EC2 instances were initiated and confirmed to be running.}\\
\small{Name} & \small{The name of the experiment so that it can be identified.}\\
\small{Number} & \small{The number of web servers to run per type. Webservers are configured in pairs, so if a DIT web server is selected then either 1 or 2 will be allowed to run while 1-4 was the range allowable for AWS based web servers. }\\
\small{Size} & \small{The size of the AQS EC2 instance to run. The parameter conforms to the specific reference name that AWS uses for its instances. Most experiments used either the \emph{t1.micro}  the \emph{m1.large } or in some cases the \emph{m3.2xlarge }}\\

  \bottomrule
\small{Package Payload} & \small{In addition to changing the experimental options above, the package used by a worker node can also be reconfigured. The most significant change is specifying the number of threads a worker instance is allowed to initiate when running. This ranged from $1$ to $100$. The number of files downloaded by a worker before it begins processing the can also be specified. The processing batch size for nearly all experiments was set to $10$.}\\
\hline
  \bottomrule
\end{tabular}
\caption{Experimental execution options for the NIMBUS pipeline.}
\label{tab:run-experiment}
\end{table}

 \tikzstyle{block1} = [rectangle, draw,fill=white,text centered, text width=7em, minimum height=2em]
\tikzstyle{disk} =   [cylinder, draw,fill=white, text width=3em, text centered,shape border rotate=90, shape aspect=0.5, inner sep=0.3333em,  minimum width=2cm, minimum height=3em]
\tikzstyle {data} = [draw,trapezium,trapezium left angle=70,trapezium right angle=-70,node distance=2.5cm,minimum height=2em]

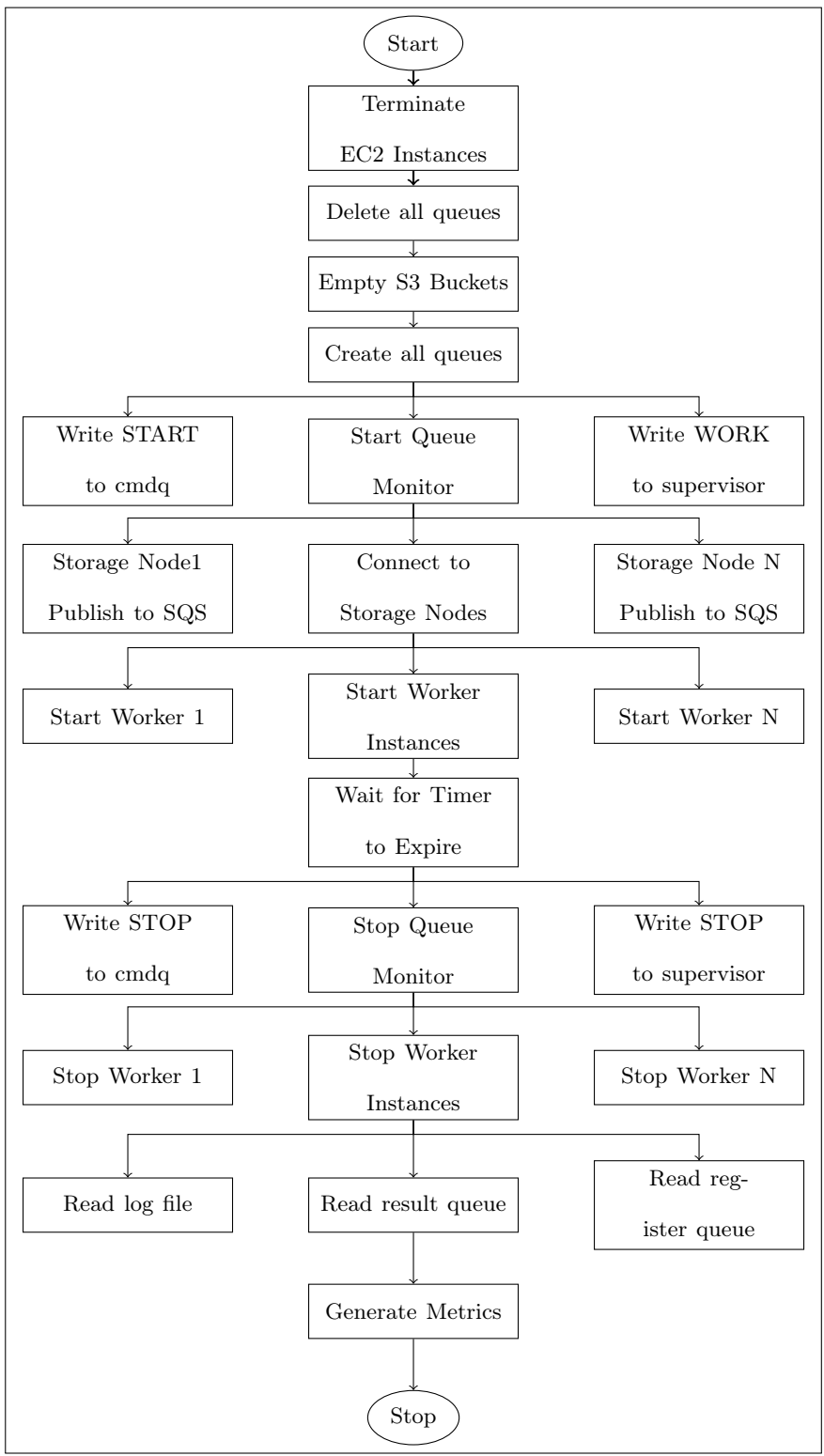
\begin{figure}[htbp]
 \begin{center}
\fbox { 

    \begin{tikzpicture}

   \node[cloud] (a)                                              {\footnotesize Start};
   \node[block1] (b)  [below of=a,node distance=1.2cm]                       {\footnotesize Terminate EC2 Instances};
   \node[block1] (c)  [below of=b,node distance=1.2cm]                       {\footnotesize Delete all queues};
   
   \node[block1] (d)  [below of=c,node distance=1cm]                       {\footnotesize Empty S3 Buckets  };
   \node[block1] (e)  [below of=d,node distance=1cm]                       {\footnotesize Create all queues };
   \node[block1] (h)  [below of=e,node distance=1.5cm]                       {\footnotesize Start Queue Monitor} ;

   \node[block1] (f)  [left of=h, node distance =4cm ]                         {\footnotesize Write START to cmdq };
 
   \node[block1] (g)  [right of=h, node distance=4cm]                       {\footnotesize Write WORK to supervisor};

 \node[block1] (i)  [below of=h,node distance=1.8cm]                       {\footnotesize Connect to Storage Nodes};
 \node[block1] (j)  [right of=i,node distance=4cm]                        {\footnotesize Storage Node N Publish to SQS };
  \node[block1] (k)  [left of=i,node distance=4cm]                          {\footnotesize Storage Node1 Publish to SQS };
   
 \node[block1] (l)  [below of=i,node distance=1.8cm]                       {\footnotesize Start Worker Instances};
 
  \node[block1] (l1)  [right of=l,node distance=4cm]                        {\footnotesize Start Worker N };
  \node[block1] (l2)  [left of=l,node distance=4cm]                          {\footnotesize Start Worker 1};

 \node[block1] (m)  [below of=l,node distance=1.5cm]                       {\footnotesize Wait for Timer to Expire};
 \node[block1] (n)  [below of=m,node distance=1.8cm]                       {\footnotesize Stop Queue Monitor};
 
   \node[block1] (o)  [left of=n, node distance =4cm ]                         {\footnotesize Write STOP to cmdq };
 
   \node[block1] (p)  [right of=n, node distance=4cm]                       {\footnotesize Write STOP to supervisor};

 \node[block1] (q)  [below of=n,node distance=1.8cm]                       {\footnotesize Stop Worker Instances};
 
  \node[block1] (q1)  [right of=q,node distance=4cm]                        {\footnotesize Stop Worker N };
  \node[block1] (q2)  [left of=q,node distance=4cm]                          {\footnotesize Stop Worker 1};

 \node[block1] (r)  [below of=q,node distance=1.8cm]                       {\footnotesize Read result queue};
 
  \node[block1] (r1)  [right of=r,node distance=4cm]                        {\footnotesize Read register queue };
  \node[block1] (r2)  [left of=r,node distance=4cm]                          {\footnotesize Read log file};

 \node[block1] (s)  [below of=r,node distance=1.5cm]                       {\footnotesize Generate Metrics};
 \node[cloud] (t) [below of=s,node distance=1.5cm]                                             {\footnotesize Stop};


   \draw[->, thick] (a) -- (b);
   \draw[->, thick] (b) -- (c);
   \draw[->] (h.south) -- (i);
   \draw[->] (c.south) -- (d);
   \draw[->] (d.south) -- (e);
      \draw[->] (e.south) -- (h);
            \draw[->] (l.south) -- (m);
            \draw[->] (m.south) -- (n);
   \draw[->] (n.south) -- (q);
   \draw[->] (q.south) -- (r);
      \draw[->] (r.south) -- (s);

    \draw[->] (s.south) -- (t);

   \draw[->] (e.south)  -- +(0mm,-2mm)  -| (f.north)  ; 
   \draw[->] (e.south)  -- +(0mm,-2mm)  -| (g.north)  ;   
   
   \draw[->] (i.south)  -- +(0mm,-2mm)  -| (l1.north)  ; 
   \draw[->] (i.south)  -- +(0mm,-2mm)  -| (l2.north)  ;   
   
   \draw[->] (m.south)  -- +(0mm,-2mm)  -| (o.north)  ; 
   \draw[->] (m.south)  -- +(0mm,-2mm)  -| (p.north)  ;  
   
   \draw[->] (n.south)  -- +(0mm,-2mm)  -| (q1.north)  ; 
   \draw[->] (n.south)  -- +(0mm,-2mm)  -| (q2.north)  ;   

   \draw[->] (q.south)  -- +(0mm,-2mm)  -| (r1.north)  ; 
   \draw[->] (q.south)  -- +(0mm,-2mm)  -| (r2.north)  ; 

   \draw[->] (h.south)  -- +(0mm,-2mm)  -| (j.north)  ; 
   \draw[->] (h.south)  -- +(0mm,-2mm)  -| (k.north)  ;   
   



   \draw[->] (i.south) -- (l);

  \end{tikzpicture} }
  \caption{NIMBUS Experimental run script flow chart}
  \label{flowchartrunexp}
 \end{center}
\end{figure} 

\par The experimental run script is a Unix bash script which call a series of Python scripts on remote system to control a specific experiment. The Python programming API used to control AWS services is BOTO, which provides access to primary services such as S3, EC2 and SQS. To ensure that an experiment starts consistently each time it is run efforts are made to ensure there is little or no system caching being performed by any system.  The S3 buckets are emptied to ensure that no previous results are being counted for the experiment. Queues are deleted and recreated to eliminate the chance of old messages contaminating an experiment. All worker instances are created from a standard AMI instance and are deleted at the end of each experimental run. 

\par The experimental pipeline performance has also been considered where possible to ensure that steps in the process for creation or tear down of the system are as efficient as possible. Some of the steps in the experiments are primarily focused on experimental integrity but are not essential to a production pipeline, such as queue deletion, and S3 result deletion.  Where large number of entries are placed on queues a multi-threaded queue reader is employed to extract all messages. As previously discussed the logfile contains the largest number of entries within an experiment and as such the reader for this queue downloads messages via multiple  threads. The process of recombining these are sorting data into a sequential stream is the final generate metrics step in the flow chart.

\par In preparation for the experiment a science payload  is created and pushed to a web server as shown in Figure \ref{flowchartpayload}. The payload will include details on how many independent workers are to be created on the instance and what work each of these workers should perform and how it should perform that work. The science payload consists of the following components:

\begin{itemize}
\item  Datafiles directory. The specific location to download batches of files.
\item  Results directory. The location where results files are held until they are uploaded to the S3 service.
\item  Masterfiles directory. Contains the mastervbias and the master flat images for use in image cleaning.
\item  bin directory. Contains the acn-aphot compiled utility used to clean image files and the funpack utility to uncompress images. 
\item  scripts directory. Contains the do\_work.sh script which performs the image uncompress and calls the acn-aphot program to perform cleaning, and the start-worker.sh script which installs the required number of worker threads and starts them. 
\item  sql-reader.py. The central control script which monitors queue, downloads images, passes them to the do\_work script, and uploaded results.
 \end{itemize}

\tikzstyle{block1} = [rectangle, draw,fill=white,text centered, text width=7em, minimum height=2em]
\tikzstyle{disk} =   [cylinder, draw,fill=white, text width=3em, text centered,shape border rotate=90, shape aspect=0.5, inner sep=0.3333em,  minimum width=2cm, minimum height=3em]
\tikzstyle {data} = [draw,trapezium,trapezium left angle=70,trapezium right angle=-70,node distance=2.5cm,minimum height=2em]

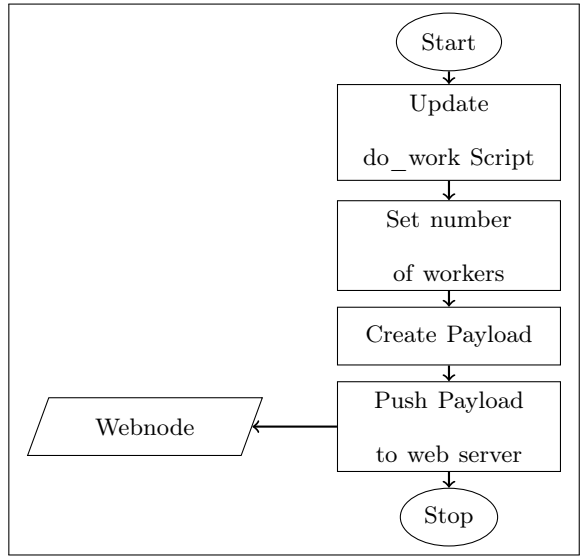
\begin{figure}[htbp]
\begin{center}
\fbox { 

    \begin{tikzpicture}[node distance=1cm]
   \node[cloud] (a)                                              {\footnotesize Start};
   \node[block1] (b)  [below of=a,node distance=1.2cm]                       {\footnotesize Update do\_work Script};
   \node[block1] (c)  [below of=b,node distance=1.5cm]                       {\footnotesize Set number of workers};
   \node[block1] (d)  [below of=c,node distance=1.2cm]                       {\footnotesize Create Payload};
   \node[block1] (e)  [below of=d,node distance=1.2cm]                       {\footnotesize Push Payload to web server};
    \node[data] (f)  [left of=e,node distance=4cm]                       {\footnotesize Webnode};

   \node[cloud] (h)  [below of=e,node distance=1.2cm]                                             {\footnotesize Stop};

   \draw[->, thick] (a) -- (b);
   \draw[->, thick] (b) -- (c);
   \draw[->, thick] (c) -- (d);
   \draw[->, thick] (d) -- (e);
   \draw[->, thick] (e) -- (f);
    \draw[->, thick] (e) -- (h);

  \end{tikzpicture} }
  \caption{Flow Chart showing creation of payload for distribution to worker nodes}
  \label{flowchartpayload}
 \end{center} 
\end{figure}

\subsection{Experiment Metrics System}
\par At the end of each experiment when the messages are written to the log file queue, the monitor queue, results queue and worker registration queue, a series of metrics are extracted to characterise the experimental behaviour. These metrics are used in the analysis of the experiment performance and to check that everything ran correctly. 

\begin{itemize}
     \item \textbf{log file queue} is a combination of all messages written by each of the workers to identify key events during the processing cycle. The metrics available from this queue are given in Table \ref{tab:logfilemetrics}. 

\item \textbf{Result queue} contains messages written by workers after a message is successfully posted to the results storage system S3. Messages in this queue are either equal to or just slightly less than then total number of files found in the results directory.  The metrics available from this queue are given in Table \ref{tab:resultmetrics}. 

\item \textbf{Worker register queue} is written to by each of the worker threads for all instances. This is used to ensure that the correct number of workers are active within the system to ensure that rate calculations are correct based on the number of workers. In most cases the workers all start, however if for any reason they fail, the registration message will not be present. The metrics available from this queue are given in Table \ref{tab:workermetrics}. 

\item \textbf{Derived Metrics} are generated from the raw data using a spreadsheet to compare the data from multiple sources. Given the distributed nature of the data and the queues,  and considering that  workers may be still processing data when an experiment is concluded differences in processing rates may be calculated using different metrics. It is also useful to measure specific attributes of the data processing performance such as  Results-Not-Posted, which shows the number of files downloaded but not yet completed processing across workers. Details of these derived metrics are shown in Tables \ref{tab:derivedmetrics1} and Table \ref{tab:derivedmetrics2}

\item \textbf{Monitor queue} is not a specific queue but rather the processing of monitoring the worker queue throughout the experiment. Once the experiments starts the monitor will keep a record on the  reported number of messages in the worker queue and perform a calculation of the rate of message processing over time. The metrics available from this queue are given in Table \ref{tab:monitormetrics} although the raw data can also be used to plot the performance over time.

 \end{itemize}

  \begin{table}[htbp]
\centering
\begin{tabular}{p{3cm} p{10cm}}
  \toprule
Metric & Description \\
  \midrule
\small{Results-Posted} & \small{ The confirmed total of all files posted successfully to the S3 bucket. This is an absolute count of work fully completed by all workers}\\

\small{Total-Downloads} & \small{The total of all files pull from all web servers.}\\
\small{Web-Server-Downloads}& \small{The total of all files pull from a specific web server.}\\
  
\small{Total-Instances} & \small{The total number of virtual or physical instances performing work during an experiment.}\\
\small{Total-Workers} & \small{The total number of worker threads running across all instances performing work during an experiment. This value is the same as instances if only 1 worker thread is running on an instance.}\\
\small{First-Start} & \small{The time that the first worker is first seen in the log file. Times are UTC based.}\\
\small{Last-Start} & \small{The time that the last worker is first seen in the logfile. Times are UTC based.}\\
\small{Last-Download} & \small{The time that the last file download occurred within a worker}\\
\small{Last-Upload} & \small{The time that the last result file was uploaded to S3  by a worker}\\

\hline
  \bottomrule
\end{tabular}
\caption{Metrics extracted from worker logfile.}
\label{tab:logfilemetrics}
\end{table}

 \begin{table}[htbp]
\centering
\begin{tabular}{p{3.5cm} p{9.5cm}}
  \toprule
Metric & Description \\
  \midrule
\small{Results-Posted} & \small{ The confirmed total of all files posted successfully to the S3 bucket. This is an absolute count of work fully completed by all workers}\\

\small{Total-Per-Thread} & \small{Each worker instance is a physical or virtual machine, and within each of the instances a number of threads can be activated to a maximum of 10. This total is the number of results posted for a specific thread across all instances}\\
  
\small{Total-Workers} & \small{The total number of virtual or physical instances performing work during an experiment.}\\
\small{Min-Worker} & \small{The minimum number of files processed by a worker thread across all instances}\\
\small{Max-Worker} & \small{The maximum number of files processed by a worker thread across all instances}\\
\small{Avg-Worker} & \small{The average number of files processed by all worker thread across all instances}\\
\small{STD Dev} & \small{Standard deviation of files processed by all worker thread across all instances}\\

\small{Total-Instances} & \small{The total number of virtual or physical instances performing work during an experiment.}\\
\small{Min-Instance} & \small{The minimum number of files processed by an instance thread across all workers}\\
\small{Max-Instance} & \small{The maximum number of files processed by an instance thread across all workers}\\
\small{Avg-Instance} & \small{The average number of files processed by an instance thread across all workers}\\
\small{STD Dev} & \small{Standard deviation  of files processed by an instance thread across all workers}\\

\hline
  \bottomrule
\end{tabular}
\caption{Metrics extracted from Results Queue.}
\label{tab:resultmetrics}
\end{table}

 \begin{table}[htbp]
\centering
\begin{tabular}{p{3.5cm} p{9.5cm}}
  \toprule
Metric & Description \\
  \midrule
\small{Results-Posted} & \small{ The confirmed total of all files posted successfully to the S3 bucket. This is an absolute count of work fully completed by all workers}\\

\small{Total-Per-Thread} & \small{Each worker instance is a physical or virtual machine, and within each of the instances a number of threads can be activated to a maximum of 10. This total is the number of results posted for a specific thread across all instances}\\
  
\small{Total-Workers} & \small{The total number of virtual or physical instances performing work during an experiment.}\\
\small{Min-Worker} & \small{The minimum number of files processed by a worker thread across all instances}\\
\small{Max-Worker} & \small{The maximum number of files processed by a worker thread across all instances}\\
\small{Avg-Worker} & \small{The average number of files processed by all worker thread across all instances}\\
\small{STD Dev} & \small{Standard deviation of files processed by all worker thread across all instances}\\

\small{Total-Instances} & \small{The total number of virtual or physical instances performing work during an experiment.}\\
\small{Min-Instance} & \small{The minimum number of files processed by an instance thread across all workers}\\
\small{Max-Instance} & \small{The maximum number of files processed by an instance thread across all workers}\\
\small{Avg-Instance} & \small{The average number of files processed by an instance thread across all workers}\\
\small{STD Dev} & \small{Standard deviation  of files processed by an instance thread across all workers}\\

\hline
  \bottomrule
\end{tabular}
\caption{Metrics extracted from worker register queue.}
\label{tab:workermetrics}
\end{table}

 \begin{table}[htbp]
\centering
\begin{tabular}{p{3.5cm} p{9.5cm}}
  \toprule
Metric & Description \\
  \midrule
\small{Results-Not-Posted} & \small{ A count of the number of files confirmed to have ben downloaded from a web server but have not been posted to the results folder on S3. This difference is present when workers are either intentionally or otherwise terminated during processing of a file downloaded. It should be closely related to the messages in flight as the message is deleted after a result it posted to S3. }\\

\small{Results-Not-Posted \%} & \small{The Results-Not-Posted metric expressed as a percentage of the total number of files downloaded}\\
  
\small{MAX-Time} & \small{The maximum amount of time that all of the workers were active. This is take as the time in seconds, between the last registration time of a worker (indicating that all workers were active) and the last download time from a web server from any worker. This is check to ensure that the approximate running time of an experiment is as expected.}\\
\small{AVG-Logfile-Rate} & \small{The total reported files uploaded to  S3 by log file queue  divided by the experimental time (1200 seconds).Setup time per worker may be included  which is evident when reviewing the MAX-Time metric above which allows for workers to register and start downloading.}\\

\small{AVG-Result-Rate} & \small{The total reported files uploaded to  S3 by result queue  divided by the experimental time.  A workers failure to write to either the log file or results queue during processing may result in differences with AVG-Logfile-Rate, however it should be a nominal difference.}\\

\small{AVG-Monitor-Rate} & \small{The total difference in the start queue size and the end queue size divided by the experimental time. This metric may  be higher than AVG-Logfile-Rate and AVG-Result-Rate as it will include any messages in flight. In cases of high volumes of workers the queue size may not be accurate at that point in time.}\\

\small{AVG-Rate} & \small{Using the averages above an average of the overall average rate across the three metrics can be obtained.}\\

  \hline
  \bottomrule \hline

\end{tabular}
\caption{Composite metrics derived from multiple raw metrics.}
\label{tab:derivedmetrics1}
\end{table}

 \begin{table}[htbp]
\centering
\begin{tabular}{p{3.5cm} p{9.5cm}}
  \toprule
Metric & Description \\
  \midrule
\small{Best-Rate} & \small{The best cleaning rate achieved using the 3 average methods. In most cases this should be the AVG-Monitor-Rate as explained above. With larger volumes of process, or larger number of workers the monitoring system may not return an accurate queue size and may return a smaller size. This is due to the \emph{eventual consistency} feature of the SQS queues.}\\

\small{Logfile-Rate-Var} & \small{The variance of the AVG-Logfile-Rate from the AVG-Rate. This can be used to determine if there are larger differences which may have skewed the calculation of the AVG-Rate.}\\
\small{Result-Rate-Var} & \small{The variance of the AVG-Result-Rate from the AVG-Rate. This can be used to determine if there are larger differences which may have skewed the calculation of the AVG-Rate.}\\

\small{Monitor-Rate-Var} & \small{The variance of the AVG-Monitor-Rate from the AVG-Rate. This can be used to determine if there are larger differences which may have skewed the calculation of the AVG-Rate.}\\

  \hline
  \bottomrule \hline

\end{tabular}
\caption{More composite metrics derived from multiple raw metrics.}
\label{tab:derivedmetrics2}
\end{table}

 \begin{table}[htbp]
\centering
\begin{tabular}{p{3.5cm} p{9.5cm}}
  \toprule
Metric & Description \\
  \midrule
\small{Start-Q-Size} & \small{ Size of the worker queue at the beginning of the experiment. If the monitor observes the queue size increasing that it waits for it to decrease before beginning.}\\

\small{End-Q-Size} & \small{Each worker instance is a physical or virtual machine, and within each of the instances a number of threads can be activated to a maximum of 10. This total is the number of results posted for a specific thread across all instances}\\
  
\small{Start-Time} & \small{The total number of virtual or physical instances performing work during an experiment.}\\
\small{End-Time} & \small{The minimum number of files processed by a worker thread across all instances}\\

\hline  \bottomrule
\end{tabular}
\caption{Metrics extracted from Monitor queue.}
\label{tab:monitormetrics}
\end{table}

\newpage
\subsubsection{Experimental Parameters}

\par Given the number of variables within the experimental setup, an exhaustive experimental approach would involve thousands of experiments. In total approximately 100 experiments were conducted with the primary aim of identifying the key variables within the system and to use that to maximise the performance of the NIMBUS pipeline within the given constraints of cost. The following seven variables were identified and where possible controlled for within each experiment.

\begin{itemize}

\item \textbf{VAR1: Web Server Location.} There are potentially significant impacts associated with the location of the web server providing files to the pipeline. While the configuration of 6 of the web servers was similar, the network between them and the workers was quite different. Network performance could have significant impact on the performance of a worker.  The locations of the web servers has already been highlighted in Figure \ref{fig:queuecontrols}. For most experiments, the AWS US East (Virginia region) web servers were used while the remaining web servers provided scaling options for larger scale experiments. The fastest performing web server was provided on loan from HEAnet \footnote{HEAnet is Irelands National Education and Research Network}, a high performance balanced web server highly tuned for large data transfer when the experimental requirement was to eliminate the web server performance from an experiment. This resource was used in a number of experiments producing the highest system performance. 
\item\textbf{VAR2: Number of web servers.} Initial testing reviewed the impact of increasing the number of servers that served files to workers. Multiple servers allows for the testing of the network performance when running small numbers of workers. There are 8 Nginx web servers and 1 FTP cluster used.  

 \item \textbf{VAR3: Number of Instances.}  An instance is a virtual or physical machine running a unix environment capable of running 1 or more workers. The instances are primarily AWS EC2 ubuntu machines and the number run in an experiment ranged from 1 to 100. 
		    
 \item\textbf{VAR4: Type of Instance.} The EC2 instance types available range in size and configuration. Experiments can be run with a number of EC2 instance types such as T1.Micro, T1.Large, M1.Large, M1.XLarge and C1.Xlarge. For a subset of experiments two physical machines were also used. A Sun/Oracle x4150 and an IBM i326e Server.  Details of the specification of these given in Appendix Table \ref{tab:instancespecs}

\item\textbf{VAR5: Workers per Instance.} For both single and multicore systems the use of multiple threads running independent workers allows for an investigation of the balance of CPU and Networking resources on an instances. Instances with more CPU cores should potentially improve the performance of the system if multiple worker instances are run. 

  \item\textbf{VAR6: Experimental Time.} Most  experiments ran for 20 minutes which was deemed long enough for the processing rate to be determined, and to ensure that a large volume of data could be processed. Some exceptions to this occurred near the final experiments to demonstrate that the processing rate could indeed be maintained over longer periods of time. 
   \item\textbf{VAR7: Batch Size.} The C program which processes images can process all files it finds in a directory. If there is a single data file found then the file is read in along with the Master Files. The images are processed, the results posted and the C program terminates and is restarted for the new batch. If the batch size is 10 then the Master Files are read in and held in memory while each of the 10 files are processed. This reduces the number of times the C program is started and stopped and reduces the file I/O by only reading the Master Files once per batch instead of once per data file.

\end {itemize}

\section{Results and Discussion}

\par This Chapter defined and executed a series of experiments to determine the overall performance of the NIMBUS system architecture presented in this chapter. The results of each experiment are broken down and analysed to provide a comprehensive view of the system. In some cases the experiments cover multiple components of the system, but in all cases the context of the results and their contribution to the overall pipeline are discussed. There are four basic sets of experiments performed, message queue performance, single and multi node instance scaling, and pipeline limit testing. Table \ref{tab:nimbusexp} shows the high-level experiments performed which are further broken down within this chapter. The web queues must be fast enough to service high levels of concurrent requests and the globally distributed computing nodes should be able to scale linearly until bottlenecks are observed. If bottlenecks are found then the architecture should be flexible enough to work around them.  All data sources for all experiments and graphs are identified in Appendix Table \ref{tab:datasets} which references an accompanying supplementary USB disk which contains raw and processed data relevant to these experiments.

  \begin{table}
\centering
\begin{tabular}{p{2cm} p{3cm} p{9cm}}
  \toprule
Reference & Measure & Description \\
  \midrule
\textbf{ \small{Exp:NIM1}}& \small{SQS Performance.} & \small {Testing the read and writing times of the web message queues}\\
\textbf{ \small{Exp:NIM2}}& \small{Single Instance Node Performance.} & \small{ Determine the variables which affect the performance of the overall processing power of a single instance.} \\
\textbf{ \small{Exp:NIM3}}& \small{Multi-Instance Node Performance.} & \small{ Focus on scaling the number of instances up to 100 looking for factors which could affect the scalability of the system.} \\
\textbf{ \small{Exp:NIM4}}& \small{System Limits.} & \small{ Identify the full scalability of the pipeline and to identify strategies to continue improving the system performance} \\

\hline
  \bottomrule
\end{tabular}
  \caption{NIMBUS Experiments Overview }
\label{tab:nimbusexp}
\end{table}

\par Limits imposed on the experiments were based on limits of available resources, although where possible indications of scaling opportunities were identified. For the pipeline to be active a minimum of $1$ worker is required to perform image cleaning and reduction. Multiple workers processes can run on a worker node/instance, which is typically a virtual AWS instance. The maximum number of instances activated within the experiments was $100$, but the maximum number of workers was $10,000$.  In some cases multiple runs of the same experiment were performed to ensure results were repeatable. 

\par It is required that the processing rates used within these experiments are expressed correctly and consistently. While the original data set is stored as multiple images per file, and the raw results from the experiments were measured as files per second, or images per second, a more useful representation of the processing rates is the amount of data processed over time. The conversion from files to bytes also needs to take into account that the data being processed is compressed so the concept of equivalent uncompressed data rate is also given. The following values and calculations are central to correctly determine processing rates. 

\begin{itemize}

\item \textbf{ $F$} Uncompressed File Size. An unprocessed image data cube is $7.297920$ MB.
\item \textbf{ $F_{c}$} Compressed File Size. An unprocessed compressed data cube is $1.247040$ MB.
\item \textbf{$F_{i}$} Images per File. The number of images within an image data cube is $10$. 
\item \textbf{$I$} Images size. An unprocessed compressed image.
 	\begin{equation}
  		I = \frac{F_{c}}{F_{i}}
	\end{equation}

\item \textbf{$C$} Compressed Rate. The size reduction of an image using the \emph{fpack} utility. 
	\begin{equation}
		C= \frac {F_{c}}{F}  
	\end{equation}
\item \textbf{  $P_{fps}$} Processing rate in files per second. The number of files processed per second using the NIMBUS pipeline. 
\item \textbf{ $P_{gps}$} Processing rate in GB per second.The number of gigabytes processed per second of compressed data using the NIMBUS pipeline. 
\begin{equation}
P_{gps} = \frac {P_{fps} *F_{i} * I }{1024}  
\end{equation}
\item \textbf{$\widehat{P}_{gps}$} Processing rate in equivalent uncompressed GB per second.  The number of gigabytes processed per second of equivalent uncompressed data using the NIMBUS pipeline. 
\begin{equation}
\widehat{P}_{gps} =  \frac{ P_{gps}}{C}
\end{equation}
\item \textbf{ $\widehat{P}_{gph}$} Processing rate in equivalent uncompressed GB per hour. The number of gigabytes processed per hour of equivalent uncompressed data using the NIMBUS pipeline. 
\begin{equation}
\widehat{P}_{gph} =  \widehat{P}_{gps} * 3600 
\end{equation}
\item \textbf{ $\widehat{P}_{tph}$} Processing rate in equivalent uncompressed TB per hour. The number of terabytes processed per hour of equivalent uncompressed data using the NIMBUS pipeline. 
\begin{equation}
\label{EQ:etbh}
\widehat{P}_{tph} =  \frac{\widehat{P}_{gph} }{1024}
\end{equation}
\end{itemize}


\subsection{Simple Queue Service (SQS) Performance}
\par To achieve a data cleaning rate of terabytes per hour it is essential that the queuing mechanism is able to advertise data sufficiently quickly to present work at a rate higher that the expected cleaning rate, and to ensure that work creation rates are expandable as the number of files to be cleaned increases. This requires that the storage nodes within the NIMBUS architecture can collectively create messages on the SQS worker queue at a rate of over 100 messages per second\footnote {100 messages for a 10 image data cubed file represents, in this system, 700Megabytes of raw data where each message points to a file of 7MB. 700MB per second $\approx$ 2.4 Terabtyes per hour}.  In addition to writing messages to the queue to generate work, the architecture of the system requires that queues are also used for monitoring and obtaining the results of an experiment. Experiments were devised to determine the sqs queue read performance. A full list of the sqs experiments are given in Table \ref{tab:queueexp}. 

  \begin{table}
\centering
\begin{tabular}{p{2.5cm} p{3.8cm} p{7.8cm}}
  \toprule
Reference & Measure & Description \\
  \midrule
\textbf{ \small{Exp:NIM1-1}}& \small{SQS Write Performance Single Node} & \small {Testing the writing time of a single storage node using a series of threaded applications  }\\
\textbf{ \small{Exp:NIM1-2}}& \small{SQS Write Performance Multi-Node} & \small {Testing the performance of the queue when multiple sources writing to it  }\\
\textbf{ \small{Exp:NIM1-3}}& \small{SQS Distributed Read Performance} & \small {Testing the distribution of messages from multiple nodes and the impact this has on queue read performance}\\
\textbf{ \small{Exp:NIM1-4}}& \small{SQS Queue Read Rates} & \small {Testing the read speed of an SQS queue using  a series of threaded approaches }\\

\hline
  \bottomrule
\end{tabular}
  \caption{NIMBUS SQS Performance Experiment Overview }
\label{tab:queueexp}
\end{table}

\subsubsection{Exp:NIM1-1 SQS Write Performance Single Node}

This test investigates the performance of a single server writing to an SQS queue to understand how quickly a storage node can advertise files to be cleaned. A series of experiments was executed on a single storage node with modifications to the utility used to write to an SQS queue. The storage node platform used was the IBM i326 server with 4GB of RAM and a 1 Terabyte remote mounted iSCSI storage drive. 

\par
\begin{figure}[htbp]
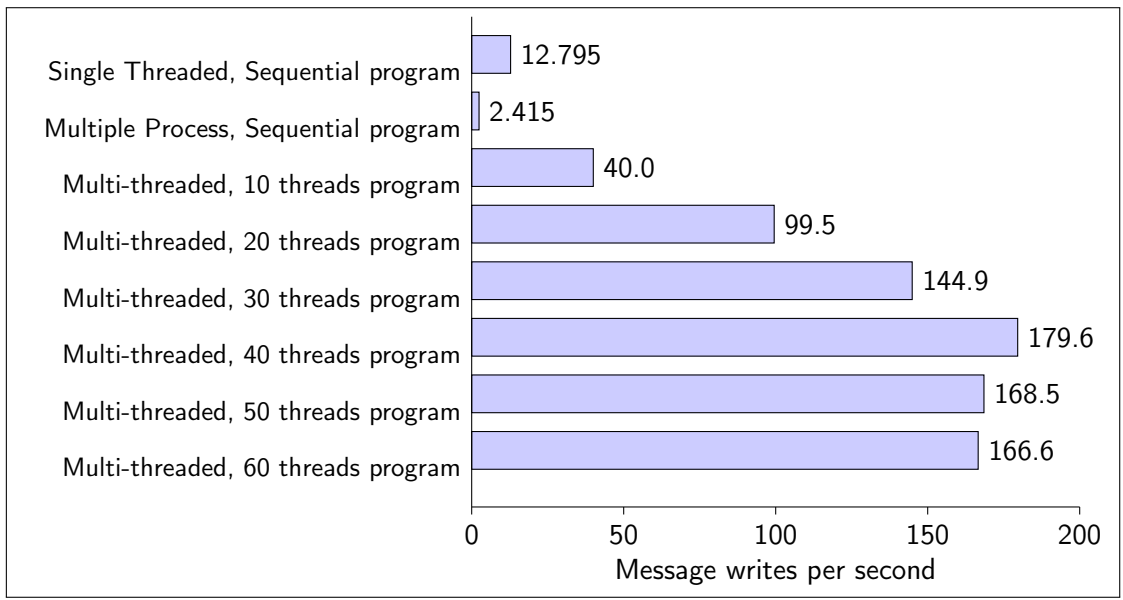
 
\centering
\fbox{      \begin{bchart}[step=50,max=200]
        \bcbar[]{12.795}
        \bclabel{\small{Single Threaded, Sequential program}}
        \smallskip
 
         \bcbar[]{2.415}
        \bclabel{\small{Multiple Process, Sequential program}}
        \smallskip

         \bcbar[]{40.0}
        \bclabel{\small{Multi-threaded, 10 threads program}}
        \smallskip
        
          \bcbar[]{99.5}
        \bclabel{\small{Multi-threaded, 20 threads program}}
        \smallskip
 
         \bcbar[]{144.9}
        \bclabel{\small{Multi-threaded, 30 threads program}}
        \smallskip

         \bcbar[]{179.6}
        \bclabel{\small{Multi-threaded, 40 threads program}}
        \smallskip

          \bcbar[]{168.5}
        \bclabel{\small{Multi-threaded, 50 threads program}}
        \smallskip
  
            \bcbar[]{166.6}
        \bclabel{\small{Multi-threaded, 60 threads program}}
        \smallskip

 \bcxlabel{Message writes per second}

    \end{bchart}}
    \caption{Exp:NIM1-1 SQS Write Performance Single Node. Average message writing time per second from a single web server node using varying methods}
  \label{fig:writetime1}

\end{figure}

 Four different approaches were used for writing messages. Sequential writing using a single threaded application, sequential writing spawning a new process per message write, multi-threaded processing for varying numbers of threads and connections to the SQS system.  The results are given in Figure \ref{fig:writetime1}.

\par The multi-threaded application used can be found in the Appendix \ref{code:sqs-writer}. Further description of the various writing approaches are described below. 

	\begin{description}
	\item [Single Threaded, Sequential program]  \hfill \\
	This approach uses a single Python program which connects to the SQS queue and writes, in strict sequence, 1,000 messages. The message writing rate is the time on average for a single message to be written. This time takes into account the connection time which is spread evenly across all of the writes. As more messages are written the connect time to the queue would become less significant. 
	\item [Multiple Process, Sequential program]  \hfill \\
	Using the Linux environment to create multiple processes, a program was written which forked 1 processes per message. The poor performance of this approach can be attributed to the creation time of the process and the fact that every single message requires a new connection to be made to the SQS queue. Each forked process also requires its own memory footprint within the server. 
	\item [Multi-threaded]  \hfill \\
	A multi-threaded program where 10 threads are created, each of which writes $\frac{1}{10}$ of the messages to the queue. In this case there is no cost for creating a full forked process for the entire program but 10 connections are set-up to the SQS queue, one per thread. 
	\item [Multi-threaded single connection]  \hfill \\
	Further optimisation of the Python program allows for the multi-threaded system to spawns an arbitrary number of processes. For the purposes of this experiment a set was selected ranging from 20-60 threads which the results posted. In this program there is a single connection made to the SQS which is shared by each of the threads. A Python queue is created which loads all image filenames into it and allows the threads to use that queue to generate messages. Threads now balance their work rate using the Python queue, with faster threads writing more messages. This modification also allows the system to take arbitrarily large numbers of files to the Python program. (see Code Listing in  Appendix: \ref{code:sqs-writer}. Results for varying threaded values are shown in Figure \ref{fig:writetime1} with the optimal performance shown around 40 threads. For the system tested this result provided the fastest message writing rate but this thread number is not the optimal number of threads for all systems, as that would be a factor of the number and performance of CPUs in the system.  	
	\end{description}

\par The single connection with multiple threads allowed for almost 180 messages written per second which on its own is greater that the required writing performance to generate a work list of over 3TB of data per hour. While a larger server may perform even faster, the next experiment focused on having multiple servers write to the queue at the same time from different locations. 

\subsubsection{Exp:NIM1-2 SQS Write Performance Multi-Node}

\par The results from Exp:NIM1-1 show a message creation rate of approximately  180 messages per second, however the testing environment relied upon a local data store from a single node. Further experimentation was required to consider the effect of  different network storage configurations and server types within the NIMBUS architecture, and the effect of multiple nodes concurrently writing to the SQS queue form different locations. This experiment is designed to test the scalability of the SQS queue to determine what message writing limitations may exist or may be relevant to this pipeline. 

\par The previous approaches implemented strict sequential message writing per thread, and due to the fact that the threads were on the same system, an element of performance balancing between threads is to be expected by the underlying operating system. If the threads were running on different storage nodes then a better understanding of the SQS central queue performance can be obtained. Of primary concern is the rate of message writing performance when the number of overall processes writing messages to a single queue is increased.   Using our multi-threaded Python message writing script on a number of different web nodes two things are testable. The first is to see if the overall message write rate for the queue will increase as the number of nodes writing is increased, and secondly will the rate of increase be a factor of the number of nodes running.

\par To exploit the distributed nature of the SQS queue in Exp:NIM1-2, 8 web servers were configured, each containing image data which can be advertised to the worker queue. This experiment attempts to see if a maximum write time could be obtained for the queue. Each of the 8 web servers contained approximately $73,000$ raw image files. Initially they were run in sequence to get a baseline of the speed at which they could write messages to the queue, next all of the nodes were run in parallel with the primary objective of all 8 writing messages onto the SQS queue concurrently. Given that each of the configurations and locations of the web servers were different as shown in Figure \ref{fig:queuecontrols}, the network connectivity to the SQS queue was varied. It would be expected that the difference in the performance of writing would arise as a result of issues such as network latency and processor performance.

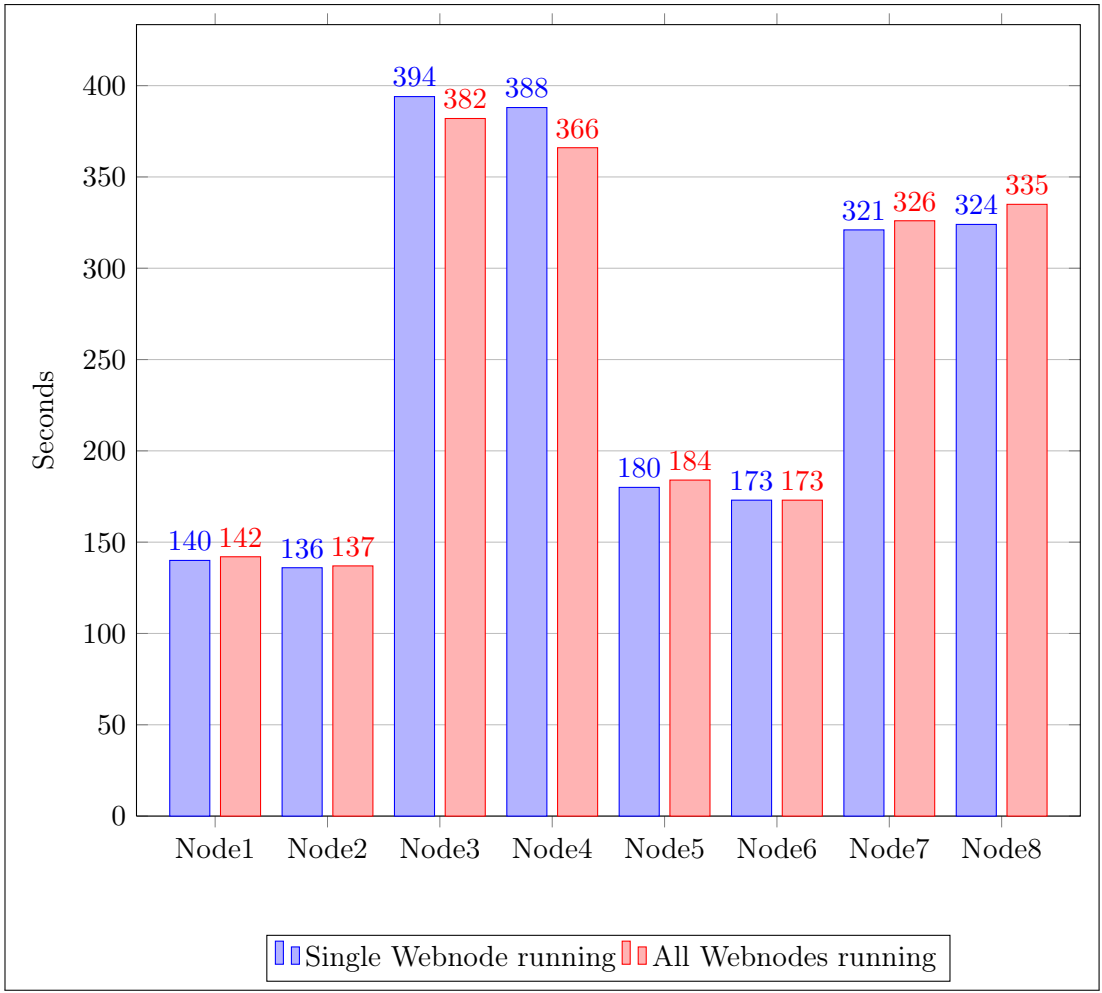
\begin{figure}[htbp] 
\centering
\fbox{     
\begin{tikzpicture}
\begin{axis}[
ylabel=Seconds,
enlarge x limits=0.1,
legend style={
    at={(0.5,-0.15)},
    anchor=north,legend columns=-1 },
ymin=0,
ybar=4pt,
xtick=data,
symbolic x coords={Node1, Node2, Node3,Node4,Node5,Node6,Node7,Node8},
grid=major,
 bar width=15pt,
width=14cm,
xmajorgrids=false,nodes near coords,
every node near coords/.append style={anchor=mid west,rotate=70,font=\tiny}
]

\addplot coordinates {(Node1,140) (Node2,136) (Node3,394) (Node4,388) (Node5,180) (Node6,173) (Node7,321) (Node8,324 )};

\addplot coordinates {(Node1,142) (Node2,137) (Node3,382) (Node4,366) (Node5,184) (Node6,173) (Node7,326) (Node8,335)};

\legend{Single Webnode running,All Webnodes running}
\end{axis}
\end{tikzpicture}
}
    \caption{Exp:Nim1-2 Comparison of sqs queue write rates per second web node  for standalone or multi-node writing.  Source Data in Appendix Table \ref{tab:queuewrites}}
  \label{fig:multiplewebnodes}

\end{figure}

\par In Figure \ref{fig:multiplewebnodes} the message write time is shown for a storage node run in isolation, and when all of the storage nodes are running. It can clearly be seen that the performance of the SQS queue was a function of the number of nodes. As the number of nodes increased, the number of messages written also increased. Running in isolation, or with other storage nodes running and writing messages, the performance of storage nodes writing was relatively unchanged.  Using the longest write time for a single node, an estimate of the average message write performance $W_{msg}$ can be obtained where $n$ is the number of nodes used, and $\max T$ is the maximum time in seconds for one of the nodes to complete writing. The total number of messages $M$ written by all 8 nodes in this example is 596,3658 with the last node finishing after 9 mins giving a message processing rate of 1104.5 messages per second (See Appendix Table \ref{tab:queuewrites} for data details) which is equivalent to 26TB of data advertised per hour.

\begin{equation}
\widehat{W}_{msg} = \frac{M} {\max T}
\end{equation}

\pgfplotstableread{Data/CanaryQ/writerate-cumulative.prn}
\datatable

\begin{figure}[htbp] 
\centering
\fbox {
\begin{tikzpicture}
    \begin{axis}[xmin=0,
    ymin=0,
    width=0.8\textwidth,
    legend style={at={(0.8,0.7)},anchor=north,legend cell align=left},
        xlabel=$Seconds$,
        ylabel=$Cumulative\ Messages\ Written$, no markers]

\addplot table[y = heanetC] from \datatable ;
\addplot table[y = webnode1C] from \datatable ;
\addplot table[y = webnode2C] from \datatable ;

\addplot table[y = webnode3C] from \datatable ;
\addplot table[y = webnode4C] from \datatable ;
\addplot table[y = webnode5C] from \datatable ;
\addplot table[y = webnode6C] from \datatable ;
\addplot table[y = webnode7C] from \datatable ;
\addplot table[y = webnode8C] from \datatable ;

\legend{$HEANet$,$Webnode1$,$Webnode2$,$Webnode3$,$Webnode4$,$Webnode5$,$Webnode6$,$Webnode7$,$Webnode8$}   

    \end{axis}

\end{tikzpicture} }
  \vskip -0.8em
    \caption{Exp:NIM1-2 Messages written to the queue over time for each storage node.}
  \label{fig:canaryqmsgovertime}
\end{figure}
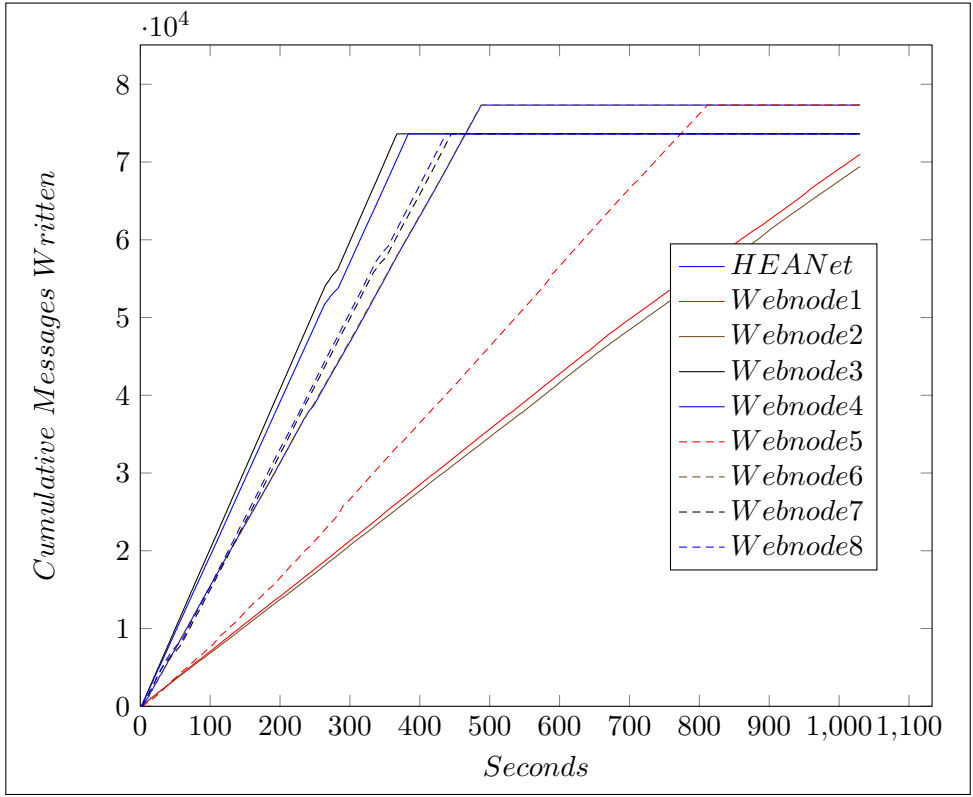

\par On consideration of the results presented in Figure \ref{fig:multiplewebnodes} there are clear variations in storage servers  writing rates but that introduction of additional servers writing to the queue does not impact the rate of writing for individual nodes. The message write rates for each server are shown in Figure \ref{fig:canaryqmsgovertime}.  The rate of writing for each node can be expressed as the number of messages $m$ written over time $t$. 

\begin{equation}
\overline{w} = \frac{m}{t} 
\end{equation}

A more accurate estimate of the system queue write rate is  the sum of $n$ storage nodes individual write rates, $\overline{W}$ as shown in equation \ref{eq:canaryqwriterate}.

\begin{equation}
\label{eq:canaryqwriterate}
\overline{W} = \sum_{1}^{n} \overline{w}
\end{equation}

\par The data used in these graphs was based on the retrieval of messages from the SQS canary queue, which was written to by the web nodes at the same time as messages are written to the SQS worker queue. This data was taken from an experimental run where data was being processed using 9 web nodes. Because  worker queue messages are deleted as they are processed the canary queue offered a mirror of the queue. When reading the canary queue the SQS attribute \emph{SentTimestamp} was requested which provides the \gls{utc}  for the message write time to the queue. Message write order is reconstructed  by combining all messages and sorting them into time buckets (in this case seconds), and for that time period determine the number of messages written from each web node. See code listing \ref{code:sqsmetric}. Figure \ref{fig:canaryqMA} plots the same data with a simple moving average with an interval of 9 using a central moving average. The raw data plot for this image can be seen in Appendix Figure \ref{fig:canaryqcumulativeraw}.  The reduction in the total write rate over time is due to the fact that the node writing faster have run out of messages, so  are no longer contributing to the system. 

\pgfplotstableread{Data/CanaryQ/writerate-MA9.prn}
\datatableA

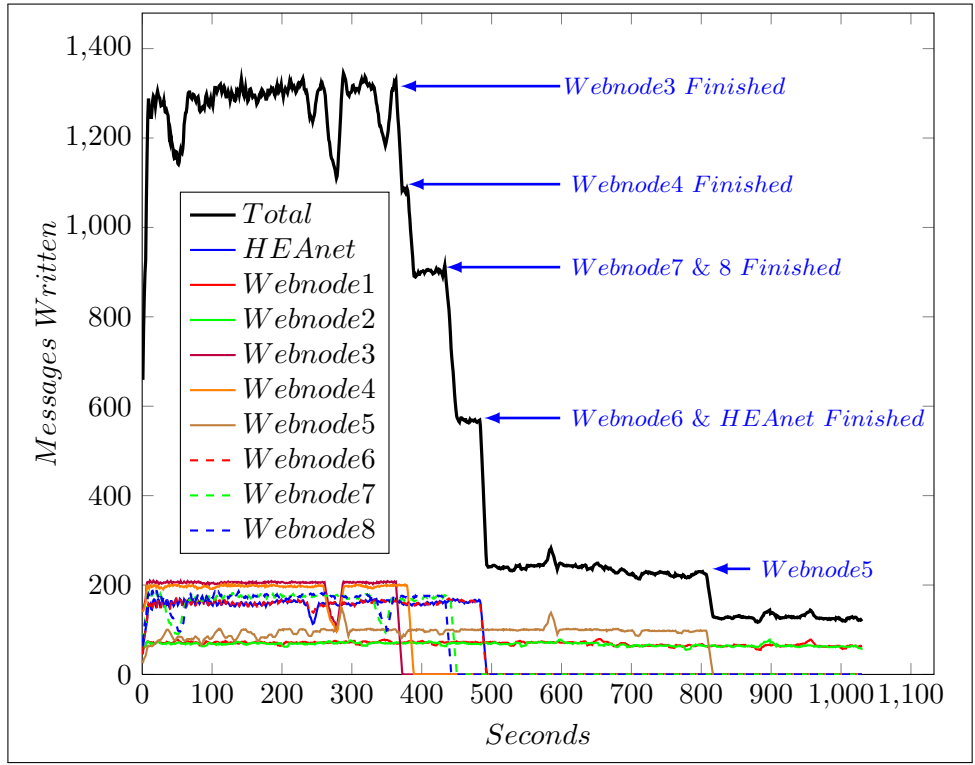
\begin{figure}[htbp] 
\centering
\fbox {
\begin{tikzpicture}
    \begin{axis}[xmin=0,
    ymin=0,
    width=0.8\textwidth,    legend style={at={(0.18,0.73)},anchor=north,legend cell align=left},
        xlabel=$Seconds$,
        ylabel=$Messages\ Written$, no markers]

\addplot[black,very thick] table[y = totalMA9] from \datatableA ;
\addplot [blue,thick] table[y = ftpMA9] from \datatableA ;
\addplot [red,thick] table[y = webnode1MA9] from \datatableA ;
\addplot [green,thick] table[y = webnode2MA9] from \datatableA ;
\addplot [purple,thick] table[y = webnode3MA9] from \datatableA ;
\addplot [orange,thick] table[y = webnode4MA9] from \datatableA ;
\addplot [brown,thick] table[y = webnode5MA9] from \datatableA ;
\addplot [red,thick,dashed] table[y = webnode6MA9] from \datatableA ;
\addplot [green,thick,dashed] table[y = webnode7MA9] from \datatableA ;
\addplot [blue,thick,dashed] table[y = webnode8MA9] from \datatableA ;

\legend{$Total$,$HEAnet$,$Webnode1$,$Webnode2$,$Webnode3$,$Webnode4$,$Webnode5$,$Webnode6$,$Webnode7$,$Webnode8$}   

    \end{axis}
\draw[-latex,blue,line width = 1pt] 
  (5.5,7.8) -- (3.4,7.8) 
  node[anchor = west, font=\footnotesize,xshift=2cm] {$Webnode3\ Finished$};
  
\draw[-latex,blue,line width = 1pt] 
  (5.5,6.5) -- (3.5,6.5) 
  node[anchor = west, font=\footnotesize,xshift=2cm] {$Webnode4\ Finished$};

\draw[-latex,blue,line width = 1pt] 
  (5.5,5.4) -- (4,5.4) 
  node[anchor = west, font=\footnotesize,xshift=1.5cm] {$Webnode7\ \&\ 8\ Finished$};

\draw[-latex,blue,line width = 1pt] 
  (5.5,3.4) -- (4.5,3.4) 
  node[anchor = west, font=\footnotesize,xshift=1cm] {$Webnode6\ \&\ HEAnet\ Finished$};

\draw[-latex,blue,line width = 1pt] 
  (8,1.4) -- (7.5,1.4) 
  node[anchor = west, font=\footnotesize,xshift=0.5cm] {$Webnode5$};

\end{tikzpicture} }
  \vskip -0.8em
    \caption{Exp:NIM1-3 Simple moving average of canary queue SQS message \emph{write rate} for all storage nodes. }
  \label{fig:canaryqMA}
\end{figure}

\subsubsection{Exp:NIM1-3 SQS Distributed Read Performance}

\par As the SQS queue does not guarantee any specific order to message delivery it is  important to understand the mix of messages on the queue as presented to multiple readers, and to have an appreciation of how that can affect a running system. From results presented in Figure \ref{fig:canaryqMA} it can be seen that the system wide message write rate reduces for the experiment as faster web nodes exhaust the number of messages they have to put on the queue. If the queue delivered messages in the same order as received then the same distribution of messages would be expected when the queue is being read. A complication is that there is no single read point for this queue as messages are read by multiple nodes. Given the distributed nature of the queue the message order will now be split over these reading nodes. The SQS system does not provide a message read timestamp so this cannot be used to reconstruct message read order. It is possible however to reconstruct an approximation  of the message delivery order to the reading worker nodes. 

\par To simulate the worker node read behavior all messages are read using a multithreaded Python program\footnote{Code Listing \ref{code:sqs-canary-reader}} which spawn 40 threads, each of which creates it's own log file that preserves the order in which the messages were read.  The population standard deviation of the number of messages read by each thread is calculated \footnote{$\sigma = \sqrt{\frac{1}{N} \sum_{i=1}^{N}(x_{i}-\mu)^{2}}$} and the relative standard deviation is obtained, which in this case is $0.25\%$. It can be assumed that each of the threads is downloading messages at a similar rate, so by grouping each of the files into bins of 100 messages and combining them\footnote{Code listing \ref{code:sqs-canary-metrics}}, It can be determined for each bin how many messages were downloaded from each storage node.  Figure \ref{fig:canaryqreads} plots this data using a simple moving average with an interval of 17 with a central moving average. Consider webnode3 which wrote messages to the queue faster than any other node, it provides consistently higher rate of messages early on in the graph and has all messages read before other nodes. The slower webnode1 starts slower and finishes last.  There is evidence of more random behaviour midway through the read process. This may be attributed to a combination of the distributed nature of the SQS queue and the method used for combining messages logs. 

\par While it has been shown that the read order is not identical to the write order it is reasonable to assume that nodes which contribute messages to the queue at a higher rate than other storage nodes will have their messages presented to worker nodes more frequently than those that contribute at a slower rate.

\pgfplotstableread{Data/CanaryQ/readdistrib.prn}
\datatable

\begin{figure}[htbp] 
\centering
\fbox {
\begin{tikzpicture}
    \begin{axis}[xmin=0,
    ymin=0,
    width=0.8\textwidth,
    legend style={at={(0.2,0.9)},anchor=north,legend cell align=left},
        xlabel=$Slot$,
        ylabel=$Messages\ Read$, no markers]

\addplot[blue] table[y = ftpMA9] from \datatable ;
\addplot [red, very thick] table[y = webnode1MA9] from \datatable ;
\addplot table[y = webnode2MA9] from \datatable ;

\addplot[black,very thick] table[y = webnode3MA9] from \datatable ;
\addplot table[y = webnode4MA9] from \datatable ;
\addplot table[y = webnode5MA9] from \datatable ;
\addplot table[y = webnode6MA9] from \datatable ;
\addplot table[y = webnode7MA9] from \datatable ;
\addplot table[y = webnode8MA9] from \datatable ;


\legend{$HEANet$,$Webnode1$,$Webnode2$,$Webnode3$,$Webnode4$,$Webnode5$,$Webnode6$,$Webnode7$,$Webnode8$}   

    \end{axis}

\end{tikzpicture} }
  \vskip -0.8em
    \caption{Exp:NIM1-3 Simple moving average representation of messages read within 350 sequential time slots with 40 threads reading the SQS canary queue.}
  \label{fig:canaryqreads}
\end{figure}
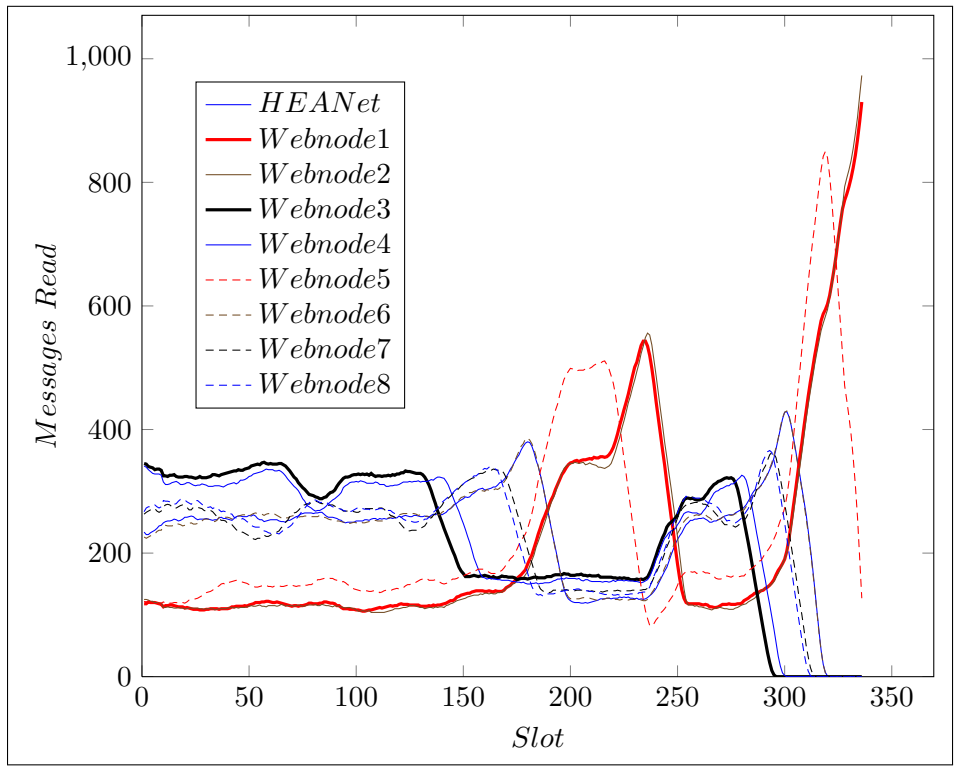

\subsubsection{Exp:NIM1-4 SQS Queue Read Rates}

The issues of reading queue message in a none distributed or parallel manner are consistent will the issues associated with writing messages to a queue. For each message read there is a connection overhead, but in addition to this there is also the overhead of deleting the message from the queue creating a double cost to message reading. When a message is read, a visibility timeout it set leaving the message invisible to others. Messages within the visibility timeout are referred to as a messages in flight, and the limit for the number of messages in flight is set currently at $120,000$. This provides a potential  limit on the performance of a system using a single worker queue. A system which has the capacity to read more that that number of messages at one time, will be blocked reading messages until the total number of messages in flight is reduced below that threshold. This requires that either a message is returned to the queue or deleted by the reading worker. 

\par While this is a limitation of the AWS SQS service it is not necessarily a limiting factor on the pipeline as it is possible to operate a larger number of queues, each containing file information for processing. If a pipeline was limited to operating at $120,000$ messages per second this would equate to $1.2 Million$ images per second using image data cubes of 10 stacked images, which in this pipeline would represent about 1 Terabyte per second, or 2.8 Petabytes per hour. 

\par The issue of reading queues quickly is more related to the processing of post experiment analysis than with the running of the system. With the pipeline's distributed nature, and the use of batch downloading per worker node, the processing time of the images is the limiting factor for the image processing rate, not  the message download time for each worker.  The use of the queues for distributed sharing of log files and result files however does require some thought on queue read performance. The reason  a queue system is used to record result information and log file information is that it allows the worker nodes to be more independent. By centralising key information about their processing to a central queue, all of the pipeline log messages are available in a single queue, although there is no specific reason why multiple queues could not be used. 
\par If the log files are being monitored for issues with workers or with processing then they need to be constantly read and monitored with key value pairs \footnote{A key/value pair could be a specific pattern within the log file data indicting an unusual or important state for a worker, such as WorkerID:STOP} being sought to identify issues. The read rate of a log file queue must therefore have a similar read rate to the worker processing rate. A monitoring process reading the queue will be doing considerably less processing that a worker node, so it should be able to process messages considerably faster, or allow for multiple monitors to operate simultaneously (which is the case). A single monitoring server reading queue messages can read approximately $100$ messages per second when running multiple threads, as shown below in Figure \ref{fig:msgreader}. Multiple monitoring servers could therefore reasonably be assumed to be able to read all log or result messages produced by a fully functional  operating pipeline. Using Equation \ref{eq:lograte} it can be estimated that the message read rate would have to be slightly greater than twice the message processing rate of a system since each worker generates just over 2 messages per image file processed. If a single monitoring system can read 100 images per second, then the total monitoring node requirements is the message processing rate of the system per second divided by 100.

\begin{figure}[htbp]
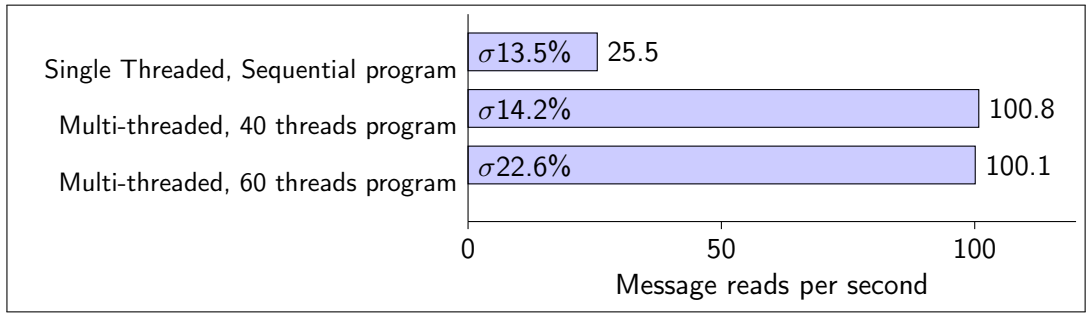
 
\centering
\fbox{      \begin{bchart}[step=50,max=120]
        \bcbar[text=$\sigma$13.5\%]{25.5}
        \bclabel{\small{Single Threaded, Sequential program}}
        \smallskip

         \bcbar[text=$\sigma$14.2\%]{100.8}
        \bclabel{\small{Multi-threaded, 40 threads program}}
        \smallskip
        
          \bcbar[text= $\sigma$22.6\%]{100.1}
        \bclabel{\small{Multi-threaded, 60 threads program}}
        \smallskip
 
 \bcxlabel{Message reads per second}

    \end{bchart}}
    \caption{Exp:NIM1-4. Messages read per second from a single monitor server node using varying levels of threads running with the standard deviation shown.}
  \label{fig:msgreader}

\end{figure}

\par A number of strategies were considered for log file message reading which are briefly discussed. 

\begin {itemize}

\item \textbf{Serial Download} requires a single-threaded application to read messages from the queue one at a time. It is possible to increase the number of messages taken from the queue at a single read, however this did not provide any significant performance improvements on the message read performance. This approach, as shown in Figure \ref {fig:msgreader} is not an optimal solution. 
\item \textbf{Multi-Threaded Download} provided considerable improvements in the overall message download rate with the number of threads being a configurable number similar to the Python listings showing message writing. In this case messages are deleted once they have been read so the queue is constantly reducing over time. 
\item \textbf{Multi-Threaded Download with messages in flight} is a faster message read given that the message is not deleted, but rather messages are given an exceptionally long visibility read time when downloaded. When a message is read, the time it remains off the queue can be set. If no messages are deleted then the message limit of $120,000$ is a system bottleneck after which no messages can be read until that number is reduced. This method only works for queues which are relatively small. In the experiments performed, queue messages can reach over $1$ Million messages. 
\item \textbf{Multi-Threaded Download with messages in flight hybrid} is a compromise solution which estimates the number of messages within the queue and has a policy of deleting a proportion of them to take advantage of the messages-in-flight mechanism, while never allowing the maximum number of messages to be in flight. For large queues however, the advantage is diminished over time. 
\end {itemize}

\newpage
\subsubsection{Analysis}
\par The message queuing system provides a number of advantages to the overall pipeline. 

\begin {itemize}

\item  \textbf{Exp:NIM1-1}. A single web node can advertise data at a rate of approximately 3TB per hour. 
\item \textbf{Exp:NIM1-2}. Using multiple web nodes writing at the same time, the advertising rate for the pipeline is over 26TB per hour, although this is unlikely the limit as write rates were linear with the number of web nodes added. 
\item \textbf{Exp:NIM1-3}. Due to the fact that some nodes may write to the queue faster than others, in the experiments where there are fixed numbers of messages written by each worker node, the faster nodes will initially contribute more to the queue and reduce over time. This may cause the process rate to appear to slowdown over time. 
\item \textbf{Exp:NIM1-4}. A single node read performance for messages is similar to the single node write performance. Downloading of messages is naturally distributed for  the pipeline. A limit per queue existing which is equivalent to a processing rate of 2.8 PB per hour. All that is required to overcome this is to increase the number of queues being used for reading. 
\end {itemize}

\subsection{Single Instance Node Performance}

\par This group of experiments focuses on understanding the behaviour of a single instance by looking at variables within the pipeline to determine how to optimise the instance performance. Give the low volume of data being processed the results of the baseline experiments are shown as  $P_{fps}$, files processed per second. 

\par There are three different variables which are tested in these experiments to determine their impact on a single instance performance, the size of the instance (Memory, Network, CPU), the number of worker threads running on the instance, and the web servers being used to provide images to clean.  All experiments run batches of size ten (downloading 10 messages before processing begins), and all work performed by a worker is identical. Experiments are broken down by variable under test and shown in Table \ref{tab:nim2-exp}

 \begin{table}
\centering
\begin{tabular}{p{2.5cm} p{3.2cm} p{8.3cm}}
  \toprule
Reference & Measure & Description \\
  \midrule
\textbf{\small{Exp:NIM2-1}}& \small{Single instance webserver performance} & \small {Testing the impact of different web server configurations to service a single worker instance.  }\\
\textbf{\small{Exp:NIM2-2}}& \small{Single instance performance by type} & \small {Testing the impact of selecting different instance machine types.   }\\
\textbf{\small{Exp:NIM2-3}}& \small{Single instance multi-worker performance} & \small {Testing the impact of increasing the number of worker threads run on an worker instance.}\\

\hline
  \bottomrule
\end{tabular}
  \caption{NIMBUS Single Instance experiments}
\label{tab:nim2-exp}
\end{table}

\subsubsection{Exp:NIM2-1 Single Instance Webserver Performance}

\par
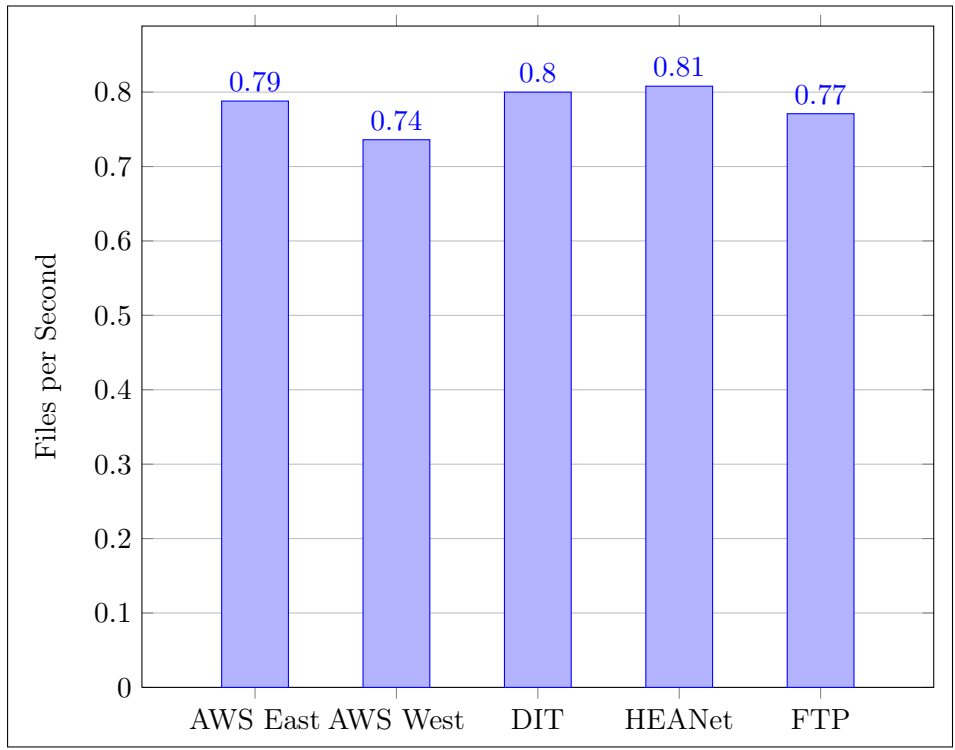
\begin{figure}[htbp]
\centering
\fbox{     

\begin{tikzpicture}
\begin{axis}[
ylabel=Files per Second, enlarge x limits=0.2,
legend style={ at={(0.5,-0.15)},  anchor=north,legend columns=-1 }, ymin=0, ybar, xtick=data, symbolic x coords={AWS East, AWS West, DIT, HEANet,FTP },
grid=major, 
bar width=25pt,
width=12cm, xmajorgrids=false,nodes near coords,every node near coords/.append style={anchor=mid west,rotate=70}]
\addplot coordinates {(AWS East,0.788) (AWS West,0.736)  (DIT,0.8) (HEANet,0.808) (FTP,0.771) };
\end{axis}
\end{tikzpicture}
}
  \caption{Exp:NIM2-1 Files Per Second: Varying Web servers and their impact on T1.Micro Instance performance }
  \label{fig:singleinstances2}

\end{figure}

 For this experiment the web servers used are those identified earlier in this chapter, and shown in Figure \ref{fig:queuecontrols}. NginX web servers were run in multiple location, on both physical (DIT \& HEANET) and virtual environments (AWS EAST containing 1 webserver, and AWS WEST which had 4 webservers), in addition to a high performance FTP server from HEAnet which also serviced HTTP requests. The worker instance was run on an AWS EC2 instance and the instance type was T1.Micro, which is the smallest of the free tier system available. The worker instance ran 5 worker threads. Figure \ref{fig:singleinstances2} show the files per second processing rate of the instance (combining the processing of each of the individual workers). From this graph it would appear that the web server location, number of web servers and type has little impact on the performance of the server instance. 

\par
\begin{figure} [!ht]
\centering
\fbox{   \includegraphics[width=0.95\textwidth] {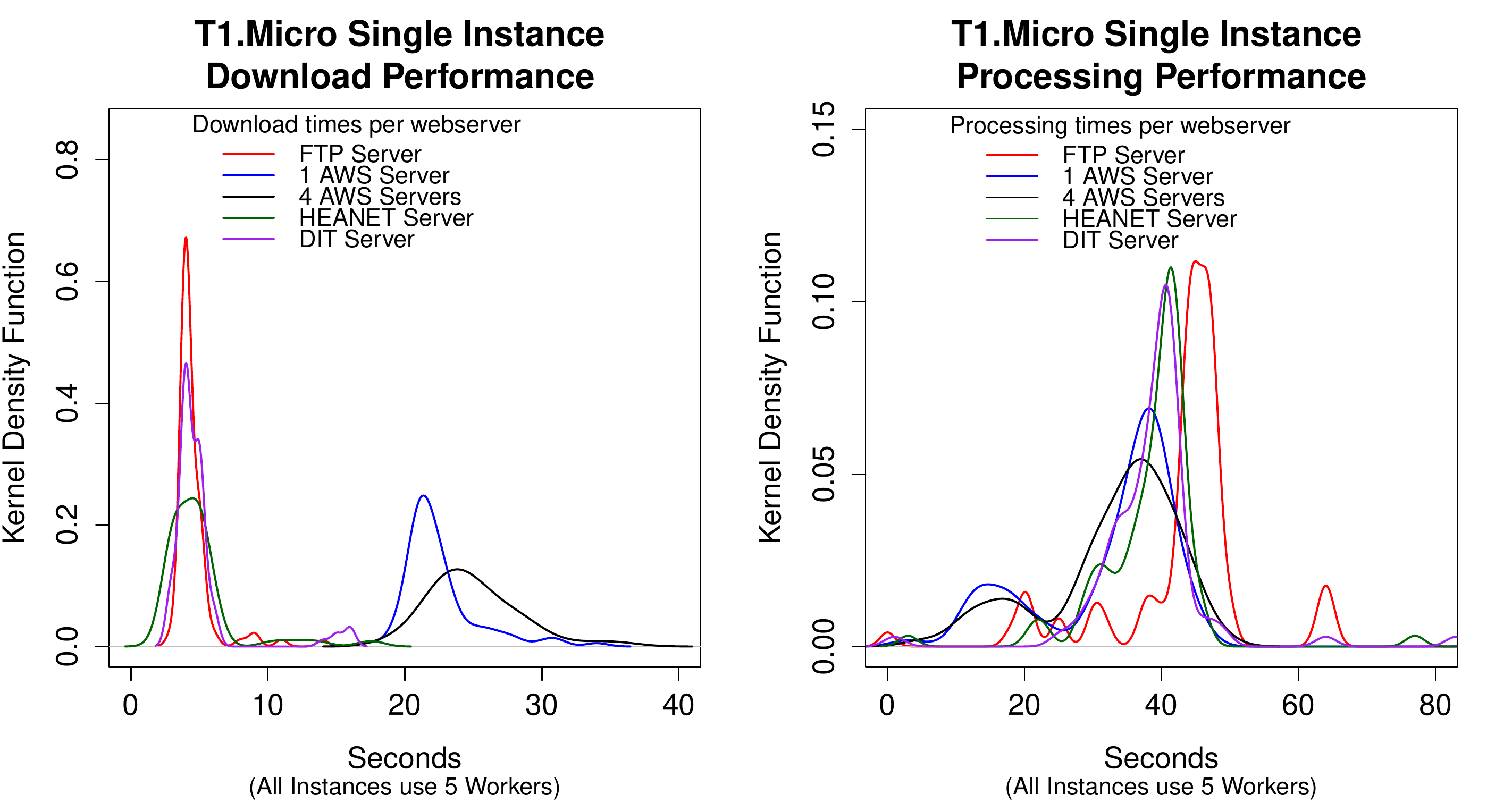}  }
  \vskip -0.8em
    \caption{Exp:NIM2-1 Single Instance Performance breakdown for different Web Servers, with 5 workers.}
  \label{fig:si-instances-breakdown-5w}
\end{figure}

 A more detailed look at the breakdown between the time to download and the time to process provides further information as shown in Figure \ref{fig:si-instances-breakdown-5w}. Using a kernel density function to plot the download times for each worker during an experiment, it is clear that the AWS web servers takes longer to download files compared to the DIT, HEANet or FTP web servers. Given that there are no other demands being placed on the web servers these results are good indicators of the maximum performance capabilities of the worker instance. In Table \ref{tab:t1.microresult1-sd-var}  the mean, standard deviation and variance for file downloads is presented,  showing a clear difference in the mean for the AWS servers. It is also evident from the low variance that the web server response times are reasonably consistent. In Table \ref{tab:t1.microresult1-sd-var-2} the mean processing times of the  workers is reasonably similar, but the variance is quite high. Given that amazon report that the T1.micro instance has CPU throttling, this would appear to be evident with some workers experiencing considerable longer delays than the mean.  With the mean values for processing times higher for the experiments where faster downloads occur it would seem to indicate that all workers are more likely to be processing data at the same time. As processing times become longer, then the CPU is most likely closer to being close to a bottleneck.   

\par It can be concluded from these experiments that there is a performance benefit from using the non-AWS web servers for single instance processing. With limits inherent within the capability of the T1.Micro instance, these benefits my not be evident, however using faster performing server instances, improvements in file processing rates could be expected. This will be explored further in Exp:NIM2-2.

\begin{table}[htbp]
\centering
  \begin{tabular} {cccc}
  \toprule

  \multicolumn{1}{b{3.5cm}}{\begin{center}Web Server\end{center}} &
   \multicolumn{1}{b{3cm}}{\begin{center}Std. Dev.\end{center}} &
   \multicolumn{1}{b{3cm}}{\begin{center}Mean\end{center}} &
   \multicolumn{1}{b{3cm}}{\begin{center}Variance\end{center}} \\

\addlinespace[-5mm]

  \midrule
  FTP 		& 1.565   		& 4.59  & 2.451      \\
  AWS East (1)		& 2.759  		& 22.49  & 7.616     \\
  AWS West (4) 		& 3.278   		& 25.00  & 10.746      \\
  HEANet 	& 2.775   		& 4.92  & 7.704    \\
  DIT 		& 2.791   		& 5.05  & 7.795     \\

\hline  \bottomrule

  \end{tabular}
  \caption{Exp:NIM2-1 T1.Micro Single Instance 5 Worker statistics for image downloads per web server }

\label{tab:t1.microresult1-sd-var}
\end{table}

\begin{table}
\centering
  \begin{tabular} {cccc}
  \toprule

  \multicolumn{1}{b{3.5cm}}{\begin{center}Web Server\end{center}} &
   \multicolumn{1}{b{3cm}}{\begin{center}Std. Dev.\end{center}} &
   \multicolumn{1}{b{3cm}}{\begin{center}Mean\end{center}} &
   \multicolumn{1}{b{3cm}}{\begin{center}Variance\end{center}} \\

\addlinespace[-5mm]

  \midrule
  FTP 		& 19.59   		& 47.05  & 383      \\
  1 x AWS 		& 18.66  		& 36.17  &  348    \\
  4 x AWS 		& 9.70   		& 32.94  & 94      \\
  HEANeet 	& 23.05   		& 45.38  & 531    \\
  DIT 		& 23.18   		& 45.65  & 537     \\

\hline  \bottomrule

  \end{tabular}
  \caption{Exp:NIM2-1 T1.Micro Single Instance 5 Worker statistics for image processing per web server }

\label{tab:t1.microresult1-sd-var-2}
\end{table}

\subsubsection{Exp:NIM2-2 Single Instance Performance by Type}

\par
\begin{figure} [htbp]
\centering
\fbox{   \includegraphics[width=0.95\textwidth] {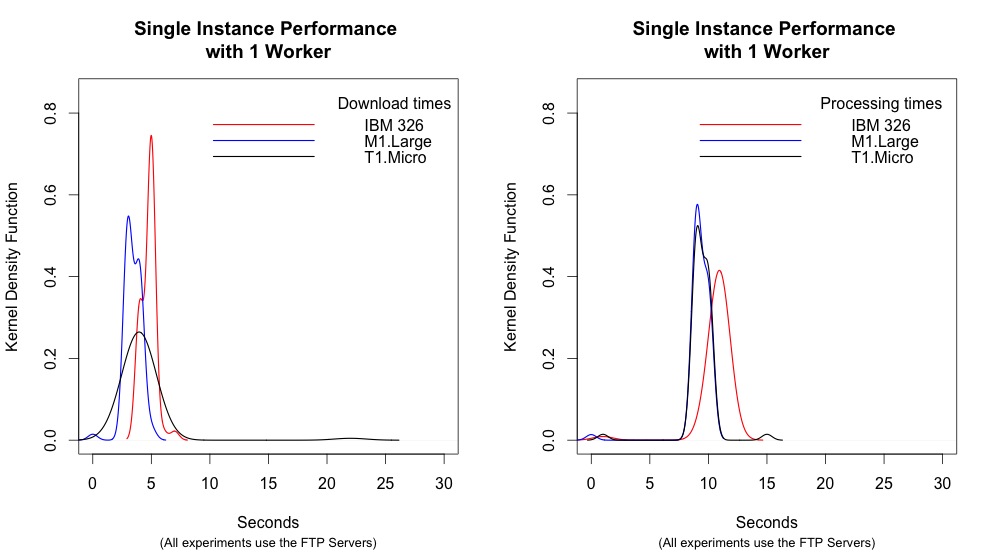}  }
  \vskip -0.8em
    \caption{Exp:NIM2-2 Single Instance 1 Worker Performance breakdown for different Instance types, using FTP Webserver}
  \label{fig:si-instances-breakdown-1w}
\end{figure}

 Given that the FTP web server from HEANet provides  a high performance load balanced web server,  that variable can remain constant for this next set of experiments. Looking at the performance of a single instance running a single worker in Figure \ref{fig:si-instances-breakdown-1w} it is clear that varying the type of server instance makes little difference. In this experiment 1 physical server was used (the IBMe326) and two different AWS instance types (T1.Micro, and M1.Large). There is a slightly longer delay in download times from the IBM server but this is likely to do with the network differences between the location of the virtual amazon instances and the physical IBM server instance. If the differences between the web servers is processing power then further experiments are required which increases the processing demand of the webserver. In Exp:NIM2-1 there were 5 workers running per server, while in this experiment there was only 1 worker running. 

\par
\begin{figure} [htbp]
\centering
\fbox{   \includegraphics[width=0.95\textwidth] {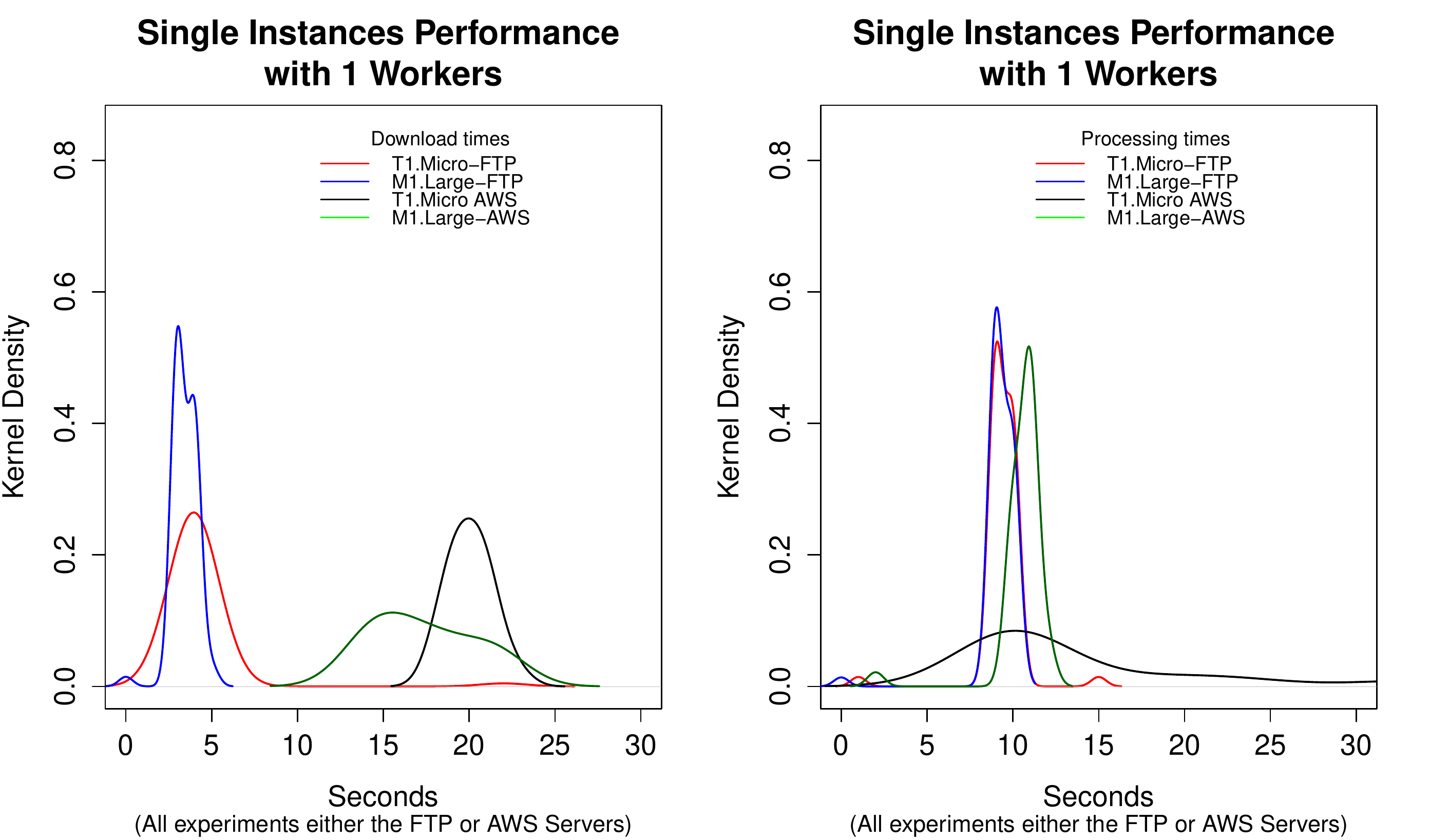}  }
  \vskip -0.8em
    \caption{Exp:NIM2-2 Single Instance 1 Worker Performance breakdown, varying web server and instance type }
  \label{fig:sip-breakdown-1w}
\end{figure}

 To verify that the web server would have an impact on the instance, a comparison of the T1.Micro and the M1.Large virtual instances performance using both the FTP and the AWS web server is shown in Figure \ref{fig:sip-breakdown-1w}. In this diagram the processing remains similar but the networking time for the downloads is clearly slower when using the AWS web servers. To further explore the instance type performance  the number of workers running to determine how to take advantage of the additional server performance must be considered. This will be explored further in Exp:NIM2-3.

\subsubsection{Exp:NIM2-3 Single Instance Multi-worker Performance}

\par  For this group of experiments a variety of physical and virtual machine instances are used to look at the impact of running multiple workers on the same instance. The assumption is that if an instance is busy downloading an image then the CPU resource is not being used. To fully utilise the CPU, additional workers can run to balance the load of the CPU over time. Workers are designed to cycle through downloading batches of files, processing them, and then uploading them. A single worker will not use all of the instance resources fully at the same time. By increasing the number of workers it would be reasonable to assume that the overall resources are being more fully used, but that there is  a point beyond which the number of workers being added does not increase the performance of the instance. 

\par
\begin{figure}[htbp] 
\centering
\fbox{     
\begin{tikzpicture}
\begin{axis}[
ylabel=Files Per Second, width=10cm,
enlarge x limits=0.25, bar width=0.75cm,
legend style={
    at={(0.5,-0.15)},
    anchor=north,legend columns=-1 },
ymin=0, ybar=0pt, xtick=data,
symbolic x coords={T1.Micro, M1.Large,IBM326e},
grid=major,  xmajorgrids=false,nodes near coords,
every node near coords/.append style={anchor=mid west,rotate=70,font=\tiny}
]
\addplot coordinates {(T1.Micro,0.511) (M1.Large,0.552)  (IBM326e,0.419) };
\addplot coordinates {(T1.Micro,0.735) (M1.Large,1.904)  (IBM326e,1.606) };
\addplot coordinates {(T1.Micro,0.771) (M1.Large,2.029)  (IBM326e,1.671) };

\legend{1 Worker, 5 Workers, 10 Workers}
\end{axis}
\end{tikzpicture}
}
  \caption{Exp:NIM2-3: Single instance performance by type, running multiple number of workers.}
  \label{fig:singleinstances-multipleworkers}

\end{figure}
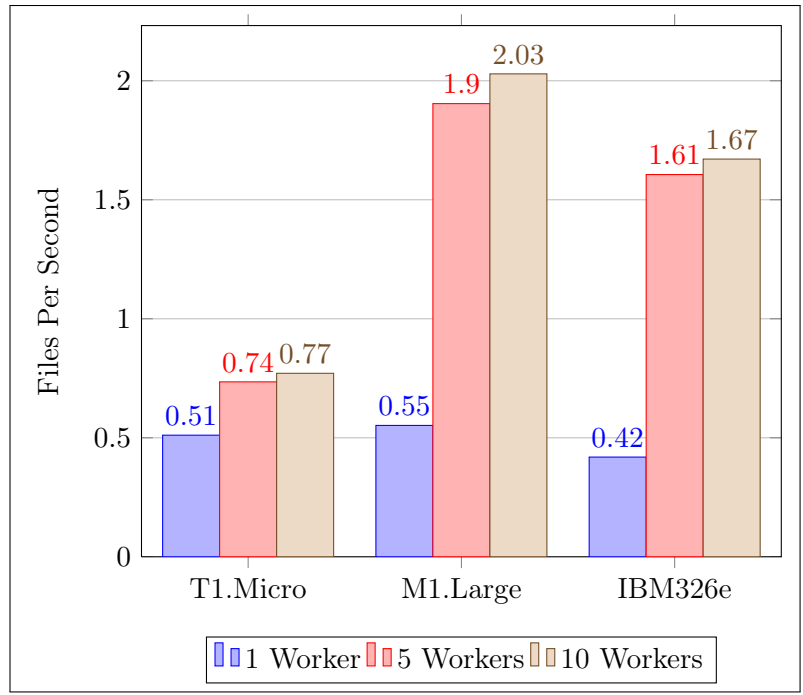

 Since different instances have different performance characteristics such as CPU, RAM and Networking, the performance characteristics should be varied across different server types and configurations. Using the T1.Micro, M1.Large and the IBM 326 physical server  for 1, 5, and 10 workers the processing rate can be drawn as shown in Figure \ref{fig:singleinstances-multipleworkers}. An increase in the rate of files processed is seen across all instance types as the number of workers increases from 1, however as the rate goes from 5 to 10, the effect is less significant in all cases. The instances with more resources show the greater gain.  

\par
\begin{figure}[htbp]
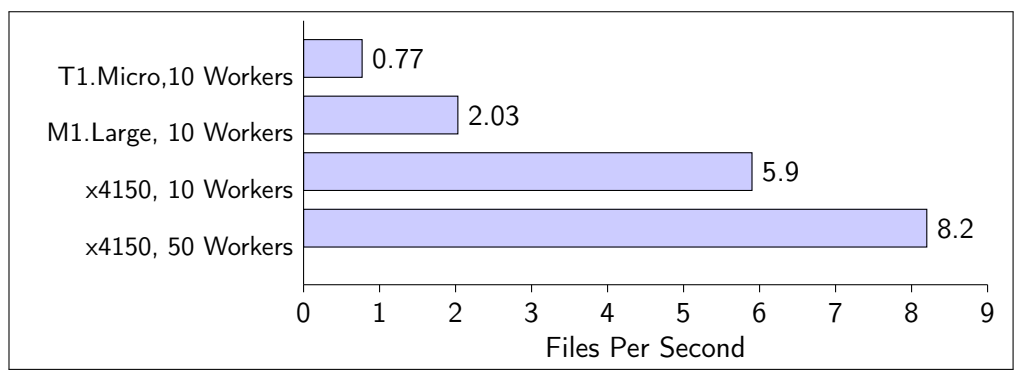
 
\centering
\fbox{      \begin{bchart}[step=1,max=9,width=9cm]
 
         \bcbar[]{0.77}
        \bclabel{\small{T1.Micro,10 Workers}}
        \smallskip
        
        \bcbar[]{2.03}
        \bclabel{\small{M1.Large, 10 Workers}}
        \smallskip

         \bcbar[]{5.9}
        \bclabel{\small{x4150, 10 Workers}}
        \smallskip
 
         \bcbar[]{8.2}
        \bclabel{\small{x4150, 50 Workers}}
        \smallskip

 \bcxlabel{Files Per Second}

    \end{bchart}}
  \caption{Exp:NIM2-3 Increasing the number of workers on faster CPU servers}
  \label{fig:largerCPUS}

\end{figure}

 To verify the relationship between increased CPU capacity and an increase in the capacity to run more worker threads, an additional experiment (see Figure \ref{fig:largerCPUS} ) was run on an x4150 server which has 4 dual quad core CPUs compared to the M1.Large which has 2 CPU cores. The x4150 was run with 10 and 50 workers. The larger server runs significantly faster using 10 workers compared to the M1.Large, but  while running 50 workers there is a levelling off of performance. Assuming that the network and the web server were not a bottleneck at this point, then the CPU is the most likely limiting factor.

\begin{figure}[htbp]
  \begin{minipage}[b]{0.5\linewidth}
    \centering
    \begin{tikzpicture}
    \begin{axis}[width=0.95\textwidth,
    legend style={at={(0.5,1.2)},anchor=north,legend cell align=left},
        xlabel=$T1.Micro\ Experiment\ Number$,
        ylabel=$Files Per Second$, xmin=0,ymax=2.2,xtick=data,ytick={0.2,0.4,0.6,0.8,1.0,1.2,1.4,1.6,1.8}]]
    \addplot plot coordinates {
    (1,0.25) (2,0.27)  (3,0.51) (4,0.52)  (5,0.53)    (6,0.78)  (7,0.73) (8,0.76) (9,0.68) (10,0.68) (11,0.71) };
     
       \legend{$Processing\ rate$\\}   
    \end{axis}

   \end{tikzpicture}

    \caption{Exp:NIM2-3 Single T1.Micro Instance file cleaning rate per second.  }
    \label{fig:t1-rates}
  \end{minipage}
  \hspace{0.5cm}
  \begin{minipage}[b]{0.5\linewidth}
  
    \centering
     \begin{tikzpicture}
    \begin{axis}[width=0.95\textwidth,
    legend style={at={(0.5,1.2)},anchor=north,legend cell align=left},
        xlabel=$M1.Large\ Experiment\ Number$,
        ylabel=$Files Per Second$, xmin=0,ymax=2.2,xtick=data,ytick={0.2,0.4,0.6,0.8,1.0,1.2,1.4,1.6,1.8}]
    \addplot plot coordinates {
     (1,0.31) (2,0.33)  (3,0.55)    (4,0.56)  (5,0.54) (6,1.43) (7,1.31) (8,1.92) (9,1.64) (10,1.44) (11,2.07) };
     
       \legend{$Processing\ rate$\\}   
    \end{axis}

   \end{tikzpicture}     
    
    \caption{Exp:NIM2-3 Single M1.Large Instance file cleaning rate per second.  }
    \label{fig:m1-rates}
  \end{minipage}
\end{figure}

\begin{table}
\centering
  \begin{tabular} {cccccc}
  \toprule

  \multicolumn{1}{b{1.5cm}}{\begin{center}Exp. Num.\end{center}} &
  \multicolumn{1}{b{1.5cm}}{\begin{center}Web Server\end{center}} &
   \multicolumn{1}{b{1.5cm}}{\begin{center}Num. Workers\end{center}} &
   \multicolumn{1}{b{2cm}}{\begin{center}Num. Web Servers\end{center}} &
   \multicolumn{1}{b{1.8cm}}{\begin{center}t1.micro fps\end{center}} &
   \multicolumn{1}{b{1.5cm}}{\begin{center}m1.large fps\end{center}} \\

\addlinespace[-5mm]

  \midrule
  1 & AWS East 		& 1   		& 1  & 0.25 & 0.31     \\
  2 &  AWS East 	& 1   		& 2  & 0.27 & 0.33     \\
  3 &  FTP 			& 1   		& 1  & 0.51 & 0.55     \\
   4 & DIT 			& 1   		& 1  & 0.52 & 0.56     \\
  5&  DIT 			& 1   		& 2  & 0.53 & 0.54     \\
 6 &   AWS East 	& 5   		& 1  & 0.78 & 1.43     \\
  7 &  AWS East 	& 5   		& 2  & 0.73 & 1.31     \\
  8 &  FTP 			& 5   		& 1  & 0.76 & 1.92     \\
  9 &  AWS East 	& 10   	& 1  & 0.68 & 1.64     \\
  10 &  AWS East 	& 10   	& 2  & 0.68 & 1.44     \\
  11 &  FTP 		& 10   	& 1  & 0.71 & 2.07     \\

\hline  \bottomrule

  \end{tabular}
  \caption{Single Instance Experimental Results Table for Fig \ref{fig:t1-rates}  and \ref{fig:m1-rates}  }

\label{tab:t1.microresult1}
\end{table}

\par Figures \ref{fig:t1-rates} and \ref{fig:m1-rates} compare, at a high level, the rate of cleaning of files per second by both the T1.micro and the M1.large instances, varying web servers and the number of workers per instance, with the data shown in Table \ref{tab:t1.microresult1}.  As expected the AWS versus the non-AWS web servers have an overall impact on the instance performance, but the increase in the number of workers is initially significant for an increase to 5 servers, but as the number of workers grows the performance improvement does not improve in a linear manner.  
\newpage
\par
\begin{figure}[htbp] 
\centering
\fbox{   \includegraphics[width=0.95\textwidth] {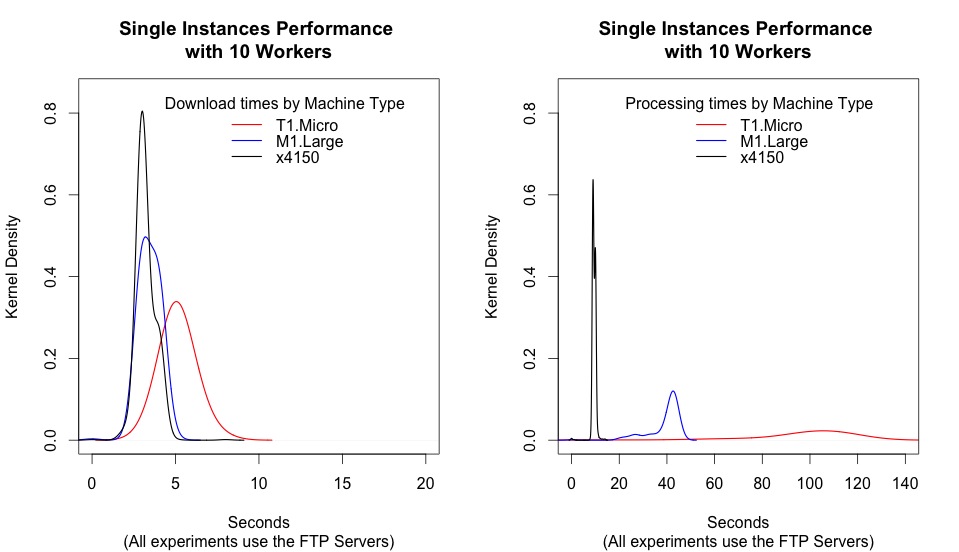}  }
  \vskip -0.8em
    \caption{Exp:NIM2-3 Single Instance Performance breakdown with increasingly powerful servers }
  \label{fig:sip-cpu-growing}
\end{figure}

 A more detailed comparison of the experiments is shown in Figure  \ref{fig:largerCPUS} where the same web server is used in all cases, and selecting the first three experiments where the same number of  worker threads are running. This ensures that  the primary variation is the CPU capability of the server. In Figure \ref{fig:sip-cpu-growing} a density probability plot of time is shown, where  the worker performance is split into download times and processing times. The processing power of the instance becomes the primary bottleneck when the web server and network are relatively unconstrained as with the t1.micro, but not for the x4150. It is clear that each instance will require a different number of workers to maximise their performance.

\par
\begin{figure}[htbp] 
\centering
\fbox{   \includegraphics[width=0.95\textwidth] {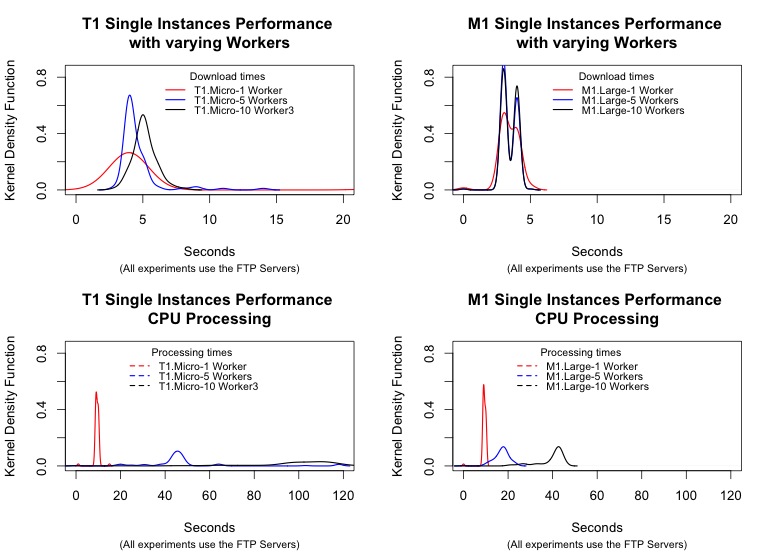}  }
  \vskip -0.8em
    \caption{Exp:NIM2-3 Single Instance Performance breakdown with increasing number of workers }
  \label{fig:sip-worker-growing}
\end{figure}

 Finally a comparison between the T1.Micro and the M1.Large is shown in Figure \ref{fig:sip-worker-growing}, breaking down the performance of each worker by time to download files and time to process them,  showing results for 1, 5 and 10 worker experiments. 

\par From these experiments it has been shown that the number of workers on a server instance has the ability to grow the overall performance of the server w.r.t file processing rate. The rate of increase is dependant on the exact configuration and  capabilities of the server, but number of CPU cores does appear to be relevant. There is a point beyond which the number of workers added will cause a negative impact on the servers processing rate, For future work, it should be possible for an instance to modify the number of worker threads it runs to maximise the overall instance processing rate. The would require a local instance monitor to operate with the ability to start-up new workers or shut them down. This could be accomplished by either multi-threaded modifications to the existing workers or to create additional sandboxes for the single threaded version to run inside as it currently the case. For single instance systems the web server is unlikely to create a bottleneck, however faster web servers will ensure that the optimal number of workers is reached sooner as the CPU is kept busy due to consistently quick file downloads.    

\subsubsection{Analysis}
\par These experiments provide basic information regarding the performance of a single instance within the pipeline architecture.

\begin {itemize}

\item  Exp:NIM2-1. For a single worker instance, running a single worker there is clearly a difference observable in the download time from the AWS based web servers used.   
\item Exp:NIM2-2.  For different worker server types, different worker processing rates are observed.  As the instance type CPU processing capability is increased, the rate CPU quickly becomes a bottleneck. By increasing the CPUs on the server the number of workers can continue to grow. If the worker nodes are increased the is a point of diminishing returns.  
\item Exp:NIM2-3. Each worker type will contain different characteristics such as CPU performance. If the number of workers is increased then providing there are sufficient CPU resources, the processing rate will improve. The overall pipeline will therefore run faster as more powerful servers are utilised until the ability to download becomes a bottleneck. 
\end {itemize}

\subsection{Multi-Instances Node Performance}

The experiments outlined in Table \ref{tab:nim3-exp} are focused on understanding the effect of increasing the number of server instances in an experiment. Experiments range from 1 to 100 EC2 amazon instances. Building on the observations of running a single instance, the aim is to determine if the increase in instances results in an overall improved performance of the system and to determine if any specific limits have been reached. 
\begin{table}[htbp]
\centering
\begin{tabular}{p{2.5cm} p{3cm} p{8.3cm}}
  \toprule
Reference & Measure & Description \\
  \midrule
\textbf{\small{Exp:NIM3-1}}& \small{Multi instance webserver performance} & \small {Testing the impact of different web server configurations to service multiple worker instances.  }\\
\textbf{\small{Exp:NIM3-2}}& \small{Multi instance performance by type} & \small {Testing the impact of selecting different instance machine types for multiple worker instances.   }\\
\textbf{\small{Exp:NIM3-3}}& \small{Multi Instance Scaling Analysis} & \small {Analysis of the rate of scaling from a single instance to 100 instances.}\\
\textbf{\small{Exp:NIM3-4}}& \small{Limit Testing} & \small {Investigating the limits of the experimental setup}\\

\hline
  \bottomrule
\end{tabular}
  \caption{NIMBUS Multi Instance experiments}
\label{tab:nim3-exp}
\end{table}

\par The three variables being tested in these experiments are Web Servers, number of instances and worker impact. Experiments are shown where the number of workers per server is consistently set to 5 for T1.Micro server instances, and 10 for M1.Large server instances, batch downloads are set to ten and all work performed  is the same. Experiments are broken down by the variable under test.

\subsubsection{Exp:NIM3-1 Multi Instance Webserver Performance}

\begin{figure}[htbp] 
\centering
\fbox{     

\begin{tikzpicture}
\begin{axis}[
ylabel=Files per Seconds, enlarge x limits=0.2,
legend style={ at={(0.5,-0.15)},  anchor=north,legend columns=-1 }, ymin=0, ybar, xtick=data, symbolic x coords={AWS East, AWS West, DIT, HEANet,FTP },
grid=major, bar width=25pt,width=12cm, xmajorgrids=false,nodes near coords,every node near coords/.append style={anchor=mid west,rotate=70}]
\addplot coordinates {(AWS East,34.3) (AWS West,74.9)  (DIT,39.9) (HEANet,80.3) (FTP,78.3) };
\end{axis}
\end{tikzpicture}
}
  \caption{Exp:NIM3-1 Files Per Second: Varying Web servers and their impact on T1.Micro 100 Instance performance }
  \label{fig:100instances2}

\end{figure}
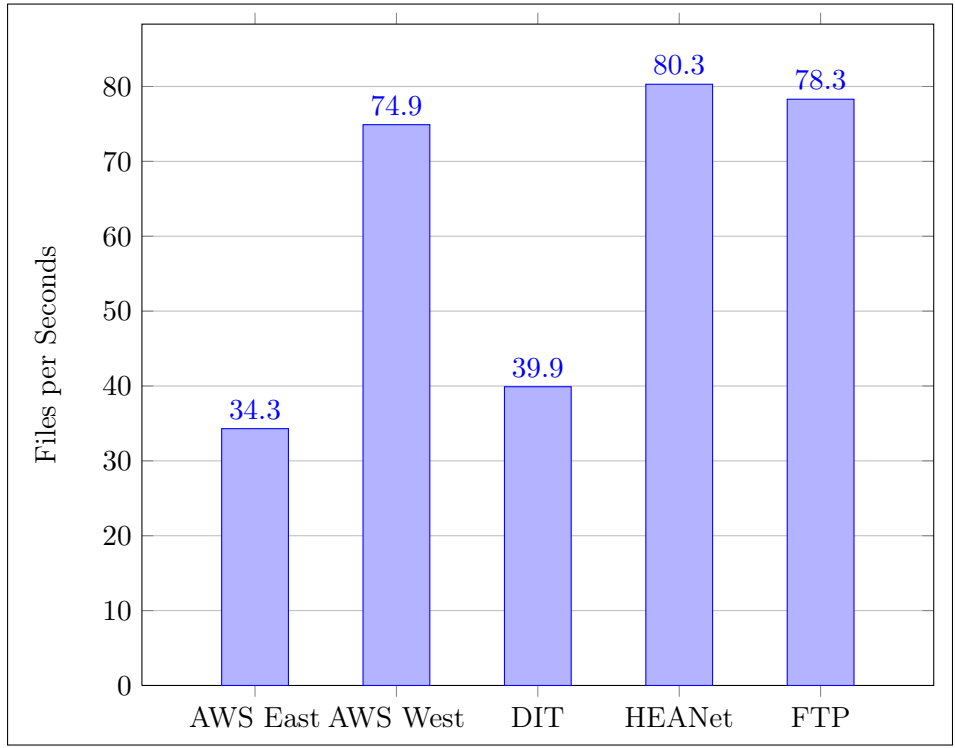

When dealing with a single instance it was clear the the AWS web server delivered files to the worker nodes slightly slower than the FTP, DIT and HEANet based web servers. As the number of instances running is increased,  the AWS versus Non-AWS web server performance are  compared to determine if the performance difference is sustained over larger number of file requests per second.  Figure \ref{fig:100instances2} shows a histogram of the overall file processing rate for 100 T1.Micro servers running 5 workers against different web servers. It is clear that the DIT and single AWS East server do not appear to scale with the increases in requests while the 4 AWS West web server combination, the HEANet Servers and the FTP servers seem to be performing well. If the cleaning rates of a single instance running 5 workers as given in Figure \ref{fig:singleinstances2} are multiplied by 100, the cleaning rates where the web server is not a bottleneck scale almost linearly, based on observation.  It is interesting to note that the web servers in the DIT and HEANet are identical in hardware and software, possibly indicating that the network is a factor in the experiment.  

\par
\begin{figure} [htbp]
\centering
\fbox{   \includegraphics[width=0.95\textwidth] {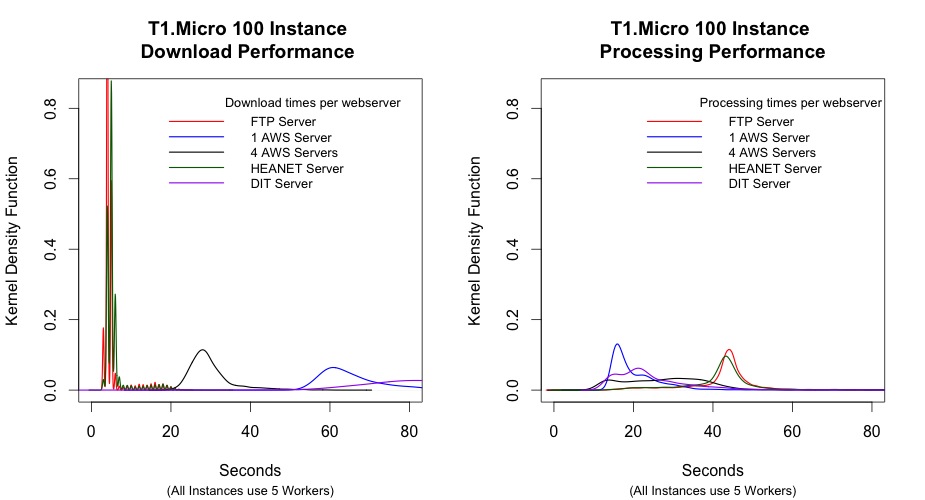}  }
  \vskip -0.8em
    \caption{Exp:NIM3-1 100 Instance Performance breakdown for different Web Servers, with 5 workers.}
  \label{fig:T1-100-instances-breakdown-5w}
\end{figure}

 Looking at the breakdown of the experiments based on download and processing  time in Figure \ref{fig:T1-100-instances-breakdown-5w}, the FTP, HEANet server configuration gives  consistent download times for data, while the other servers are consistently slower. The mean, standard deviation and variance in Table \ref{tab:t1.microresult100-sd-var} from the FTP and HEANet servers indicate consistent reliable network performance throughout the experiment. The processing performance in Table \ref{tab:t1.microresult100-sd-var-2} again shows similar behaviour for the experiments with fast downloads, while the slower web server based experiments have lower variance, standard deviation and mean for the overall slower experiments as the CPU is less loaded due to slower download of data. 

\begin{table} [htbp]
\centering
  \begin{tabular} {cccc}
  \toprule

  \multicolumn{1}{b{3.5cm}}{\begin{center}Web Server\end{center}} &
   \multicolumn{1}{b{3cm}}{\begin{center}Std. Dev.\end{center}} &
   \multicolumn{1}{b{3cm}}{\begin{center}Mean\end{center}} &
   \multicolumn{1}{b{3cm}}{\begin{center}Variance\end{center}} \\

\addlinespace[-5mm]

  \midrule
  FTP 		& 3.148   		& 5.12  & 9.91      \\
  1 x AWS 		& 10.813  		& 66.62  & 116.90     \\
  4 x AWS 		& 5.509   		& 29.72  & 30.35      \\
  HEANeet 	& 3.065   		& 5,72  & 9.39    \\
  DIT 		& 15.585   	& 84.35  & 242.91     \\
\hline  \bottomrule

  \end{tabular}
  \caption{T1.Micro 100 Instance 5 Worker statistics for image downloads per web server }

\label{tab:t1.microresult100-sd-var}
\end{table}

\begin{table}
\centering
  \begin{tabular} {cccc}
  \toprule

  \multicolumn{1}{b{3.5cm}}{\begin{center}Web Server\end{center}} &
   \multicolumn{1}{b{3cm}}{\begin{center}Std. Dev.\end{center}} &
   \multicolumn{1}{b{3cm}}{\begin{center}Mean\end{center}} &
   \multicolumn{1}{b{3cm}}{\begin{center}Variance\end{center}} \\

\addlinespace[-5mm]

  \midrule
  FTP 		& 28.67   		& 52.85  & 822      \\
  1 x AWS 		& 13.63  		& 22.35  &  186    \\
  4 x AWS 		& 20.15   		& 32.27  & 405      \\
  HEANeet 	& 28.06   		& 51.54  & 787    \\
  DIT 		& 13.24   		& 25.45  & 175     \\

\hline  \bottomrule

  \end{tabular}
  \caption{T1.Micro 100 Instance 5 Worker statistics for image processing per web server }

\label{tab:t1.microresult100-sd-var-2}
\end{table}

\

\par It can be concluded from these experiments that there is a performance benefit from using the non-AWS web servers when scaling worker instances. When using the FTP HEANet server the increase in processing rates has been almost linear. Clearly there are also network effects given  different network configurations but identical hardware configurations for the DIT and HEANet web servers. Combinations of multiple web servers such as the AWS West 4 server setup will also allow the performance of the pipeline to continue to grow with the number of instances. 

\par Given the limited processing performance of the T1.Micro instance, it is possible that bottlenecks in the web servers have not yet been reached. As the larger M1.Large instance type puts higher demands on the web servers due to the higher processing rates,  additional experiments are considered in Exp:NIM3-2 comparing the performance of the two instance types.

\subsubsection{Exp:NIM3-2 Multi Instance  Performance by Type}

\par These experiments look at the ability of the system to continue increasing performance as the number of instances is increased. The different instance types are explored to see if the increase in performance is linear for all instance types using the FTP web server to service all requests. 

\par
\begin{figure}[htbp] 
\centering
\fbox{   \includegraphics[width=0.95\textwidth] {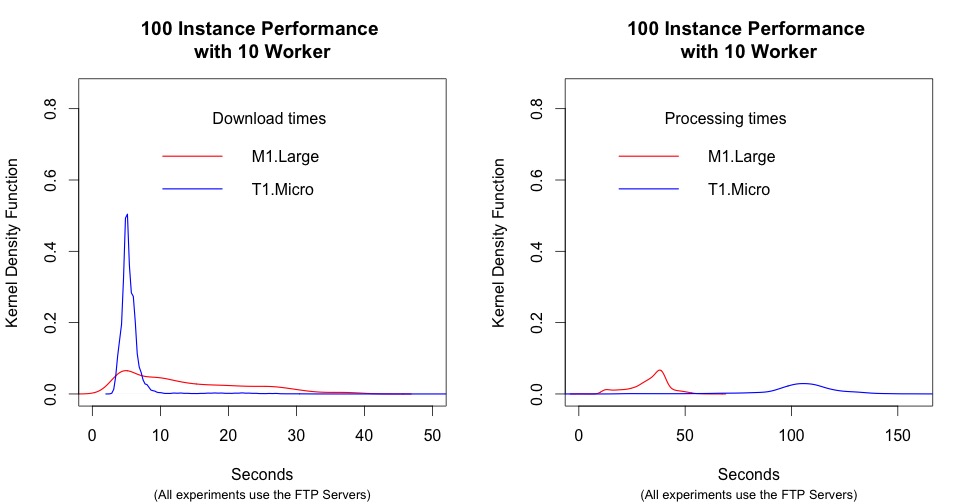}  }
  \vskip -0.8em
    \caption{Exp:NIM3-2 100 Instance Performance breakdown for different instance types running 10 workers using the FTP web server.}
  \label{fig:100-instances-breakdown-10w}
\end{figure}

\begin{figure} [!ht]
\centering
\fbox{   \includegraphics[width=0.95\textwidth] {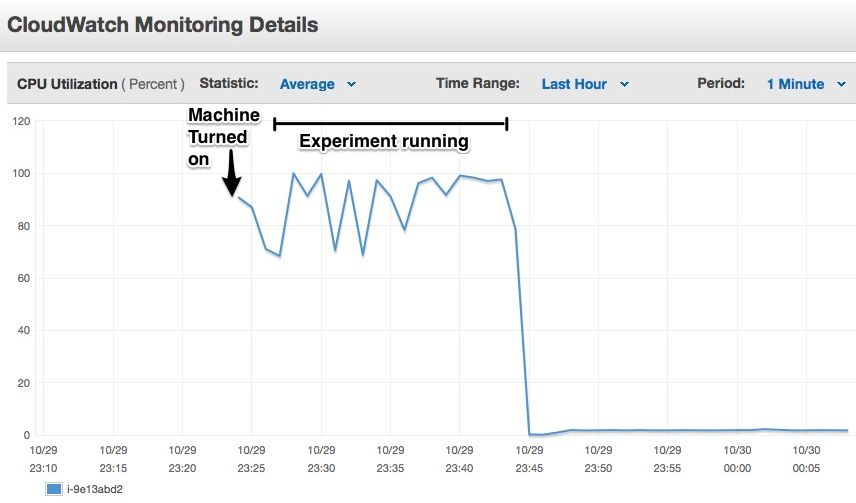}  }
  \vskip -0.8em
    \caption{Exp:NIM3-2 CPU Performance for M1.Large canary instance during 100 instance run, with 10 workers.}
  \label{fig:canarycpu1}
\end{figure}

\begin{figure} [!ht]
\centering
\fbox{   \includegraphics[width=0.95\textwidth] {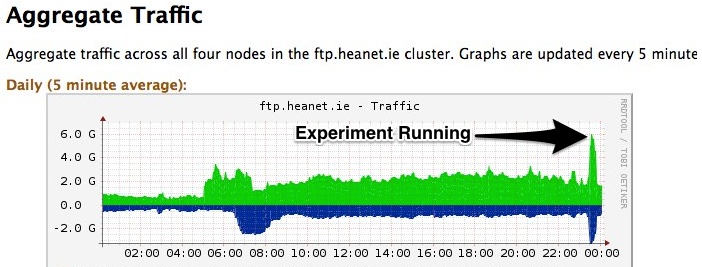}  }
  \vskip -0.8em
    \caption{Exp:NIM3-2 Network Performance for FTP.heanet.ie web server during M1.Large 100 instance run, with 10 workers per instance.}
  \label{fig:ftpnetwork1}
\end{figure}

 In Figure \ref{fig:100-instances-breakdown-10w} the T1.Micro and the M1.Large instance types are compared while running 10 workers per instance. If the mean time for workers to download and process data to those of a single instance is compare in Table \ref{tab:compare-1-100-sd-var-0} and Table \ref{tab:compare-1-100-sd-var-2} it is clear that the T1.Micro network mean is similar, but there is more fluctuation in the download times, while the T1.Large mean is increasing.  This might suggest that the increase in demand on the FTP server is starting to indicate some pressure on its ability to service file requests. The CPU processing mean remains similar for the T1.Micro and the M1.Large, but there is considerably more variance introduced into both instance types. This could be due to the larger sample size. In Figure \ref{fig:canarycpu1} the CPU of one of the running instances of the  M1.Large instance type is shown, indicating that the CPU of the M1.Large is close to maximum for the length of the experiment while Figure \ref{fig:ftpnetwork1} shows the additional load generated by the experiment on the ftp.heanet.ie web server. 

\begin{table}[htbp]
\centering
  \begin{tabular} {ccccc}
  \toprule

  \multicolumn{1}{b{3cm}}{\begin{center}Instance Type\end{center}} &
    \multicolumn{1}{b{3cm}}{\begin{center}Num. Instances\end{center}} &
   \multicolumn{1}{b{2cm}}{\begin{center}Std. Dev.\end{center}} &
   \multicolumn{1}{b{2cm}}{\begin{center}Mean\end{center}} &
   \multicolumn{1}{b{2cm}}{\begin{center}Variance\end{center}} \\

\addlinespace[-5mm]

  \midrule
  T1.Micro 		& 1   		&  0.8 &   5.1 &  0.7   \\
  T1.Micro 		& 100  	& 3.2  &  5.8  & 10.4 \\
  M1.Large 		& 1   		&  0.6 &   3.5 &  0.3\\
  M1.Large 	        & 100   	& 8.9  & 13.8    & 79.4\\

 \hline \bottomrule

  \end{tabular}
  \caption{Comparing 1 vs 100 Instance 10 Worker network statistics using the FTP web server }

\label{tab:compare-1-100-sd-var-0}
\end{table}

\begin{table}[htbp]
\centering
  \begin{tabular} {ccccc}
  \toprule

  \multicolumn{1}{b{3cm}}{\begin{center}Instance Type\end{center}} &
    \multicolumn{1}{b{3cm}}{\begin{center}Num. Instances\end{center}} &
   \multicolumn{1}{b{2cm}}{\begin{center}Std. Dev.\end{center}} &
   \multicolumn{1}{b{2cm}}{\begin{center}Mean\end{center}} &
   \multicolumn{1}{b{2cm}}{\begin{center}Variance\end{center}} \\

\addlinespace[-5mm]

  \midrule
  T1.Micro 		& 1   		&   21.9&  97.0  &  479.3 \\
  T1.Micro 		& 100  	&   57.7 &  120.8  &  3338 \\
  M1.Large 		& 1   		&   6.6 &    39.3 &  44.1 \\
  M1.Large 	        & 100   	&  9.0 &  33.2  & 81.8  \\

 \hline \bottomrule

  \end{tabular}
  \caption{Comparing 1 vs 100 Instance 10 Worker CPU statistics using the FTP web server }

\label{tab:compare-1-100-sd-var-2}
\end{table}

\par The overall processing rate of the T1.Micro is just over 72 files per second, while the M1.Large is almost 192 files per second. From Table \ref{tab:t1.microresult1} the single instance processing times for 10 workers for the T1.Micro was 0.71 files per second, and the M1.Large was 2.07 files per second indicating that the increases in instances is indeed scaling linearly however fluctuations in the performance of the workers is starting to occur.

\subsubsection{Exp:NIM3-3 Multi Instance Scaling Analysis}

\par From the previous section it can be seen that the increase from 1 to 100 instances appears to provide an almost linear increase in performance for both the T1.Micro and the M1.Large instance types, although the larger instance experiments are seeing more fluctuations in the worker performance. To test the statistical significance of the increase in overall system performance a set of statistical tests have been run on the M1.Large instance type. This instance type was selected as it does not have the reported T1.Micro CPU throttling issues reported by AWS, which eliminates yet another variable from the results.  As already shown, the web server performance has an effect on the outcome of the experiments for larger numbers of instances. To eliminate this bottleneck from experiments when seeking a correlation between the experiment performance and the number of instances, the selection of experiments for larger instances was restricted to the faster web servers configurations, (FTP, 4xAWS servers, HEANEet web servers). Experiments with either 5 or 10 workers were included. So only for the larger instances have the single AWS and DIT web servers been removed.     

\par
\begin{figure}[htbp]
\centering
\fbox{   \includegraphics[width=0.95\textwidth] {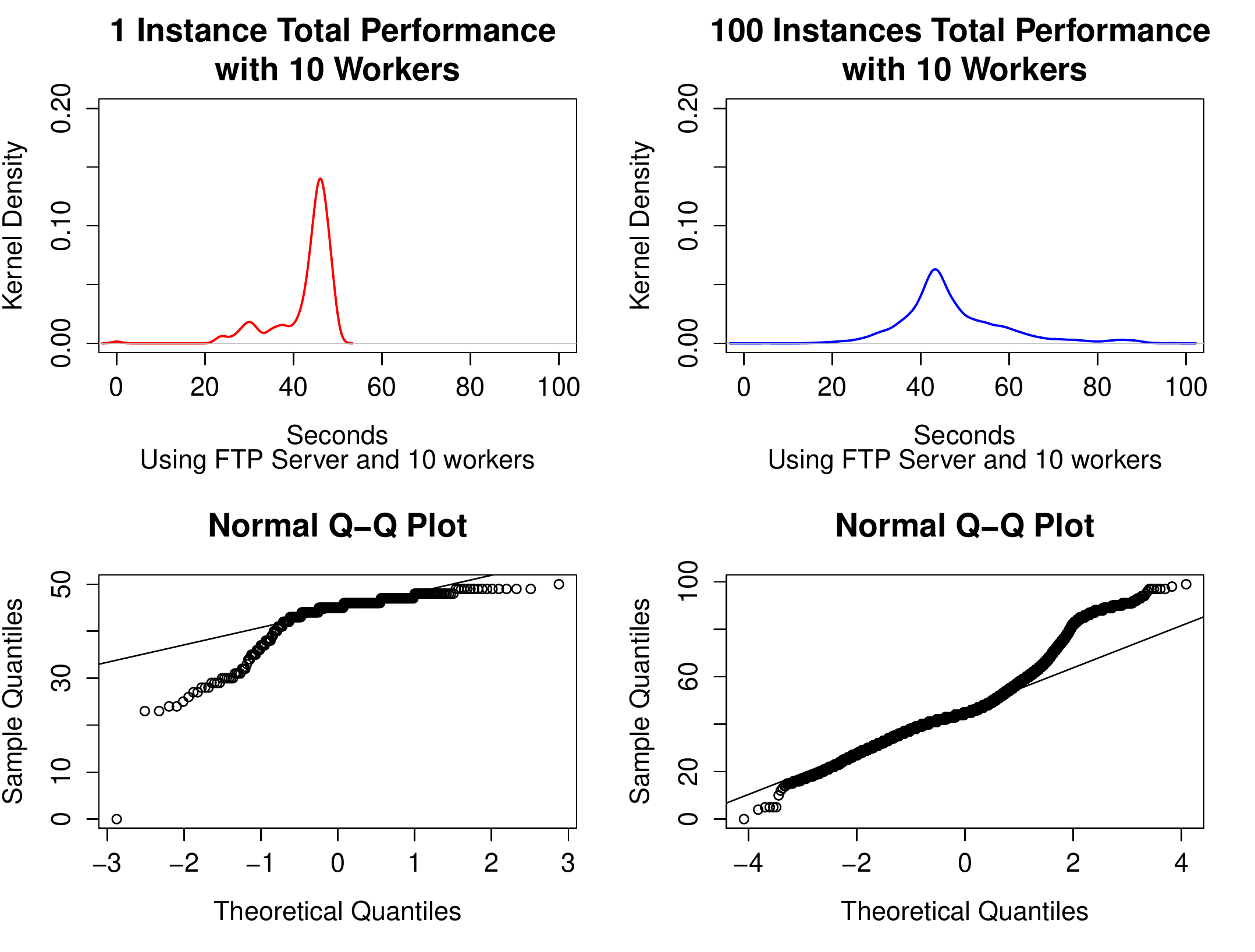}  }
  \vskip -0.8em
    \caption{Exp:NIM3-3 Testing for normal distribution of worker performance for 1 and 100 M1.Large instance experiments}
  \label{fig:testfornormal1}
\end{figure}

 Before running a correlation or a T-test a test for normality of the data must first be performed. Taking two experiments,  both using the FTP server and 10 workers per instance, where the first has a single instance running and the second has 100 instances running, a density plot and the corresponding Normal Q-Q plot is show in Figure \ref{fig:testfornormal1}. From this it can be assumed that the data is reasonably normally distributed and it is appropriate to run a correlation test and T-test.

\par
\begin{figure}[htbp]
\centering
\fbox{   \includegraphics[width=0.8\textwidth] {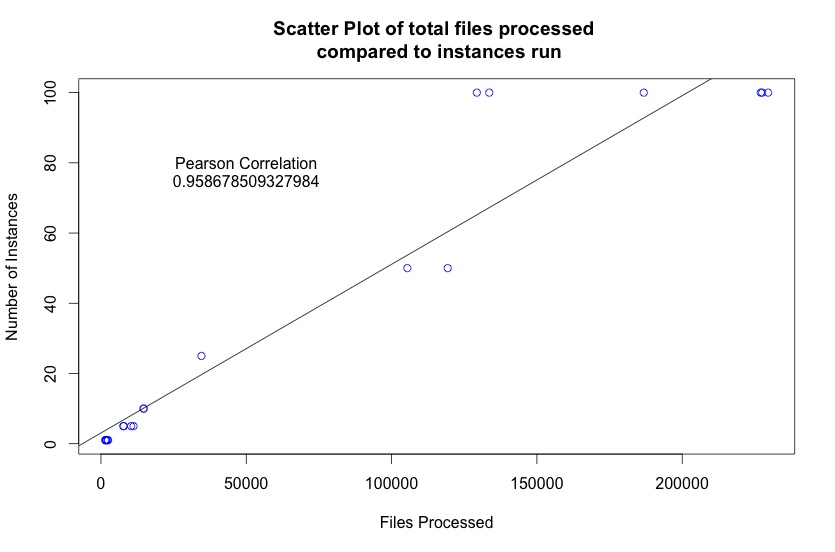}  }
  \vskip -0.8em
    \caption{Exp:NIM3-3 Testing for a correlation between instances and files process. Data for the scatter plot given in Appendix Table \ref{tab:pearsons1}}
  \label{fig:pearson1}
\end{figure}

 The Pearson product-moment correlation coefficient is used to measure the dependance  between instance numbers and files processed and the scatter plot along with the pearson coefficient is given in Figure \ref{fig:pearson1}. With a value of $\rho = 0.95$ it can be determined that there is a \emph {strong and positive} correlation between the number of instances and the number of files process, for the M1.Large instance type experiments. It is important to note that $\rho$ provides a measure of the linear relationship between these variable, however it does not in itself indicate a causal relationship. 

\par
\begin{figure}[htbp]
\centering
\fbox{   \includegraphics[width=0.8\textwidth] {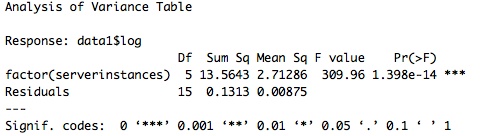}  }
  \vskip -0.8em
    \caption{Exp:NIM3-3 One Way ANOVA  Table for increasing instances}
  \label{fig:anova1}
\end{figure}

\begin{figure}[htbp]
\centering
\fbox{   \includegraphics[width=0.8\textwidth] {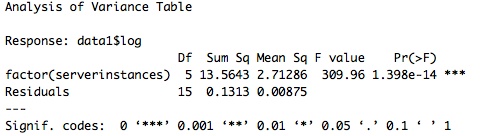}  }
  \vskip -0.8em
    \caption{Exp:NIM3-3 Pairwise T Test}
  \label{fig:pairwise1}
\end{figure}

 Assuming the the data is normally distributed, a one-way ANOVA test could be run to perform an analysis of the variance, however  with a skewness of $0.859$ (skewed to the right) and a kurtosis of $2.149$, implying the data is platykurtic, it is prudent to transform the total number of files process using a Log10 function. Testing can now be done testing the null hypothesis, that the means are all equal. The P-value is calculated to help determine if the null hypothesis should be rejected. The result of the one way ANOVA in Figure \ref{fig:anova1} is considered significant with a P value < 0.001, so a pairwise comparison is performed to test if the differences are statistically significant, while adjusting for Type 1 errors. The results of the pairwise test are shown in Figure \ref{fig:pairwise1} and with p-values < 0.001 in most cases, it can be concluded that there is a statistically significant difference comparing instance numbers to files processed.

\subsubsection{Exp:NIM3-4 Limit Testing}

\par The majority of experiments run and presented so far have been based on the  T1.Micro instance and M1.Large instance servers running within the amazon AMS environment. Due to physical hardware limitations, some additional experiments were completed that have shown the ability of the system to process higher rates of data when using more powerful servers, but the number of physical servers was limited to 1.  
\par It has also been shown that the impact of the number of workers and the web servers serving data provides a dynamic which can often have smaller number of workers performance faster, or comparable to those with higher numbers depending on how fast data is available to the server and the CPU load generated on the instance. The T1.Micro and the L1.Large performances for 100 instances can be summarised in Figures \ref{fig:t1-100-rates} and \ref{fig:m1-100-rates}, with the data for the figures given in Table \ref{tab:M1.mediumresult1}. In line with previous observations,  experiment 6, which used the DIT web servers has reduced performance across instance types, and while the T1.Micro instances have a CPU bottleneck around 80 files per second, the M1.Large instances continue to increase the processing rate as more workers are run per instances, and the web server capacity is increased. 

\begin{figure}[htbp]
  \begin{minipage}[b]{0.5\linewidth}
    \centering
    \begin{tikzpicture}
    \begin{axis}[width=0.95\textwidth,
    legend style={at={(0.5,1.2)},anchor=north,legend cell align=left},
        xlabel=$T1.Micro\ Experiment\ Number$,
        ylabel=$Files Per Second$, xmin=0,ymax=220,xtick=data,ytick={20,40,60,80,100,120,140,160,180}]]
    \addplot plot coordinates {
    (1,27.4) (2,26.1)  (3,55.8) (4,64.7)  (5,75)    (6,39.9)  (7,75.1) (8,78.3) (9,79) };
     
       \legend{$Processing\ rate$\\}   
    \end{axis}

   \end{tikzpicture}

    \caption{Exp:NIM3-4 100 T1.Micro Instances, file cleaning rate per second.  }
    \label{fig:t1-100-rates}
  \end{minipage}
  \hspace{0.5cm}
  \begin{minipage}[b]{0.5\linewidth}
  
    \centering
     \begin{tikzpicture}
    \begin{axis}[width=0.95\textwidth,
    legend style={at={(0.5,1.2)},anchor=north,legend cell align=left},
        xlabel=$M1.Large\ Experiment\ Number$,
        ylabel=$Files Per Second$, xmin=0,ymax=220,xtick=data,ytick={20,40,60,80,100,120,140,160,180}]
    \addplot plot coordinates {
     (1,28.6) (2,28.9)  (3,58.1)    (4,56.5)  (5,107.8) (6,39.1) (7,111.3) (8,191.3) (9,189.5) };
     
       \legend{$Processing\ rate$\\}   
    \end{axis}

   \end{tikzpicture}     
    
    \caption{Exp:NIM3-4 100 M1.Large Instance file cleaning rate per second.  }
    \label{fig:m1-100-rates}
  \end{minipage}
\end{figure}

\begin{table}
\centering
  \begin{tabular} {clcccc}
  \toprule

  \multicolumn{1}{b{1.5cm}}{\begin{center}Exp. Num.\end{center}} &
  \multicolumn{1}{b{1.5cm}}{\begin{center}Web Server\end{center}} &
   \multicolumn{1}{b{1.5cm}}{\begin{center}Num. Workers\end{center}} &
   \multicolumn{1}{b{2cm}}{\begin{center}Num. Web Servers\end{center}} &
   \multicolumn{1}{b{1.8cm}}{\begin{center}t1.micro fps\end{center}} &
   \multicolumn{1}{b{1.5cm}}{\begin{center}m1.large fps\end{center}} \\

\addlinespace[-5mm]

  \midrule
  1 & AWS East 		& 1   		& 1  & 27.4 &  28.6    \\
  2 &  AWS East 		& 1  		& 2  & 26.1 & 28.9    \\
  3 &  FTP 			& 1   		& 1  & 55.8 & 58.1     \\
   4 & AWS East 		& 5/10   	& 2  & 64.7 & 56.5     \\
  5&  AWS East and West 	& 5/10   	& 4  & 75.0 & 107.8     \\
 6 &   DIT 				& 5/10   	& 2  & 39.9 & 39.1   \\
  7 &  HEANET 			& 5/10   	& 2  & 75.1 & 111.3     \\
  8 &  FTP 			& 5/10   	& 1  & 78.3 & 191.3    \\
  9 &  ALL 				& 10/10   	& 10  & 79.0 & 189.5     \\

\hline  \bottomrule

  \end{tabular}
  \caption{100 Instance Experimental Results Table for Figure \ref{fig:t1-100-rates}  and \ref{fig:m1-100-rates}  }

\label{tab:M1.mediumresult1}
\end{table}

\par
\begin{figure}[htbp] 
\centering
\fbox{     

\begin{tikzpicture}
\begin{axis}[
ylabel=Files Per Seconds, enlarge x limits=0.2,
legend style={ at={(0.5,-0.15)},  anchor=north,legend columns=-1 }, ymin=0, ybar, xtick=data, symbolic x coords={T1.Micro-5w, M1.Large-10w,M1.XLarge-20w, M1.XLarge-50w, C1.XLarge-100w},
grid=major, bar width=10mm,width=12cm, xmajorgrids=false,nodes near coords,every node near coords/.append style={anchor=mid west,rotate=70},xticklabel style={
            inner sep=0pt,
            anchor=north east,
            rotate=45 } ]
\addplot coordinates {(T1.Micro-5w,80) (M1.Large-10w,191)  (M1.XLarge-20w,223) (M1.XLarge-50w,174) (C1.XLarge-100w,234) };
\end{axis}
\end{tikzpicture}
}
  \caption{Exp:NIM3-4 Best file processing rates per second, for different instance types using 100 Instance experiments. See Table \ref{tab:bestperfacrossinstances} }
  \label{fig:bestperformancerates}

\end{figure}
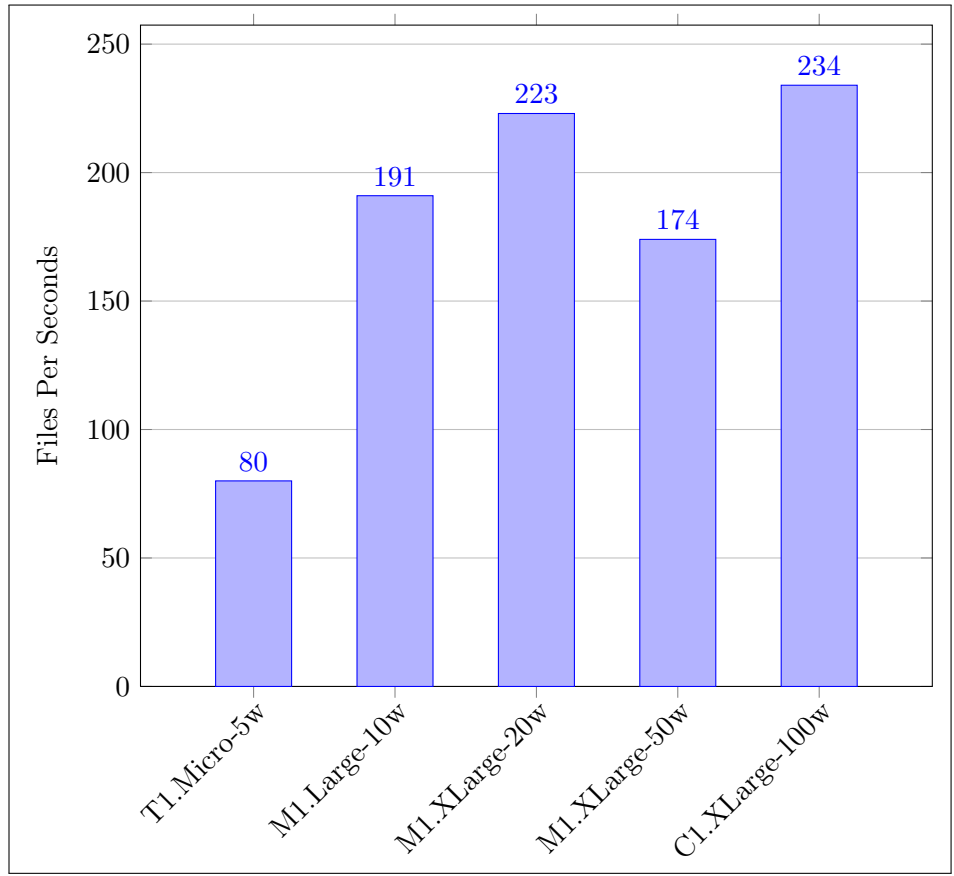

 To determine if the observations about the increased performance of the M1.Large can continue as larger instances are available (as shown with the x4150 server instance previously), two additional AWS instance types were selected to continue a limited set of additional experiments. This limitation was primary due to the cost of these systems.  If the best performance obtained using the T1.Micro instance and the M1.Large instance it taken, then as an instance type increases its capacity to run more workers it can affect the overall performance of the instance. Where the number of workers increases beyond the capacity of the CPU, a processing bottleneck is created and the instance slows down as evidenced by the T1.Micro having poorer performance when running 10 workers per instance compared to 10 instances. It is also clear that the  FTP.heanet.ie server acts as a high-performance web server for these experiments without providing a significant bottleneck on file availability for workers. By continuing to use the ftp.heanet.ie server and by running additional experiments, all of which execute 100 instances, it can be shown that larger capacity instances can yield increased performance. Regardless of how large an instance is however there should be a point at which a CPU bottleneck is observed. Figure \ref{fig:bestperformancerates} provides a histogram showing the best processing rates obtained for all 100 instance experiments including the M1.XLarge and the C1.XLarge instance types showing the number of workers per instance. All experiments used the ftp.heanet.ie web server. As can be seen in Experiment 5 in Table \ref{tab:bestperfacrossinstances}, there is reducing performance for the M1.Xlarge instance type as the number of workers is increased from 20 to 50. 

\begin{table}
\centering
  \begin{tabular} {cccccc}
  \toprule

  \multicolumn{1}{b{1.5cm}}{\begin{center}Exp. Num.\end{center}} &
  \multicolumn{1}{b{1.5cm}}{\begin{center}Web Server\end{center}} &
    \multicolumn{1}{b{1.5cm}}{\begin{center}Instance\end{center}} &
   \multicolumn{1}{b{1.5cm}}{\begin{center}Num. Workers\end{center}} &
   \multicolumn{1}{b{1.8cm}}{\begin{center}Files per Sec\end{center}} &
   \multicolumn{1}{b{1.5cm}}{\begin{center}GB per Hr\end{center}} \\

\addlinespace[-5mm]

  \midrule
  1 & FTP 	& T1.micro	& 5    	  & 78.3 &  343.1    \\
  2 & FTP 	& M1.Large	& 10    	  & 191.3 &  838.7    \\
  3 &  FTP & M1.XLarge	& 20  	 & 193.3 & 847.4    \\
  4 &  FTP & M1.XLarge	& 20   	 & 223.1 & 978.0     \\
   5 & FTP & M1.XLarge	& 50   	 & 173.6 & 761.0     \\
  6&  FTP 	& C1.XLarge	& 100   	 & 212.2 & 930.5     \\
 7 &   FTP & C1.XLarge	& 100   	 & 233.8 & 1024.1   \\

\hline  \bottomrule

  \end{tabular}
  \caption{100 Instance Experimental Results for different instance types using the FTP web server. Table for Fig \ref{fig:bestperformancerates}  }

\label{tab:bestperfacrossinstances}
\end{table}

\par To better understand the behavior of the running experiment, the processing performance of each experiment shown in Table \ref{tab:bestperfacrossinstances} was plotted over time, showing the cumulative total of result files written to the S3 Storage. Each experiment was 1,200 seconds in duration, and each ran 100 instances using the FTP server. Figure \ref{fig:100instances} shows the cumulative number of files posted over time. From the graph, the throttle of the T1.Micro CPU is evident over time as the rate of processing actually slows down after a period of time. The same effect is not seen by the other instance types. 

\par As the number of workers increases there is a significant delay in starting the C1.XLarge instances as there is a sequential process for copying and initialising each of the workers. This minor sequential process causes no significant delays then the number of workers is small, however it is increasingly significant as the number of workers increases. This would require additional modification of the worker software to eliminate this behaviour. The C1.XLarge processing rate however, once established is slightly better that the other workers. The minor tail at the end of the experiments is due to either the message queue becoming empty or the experiment concluding.

\pgfplotstableread{Data/Best-Result-RawData/Combined-Group2-Graphs.prn}
\datatable

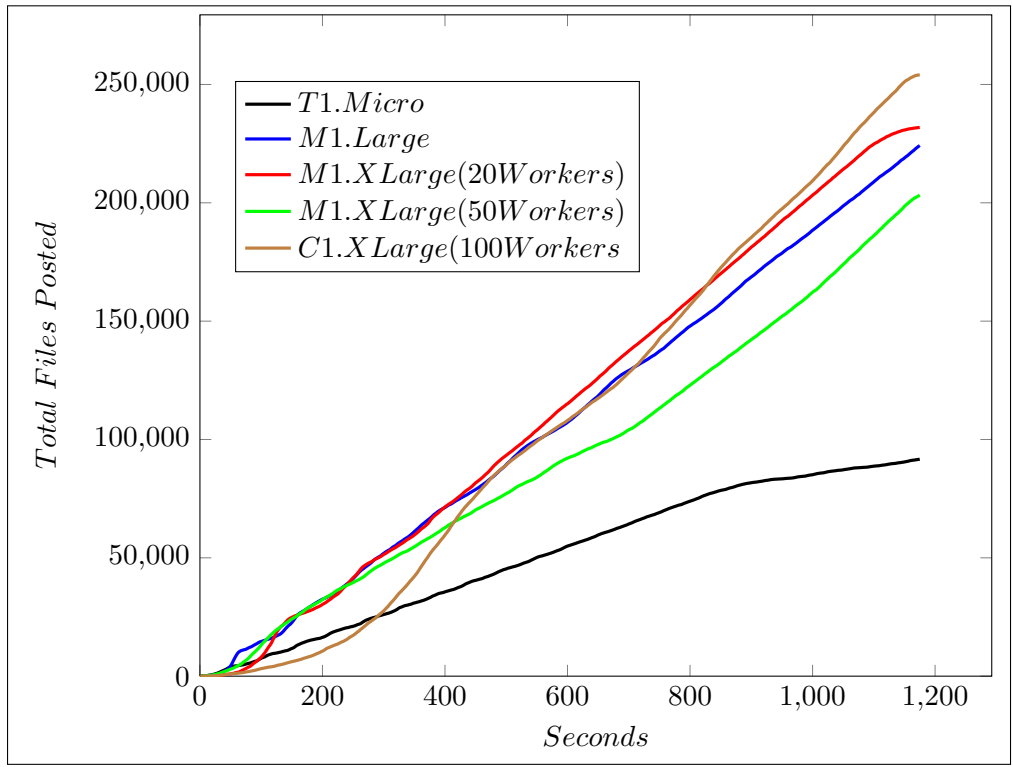
\begin{figure}[htbp] 
\centering
\fbox {
\begin{tikzpicture}
    \begin{axis}[xmin=0,
    ymin=0, scaled ticks=false, yticklabel style={/pgf/number format/fixed},
    width=0.8\textwidth, ylabel style={yshift=0.8cm},
    legend style={at={(0.3,0.9)},anchor=north,legend cell align=left},
        xlabel=$Seconds$,
        ylabel=$Total\ Files\ Posted$, no markers]

\addplot[black,very thick] table[y = T1.Micro] from \datatable ;
\addplot [blue,very thick] table[y = M1.Large] from \datatable ;
\addplot [red, very thick] table[y = M1.Xlarge-20workersA] from \datatable ;
\addplot [green, very thick] table[y = M1.Xlarge-50Workers] from \datatable ;
\addplot [brown, very thick] table[y = C1.Xlarge-100Workers] from \datatable ;

\legend{$T1.Micro$,$M1.Large$,$M1.XLarge(20 Workers)$,$M1.XLarge(50 Workers)$,$C1.XLarge(100 Workers$}   

    \end{axis}

\end{tikzpicture} }
  \vskip -0.8em
    \caption{Exp:NIM3-4 Cumulative files processed over time.}
  \label{fig:100instances}
\end{figure}

\par To ensure that the processing rates seen in Figure \ref{fig:100instances} are sustainable, the best performing experiment, the C1.XLarge with 100 workers, was selected to run for 3,000 seconds. Figure \ref{fig:3000seconds} shows a consistent processing rate once all of the workers have started, with no indication of an unsustainable process rate. The sequential initialisation process is evident as the instances start up.

\par  To investigate the slow start to the processing rate, further analysis was performed. The time stamp used per instance which determine the time at which a result was posted to S3, are based on the internal time of the instance, which may be slightly out of sync with other instances.  As it is possible that these times may be different across instances, two different plots are shown indicating the time a worker first posts a result using two different times. The first is the time as recorded by the instance, the second is an offset from when the instance initially registers itself.  So in the case where the actual time recorded by the instance is used, each worker first data publish time is shown in black in Figure \ref{fig:workerstarttimes}. An alternative plot is also shown in red, where the earliest result across all of the instances is used, and using an assumption that all instances started approximately the same time, each timestamp from each instance is calibrated back to align with the start time of the earliest result posted. 
\par Given that the Figure \ref{fig:3000seconds} uses the worker time and not a calibrated time, a more elongated start time for workers would be expected in Figure \ref{fig:workerstarttimes}, which is in fact observed.  
 
\pgfplotstableread{Data/Best-Result-RawData/CX1-Graph.prn}
\datatable

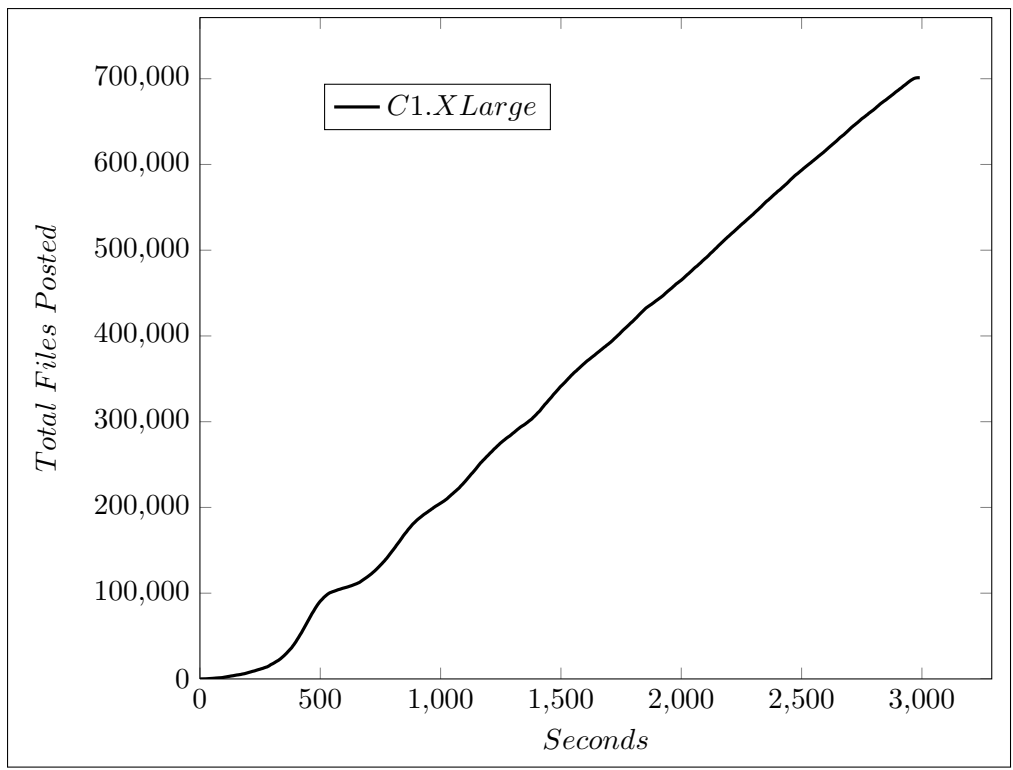
\begin{figure}[htbp] 
\centering
\fbox {
\begin{tikzpicture}
    \begin{axis}[xmin=0,
    ymin=0, scaled ticks=false, yticklabel style={/pgf/number format/fixed},
    width=0.8\textwidth, ylabel style={yshift=0.8cm},
    legend style={at={(0.3,0.9)},anchor=north,legend cell align=left},
        xlabel=$Seconds$,
        ylabel=$Total\ Files\ Posted$, no markers]

\addplot [black, very thick] table[y = C1.Xlarge-100Workers-50Mins] from \datatable ;

\legend{$C1.XLarge$}   

    \end{axis}

\end{tikzpicture} }
  \vskip -0.8em
    \caption{Exp:NIM3-4 Cumulative files processed over time for 100 C1.XLarge Instances, running 100 workers for 3,000 seconds.}
  \label{fig:3000seconds}
\end{figure}

\pgfplotstableread{Data/Best-Result-RawData/06-FTP-100wXL-c1.xlarge/2013-10-31-01-50-55-50mins/starttimes.prn}
\datatable

\begin{figure}[htbp] 
\centering
\fbox {
\begin{tikzpicture}
    \begin{axis}[xmin=0,
    ymin=0,ymax=50,
    width=0.8\textwidth,
    legend style={at={(0.5,0.9)},anchor=north,legend cell align=left},
        xlabel=$Time\ in\ Seconds$,
        ylabel=$Results\ Posted$, no markers]

\addplot[black,very thick] table[y = start1] from \datatable ;
\addplot [red, very thick] table[y = start2] from \datatable ;

\legend{$Worker\ Threads\  Start\ Time$,$Worker\ Threads\ Aligned$}   

    \end{axis}

\end{tikzpicture} }
  \vskip -0.8em
    \caption{Exp:NIM3-4 Start times for each threaded worker across all Instances. }
  \label{fig:workerstarttimes}
\end{figure}
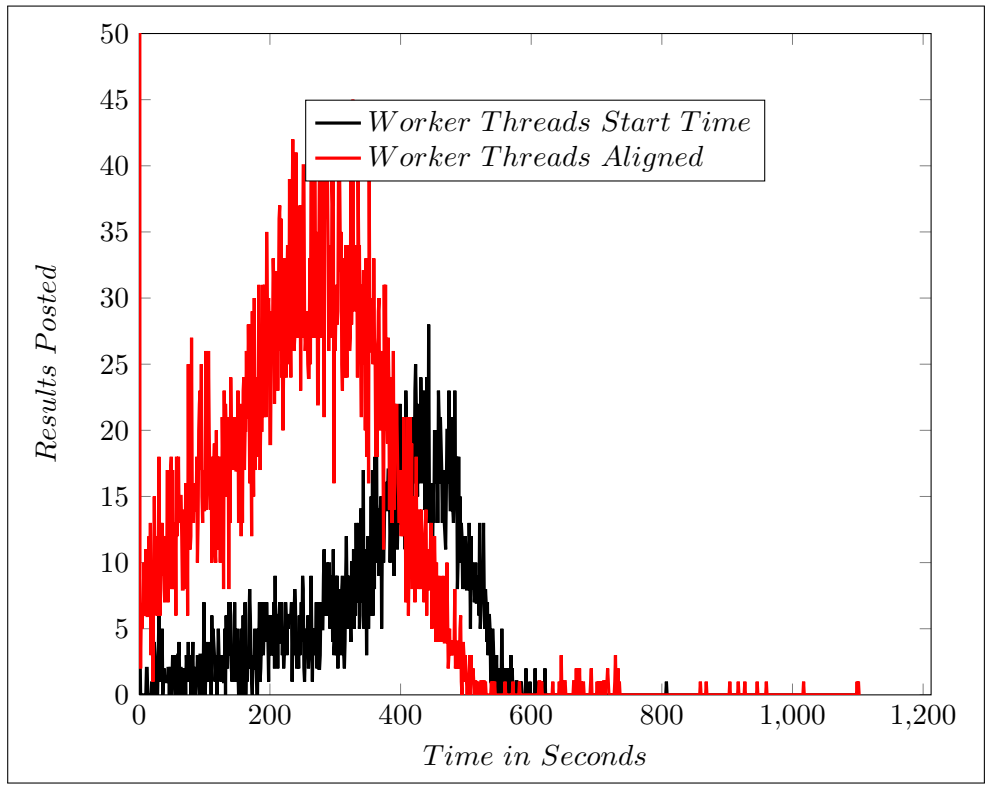

\par Table \ref{tab:bestperfacrossinstances} gives the details for Figure \ref{fig:bestperformancerates} which presents the most significant results of these experiments. Further increases in either the instance types or the number of instances requires additional funding and investment. As the instance type is increased, so is the overall performance. At some point the web server providing the data will become the bottleneck for the system, however this is not an issue once the data being processed is  distributed  across different locations.

Data generation rate must be slower than the data publication rate to ensure that work can be advertised as quickly as it is generated. Work must also be processed faster then the generation rate. So simply put $D < W < P$ where D is the rate at which data is generated, W is the rate at which work is processed and P is the rate at which work can be published.

For all experiments which have 100 instances running, the FTP web server is required or there is a noticeable decrease in the performance of the overall system as the web server becomes the bottleneck. For the system to scale, the source of the data being processed must be distributed. Lower performance web servers which grouped together can match the high performance web server from HEANet for example.

\subsubsection{Analysis}
\par The pipeline allows for instance types to vary, for the number of instances to increase and for the number of web servers to be extended. Each of the these components contribute to the overall performance of the system, and these experiments review the core factors as each of these components in increased. The key findings of these experiments are summarised below. 

\begin {itemize}

\item  Exp:NIM3-1. As the number of instances increases some of the web servers failed to scale and download time increased on average.  Where multiple slower web servers were used, they maintained good performance due to the load balancing across them.  
\item Exp:NIM3-2.  The differences in machine types performance continues as the number instances increases, but there is a higher variance observerd most likely due to the increase in the number of overall requests to the server. The M1.Large instance type is still maintaining faster processing rates consistently compared to the T1.Micro.  
\item Exp:NIM3-3. As the number of instances increases so does the overall processing of the pipeline. It is shown the this increase is statistically significant, and there is a high correlation between the number of worker instances and the performance of the pipeline. 
\item Exp:NIM3-4. Limit based experiments demonstrate the core factors in system scaling. The optimal number of worker threads on an instance is different for each instance type used, and the web server ability to respond to increases in the number of requests varies by configuration. If more web servers are used the load is spread out allowing the processing rates to continue to grow literally with number of instances. Using the larger machine types and the more powerful web servers high processing rates are observed, but the similarity in resuls may indicate that despite the high processing rate, a limit may be emerging within the system. 

\end {itemize}

\subsection{System Limits }

\par
\begin{figure}[htbp]
\centering
\fbox{   \includegraphics[width=0.95\textwidth] {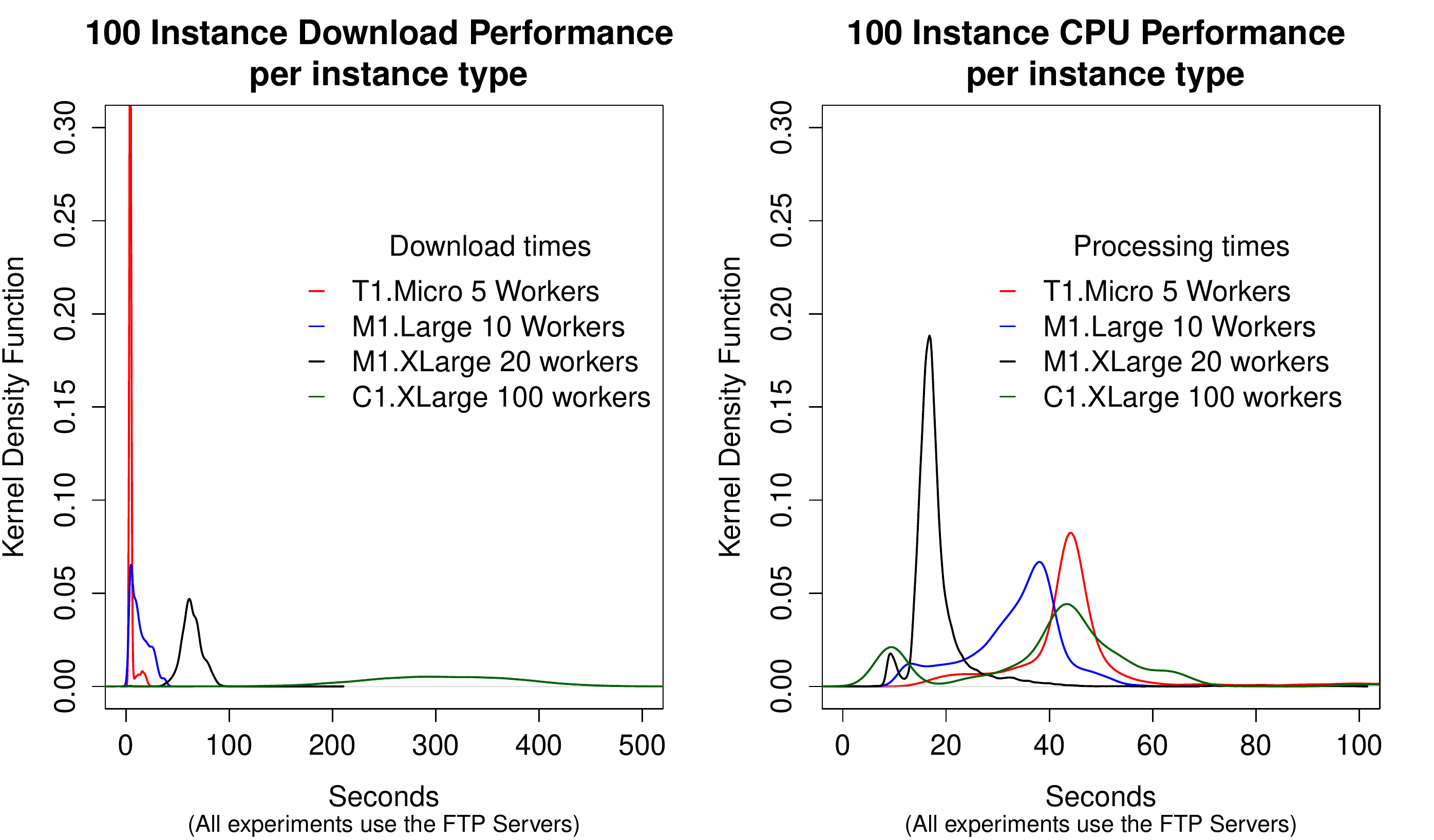}  }
  \vskip -0.8em
    \caption{Breakdown of CPU and Network performance on fastest experiments by instance type.}
  \label{fig:breakdown-best}
\end{figure}

 Of interest in Figure \ref{fig:100instances} is the similarity of the processing rates across machine types. To determine if this is a CPU bottleneck or a web sever download issue, a more detailed look at the split between these two activities for the worker node within each experiment is shown in Figure \ref{fig:breakdown-best}. As expected the large amazon EC2 instances process data the fastest, however considering that they are each using the same web server for download, there are clearly difference in download times for each experiment. Given an overall lower processing rate for the T1.Micro experiment, the FTP web server performance consistently well with fast download times, while the M1.XLarge running 20 workers experiences considerably slower download rates. The M1.XLarge processing times are faster than the download times, leaving the primary bottleneck as the web server. This is reversed for the T1.Micro which has fast downloads but slower processing times.  Of primary concern in this graph is the increase in the download times for the C1.XLarge instances. As the machine times get faster the mean download time gets longer, going from 5 seconds to 300 seconds. The reduction in the download time could be an indication of a bottleneck with the overall system. To investigate this issue, the following possible restrictions were considered and experiments designed to identify and eliminate the bottleneck to ensure that there were no underlying scalability issues within the architecture. Table \ref{tab:nim4-exp} contains details of these experiments.

\begin{table} [htbp]
\centering
\begin{tabular}{p{2.5cm} p{3.5cm} p{8cm}}
  \toprule
Reference & Title & Experimental Objective \\
  \midrule
\textbf{\small{Exp:NIM4-1}}& \small{Instance limitations based on workers} & \small {As the number of workers increases on a server instance, even a larger server, at some point there may be diminishing rates of returns for processing and/or downloading files. }\\
\textbf{\small{Exp:NIM4-2}}& \small{Virtual Machine sharing as a bottleneck} & \small {Depending on the configuration and deployment of the virtual machine instance it is possible that the service of the network could degrade as virtual machine instances shared physical networks. While is is not knowable exactly how virtual instances are deployed on  physical machines, the basic transfer rates can be looked at.  }\\
\textbf{\small{Exp:NIM4-3}}& \small{Bandwidth as a Bottleneck} & \small {This would indicate that the network either had bandwidth throttles or limited bandwidth available, and that the overall experiments were exceeding these values.}\\
\textbf{\small{Exp:NIM4-4}}& \small{Web Server as a bottleneck} & \small {Determine if the service form the web server had some hard limit for servicing file requests.}\\
\textbf{\small{Exp:NIM4-5}}& \small{System Scalability.} & \small{ The scalability and flexibility of the system is tested taking into account any limits observed in previous experiments. } \\

\hline
  \bottomrule
\end{tabular}
  \caption{NIM4: Testing the pipeline architecture for system bottlenecks}
\label{tab:nim4-exp}
\end{table}

\subsubsection{Exp:NIM4-1 Instance limitations based on workers }
\par For these experiments a single M3.2XLarge instance type was chosen. Comparable to the C1.XLarge instance type, this is an 8 VCPU instance with 30GB of RAM using SSD drives. This instance type is described as a balance between compute, memory and network performance with the CPU usually a high frequency Intel Xeon E5-2670 v2 processor. 

\begin{figure}[htbp]
\centering
\fbox{   \includegraphics[width=0.95\textwidth] {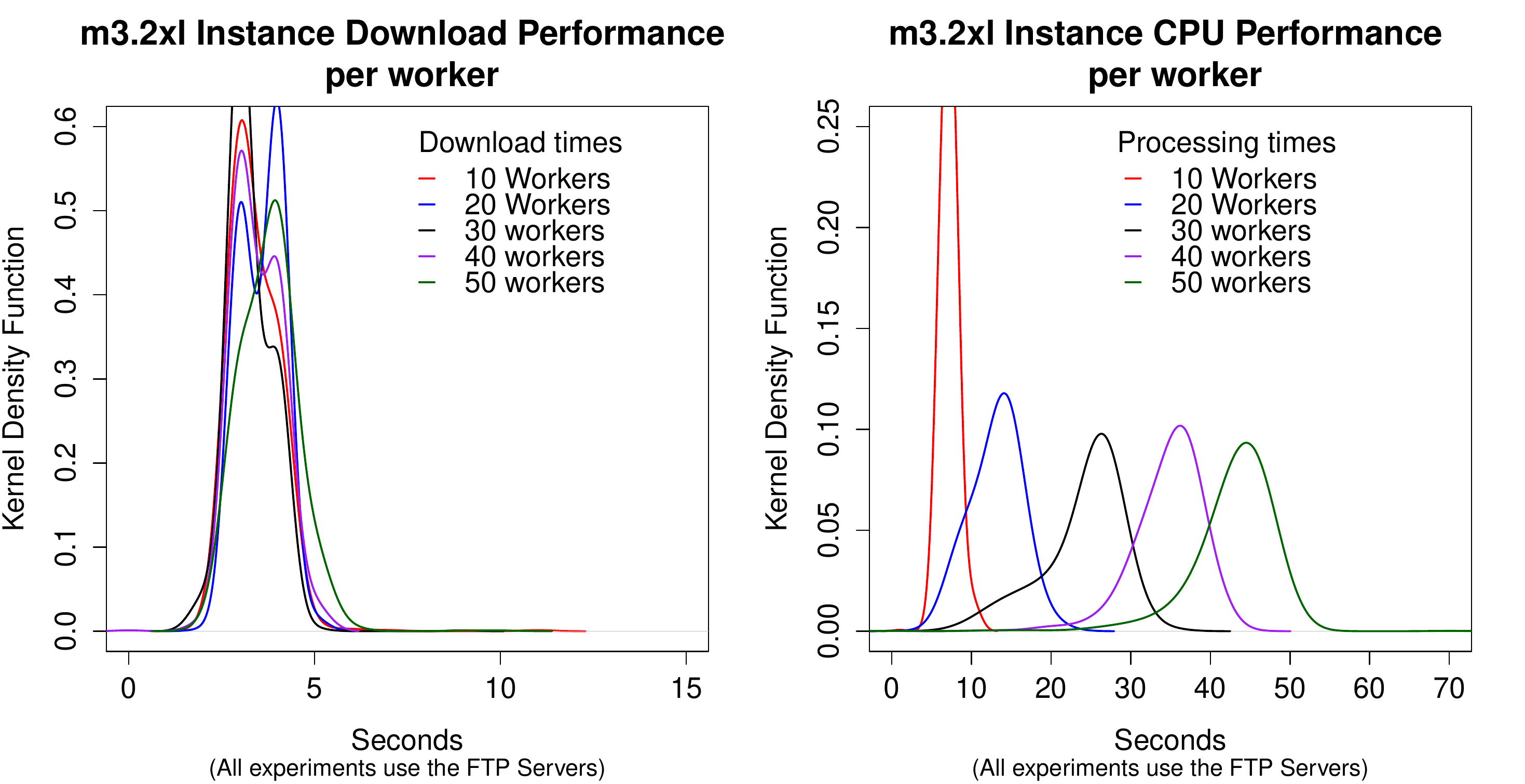}  }
  \vskip -0.8em
    \caption{Exp:NIM4-1 Breakdown of CPU and Network performance by number of workers }
  \label{fig:breakdown-scalingworkers}
\end{figure}

\par In Figure \ref{fig:breakdown-scalingworkers} the breakdown of network download and performance for a single instance is shown. The web server used was the ftp.heanet.ie server and as can be seen quite clearly, the web server offers consistent performance, with the mean download time ranging from 3.39 to 3.76 seconds for a batch of 10 images with a standard deviation of less than 0.7 seconds (Table: \ref{tab:compare-1-100-sd-var-4}). The processing rate in seconds, for cleaning the downloaded batch of 10 images, demonstrates a different behaviour showing a clear relationship between the number of workers and the processing time for a batch of images. As the processing times increase as more workers are added, the overall processing rate remains about the same for the experiment run.  This relationship will be dependant on the system configuration used, and may require additional monitoring and adjustment to obtain an optimal rate over time. Assuming similar performance, less workers per system is preferred purely for the reduction in complexity in post experimental analysis to provide a smaller data set per instance for analysis. The file processing rate  per second $P_{fps}$ is calculated using Equation \ref{eq:fprps}, where the experimental time $E_{time}$ was limited to $1200$ seconds, and $F_{total}$ is the total number of files processed by the experiment. 

\begin{equation}
\label{eq:fprps}
         {P}_{fps}=\frac {F_{total}}{E_{time}}
\end{equation}

\begin{table} [htbp]
\centering
  \begin{tabular} {cccccccc}
  \toprule
  \addlinespace[-5mm]
  \multicolumn{1}{b{1.5cm}}{\begin{center}Num Workers\end{center}} &
    \multicolumn{1}{b{1.5cm}}{\begin{center}Download Mean\end{center}} &
   \multicolumn{1}{b{1.5cm}}{\begin{center}Download Std. Dev.\end{center}} &
   \multicolumn{1}{b{1.5cm}}{\begin{center}Download Variance\end{center}} &
   \multicolumn{1}{b{1.4cm}}{\begin{center}CPU Mean\end{center}} &
   \multicolumn{1}{b{1.4cm}}{\begin{center}CPU Std. Dev.\end{center}} &
   \multicolumn{1}{b{1.4cm}}{\begin{center}CPU Variance\end{center}} &
   \multicolumn{1}{b{1.4cm}}{\begin{center}fps\end{center}} \\
\addlinespace[-5mm]
  \midrule
  10 		& 3.39   	& 0.59  &  0.36 &  7.1   & 1.0 & 1.1   & 6.4\\
  20 		& 3.57  	& 0.52  &  0.27 & 13.1  & 2.9 &  8.7 & 9.0\\
  30 		& 3.26   	& 0.54  &  0.29 	&  24.0 & 4.8 &  23.4 & 9.0 \\
  40 	        & 3.45 	& 0.58  &  0.34  & 34.7 & 4.3 &  18.2 & 9.0 \\
  50 	        & 3.76   	& 0.69  &  0.47  & 43.0 & 5.4 &  28.9 & 9.3\\

 \hline
 \bottomrule
  \end{tabular}
  \caption{Comparing single instance M3.2XLarge performance statistics and file processing rate per second for different workers. }

\label{tab:compare-1-100-sd-var-4}
\end{table}

\subsubsection{Exp:NIM4-2 Virtual Machine sharing as a bottleneck }
The aim of this experiment is to scale the number of instances until a bottleneck is evident using instances from a single physical AWS location and to then determine if the limitation can be overcome by ensuring that additional instances are created which do not share physical hardware. To accomplish this multiple AWS server locations around the world are used, and in this way any resource sharing at the network or physical machine level are eliminated.  While the advertised performance of the M3.2XLarge instance is set to "Fast" there is no specific performance service level agreement given. The expectation is that the network performance of the instance should be approximately 1GBits per second. From the specifications of the AWS services it is unclear if this is a maximum burst capacity, or a sustainable data transfer rate. The possibility exists that virtual machine deployment could dictate that sharing of network capacity exists to some extent. From the experiments on a single instance with 20 workers, as shown in Table \ref{tab:compare-1-100-sd-var-4}, and using the files processed per second $P_{fps}$  for a single instance, the data throughput for the network  in Gigabits per second, $B_{Gbps}$ is calculated as follows, where $n$ is the number of instances being run,  and $D$ is the size of a data cubed image file in Megabytes. To convert from bytes to bits, the result is multiplied by 8.  

\begin{equation}
 B_{Gbps} = \frac{D}{1024}  \sum_{i=1}^{n}P_{fps} * 8
  \label{eq:Expgrp3-2}
  \end{equation}
 
 \par
\begin{figure}[htbp] 
\centering
\fbox{     

\begin{tikzpicture}
\begin{axis} [
ylabel=Files Per Second,
enlarge x limits=0.2,
legend style={ at={(0.5,-0.25)},  anchor=north,legend columns=-1 }, 
ymin=0, ybar, 
xtick=data, 
symbolic x coords={1 Instance, 10 Instances,20 Instances, 25 Instances, 35 Instances, 40 Instances,50 Instances, 50 Instances*},
grid=major, bar width=5mm,
width=15cm, 
xmajorgrids=false, nodes near coords,
every node near coords/.append style={anchor=mid west,rotate=70,font=\tiny},
xticklabel style={
            inner sep=0pt,
            anchor=north east,
            rotate=45 } 
]
\addplot coordinates {(1 Instance,9) (10 Instances,88)  (20 Instances,185) (25 Instances,207) (35 Instances,222) (40 Instances,164) (50 Instances,197)  (50 Instances*,227) };
\addplot coordinates {(1 Instance,9) (10 Instances,90)  (20 Instances,180) (25 Instances,225) (35 Instances,315) (40 Instances,360) (50 Instances,450)  (50 Instances*,450) };
\legend{Actual Result, Linear Result}
\end{axis}
\end{tikzpicture}
}
  \caption{Exp:NIM4-2 File processing rates for different numbers of Instances using the M3.2XLarge instance type using 20 workers. }
  \label{fig:Expgrp3-2}

\end{figure}
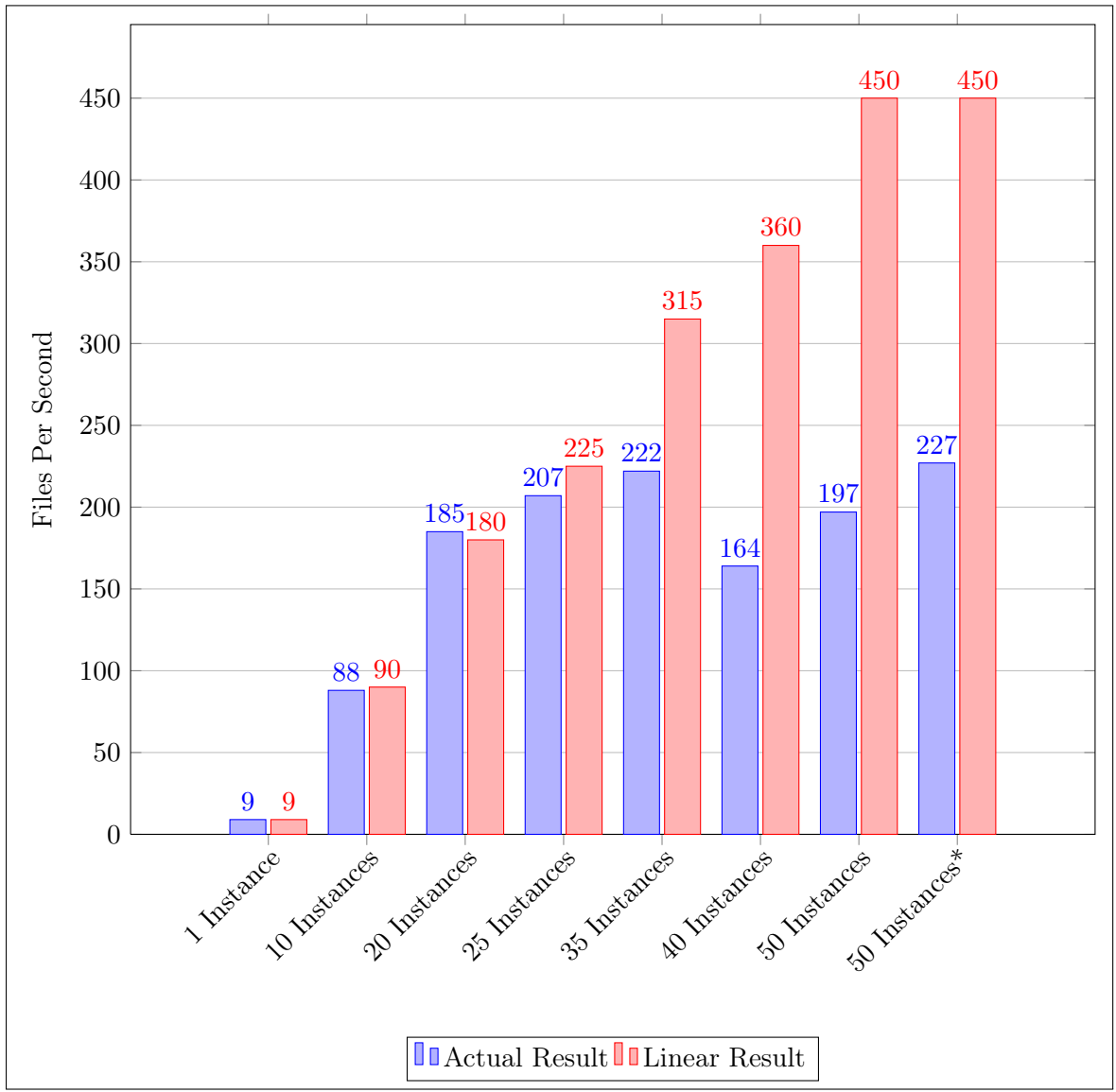

 For the images used in these experiments the size is 2.4Mbytes, and unless otherwise stated, experiments run for 1200 seconds. While the file process rate for an instance may vary to some extent, it can be estimated that  $P_{fps}$ for 50 instances should be approximately 8.4 Gbps. This should be well within the limits of the individual instance specifications which  is 1 Gbps per server and where a potential maximum network performance of 50Gbps (1 Gbps per instance). There is a possibility that the virtual instances may not be able to perform at sustained data transfer speeds or that the network connection between the instances and the HEANnet ftp server are not sufficient to sustain these transfer rates. To determine if a limit of sustainable Gbps transfer speed exists  a set of experiments were run using varying numbers of M3.2XLarge instances each running 20 workers. Figure \ref{fig:Expgrp3-2} clearly shows a roughly linear increase in files processed per second as instances are initially scaled, however as the number of instances rises, this linear increase in performance does not continue after about 25 instances. 
 
 \par If this is a bottleneck due to physical resource sharing by virtual instances, the bottleneck can be circumvented by running 25 of the instances from the initial location, and 25 additional instances from alternative locations. If the linear scaling of the architecture is to continue,  file processing rates for 50 Instances should be twice that of 25 instances. Looking at Figure \ref{fig:Expgrp3-2}, the label 50 Instances* represents an experiment where half of the instances were run in the primary AWS location used for all experiments, while the remaining 25 instances were run in a different physical location. The bottleneck observed after 25 instances is still evident in the result, and as such the sharing of virtual resources as the cause for this can be ruled out based on these results.  

\subsubsection{Exp:NIM4-3: Bandwith as a Bottleneck }

\par The issue of a bottleneck existing as a result of the shared physical network between the AWS and the HEANet needs also to be considered. If the maximum connection speed between AWS and HEAnet was reached after processing at a rate of 230 Files per second then no additional increase in instances would affect the overall result. Using equation \ref{eq:Expgrp3-2}, the bandwidth required, in Gbps, for 300 file per second for files of size 2.4Mbytes, is less than 4.5 Gbps. Further investigation revealed that the actual connection between these two networks at the edge for the AWS networking within Ireland, was 10Gbps. Data was limited to 1Gbps from AWS regions outside of Ireland which in total 11Gbps.  While network sharing would be expected to occur, experiments using the HEAnet FTP server consistently showed a limit of approximately 4 to 4.5 Gbps regardless of the HEANet FTP server load. The network bandwidth limitation between the instances and the server would appear not be responsible for the bottleneck in performance.

\subsubsection{Exp:NIM4-4: Web Server as a Bottleneck }

\par Considering that experiments have shown that moving instances to different regions around the AWS network allowed the architecture to use additional bandwidth and to ensure that machine resource sharing was not an issue, the final set of tests considers the web server as the primary bottleneck. As the number of requests increases it is possible that the web server's ability to service the requests is limited. The ftp.heanet.ie service uses a load balancing service which is supported by four apache web servers.  As requests are made to the ftp.heanet.ie url the load balancer will offload requests to one of four web servers each of which has a replicated copy of the data. The configuration of each of these web servers should allow for more concurrent connections than requested from the processing pipeline. To eliminate the load balancer as an issue, tests were run against one of the four web servers. With 50 instances running, the test used the same network and instances as pervious tests. The result was that the total number of files processed in 1200 seconds was 66 thousand, which is approximately a quarter of the total processed when all four of the web servers are utilised by the load balancer. From this observation it would appear that the bottleneck in the experiments provided is the ftp.heanet.ie web server. While its performance was superior to the other web servers used, its ability to process more that 250 file per second would appear to be limited. As the 100 instances used in earlier experiments with the C1.2XLarge virtual machine type was about the limit of the entire system, this bottleneck was not initially evident. Only by increasing the instance performance, and splitting the download versus processing time was this bottleneck evident.

\subsubsection{Exp:NIM4-5 System Scalability}
\par To ensure that the proposed architecture is scalable, modifications were made to the system to address the possible bottlenecks die tidied and the actual web server bottleneck already discussed. Three key changes were introduced. 

 \begin{itemize}

	\item Diversity in the location of the virtual machines to eliminate resource sharing concerns. In this case 4 AWS regions were chosen to run virtual machines. Ireland, Virginia, Tokyo and San Paolo. 
	\item Increase in the number of web servers. In the event of a single web server being a bottleneck due to the high rate of request, data should be spread across multiple servers. In this case two additional servers were used which were also within the HEANet network. 
	\item Increase the number of SQS queues.  Although there was no reduction in the performance of the get message requests, additional worker queue were created to demonstrate the flexible nation of this approach. If worker instances have a choice of sqs queue to read, then they could in theory self load balance by moving queues if the response time for a message started to degrade. 
	
\end{itemize}

\par
\begin{figure}[htbp]
\centering
\fbox{   \includegraphics[width=0.8\textwidth] {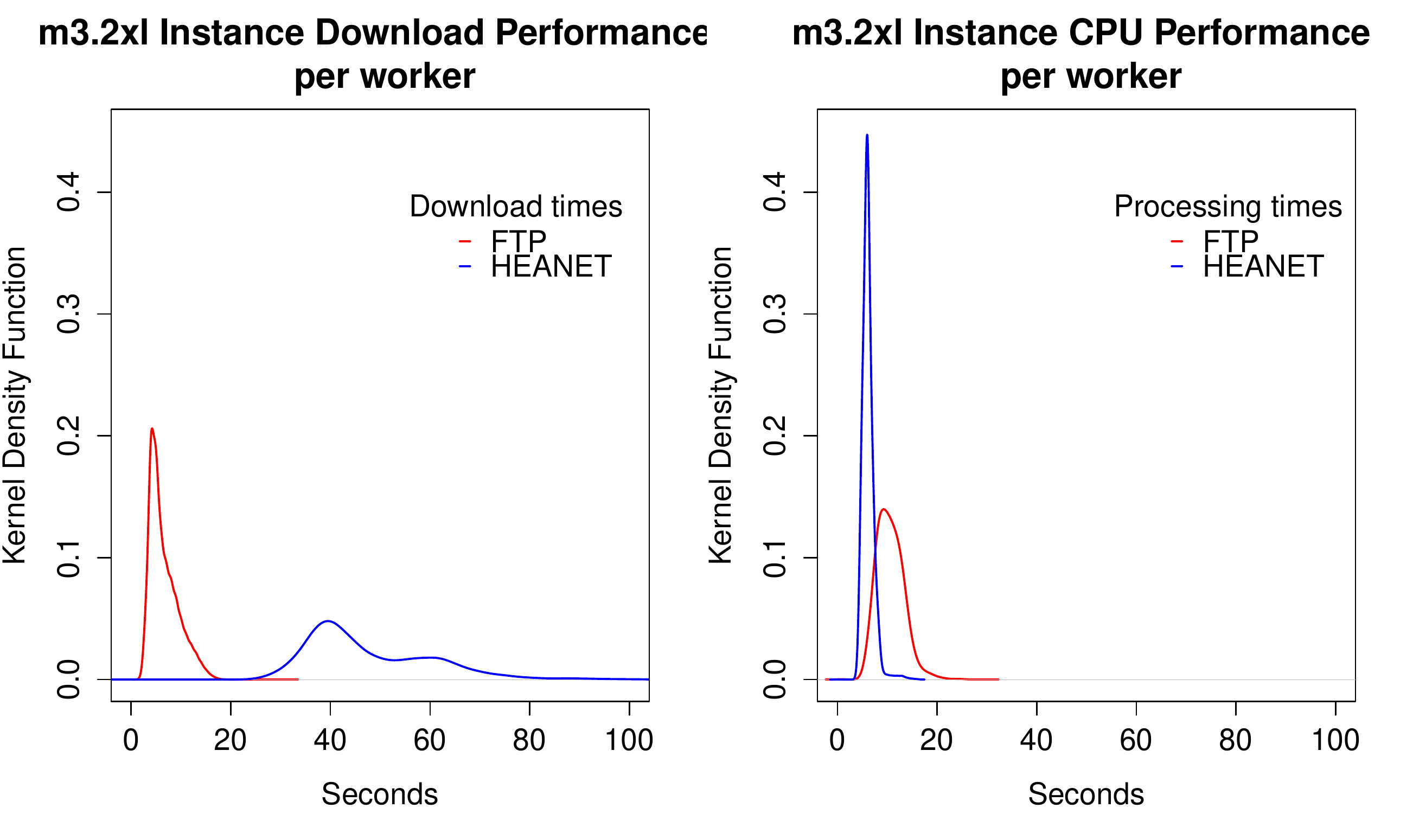}  }
  \vskip -0.8em
    \caption{Exp:NIM4-5 Breakdown of CPU and Network performance for 20 worker per Instance experiment where  Instances are located in multiple AWS regions and use multiple web servers.  }
  \label{fig:breakdown-international}
\end{figure}

 Using these modifications, a final experiment was run which consisted of 25 instances in the Ireland AWS region and 15 instances in the three additional AWS regions mentioned above. Each instance was of type M3.2XLarge, and ran 20 workers threads each. One workerq contained work for the FTP.HEANet.ie web server which was used by the Ireland based instances, and a second wokerq pointing and two additional web servers which was used by the 3 non-Irish based locations, each of which ran 15 instances for a total of 45 instances, giving 70 instances in all for the experiment.  Figure \ref{fig:breakdown-international} shows consistent network download performance and CPU performance. Within the experiment time a total of almost 400,000 files were processed, which equates to 322.9 files per second. Using the formula \ref{EQ:etbh} this is an equivalent processing rate of 194 TB per 24 hours. 

\pgfplotstableread{Data/Best-Result-RawData/bigpicture-sorted.prn}
\datatable

\begin{figure} [htbp]
\centering
\fbox {
\begin{tikzpicture}
    \begin{axis}[xmin=0,ymax=350,
    ymin=0, yticklabel style={/pgf/number format/fixed},
    width=0.75\textwidth,ybar,bar width=2pt, height=4.5cm,
    legend style={at={(0.3,0.9)},anchor=north,legend cell align=left},
        xlabel=$Experiment\ Number$,
        ylabel=$Total\ Files\ Per\ Second$]

\addplot table[y = fps] from \datatable ;
    \end{axis}
\end{tikzpicture} }
  \vskip -0.8em
    \caption{Exp:NIM4-5 Files Per Second. See Appendix Table \ref{tab:longtable2}.}
  \label{fig:bigpicture}
\end{figure}
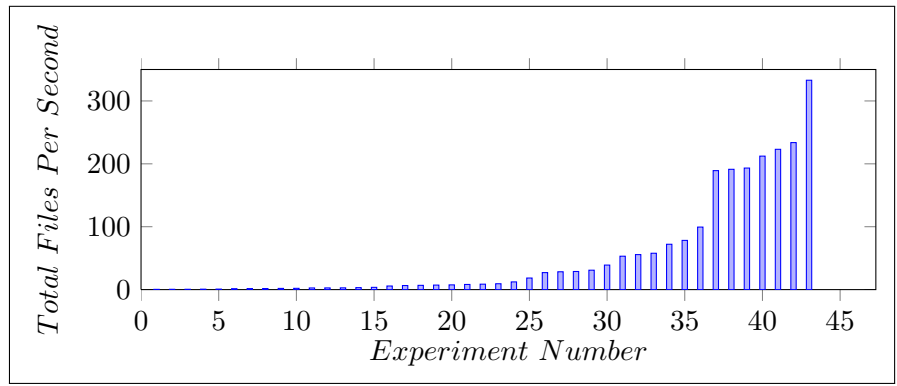

%
%
\pgfplotstableread{Data/Best-Result-RawData/bigpicture-sorted-tw.prn}
\datatable

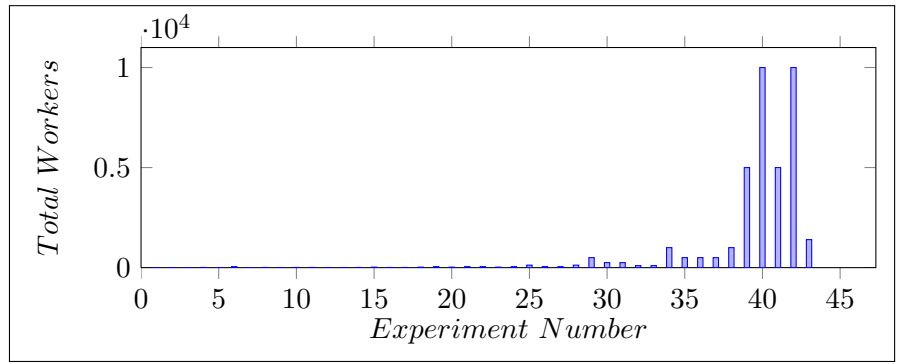
\begin{figure}[htbp]
\centering
\fbox {
\begin{tikzpicture}
    \begin{axis}[xmin=0,ymax=11000,
    ymin=0, yticklabel style={/pgf/number format/fixed},
    width=0.75\textwidth,ybar,bar width=2pt, height=4.5cm,
    legend style={at={(0.3,0.9)},anchor=north,legend cell align=left},
        xlabel=$Experiment\ Number$,
        ylabel=$Total\ Workers$]

\addplot table[y = tw] from \datatable ;

    \end{axis}

\end{tikzpicture} }
  \vskip -0.8em
\caption{Exp:NIM4-5 Total Workers Per Experiment. See Appendix Table \ref{tab:longtable2}.}

  \label{fig:bigpicture1}
\end{figure}

\pgfplotstableread{Data/Best-Result-RawData/bigpicture-sorted-ti.prn}
\datatable

\begin{figure}[htbp]
\centering
\fbox {
\begin{tikzpicture}
    \begin{axis}[xmin=0,ymax=110,
    ymin=0, yticklabel style={/pgf/number format/fixed},
    width=0.75\textwidth,ybar,bar width=2pt, height=4.5cm,
    legend style={at={(0.3,0.9)},anchor=north,legend cell align=left},
        xlabel=$Experiment\ Number$,
        ylabel=$Total\ Instances$]

\addplot table[y = Instances] from \datatable ;
    \end{axis}

\end{tikzpicture} }
  \vskip -0.8em
\caption{Exp:NIM4-5 Total Instances Per Experiment. See Appendix Table \ref{tab:longtable2}.}
  \label{fig:bigpicture2}
\end{figure}
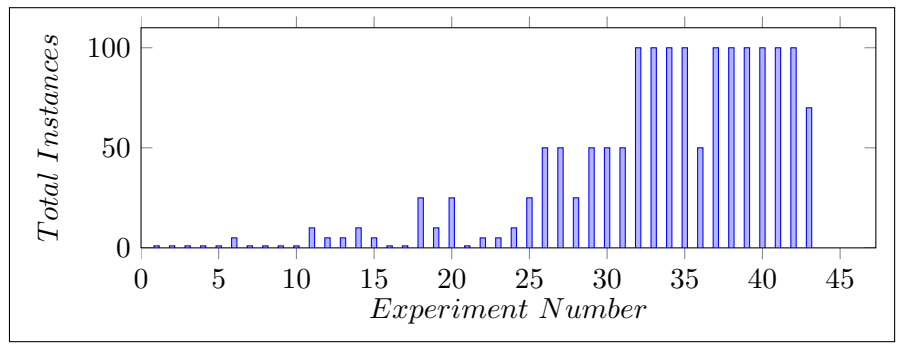

\par Figures \ref{fig:bigpicture}, \ref{fig:bigpicture1}, and \ref{fig:bigpicture2} plot different aspects of 43 experiments. Figure  \ref{fig:bigpicture} shows the increase in files per second for each experiment, with the final experiment giving the fastest processing rate of 322.9 files per second. If this is cross referenced to Figure \ref{fig:bigpicture1}, which shows the total number of workers operating within each experiment, it can be shown that the total number of workers does not necessarily correspond to faster process rates. The fastest experiment for example ran 1400 worker instances, while previous experiments ran as many as 10,000. 

\par Figure \ref{fig:bigpicture2} shows the number of instances run per experiment and again the fastest experiment had only 70 instances. While the number of instances was important, the number of workers per instance was a function of the power of the server, with more powerful servers being able to run more workers. The major difference however between the fastest experiments was that with the introduction of additional web servers, the pipeline could continue to increase its processing rate, whereas previous experiments were hitting a web server bottleneck. So it is shown that by altering certain parameters of the pipeline, but without changing the underlying architecture, an experiment with 70 globally distributed worker instances, running a total of 1400 worker threads, outperformed all other experiments.

\subsubsection{Analysis}
\par The pipeline allows for an instance type to vary, for the number of instances to increase, the number of worker queues expanded and for the number of web servers to be extended. Each of the these components contribute to the overall performance of the system, and these experiments review the core factors as each of these components in increased. The key findings of these experiments are summarised below. Ultimately when all parameters were reviewed any observable limitations were identified and the architecture of the pipeline allowed these to be eliminated. 

\begin {itemize}

\item Exp:NIM4-1. As the number of worker threads per instances increases these experiments show that the web server delivers images at a consistent rate, but the processing time as the workers increase goes up.  If the web server is not a limit then worker threads will eventually become their own bottleneck if they continue to increase.  
\item Exp:NIM4-2. To eliminate the possibility that the virtual machine instances used with the AWS, which were based in the same geographical region had limits on their collective shared bandwidth, experiments were run with EC2 instances starting in multiple geographical regions. Using tests with 50 instance, there was no significant difference on the overall result based on the location of the virtual machines. 
\item Exp:NIM4-3. The bandwidth required for a processing rate of 300 files per second is less then 5GBs, the bandwidth connecting AWS to HEANet is 10Gbs. Experiments were run when the HEANet server was both heavily and lightly loaded, and in all cases the data transfer from the web servers was less 4Gbps.  The HEANet bandwidth clearly could operate at a higher rate, but the pipeline could not get better performance rates that about 230fps. 
\item Exp:NIM4-4. The HEAnet FTP web server was tested to see if it was a bottleneck. The web server is load balanced, so testing was performed agains one of the servers and the processing rate dropped to one quarter of the previous results. It can be concluded that this websever had finally hit a data serving limit, possibly due to disk access read limitations.      
\item Exp:NIM4-5. The final experiment was run taking all results into account. Using 70 virtual machines distributed around the world, and using multiple web servers and worker queues the processing rate achieved was close to 200TB per day. Further scalability may be possible, but the budgets used were limited.  
\end {itemize}

\section{Conclusion}

\par The NIMBUS pipeline extends the ACN pipeline into a global environment. The use of the publicly accessibly SQS web queue service from AWS enabled computing resources from around the globe to join the processing cloud which was a key modification to the ACN architecture.  The SQS is a distributed architecture which scales horizontally allowing for a maximum of 120,000 messages in flight at a time for a single queue. NIMBUS allows for the used of multiple queues  both for advertising work, controlling the behaviour of the worker instances, and monitoring running experiments.

\par Most of the worker instances used for experiments were EC2 virtual machines within the AWS cloud. This was a convenience for experimentation rather than a primary required feature of the architecture. Physical machines were also integrated into the pipeline for various experimental purposes, specifically when reviewing maximum performance for individual servers. Experiments ranged from single instances to a maximum of 100 instances. Each worker instance had multiple threads activated to determine optimal configuration of the instance type. 

\par The experimental variables modified while testing the architectures ability to scale to terabytes per hour were, virtual machine location, virtual machine size, number of worker threads per instance, web server type and number of web servers, and number of queue for advertising work. 

\par The data processed by the system is assumed to be part of an existing data archive and is already compressed. The requirement for a datastore to participate in the pipeline is to create an SQS message for each file to be processed. 

\par The final experiment demonstrated that horizontal scaling of all primary components is possible, and that this approach will ensure system bottlenecks are overcome. 

The following is a summary of the experiments used to evaluate the performance of the pipeline. 

\begin {itemize}

\item The SQS queue performance was evaluated to ensure that its ability to have messages written was sufficiently fast to advertise work for processing, and for messages to be read by workers within the processing cloud as it scales to thousands of worker threads. Work advertising needs to be significantly faster than work processing. Multi-threaded Python programs were required to ensure that the speed was sufficiently fast.     

\item Before scaling to multiple worker instances, the performance limit of a single instance was explored. By varying the type of instance, the number of worker threads run and the web server used, a baseline was obtained against which the scaling architecture could be compared.  Multiple threads increased performance, but bottlenecks were identified after too many threads were run. With a sufficiently fast web server to service requests, higher performance from the worker instance was based on the number and speed of the CPUs used. 
 
\item Scaling from a base of good performing single instances, the experiments reviewed the ability of the web servers to maintain data rates. Multiple web servers provided the best solution to ensure that data requirements from the scaling instances could be serviced. It was shown that there was a strong correlation between instance numbers and the processing speed of the pipeline. 

\item Despite the good performance of the scaling experiments, further investigation revealed possible limitations in the processing rate. A series of tests identified the possible sources of these limits and a final experiment was run which increased the horizontal scaling of each of the primary architectural components. This resulted in an overall processing rate improvement using less instances. While funding limited the ability to run additional experiments, the result of this final experiment were such that 192TB of data per 24hours processing could be achieved, with evidence that further improvement would be possible through the use of more instance types, and more web servers providing data. 

\end {itemize}

%% file: chapter7.tex

\chapter{Conclusion}

\par The research hypothesis proposed within this thesis was to determine if a globally distributed astronomical CCD image reduction pipeline can process data at a rate which exceeds existing data processing requirements and is scalable such that it can meet future data processing challenges. The hypothesis has been proven to be true  in Chapter 4 (the ACN pipeline), and Chapter 5 (the NIMBUS pipeline). The ACN demonstrates that data can be compressed and uploaded to a central storage service at a rate higher than it can be generated. The NIMBUS pipeline demonstrates that data can be distributed and processed using a horizontally scalable elastic architecture at proven rates of 200 terabytes per day.

\par This thesis focused on the processing of CCD image reduction and photometry in a distributed global network to determine if an architecture could be devised which had the ability to process terabytes of data per day to meet the growing demands of data production for projects like the LSST which are expected to reach tens of terabyte per day within the next few years. It has been shown that a distributed horizontally scalable cloud based architecture can process hundreds of terabytes of data per day, and that the architecture is flexible enough to continue to scale to petabytes per day by allowing individual processing nodes to join and leave the pipeline without impacting the integrity of the system. The form of CCD data used is such that it is suitable for parallel processing, and the pipeline developed can be applied to alternative science areas  with similar characteristics. 

\par Whilst existing solutions to CCD data processing for large scale projects, such as space crafts or large ground-based observatories, are processed either serially or using large computing data centres, much of the archive data made available  to researchers still requires reprocessing. A solution, such as NIMBUS, which is flexible and scalable, is needed to assist with this requirement as the volume of data increases. With the increased availability and sensitivity of CCD or CMOS devices, further innovative research opportunities exist to support high-speed photometry,  and robotic telescope farms. Each of these areas will require data processing rates to keep pace with data generation rates.    

\par The NIMBUS processing pipeline architecture is presented as the final stage in an iterative process, where a series of architectures were proposed and experiments designed to evaluate  limits of each of the bottlenecks identified through the research. The initial stage focused on a pilot system called FEBRUUS, which performed basic calibration of CCD images to establish a benchmark of performance for a sequential processing pipeline. Work on this pipeline clearly demonstrated that parallel  processing was a suitable solution that could clearly be used. The second stage was the consideration of IRAF instances implemented within the cloud. This architecture considered virtualisation instances of existing reduction software, communicating via a centralised queue. From the analysis of this potential solution, the queuing design was identified as an essential communication feature. A light-weight utility based on the FEBRUUS pilot, which also performed photometry, was implemented for the ACN pipeline that focused on compression and uploading of data to a distributed storage node, for a distributed processing private cloud to access. The private process cloud was constructed across three Irish institutes of technology. Using a private NFS queuing system limited the system to processing nodes which were logically within the same network and hence failed to incorporate globally distributed nodes. The final pipeline, NIMBUS, incorporated web based queues for advertising work and for controlling the processing cloud. With work being advertised using a globally available message queue, the number of instances, and the number of worker threads within the instances, were expanded and experiments were devised to test the scaling nature of the system.  Final experiments demonstrated the ability to process hundreds of terabytes per day with limits of budget finally restricting additional experimentation. No evidence was found that this final experiment was in any way the maximum processing rate possible by the pipeline. Evidence suggests that further expansion of the system is likely through an increase in available processing units. 

\section{Summary of Contributions and Achievements}

\par The NIMBUS and ACN \cite{doyle2012significantly} architectures forms the core contribution of this thesis. The NIMBUS architecture is focused on data processing of archive systems, while the ACN pipeline focused on the compression and uploading of data to an archive. The NIMBUS system can support both modes of operation, requiring only that all data to be processed is advertised via a central web queue. While Amazon Web Services (AWS) were used in many of the experiments, this was for convenience only and the NIMBUS architecture does not have to be AWS based.  The following are the core features of the pipeline, which form the contributions of this thesis. 

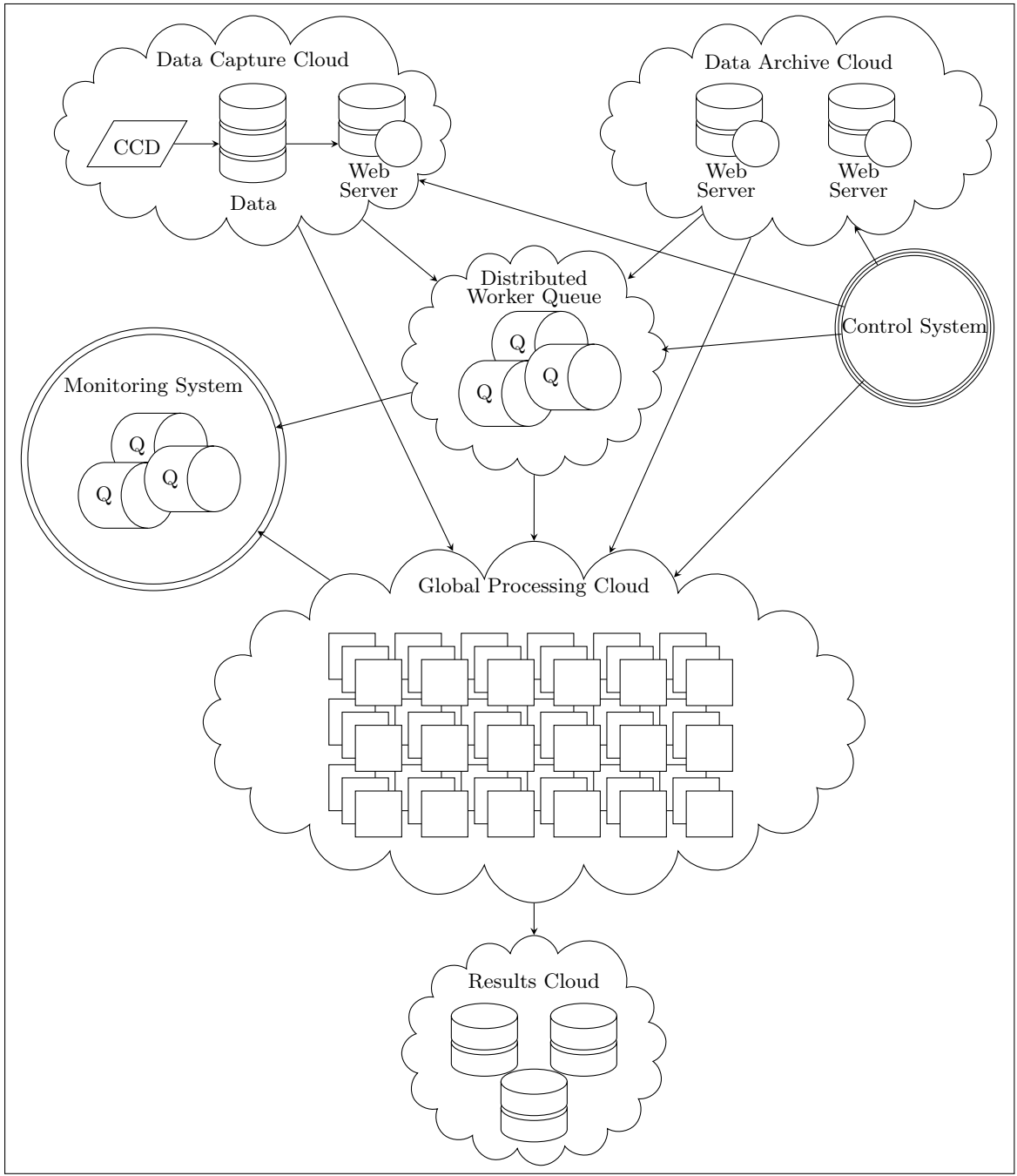
\begin{figure} [htbp]
 \begin{center}
\fbox { 

  \begin{tikzpicture}[node distance=2.5cm, auto, >=stealth,cross line/.style={preaction={draw=white,->,line width=4pt}}]

%
%
    \node[draw, cylinder, shape aspect=1.5, inner sep=0.3333em, fill=white, minimum width=1cm, minimum height=1cm] (cyl1) at (2,-1.25) {\footnotesize{Q}};
    \node[draw, cylinder, shape aspect=1.5, inner sep=0.3333em, fill=white, minimum width=1cm, minimum height=1cm] (cyl1) at (1.5,-2) {\footnotesize{Q}};

  \node[draw, cylinder, shape aspect=1.5, inner sep=0.3333em, fill=white, minimum width=1cm, minimum height=1cm] (cyl1) at (2.5,-1.75) {\footnotesize{Q}};
  \node[shape=cloud, cloud puffs=15.7, cloud ignores aspect, minimum width=4cm, minimum height=3.5cm, align=center, draw] (wrkq) at (2.25, -1.5) {};
  \draw  ($(wrkq.north)+(-0mm,-5mm)$)                     node                {\footnotesize {Distributed}};
  \draw  ($(wrkq.north)+(-0mm,-8mm)$)                     node                {\footnotesize {Worker Queue}};
 
%
%
    \node[shape=cloud, cloud puffs=15.7, cloud ignores aspect, minimum width=6cm, minimum height=3.5cm, align=center, draw] (ARCHIVECloud) at (6.25, 2) {};
    \draw  ($(ARCHIVECloud.north)+(-0mm,-7mm)$)                     node                {\footnotesize {Data Archive Cloud}};
  
    \node[draw, cylinder, shape aspect=1.5, inner sep=0.3333em, fill=white,
    rotate=90, minimum width=1cm, minimum height=0.05cm] (cyl2a) at (5.2,1.8) {};
    \node[draw, cylinder, shape aspect=1.5, inner sep=0.3333em, fill=white,
    rotate=90, minimum width=1cm, minimum height=0.05cm] (cyl3) at (5.2,2.2) {};  
    \draw  ($(cyl3.south)+(-5.5mm,-8mm)$)                     node                {\footnotesize {Web }};
    \draw  ($(cyl3.south)+(-5.5mm,-11mm)$)                     node                {\footnotesize {Server }};

      \node[shape=circle, fill=white,minimum width=.7cm, minimum height=.7cm, align=center, draw] (web) at (5.6, 1.8) {};

    \node[draw, cylinder, shape aspect=1.5, inner sep=0.3333em, fill=white,
    rotate=90, minimum width=1cm, minimum height=0.05cm] (cyl2a) at (7.2,1.8) {};
    \node[draw, cylinder, shape aspect=1.5, inner sep=0.3333em, fill=white,
    rotate=90, minimum width=1cm, minimum height=0.05cm] (cyl3) at (7.2,2.2) {};  
    \draw  ($(cyl3.south)+(-5.5mm,-8mm)$)                     node                {\footnotesize {Web }};
    \draw  ($(cyl3.south)+(-5.5mm,-11mm)$)                     node                {\footnotesize {Server }};

      \node[shape=circle, fill=white,minimum width=.7cm, minimum height=.7cm, align=center, draw] (web) at (7.6, 1.8) {};    
      
%
%
    \node[shape=circle, fill=white,minimum width=2.4cm,  align=center, draw] (controller) at (8, -1) {};    
    \node[shape=circle, fill=white,minimum width=2.3cm,  align=center, draw] (controller) at (8,-1) {};    
    \node[shape=circle, fill=white,minimum width=2.2cm,  align=center, draw] (controller) at (8, -1) {};    
  \draw  ($(controller)+(-0mm,-0mm)$)                     node                {\footnotesize {Control System}};

%
%
    \node[shape=cloud, cloud puffs=15.7, cloud ignores aspect, minimum width=6cm, minimum height=3.5cm, align=center, draw] (DCCloud) at (-2, 2) {};
    \draw  ($(DCCloud.north)+(-0mm,-7mm)$)                     node                {\footnotesize {Data Capture Cloud}};
    
     \node[shape=trapezium, minimum width=.7cm, minimum height=.7cm, align=center, draw,trapezium right angle=120,trapezium left angle=60] (CCD) at (-3.75, 1.8) {};
    \draw  ($(CCD.north)+(-0mm,-4mm)$)                     node                {\footnotesize {CCD}};

    \node[draw, cylinder, shape aspect=1.5, inner sep=0.3333em, fill=white,
    rotate=90, minimum width=1cm, minimum height=0.05cm] (cyl1) at (-2,1.4) {};
    \node[draw, cylinder, shape aspect=1.5, inner sep=0.3333em, fill=white,
    rotate=90, minimum width=1cm, minimum height=0.05cm] (cyl2-1) at (-2,1.8) {};
    \node[draw, cylinder, shape aspect=1.5, inner sep=0.3333em, fill=white,
    rotate=90, minimum width=1cm, minimum height=0.05cm] (cyl3) at (-2,2.2) {};  
  
    \draw  ($(cyl1.south)+(-5mm,-5mm)$)                     node                {\footnotesize {Data }};  
    \node[draw, cylinder, shape aspect=1.5, inner sep=0.3333em, fill=white,
    rotate=90, minimum width=1cm, minimum height=0.05cm] (cyl2a) at (-0.2,1.8) {};
    \node[draw, cylinder, shape aspect=1.5, inner sep=0.3333em, fill=white,
    rotate=90, minimum width=1cm, minimum height=0.05cm] (cyl3) at (-0.2,2.2) {};  
    \draw  ($(cyl3.south)+(-5.5mm,-8mm)$)                     node                {\footnotesize {Web }};
    \draw  ($(cyl3.south)+(-5.5mm,-11mm)$)                     node                {\footnotesize {Server }};

      \node[shape=circle, fill=white,minimum width=.7cm, minimum height=.7cm, align=center, draw] (web) at (0.2,1.8) {};    
%
 %
%
  \node[draw, cylinder, shape aspect=1.5, inner sep=0.3333em, fill=white, minimum width=1cm, minimum height=1cm] (cyla) at (-3.75,-2.8) {\footnotesize{Q}};
  \node[draw, cylinder, shape aspect=1.5, inner sep=0.3333em, fill=white, minimum width=1cm, minimum height=1cm] (cylb) at (-4.25,-3.55) {\footnotesize{Q}};
  \node[draw, cylinder, shape aspect=1.5, inner sep=0.3333em, fill=white, minimum width=1cm, minimum height=1cm] (cylc) at (-3.25,-3.3) {\footnotesize{Q}};
  \node[shape=circle, minimum width=4cm, minimum height=3.5cm, align=center, draw] (moncloud) at (-3.5, -3) {};
    \node[shape=circle, minimum width=3.8cm, minimum height=3.5cm, align=center, draw] (moncloud) at (-3.5, -3) {};

  \draw  ($(moncloud.north)+(-0mm,-8mm)$)                     node                {\footnotesize {Monitoring System}};

%
  \node[shape=cloud, cloud puffs=20, cloud ignores aspect, minimum width=10cm, minimum height=5.5cm, align=center, draw] (heanetcloud) at (2.25, -7) {};
  \draw  ($(heanetcloud.north)+(-0mm,-7mm)$)                     node                {\footnotesize {Global Processing Cloud}};

       \node[shape=rectangle, fill=white,minimum width=.7cm, minimum height=.7cm, align=center, draw] (cpu1) at (4.5, -8) {};
  \node[shape=rectangle, fill=white,minimum width=.7cm, minimum height=.7cm, align=center, draw] (cpu1) at (4.7, -8.2) {};
  \node[shape=rectangle, fill=white,minimum width=.7cm, minimum height=.7cm, align=center, draw] (cpu1) at (4.9, -8.4) {};
  
    \node[shape=rectangle, fill=white,minimum width=.7cm, minimum height=.7cm, align=center, draw] (cpu1) at (3.5, -8) {};
  \node[shape=rectangle, fill=white,minimum width=.7cm, minimum height=.7cm, align=center, draw] (cpu1) at (3.7, -8.2) {};
  \node[shape=rectangle, fill=white,minimum width=.7cm, minimum height=.7cm, align=center, draw] (cpu1) at (3.9, -8.4) {};

 \node[shape=rectangle, fill=white,minimum width=.7cm, minimum height=.7cm, align=center, draw] (cpu1) at (2.5, -8) {};
  \node[shape=rectangle, fill=white,minimum width=.7cm, minimum height=.7cm, align=center, draw] (cpu1) at (2.7, -8.2) {};
  \node[shape=rectangle, fill=white,minimum width=.7cm, minimum height=.7cm, align=center, draw] (cpu1) at (2.9, -8.4) {};

 \node[shape=rectangle, fill=white,minimum width=.7cm, minimum height=.7cm, align=center, draw] (cpu1) at (1.5, -8) {};
  \node[shape=rectangle, fill=white,minimum width=.7cm, minimum height=.7cm, align=center, draw] (cpu1) at (1.7, -8.2) {};
  \node[shape=rectangle, fill=white,minimum width=.7cm, minimum height=.7cm, align=center, draw] (cpu1) at (1.9, -8.4) {};

  \node[shape=rectangle, fill=white,minimum width=.7cm, minimum height=.7cm, align=center, draw] (cpu1) at (0.5, -8) {};
  \node[shape=rectangle, fill=white,minimum width=.7cm, minimum height=.7cm, align=center, draw] (cpu1) at (0.7, -8.2) {};
  \node[shape=rectangle, fill=white,minimum width=.7cm, minimum height=.7cm, align=center, draw] (cpu1) at (0.9, -8.4) {};
  
    \node[shape=rectangle, fill=white,minimum width=.7cm, minimum height=.7cm, align=center, draw] (cpu1) at (-0.5, -8) {};
  \node[shape=rectangle, fill=white,minimum width=.7cm, minimum height=.7cm, align=center, draw] (cpu1) at (-0.3, -8.2) {};
  \node[shape=rectangle, fill=white,minimum width=.7cm, minimum height=.7cm, align=center, draw] (cpu1) at (-0.1, -8.4) {};

        \node[shape=rectangle, fill=white,minimum width=.7cm, minimum height=.7cm, align=center, draw] (cpu1) at (4.5, -7) {};
  \node[shape=rectangle, fill=white,minimum width=.7cm, minimum height=.7cm, align=center, draw] (cpu1) at (4.7, -7.2) {};
  \node[shape=rectangle, fill=white,minimum width=.7cm, minimum height=.7cm, align=center, draw] (cpu1) at (4.9, -7.4) {};
    
        \node[shape=rectangle, fill=white,minimum width=.7cm, minimum height=.7cm, align=center, draw] (cpu1) at (3.5, -7) {};
  \node[shape=rectangle, fill=white,minimum width=.7cm, minimum height=.7cm, align=center, draw] (cpu1) at (3.7, -7.2) {};
  \node[shape=rectangle, fill=white,minimum width=.7cm, minimum height=.7cm, align=center, draw] (cpu1) at (3.9, -7.4) {};

 \node[shape=rectangle, fill=white,minimum width=.7cm, minimum height=.7cm, align=center, draw] (cpu1) at (2.5, -7) {};
  \node[shape=rectangle, fill=white,minimum width=.7cm, minimum height=.7cm, align=center, draw] (cpu1) at (2.7, -7.2) {};
  \node[shape=rectangle, fill=white,minimum width=.7cm, minimum height=.7cm, align=center, draw] (cpu1) at (2.9, -7.4) {};

 \node[shape=rectangle, fill=white,minimum width=.7cm, minimum height=.7cm, align=center, draw] (cpu1) at (1.5, -7) {};
  \node[shape=rectangle, fill=white,minimum width=.7cm, minimum height=.7cm, align=center, draw] (cpu1) at (1.7, -7.2) {};
  \node[shape=rectangle, fill=white,minimum width=.7cm, minimum height=.7cm, align=center, draw] (cpu1) at (1.9, -7.4) {};

  \node[shape=rectangle, fill=white,minimum width=.7cm, minimum height=.7cm, align=center, draw] (cpu1) at (0.5, -7) {};
  \node[shape=rectangle, fill=white,minimum width=.7cm, minimum height=.7cm, align=center, draw] (cpu1) at (0.7, -7.2) {};
  \node[shape=rectangle, fill=white,minimum width=.7cm, minimum height=.7cm, align=center, draw] (cpu1) at (0.9, -7.4) {};
  
  \node[shape=rectangle, fill=white,minimum width=.7cm, minimum height=.7cm, align=center, draw] (cpu1) at (-0.5, -7) {};
  \node[shape=rectangle, fill=white,minimum width=.7cm, minimum height=.7cm, align=center, draw] (cpu1) at (-0.3, -7.2) {};
  \node[shape=rectangle, fill=white,minimum width=.7cm, minimum height=.7cm, align=center, draw] (cpu1) at (-0.1, -7.4) {};

      \node[shape=rectangle, fill=white,minimum width=.7cm, minimum height=.7cm, align=center, draw] (cpu1) at (4.5, -6) {};
  \node[shape=rectangle, fill=white,minimum width=.7cm, minimum height=.7cm, align=center, draw] (cpu1) at (4.7, -6.2) {};
  \node[shape=rectangle, fill=white,minimum width=.7cm, minimum height=.7cm, align=center, draw] (cpu1) at (4.9, -6.4) {};
    
      \node[shape=rectangle, fill=white,minimum width=.7cm, minimum height=.7cm, align=center, draw] (cpu1) at (3.5, -6) {};
  \node[shape=rectangle, fill=white,minimum width=.7cm, minimum height=.7cm, align=center, draw] (cpu1) at (3.7, -6.2) {};
  \node[shape=rectangle, fill=white,minimum width=.7cm, minimum height=.7cm, align=center, draw] (cpu1) at (3.9, -6.4) {};

 \node[shape=rectangle, fill=white,minimum width=.7cm, minimum height=.7cm, align=center, draw] (cpu1) at (2.5, -6) {};
  \node[shape=rectangle, fill=white,minimum width=.7cm, minimum height=.7cm, align=center, draw] (cpu1) at (2.7, -6.2) {};
  \node[shape=rectangle, fill=white,minimum width=.7cm, minimum height=.7cm, align=center, draw] (cpu1) at (2.9, -6.4) {};

 \node[shape=rectangle, fill=white,minimum width=.7cm, minimum height=.7cm, align=center, draw] (cpu1) at (1.5, -6) {};
  \node[shape=rectangle, fill=white,minimum width=.7cm, minimum height=.7cm, align=center, draw] (cpu1) at (1.7, -6.2) {};
  \node[shape=rectangle, fill=white,minimum width=.7cm, minimum height=.7cm, align=center, draw] (cpu1) at (1.9, -6.4) {};

  \node[shape=rectangle, fill=white,minimum width=.7cm, minimum height=.7cm, align=center, draw] (cpu1) at (0.5, -6) {};
  \node[shape=rectangle, fill=white,minimum width=.7cm, minimum height=.7cm, align=center, draw] (cpu1) at (0.7, -6.2) {};
  \node[shape=rectangle, fill=white,minimum width=.7cm, minimum height=.7cm, align=center, draw] (cpu1) at (0.9, -6.4) {};
 
   \node[shape=rectangle, fill=white,minimum width=.7cm, minimum height=.7cm, align=center, draw] (cpu1) at (-0.5, -6) {};
  \node[shape=rectangle, fill=white,minimum width=.7cm, minimum height=.7cm, align=center, draw] (cpu1) at (-0.3, -6.2) {};
  \node[shape=rectangle, fill=white,minimum width=.7cm, minimum height=.7cm, align=center, draw] (cpu1) at (-0.1, -6.4) {}; 
  
%
%
  \node[shape=cloud, cloud puffs=15.7, cloud ignores aspect, minimum width=4cm, minimum height=3.5cm, align=center, draw] (virginiacloud) at (2.25, -12) {};
  \draw  ($(virginiacloud.north)+(-0mm,-7mm)$)                     node                {\footnotesize {Results Cloud}};

    \node[draw, cylinder, shape aspect=1.5, inner sep=0.3333em, fill=white, rotate=90, minimum width=1cm, minimum height=0.05cm] (cyl2) at (3,-12.2) {};
    \node[draw, cylinder, shape aspect=1.5, inner sep=0.3333em, fill=white, rotate=90, minimum width=1cm, minimum height=0.05cm] (cyl3) at (3,-11.8) {};      
    
    \node[draw, cylinder, shape aspect=1.5, inner sep=0.3333em, fill=white,
    rotate=90, minimum width=1cm, minimum height=0.05cm] (cyl2) at (1.5,-12.2) {};
    \node[draw, cylinder, shape aspect=1.5, inner sep=0.3333em, fill=white,
    rotate=90, minimum width=1cm, minimum height=0.05cm] (cyl3) at (1.5,-11.8) {};    
    
    \node[draw, cylinder, shape aspect=1.5, inner sep=0.3333em, fill=white,    rotate=90, minimum width=1cm, minimum height=0.05cm] (cyl2) at (2.25,-13.2) {};
    \node[draw, cylinder, shape aspect=1.5, inner sep=0.3333em, fill=white,    rotate=90, minimum width=1cm, minimum height=0.05cm] (cyl3) at (2.25,-12.8) {};  

%
%

    \draw[->] (ARCHIVECloud) -- (wrkq) ;
    \draw[->] (ARCHIVECloud) -- (heanetcloud) ;
    \draw[->] (heanetcloud) -- (virginiacloud.north) ;   
    \draw[->] (DCCloud) -- (wrkq) ;
    \draw[->] (DCCloud) -- (heanetcloud) ;
    \draw[->] (wrkq) -- (heanetcloud) ;
    \draw[->] (heanetcloud) -- (moncloud) ;
    \draw[->] (CCD) -- (cyl2-1) ;
    \draw[->] (cyl2-1) -- (cyl2a) ;

    \draw[->] (controller) -- (heanetcloud) ;
    \draw[->] (controller) -- (wrkq) ;
    \draw[->] (controller) -- (ARCHIVECloud) ;
    \draw[->] (controller) -- (DCCloud) ;

 %
    \draw[->] (wrkq) -- (moncloud);

  \end{tikzpicture} }
  \caption{NIMBUS Architecture }
  \label{fig:NIMBUSfinal}
 \end{center} 
\end{figure}

	\begin{description}
	\item [A globally distributed data processing architecture]  \hfill \\
	The global architecture is a combination of the ACN pipeline which was further extended by the NIMBUS pipeline. An overview of NIMBUS is shown in Figure \ref{fig:NIMBUSfinal}. Data was produced either within a dispersed cloud of CCD generating detectors which compress data and service the data via http web requests, or is available from a data archive which contains compressed images. Data availability was advertised via a distributed worker web queue, which is readable by all workers within the data processing cloud. Worker Instances read from the web worker queue, download data from the data provider, process it and upload to a defined storage cloud containing a distributed storage system. A central controlling server acts as a broker, publishing the location of the workers and the queues.  A series of monitoring systems are provided to help assess the overall system performance. Workers and web servers can easily be added to the overall system allowing it to scale up as required.
	\item [A self configuring, balancing system that is scalable and resilient to failure]  \hfill \\
	The use of web based message queues, as shown in the NIMBUS pipeline, decouples the processing cloud from the individual worker performance. The extent of decoupling is that different worker instance types can process work at different rates without holding up the overall system. If for any reason an instance fails, the message is not lost but restored back to the queue for another worker to process. The distributed set of web queues allows for management of thousands of workers through a series of instructions which provide instructions on where work can be found. 
	\item [Reconfigurable for multiple science payloads]  \hfill \\
	The NIMBUS pipeline works as a distributed data processing system that can be used whenever workers can process files in parallel with no specific ordering required. Once the files are advertised as available from a web server, the processing workers initialise themselves by looking for a science payload to download, install and run. The pipeline provides the necessary supporting architecture and calls the downloaded control scripts to perform the work. This pipeline could be used for any process, which simply requires files to be processed in parallel, where a worker performs a specific task on the file and uploads results when completed. 
	\item [Enabling real-time data processing]  \hfill \\
	The processing speed of the NIMBUS pipeline was tested to just almost 200TB of data per day, which equates to 3,220 images processed per second. Given multiple data sources such as a robotics telescope farm, this processing speed would enable a process where real feedback could be provided to a targeting system to modify the location they are searching, thereby facilitating a real-time feedback loop.    
	
\end {description}

		\tikzstyle {webserver} = [draw,rectangle,node distance=2.5cm,rounded corners=0.5ex,minimum height=2em]
\tikzstyle {queue} = [draw,trapezium,fill=white,trapezium left angle=70,trapezium right angle=-70,shape border rotate=180,node distance=2.5cm,minimum height=2em]

\newpage
\section{Future Work}
\par Further optimisations of the pipeline are considered as possible future areas of research. Much of the experimentation performed in this thesis demonstrated the extensive capability of a distributed system and identified the key factors within the system. It is possible to take these factors and monitor them such that a machine learning approach could be used to optimise a running system. 

\par Further investigation into data compression and clipping optimisation could reduce the amount of data to be transferred. Clipped images, if not stacked deeply enough, have considerable readwrite costs, so data cubes could be constructed at source which are optimised for data movement to reduce the data transfer requirements. Rather than providing a static measure of how many files should be contained within the data cube, a machine learning approach could be used to find optimal sizes for data given different characteristics of the processing pipeline. 

\par With the  increase in the number of survey telescopes and robotic telescope farms, data processing requirements continue to grow. It should be possible to incorporate the NIMBUS pipeline into such systems for data processing to be ongoing and automated. A full implementation of a user interface would be required to provide flexibility regarding the science payloads to be used within the pipeline. A more sophisticated identification of stars using the world coordinate system would ensure that data could be categorised for future use. These science packages would be required to extend the basic photometry of such a system, which performed fundamental analysis, from which control commands could be incorporated back to the robotics farm. 

\par Research should be performed into data storage and subsequent retrieval of processed data. Integration with the Virtual Observatory would be a requirement for both raw and processed data using search terms which were suitable for large scale data base queries such as NoSQL Column Databases \cite{han2011survey}. Objects should be identified by type, position and time index. This would allow for research to be performed on objects observed at different times and with different instruments. A review of the search capabilities and indexing of astronomical data sets would provide context for the results of this pipeline.  

\par To dramatically reduce the overall data movement where live telescopes are being used, data processing at the telescope site, using a GPU system, could result in the transmission of processed data, instead of the raw image.  Work on light curve generation within the pipeline could also be incorporated into the worker nodes. Further research would be required into data reduction at the source of data production which would ensure that the NIMBUS pipeline could increase the overall processing rates by changing the ratio between data movement and data processing. 

\par The NIMBUS pipeline is applicable to other areas within science. A data analysis of exiting science processing pipelines would be required to determine their suitability for data processing using NIMBUS. The structure and format of the dataset would require  that it facilitated parallel processing at some level. If large volumes of data could be processed independently then a mapping of data files to worker nodes is possible via the queue generation sequence, where the files are advertised by each of the web stores where the data resides. Opportunities for distributed processing also exists where data is naturally distributed and data must be categorised or processed independently.  Post processing of results similar to the map-reduce module would allow for results to be recombined to form new datasets. The migration of existing science projects such as the Solar Monitor, \url{http:\\solarmonitor.org} to the NIMBUS platform  would demonstrate the flexible nature of the architecture and its applicability to multiple scientific disciplines. The Solar Monitor was developed by the Solar Research Group, Trinity College Dublin, led by Peter Gallagher and is an IDL based system which integrates data from  multiple sources to produce near real time monitoring of active regions of solar flare activity \cite{gallagher2002active}.

\par NIMBUS and the ACN pipelines do not contain a security layer, and while the data being processed is not sensitive for the purposes of this thesis, other datasets are likely to require some form of security. Security within the pipeline should be researched, but where possible it should incorporate the cost of additional processing into the distributed components of the system. Public/private encryption could be performed with all data transferred over http port 80 or 8080 thus ensuring that firewall restrictions would not impact the overall connectedness of the  pipeline. Data could either be encrypted at source, or by the web server before the data is advertised via a web queue. 

\par While the current pipeline demonstrated that data can be processed by distributed computers, the process of preparing and advertising data also lends itself to opportunities for processing data using citizen science. For example  \url{http://galaxyzoo.org} is a galaxy classification project where volunteers are presented with images within browsers, which they must then classify. The NIMBUS pipeline could incorporate a browser based application which downloads messages from the web queue which would include both the questions to be asked about the images, and links to the images themselves. In this way a browser could act as a generic system providing a per message citizen science question for the volunteer from a generic web based application. In this system the data processing is done by the end user. 

\par A final thought to possible expansions of the NIMBUS architecture is to enable a global data processing cloud which uses the processing power of the mobile phone community. Given the proliferation of these devices a cloud could be constructed within minutes which could provide the processing power of a super computing thorough the use of millions of mobile devices. Possible options to encourage the deployment of the application on the mobile application could be the inclusion of a micro-payment scheme for applications that process specific amounts of data. Given the public nature of this approach it is also likely that additional security would be required to ensure that values returned were correct. Typical methods to ensure accurate data processing requires data to be processed by multiple clients and only accepting results which are consistent between clients.   Using this method clients could download data while the phone is being charged and within WIFI range so there are no data download charges and the performance of the phone was not compromised. A global flash cloud could be created easily which would be transient and potentially cost controlled by ensuring processing was done using a pricing module similar to the AWS spot price option where researchers could bid for computing resources at specific prices. 

\par This research set out to determine if a distributed processing pipeline could scale to meet the existing, and future demands, of image processing for astronomical photometry, which for optical photometry is projected to be tens of terabytes of CCD image data per day in the near future. NIMBUS is presented as an architecture that is truly global in scale, with a demonstrated capability for processing hundreds of terabytes of CCD data per day. The limits of the scalability of this architecture were not fully determined due to financial constraints, but evidence suggests that significant scaling is still possible, in the order of petabytes per day.   NIMBUS is  also resilient, and flexible enough to address the processing requirements of other large scientific datasets.

%% file: appendixA.tex

\appendix

\printglossary[title=Glossary]

    \chapter{Additional Material for Chapter 1}
\label{app:chapter1}    

  \begin{longtable} {ccccc}

  \toprule
    \addlinespace[-10mm]

   \multicolumn{1}{b{2cm}}{\begin{center}Time(secs)\end{center}} &
   \multicolumn{1}{b{2cm}}{\begin{center}fps\end{center}} &
   \multicolumn{1}{b{3cm}}{\begin{center}pixels\end{center}} &
   \multicolumn{1}{b{2cm}}{\begin{center}bits\end{center}} &
  \multicolumn{1}{b{2.3cm}}{\begin{center}terabytes\end{center}} \\
 
     \addlinespace[-5mm]

  \midrule

28800   &100  &  5.5 MegaPixel   &  5.07E+14  & 63.36 \\
28800   &  50  &  5.5 MegaPixel   &  2.53E+14  & 31.68 \\
28800   &  25  &  5.5 MegaPixel   &  1.27E+14  & 15.84 \\
28800   &  10  &  5.5 MegaPixel   &  5.07E+13  &  6.34 \\
28800   &    5  &  5.5 MegaPixel   &  2.53E+13  & 3.17 \\
28800   &    1  &  5.5 MegaPixel   &  5.07E+12  & 0.63 \\

28800   &100  &  4 MegaPixel   &  3.69E+14  & 46.08 \\
28800   &  50  &  4 MegaPixel   &  1.84E+14  & 23.04 \\
28800   &  25  &  4 MegaPixel   &  9.22E+13  & 11.52 \\
28800   &  10  &  4 MegaPixel   &  3.69E+13  &  4.61\\
28800   &    5  &  4 MegaPixel   &  1.84E+13  & 2.30 \\
28800   &    1  &  4 MegaPixel   &  3.69E+12  & 0.46 \\

28800   &100  &  2 MegaPixel   &  1.84E+14  & 23.04 \\
28800   &  50  &  2 MegaPixel   &  9.22E+13  & 11.52 \\
28800   &  25  &  2 MegaPixel   &  4.61E+13  & 5.76 \\
28800   &  10  &  2 MegaPixel   &  1.84E+13  &  2.30\\
28800   &    5  &  2 MegaPixel   &  9.22E+12  & 1.15 \\
28800   &    1  &  2 MegaPixel   &  1.84E+12  & 0.23 \\

28800   &100  &  1 MegaPixel   &  9.22E+13  & 11.52 \\
28800   &  50  &  1 MegaPixel   &  4.61E+12  & 5.76 \\
28800   &  25  &  1 MegaPixel   &  2.30E+12  & 2.88 \\
28800   &  10  &  1 MegaPixel   &  9.22E+12  &  1.15\\
28800   &    5  &  1 MegaPixel   &  4.61E+11  & 0.58 \\
28800   &    1  &  1 MegaPixel   &  9.22E+11  & 0.12 \\

28800   &100  &  0.24 MegaPixel   &  3.02E+12  & 3.02 \\
28800   &  50  &  0.24 MegaPixel   &  1.51E+12  & 1.51 \\
28800   &  25  &  0.24 MegaPixel   &  7.55E+11  & 0.75 \\
28800   &  10  &  0.24 MegaPixel   &  3.02E+11  &  0.30\\
28800   &    5  &  0.24 MegaPixel   &  1.51E+11  & 0.15 \\
28800   &    1  &  0.24 MegaPixel   &  3.02E+10  & 0.03 \\

\bottomrule
\caption{CCD data generation raw data for 32bit pixel precision for varying resolutions and frames per second over 8hr period} 
\label{tab:longtableccd}
\end{longtable}

\chapter{Additional Material for Chapter 3}  

\par The FEBRUUS pilot is a series of CFITSIO files which are used to perform pixel calibration on CCD images. The flowcharts for these programs are presented within this section of the appendix. 

\label{app:chapter3}    
\begin{figure} [htbp] 
\centering
\fbox{   \includegraphics[width=0.95\textwidth] {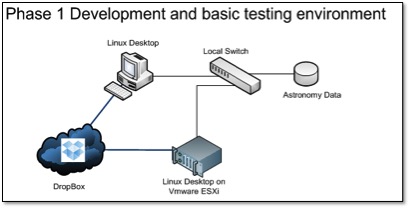}  }
  \vskip -0.8em
    \caption{FEBRUUS: Initial configuration for the Pilot Pipeline }
  \label{fig:pilotconfig}
\end{figure}
   
\begin{figure} [htbp]
\centering
\fbox{   \includegraphics[width=0.8\textwidth] {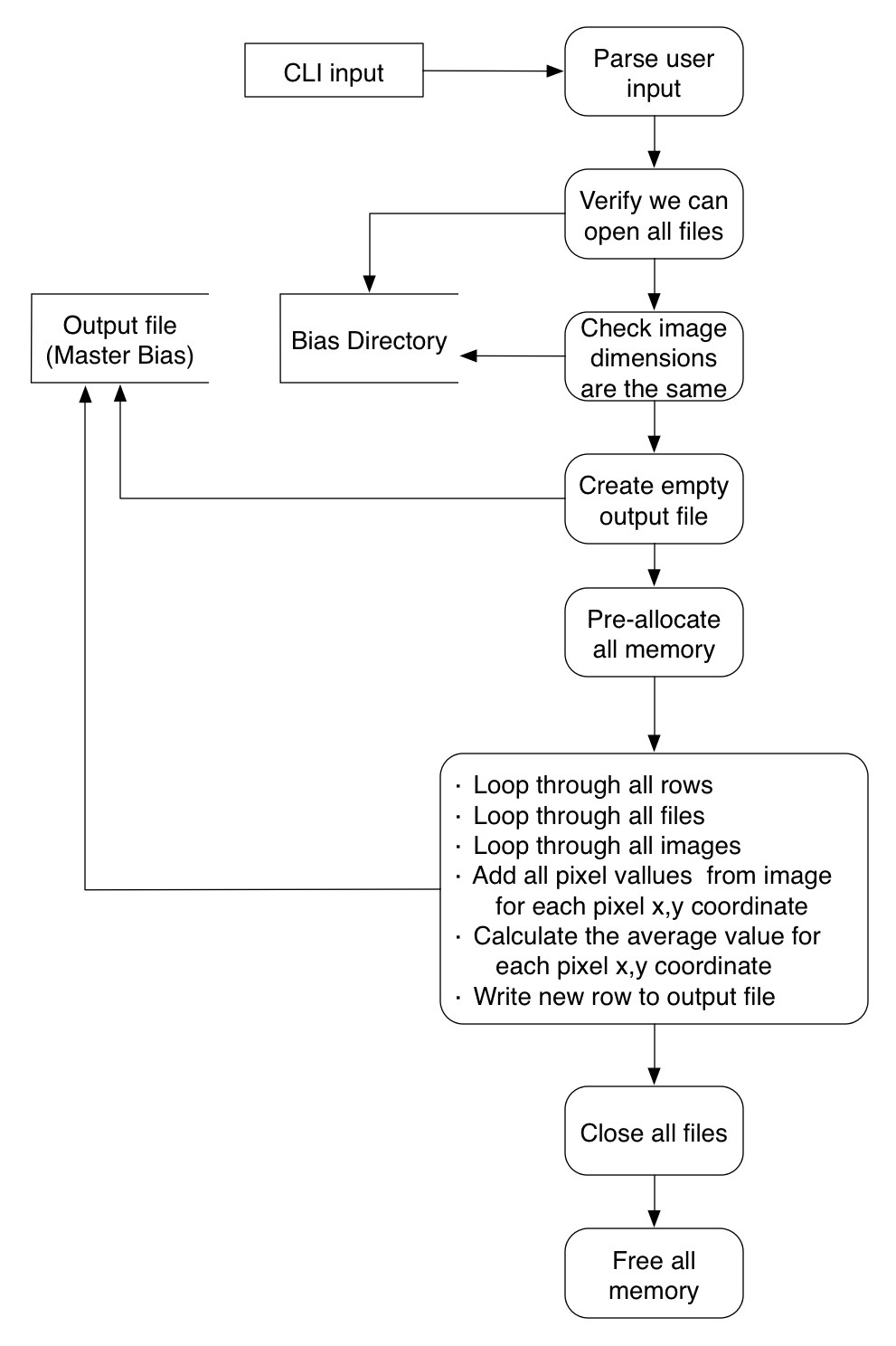}  }
  \vskip -0.8em
    \caption{FEBRUUS: The GMB.c program flowchart }
  \label{fig:gmb-flow}
\end{figure}

\begin{figure} [htbp]
\centering
\fbox{   \includegraphics[width=0.8\textwidth] {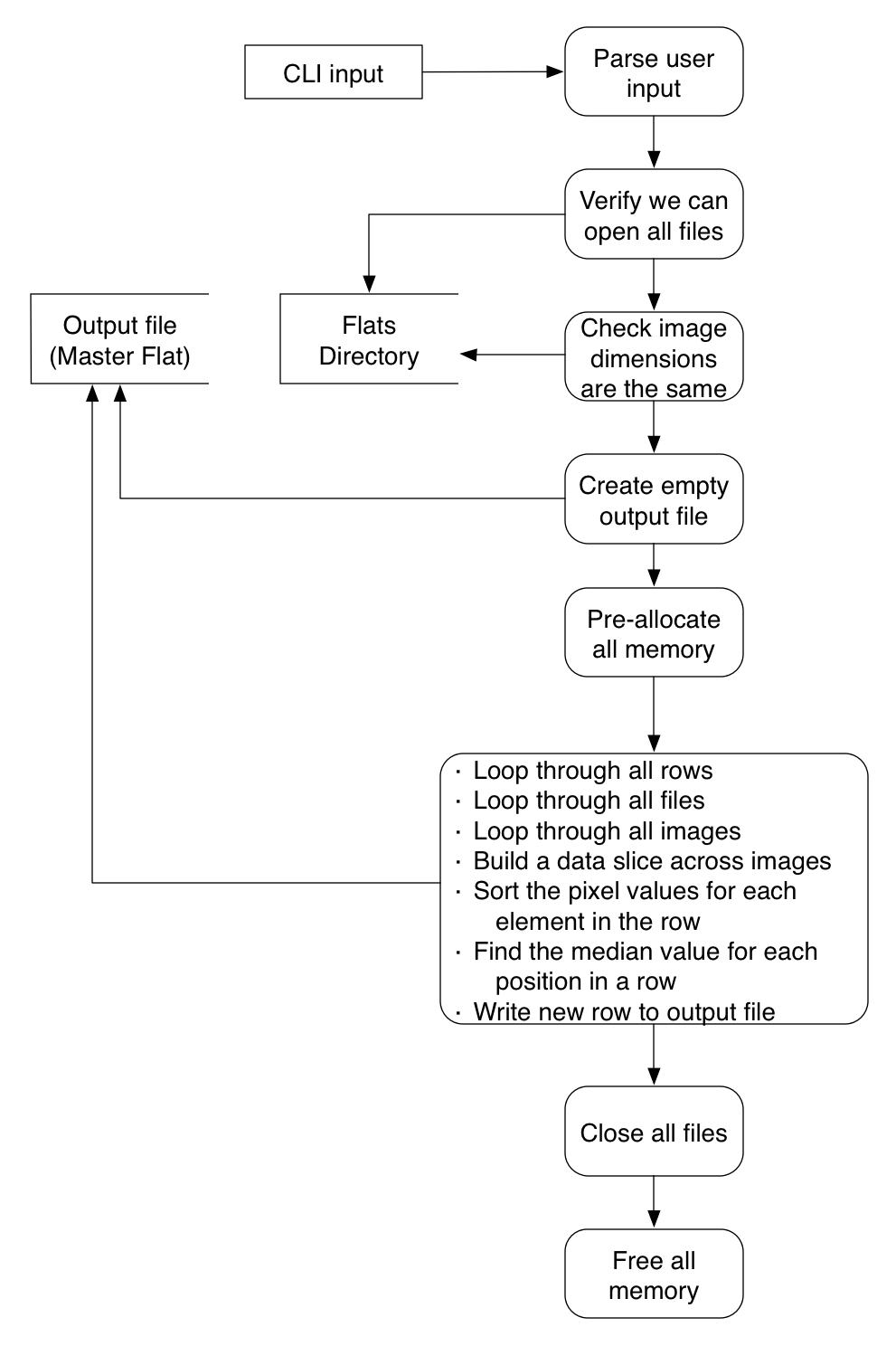}  }
  \vskip -0.8em
    \caption{FEBRUUS: The GMF program flowchart }
  \label{fig:gmf-flow}
\end{figure}

\begin{figure} [htbp]
\centering
\fbox{   \includegraphics[width=0.8\textwidth] {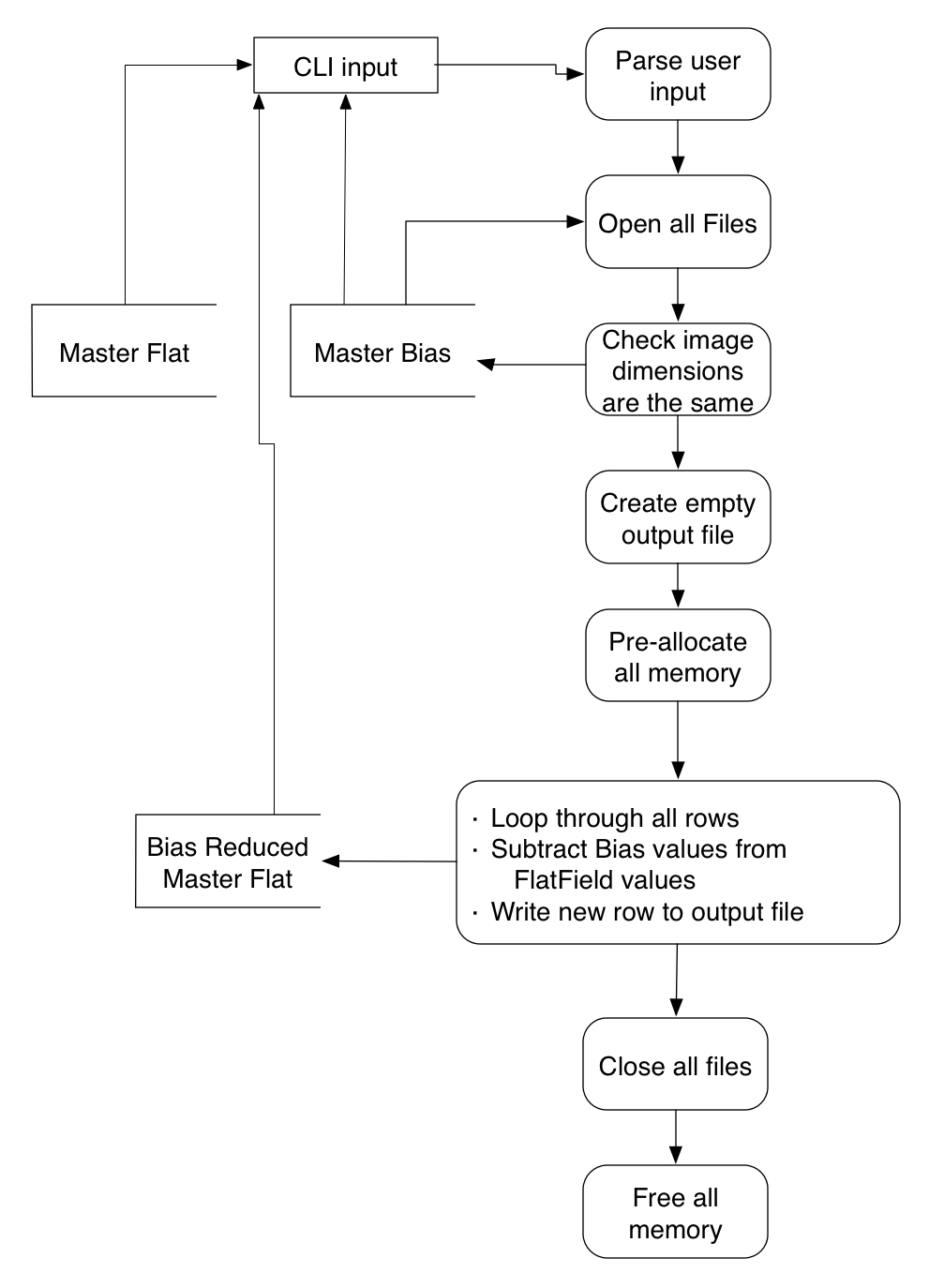}  }
  \vskip -0.8em
    \caption{FEBRUUS: The BMF program flowchart }
  \label{fig:bmf-flow}
\end{figure}

\begin{figure} [htbp]
\centering
\fbox{   \includegraphics[width=0.8\textwidth] {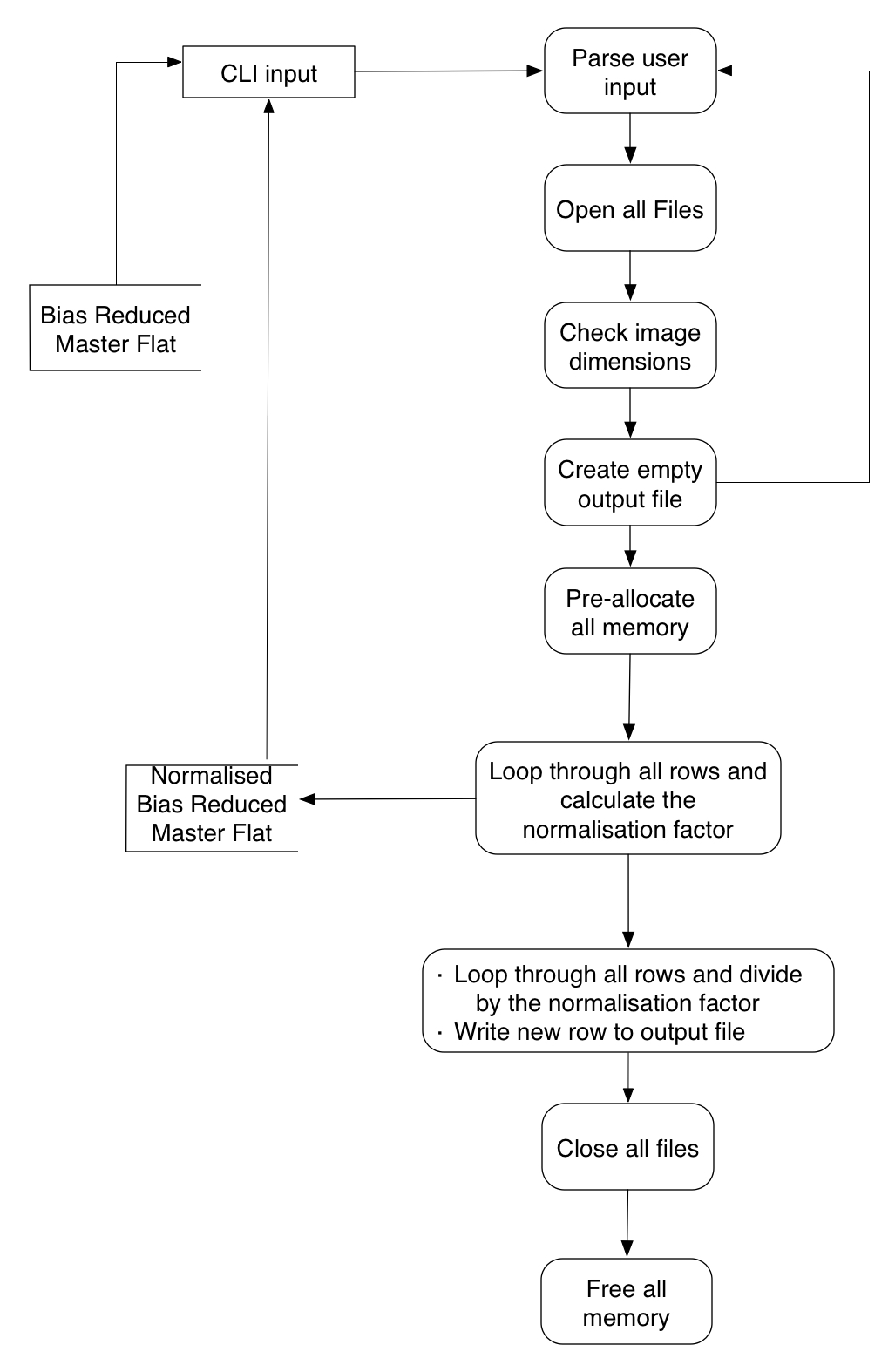}  }
  \vskip -0.8em
    \caption{FEBRUUS: The NMF program flowchart }
  \label{fig:nmf-flow}
\end{figure}

\begin{figure} [htbp]
\centering
\fbox{   \includegraphics[width=0.8\textwidth] {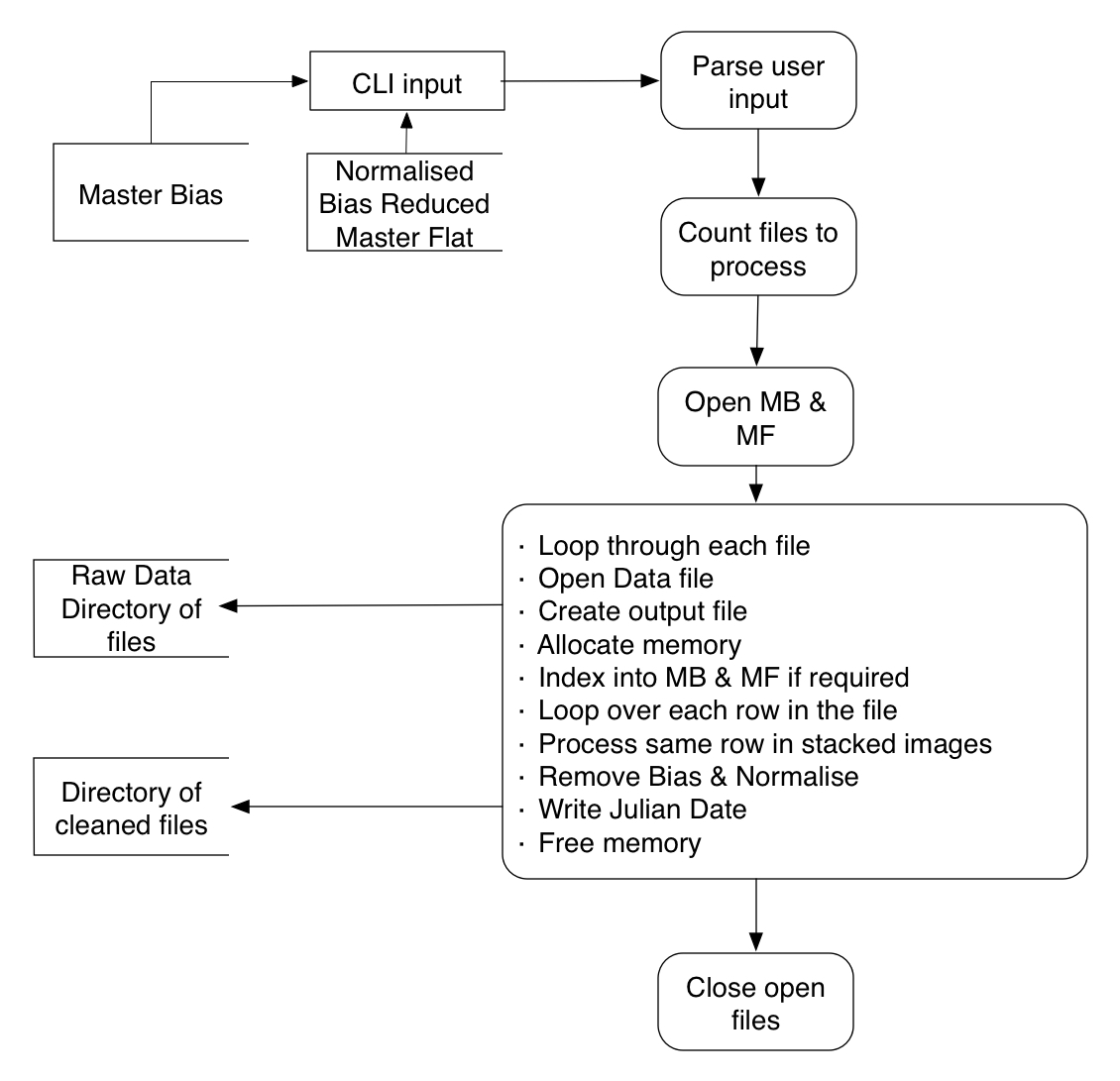}  }
  \vskip -0.8em
    \caption{FEBRUUS: The RRF program flowchart }
  \label{fig:rrf-flow}
\end{figure}

\begin{figure} [htbp]
\centering
\fbox{   \includegraphics[width=0.8\textwidth] {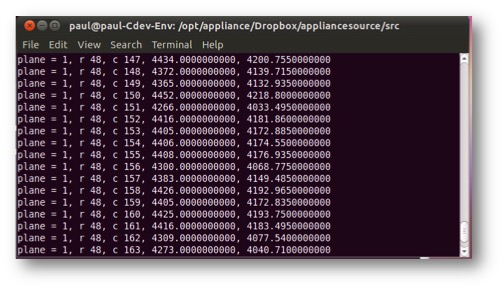}  }
  \vskip -0.8em
    \caption{FEBRUUS: bias reduced master flat output. }
  \label{fig:bmf-output}
\end{figure}

\begin{figure} [htbp]
\centering
\fbox{   \includegraphics[width=0.8\textwidth] {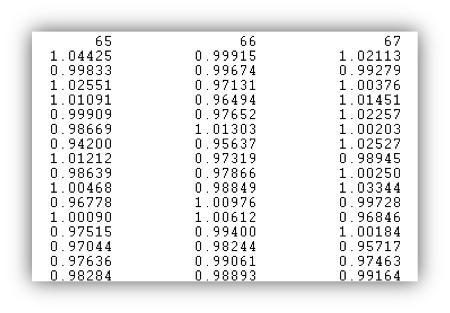}  }
  \vskip -0.8em
    \caption{FEBRUUS: Normalised master flat output. }
  \label{fig:nmf-output}
\end{figure}

  \chapter{Additional Material for Chapter 4}
 \label{app:chapter4}

  \begin{table}[htbp]
\centering
\begin{tabular}{p{6cm} p{8cm}}
  \toprule
Command & Description \\
\midrule
\small{Activate-ACN \hbox{-}r NODEFILENAME   }& \small{This command will activate all ACN Nodes in the nodes file so they start looking to a queue for image files to process. It is possible to have ACN nodes wait for a queue, but this option provides better control for experimentation so it is clean which nodes are active. Once a node is running, it checks the queue for unlocked files and locks them as shown in Figure 49. Once locked, the full image file is downloaded from an S3 bucket and cleaned using the acn-aphot program. This program once for each file downloaded. }\\
\hline
\small{Activate-ACN \hbox{-}c NODEFILENAME   }& \small{This command will reset all ACN Nodes in the nodes file so they stop running, remove all temporary data and download the latest utility files, configuration files and Master Bias/Master Flat images ready for the next round of processing.}\\
\hline
\small{Activate-ACN \hbox{-}p NODEFILENAME   }& \small{This command will Ping all of the nodes to ensure that they are accessible to the pipeline.}\\
\hline

\small{Activate-ACN \hbox{-}x NODEFILENAME   }& \small{Reboot all of the nodes to ensure that they have flushed all caches. If a node processes the same data multiple times, it may operate faster on the subsequent executions due to caching of data in memory.}\\
\bottomrule
\end{tabular}
\caption{Control Node commands.}
\label{tabla:ctrlnode}
\end{table}

  \begin{table}[H]
\centering
\begin{tabular}{p{6cm} p{8cm}}
  \toprule
Command & Description \\
  \midrule
\small{fits-compress \hbox{-}p DATADIR   }& \small{This script has the option of performing compression in parallel or in sequence. The parallel execution spawns off processes and requires a machine with good RAM and processing capabilities. A comparison of performance for this script running in both modes is given in the next chapter. }\\
\hline
\small{s3-upload \hbox{-}p DATADIR     }& \small{This script has the option of performing upload in parallel or in sequence. The parallel execution spawns off processes and requires a machine with good RAM and processing capabilities. A comparison of performance for this script running in both modes is given in the next chapter. Compressed or uncompressed data (depending on what is in the data directory) is uploaded to an S3 bucket. The data transfer rate through the server Ethernet card is increased using this approach. A comparison of performance for this script running in both modes is given in the next chapter.}\\
  \bottomrule
\end{tabular}
\caption{Storage Control commands.}
\label{tabla:strnode}
\end{table}

  \begin{table}[H]
\centering
\begin{tabular}{p{6cm} p{8cm}}
  \toprule
Command & Description \\
  \midrule
\small{Activate-ACN \hbox{-}q compressed \textbar \ standard \textbar \ clipped}& \small{This command will create a list of empty files in a directory, which can be used as a simple queue. Files are named with a prefix source and traverse a dataset creating an entry in the queue for all files found. A compressed source file of 0000001.fits.fz has a corresponding queue entry of Queued-0000001.fits.fz. When successfully locked for processing by an ACN-Node this is changed to LOCKED-0000001.fits.fz. The NFS file system ensures only one lock can be obtained.}\\
  \bottomrule

\end{tabular}
\caption{Queue Control commands.}
\label{tabla:ctrlqueue}
\end{table}

\begin{table}[htbp]
\centering
  \begin{tabular} {ccccccc}
  \toprule
  \addlinespace[-5mm]

  \multicolumn{1}{b{1.5cm}}{\begin{center}Hardware\end{center}} &
   \multicolumn{1}{b{1.5cm}}{\begin{center}Storage Type\end{center}} &
   \multicolumn{1}{b{1.5cm}}{\begin{center}Reduction Steps\end{center}} &
   \multicolumn{1}{b{1.5cm}}{\begin{center}Step 1 Time\end{center}} &
   \multicolumn{1}{b{1.5cm}}{\begin{center}Step 2 Time\end{center}} &
   \multicolumn{1}{b{1.5cm}}{\begin{center}Total Time\end{center}} &
   \multicolumn{1}{b{1.5cm}}{\begin{center}Files per Sec\end{center}} \\

\addlinespace[-5mm]

  \midrule
  Macbook Pro & SATA   & 2  & 11:21:45 & 00:54:48  & 12:16:33  & 0.8332   \\
  Macbook Pro & SATA   & 1  &                & 0055:53  & 00:55:53  & 1.09812   \\

  eServer326 & SCSI     & 2  & 00:49:00 & 01:15:23  & 02:04:23  & 0.49337   \\
  eServer326 & SCSI     & 1  &                & 01:22:30  & 01:22:30  & 0.8332   \\

  eServer326 & NFS      & 2  & 00:49:53 & 01:15:08  & 02:05:01  & 0.49087   \\
  eServer326 & NFS      & 1  & 	             & 01:17:25  & 01:17:25   & 0.79268   \\

  \bottomrule
\hline
  \end{tabular}
  \caption{Single Node performance data for 1 and 2 step reduction processing }

\label{tab:singlenode}
\end{table}

\begin{longtable} {ccccc}

  \toprule
    \addlinespace[-10mm]

   \multicolumn{1}{b{2cm}}{\begin{center}Node Type\end{center}} &
   \multicolumn{1}{b{2cm}}{\begin{center}Node Ref\end{center}} &
   \multicolumn{1}{b{3cm}}{\begin{center}Seconds running\end{center}} &
   \multicolumn{1}{b{2cm}}{\begin{center}Files Cleaned\end{center}} &
  \multicolumn{1}{b{2.3cm}}{\begin{center}Cleaning Rate\end{center}} \\
 
     \addlinespace[-5mm]

  \midrule

VM-Tokyo	&	1	&	133	&	91	&	1.4615	\\
VM-Tokyo	&	2	&	133	&	93	&	1.4301	\\
VM-Tokyo	&	3	&	133	&	92	&	1.4456	\\
VM-Tokyo	&	4	&	134	&	94	&	1.4255	\\
VM-Sydney	&	5	&	132	&	93	&	1.4193	\\
VM-Sydney	&	6	&	133	&	88	&	1.5113	\\
VM-Sydney	&	7	&	133	&	88	&	1.5113	\\
VM-Sydney	&	8	&	133	&	92	&	1.4456	\\
VM-Paris	&	9	&	128	&	48	&	2.6666	\\
VM-Paris	&	10	&	131	&	50	&	2.62	\\
VM-Paris	&	11	&	132	&	46	&	2.8695	\\
VM-Paris	&	12	&	132	&	49	&	2.6938	\\
VM-Paris	&	13	&	133	&	53	&	2.5094	\\
VM-Paris	&	14	&	133	&	52	&	2.5576	\\
VM-Paris	&	15	&	133	&	53	&	2.5094	\\
VM-Paris	&	16	&	134	&	53	&	2.5283	\\
VM-Paris	&	17	&	134	&	52	&	2.5769	\\
VM-Paris	&	18	&	134	&	52	&	2.5769	\\
VM-Paris	&	19	&	134	&	51	&	2.6274	\\
VM-Paris	&	20	&	134	&	53	&	2.5283	\\
VM-London	&	21	&	131	&	37	&	3.5405	\\
VM-London	&	22	&	132	&	37	&	3.5675	\\
VM-London	&	23	&	132	&	37	&	3.5675	\\
VM-London	&	24	&	133	&	37	&	3.5945	\\
VM-London	&	25	&	133	&	37	&	3.5945	\\
VM-London	&	26	&	133	&	37	&	3.5945	\\
VM-London	&	27	&	133	&	37	&	3.5945	\\
VM-London	&	28	&	133	&	37	&	3.5945	\\
VM-London	&	29	&	133	&	38	&	3.5	\\
VM-London	&	30	&	134	&	38	&	3.5263	\\
VM-London	&	31	&	134	&	38	&	3.5263	\\
VM-London	&	32	&	134	&	38	&	3.5263	\\
VM-London	&	33	&	134	&	38	&	3.5263	\\
VM-London	&	34	&	134	&	37	&	3.6216	\\
VM-London	&	35	&	134	&	38	&	3.5263	\\
ITTD	&	36	&	133	&	78	&	1.7051	\\
ITTD	&	37	&	134	&	71	&	1.8873	\\
ITTD	&	38	&	134	&	77	&	1.7402	\\
ITTD	&	39	&	134	&	77	&	1.7402	\\
ITTD	&	40	&	134	&	73	&	1.8356	\\
ITTD	&	41	&	134	&	78	&	1.7179	\\
ITTD	&	42	&	134	&	76	&	1.7631	\\
ITTD	&	43	&	134	&	77	&	1.7402	\\
DIT	&	44	&	133	&	69	&	1.9275	\\
DIT	&	45	&	133	&	70	&	1.9	\\
DIT	&	46	&	133	&	77	&	1.7272	\\
DIT	&	47	&	134	&	80	&	1.675	\\
DIT	&	48	&	134	&	78	&	1.7179	\\
DIT	&	49	&	134	&	71	&	1.8873	\\
DIT	&	50	&	134	&	71	&	1.8873	\\
DIT	&	51	&	134	&	70	&	1.9142	\\
BCO	&	52	&	133	&	70	&	1.9	\\
BCO	&	53	&	133	&	72	&	1.8472	\\
BCO	&	54	&	133	&	72	&	1.8472	\\
BCO	&	55	&	133	&	70	&	1.9	\\
BCO	&	56	&	134	&	79	&	1.6962	\\
BCO	&	57	&	134	&	79	&	1.6962	\\
BCO	&	58	&	134	&	72	&	1.8611	\\
BCO	&	59	&	134	&	71	&	1.8873	\\
\bottomrule
\caption{ACN Multi-Node processing raw data} 
\label{tab:longtable1}
\end{longtable}

\chapter{Additional Material for Chapter 5}
\label{app:chapter5}
\begin{table}[htbp]
\centering
  \begin{tabular} {cccccc}
  \toprule

  \multicolumn{1}{b{1.7cm}}{\begin{center}Webnode\end{center}} &
   \multicolumn{1}{b{1.5cm}}{\begin{center}No. Msgs\end{center}} &
   \multicolumn{1}{b{1.5cm}}{\begin{center}Single Node Write Time\end{center}} &
   \multicolumn{1}{b{1.5cm}}{\begin{center}Msg Write Rate\end{center}} &
     \multicolumn{1}{b{1.5cm}}{\begin{center}Multi Node Write Time\end{center}} &
   \multicolumn{1}{b{1.5cm}}{\begin{center}Msg Write Rate\end{center}} \\ 
   
\addlinespace[-5mm]

  \midrule
  Webnode1    & 73620 & 8m47.279s & 139.7  & 8m40.200s  &  141.6   \\
  Webnode2  & 73620  & 9m0.286s    &  136.3 & 8m57.302s  &  137.1   \\

  Webnode3  & 73620 & 3m7.055s     & 393.7 & 3m13.622s  &  381.5   \\
  Webnode4 & 73620 & 3m10.238s       &387.5& 3m21.267s  & 366.3   \\

  Webnode5     & 77342 & 7m11.410s   & 179.5 & 7m0.083s  & 184.2   \\
  Webnode6    & 77342 & 7m28.142s    & 172.6 & 7m26.385s  &  173.4   \\

  Webnode7     & 73620  & 3m49.490s  & 	321.4 & 3m46.775s  &  325.8   \\
  Webnode8  & 73620  & 3m47.520s     & 324.3 & 3m40.521s  &  334.7   \\

  \bottomrule
\hline
  \end{tabular}
  \caption{SQS queue write performance data }

\label{tab:queuewrites}
\end{table}

\definecolor{dkgreen}{rgb}{0,0.6,0}
\definecolor{gray}{rgb}{0.5,0.5,0.5}
\definecolor{mauve}{rgb}{0.58,0,0.82}


\lstset{
language=bash,
basicstyle=\small\sffamily,
  breaklines=true,
  breakatwhitespace=true,
numbers=left,
  commentstyle=\color{dkgreen},
numberstyle=\tiny\color{gray},
frame=tb,
 keywordstyle=\color{blue},
  identifierstyle=\bfseries\color{black},
columns=fullflexible,
showstringspaces=false
}

\begin{lstlisting}[caption=Generate Metrics: writes per node per second from RAW canaryq source files.,
  label=code:sqsmetric,
  float=t]
#!/bin/bash
# This generates a pair of values - epoch time and webserver 
SERVERLIST=$(cat SQS* |  awk -F "," '{print $5 }'  |  grep http |  awk -F "http://" '{print $2 }'  |  awk -F "/" '{ print $1 }' | sort | uniq)

# Time is rounded to nearest second
TIMELIST=$(cat SQS*  | grep SentTime | awk -F "," '{print $1 $5}' | awk -F "http://" '{print $1 "," $2 }' | awk -F "/" '{print $1}' | sed s/\{u\'SentTimestamp\'\:[[:space:]]u\'// |  awk -F "\'\}," '{print $1}' | sed 's/\(^.\{10\}\).\{3\}\(.*\)/\1\2/' | sort -n -k 1 | uniq)

for times in $TIMELIST
do
	echo -n $times ","
	for servers in $SERVERLIST
	do
		counter1=$(cat SQS*  | grep $times | grep $servers | wc -l)
		echo -n $counter1 ","
	done
	echo
done

\end{lstlisting}

\lstset{
language=python,
basicstyle=\small\sffamily,
  breaklines=true,
  breakatwhitespace=true,
numbers=left,
  commentstyle=\color{dkgreen},
numberstyle=\tiny\color{gray},
frame=tb,
 keywordstyle=\color{blue},
  identifierstyle=\bfseries\color{black},
columns=fullflexible,
showstringspaces=false
}

\begin{lstlisting}[caption=Python code extract for multi-threaded single node SQS testing.,
  label=code:sqs-writer,
  float=t]
sqs_queue = conn.get_queue(args.queuearg)
canary_queue = conn.get_queue("canaryq")
          
class Sender(threading.Thread):  
        def __init__(self):
                threading.Thread.__init__(self)
        def run(self):
                global sqs_queue,canary_queue,queue
                while True: 
                        try: 
                                msg = queue.get(True,3)
                                m = Message()
                                m.set_body(msg)
                                status = sqs_queue.write(m)
                        except Queue.Empty:
                                return 
                        except:
                                return
queue = Queue.Queue(0)

for file in sys.stdin:
        file = ipadd+file
        queue.put(file)

threads = []
for n in xrange(40):
        t = Sender()
        t.start()
        threads.append(t)

for t in threads:
        t.join()

\end{lstlisting}

\begin{lstlisting}[caption=Python code extract for multi-threaded SQS queue reading.,
  label=code:sqs-canary-reader,
  float=t]

class Sender(threading.Thread):
        def __init__(self):
                threading.Thread.__init__(self)

        def run(self):
                global sqs_queue,queue
                name = args.experiment+str(queue.get())+"-"+args.queuearg+".csv"
                f = open(name,'w')
                
                while True:
                        try:
                                m = sqs_queue.read(60)
                                m = sqs_queue.get_messages(num_messages=1, attributes='SentTimestamp')
                                print "This is the message->", m[0].attributes,m[0].get_body()
                                f.write(str(m[0].attributes)+str(m[0].get_body())+"\n")
                                sqs_queue.delete_message(m[0])
                        except:
                                if sqs_queue.count() < 1:
                                        f.write(args.queuearg + " is empty\n")
                                        return
queue = Queue.Queue(0)

threads = []
for n in xrange(40):
        queue.put(n)
        t = Sender()
        t.start()
        threads.append(t)

for t in threads:
        t.join()
        
\end{lstlisting}        
        
\lstset{
language=bash,
basicstyle=\small\sffamily,
  breaklines=true,
  breakatwhitespace=true,
numbers=left,
  commentstyle=\color{dkgreen},
numberstyle=\tiny\color{gray},
frame=tb,
 keywordstyle=\color{blue},
  identifierstyle=\bfseries\color{black},
columns=fullflexible,
showstringspaces=false
}        
\begin{lstlisting}[caption=Bash code combining canary queue messages into bins.,
  label=code:sqs-canary-metrics,
  float=t]
        
#!/bin/bash
# This generates a pair of values - epoch time and webserver
FILELIST=$(ls SQS*)
SERVERLIST=$(cat SQS* | awk -F "," '{print $5 }' | grep http |  awk -F "http://" '{print $2 }' |  awk -F "/" '{print $1 }' | sort | uniq)

for servers in $SERVERLIST
do
        echo -n $servers ","
done
echo
x=1     
y=$(($x+100))
filelen=$(cat SQS-canary9-* | wc -l)
loops=$(($filelen/100))

for (( loopcount=1; loopcount<=$loops; loopcount++ )) 
do      
        for servers in $SERVERLIST
        do      
                counter=0
                for files in $FILELIST
                do      
                        counter1=$(cat $files | sed -n "$x,$y p"  | grep $servers | wc -l)
                        counter=$(($counter+counter1))
                done    
                echo -n $counter ","
        done    
        echo    
        x=$(($x+100))
        y=$(($x+100))
done            
        
      \end{lstlisting}

\begin{lstlisting}[caption=Experimental Batch Script for running multiple experiments.,
  label=code:batchrunexperiments,
  float=t]
#!/bin/bash
#  Configuration options for running experiments
#  -a   Use Amazon Web servers
#  -d   Use DIT Web servers
#  -n	Use HEANT Web servers 
#  -x	Use ALL Web servers
#
#   Num Instances to Run
#   Time in Seconds to run the experiment
#   Name of the Experiment
#   Number of Web Servers to run (1 or 2)
#   Size of the instances to use in the experiment 
#   All experiments are formally names to correspond to a specific configuration
#   Format of the run command below is as follows. 
#
#  webservertype  instancenum seconds expname webservernum  instancetype
#
#AWS Experiments group 1, Workers= 1 per instance, BatchSize = 10
#
# The following parameters must be set in the Worker Package
# 	Workers per instance in this case is 1
# 	Batch Size per worker is set to 10
#
# 	The canary1.nightsky.ie should also be running rounding up the 
#	number of instances by 1 in all cases
./run-experiment.sh -a  9   1200 SetB5w.2 1 t1.micro   
./run-experiment.sh -a  24  1200 SetB5w.3 1 t1.micro   
./run-experiment.sh -a  49  1200 SetB5w.4 1 t1.micro   
./run-experiment.sh -a  99  1200 SetB5w.5 1 t1.micro   
./run-experiment.sh -a  4   1200 SetB5w.6 1 m1.large
./run-experiment.sh -a  9  1200 SetB5w.7 1 m1.large
./run-experiment.sh -a  24  1200 SetB5w.8 1 m1.large
./run-experiment.sh -a  49  1200 SetB5w.9 1 m1.large
./run-experiment.sh -a  99 1200 SetB5w.10 1 m1.large
      \end{lstlisting}

\pgfplotstableread{Data/CanaryQ/writerate-cumulative.prn}
\datatable

\begin{figure} 
\centering
\fbox {
\begin{tikzpicture}
    \begin{axis}[xmin=0,
    ymin=0,
    width=0.8\textwidth,
    legend style={at={(0.8,1)},anchor=north,legend cell align=left},
        xlabel=$Seconds$,
        ylabel=$Messages\ Written$, no markers]

\addplot table[y = heanet] from \datatable ;
\addlegendentry{Heanet FTP Server}
\addplot table[y = webnode1] from \datatable ;
\addlegendentry{WebNode1}
\addplot table[y = webnode2] from \datatable ;
\addlegendentry{WebNode2}

\addplot table[y = webnode3] from \datatable ;
\addplot table[y = webnode4] from \datatable ;
\addplot table[y = webnode5] from \datatable ;
\addplot table[y = webnode6] from \datatable ;
\addplot table[y = webnode7] from \datatable ;
\addplot table[y = webnode8] from \datatable ;

\legend{$HEANet$,$Webnode1$,$Webnode2$,$Webnode3$,$Webnode4$,$Webnode5$,$Webnode6$,$Webnode7$,$Webnode8$}   

    \end{axis}

\end{tikzpicture} }
  \vskip -0.8em
    \caption{Cumulative writing of messages to the canary queue}
  \label{fig:canaryqcumulativeraw}
\end{figure}
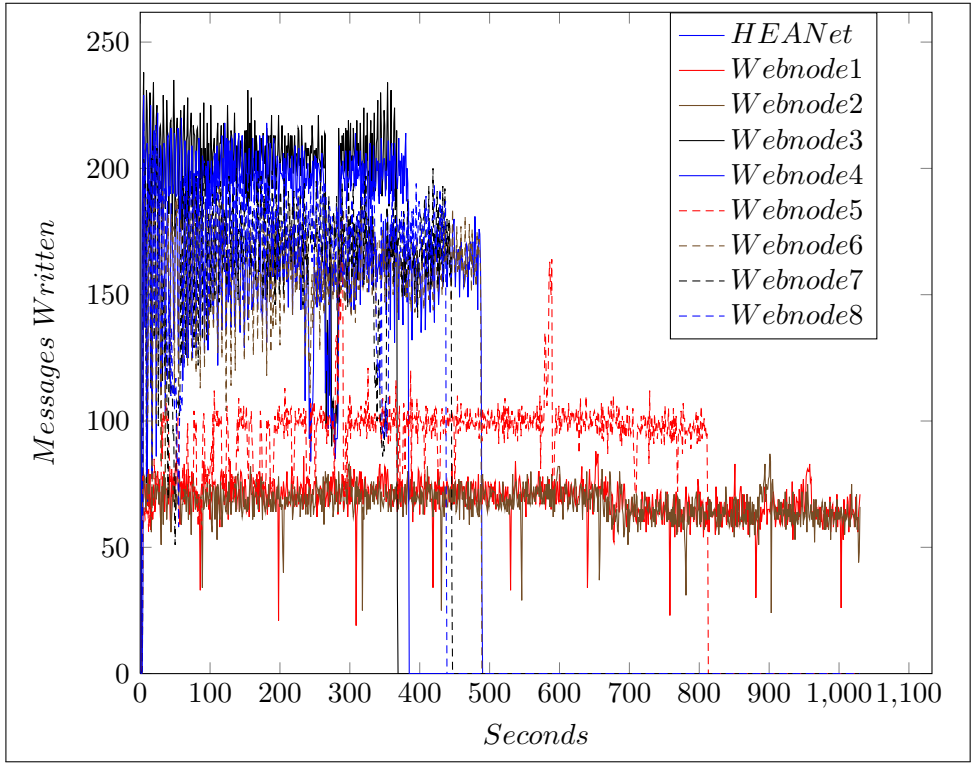

\begin{table}[htbp]
\centering
  \begin{tabular} {lll}
  \toprule

  \multicolumn{1}{b{1.5cm}}{\begin{center}Ref.\end{center}} &
  \multicolumn{1}{b{1.5cm}}{\begin{center}Directory\end{center}} &
   \multicolumn{1}{b{1.5cm}}{\begin{center}Comment\end{center}} \\
   
\addlinespace[-5mm]

  \midrule
  \small{RAW DATA}    & \small{DataSet-BCO/22-9-2003} & \small{BCO 26GB Raw CCD Data}\\
  \small{FEBRUSS}     & \small{FEBRUUS-Pilot/Source Files} & \small{C programs for processing FITS files}\\
  \small{FEBRUSS}     & \small{FEBRUUS-Pilot/FITS-Testfiles} & \small{Sample FITS files for accuracy testing}\\

  \small{ACN} & \small{ACN-Pipeline/ACN-C-Source} & \small{Core CCD Image processing C files}\\
  \small{ACN} & \small{ACN-Pipeline/ACN-Controller} & \small{BASH scripts for controlling experiments} \\
  \small{ACN} & \small{ACN-Pipeline/ACN-Worker} & \small{BASH scripts for controlling Worker Nodes}\\
  \small{ACN} & \small{ACN-Pipeline/Experimental Data} & \small{Excel Spreadsheets with Result Data}\\

  \small{NIMBUS} & \small{NIMBUS-Pipeline/ACN-APHOT-SRC} & \small{C Program for cleaning and Photometry}\\
  \small{NIMBUS} & \small{NIMBUS-Pipeline/NIMBUS Controller} & \small{Python source for experimental execution}\\
  \small{NIMBUS} & \small{NIMBUS-Pipeline/NIMBUS Worker} & \small{Python source for Worker execution}\\
  \small{NIMBUS} & \small{NIMBUS-Pipeline/acn-science.pkg.tar.z} & \small{Compressed Pkg an utilities tools}\\
  \small{NIMBUS} & \small{NIMBUS-Pipeline/Experimental Data} & \small{Raw data with R-files and Excel} \\

  \bottomrule
\hline
  \end{tabular}
  \caption{Data Sources for experiments and graphs}

\label{tab:datasets}
\end{table}

\begin{table}[htbp]
\centering
  \begin{tabular} {llcc}
  \toprule

  \multicolumn{1}{b{2cm}}{\begin{center}Family \end{center}} &
  \multicolumn{1}{b{2cm}}{\begin{center}Type\end{center}} &
  \multicolumn{1}{b{2cm}}{\begin{center}CPU\end{center}} &
   \multicolumn{1}{b{2cm}}{\begin{center}Memory (GB)\end{center}} \\
   
\addlinespace[-5mm]

  \midrule
  Virtual-Micro   			& t1.micro      & 1 & 0.65\\
  Virtual-General    		& m1.large    & 2  & 7.5\\
  Virtual-General                 & m1.xlarge  & 4 & 15\\
  Virtual-Compute               & c1.xlarge   & 8 & 7 \\
  Physical                 		& x4150	&  8 & 64\\
  Physical                 		& IMB326e  & 1 & 4\\

  \bottomrule
\hline
  \end{tabular}
  \caption{Instance Types and their Specification }

\label{tab:instancespecs}
\end{table}

\begin{longtable} {cccccc}

  \toprule
    \addlinespace[-10mm]

   \multicolumn{1}{b{1.5cm}}{\begin{center}Exp. Num.\end{center}} &
  \multicolumn{1}{b{1.5cm}}{\begin{center}Web Server\end{center}} &
   \multicolumn{1}{b{1.5cm}}{\begin{center}Instance\end{center}} &

     \multicolumn{1}{b{1.5cm}}{\begin{center}Num. Instances\end{center}} &
   \multicolumn{1}{b{1.5cm}}{\begin{center}Workers per Inst.\end{center}} &
   \multicolumn{1}{b{1.8cm}}{\begin{center}FPS\end{center}} \\

     \addlinespace[-5mm]

  \midrule

1   &  FTP  	&	IBM326	&	1	&	1 	&	0.42 \\
2   &  FTP  	&	T1.Micro	&	1	&	1 	&	0.51 \\
3   &  FTP  	&	M1.Large	&	1	&	1 	&	0.54 \\
4   &  FTP  	&	T1.Micro	&	1	&	10 	&	0.71. \\
5   &  HEANET  &	T1.Micro	&	1	&	5 	&	0.81\\
6   &  FTP  	&	T1.Micro	&	5	&	10 	&	1.45 \\
7   &  FTP  	&	IBM326	&	1	&	5 	&	1.61 \\
8   &  FTP  	&	IBM326	&	1	&	10 	&	1.66 \\§
9   &  FTP  	&	M1.Large	&	1	&	5 	&	1.91 \\
10   &  FTP  	&	M1.Large	&	1	&	10 	&	2.04 \\
11   &  AWS  	&	T1.Micro	&	10	&	1 	&	2.64 \\
12   &  FTP  	&	T1.Micro	&	5	&	1 	&	2.70 \\
13   &  FTP  	&	M1.Large	&	5	&	1 	&	2.81 \\
14   &  AWS  	&	M1.Large	&	10	&	1 	&	3.13 \\
15   &  FTP  	&	T1.Micro	&	5	&	5 	&	3.60 \\
16   &  FTP  	&	x4150	&	1	&	10 	&	5.73 \\
17   &  FTP  	&	x4150	&	1	&	10 	&	6.47 \\
18   &  AWS  	&	T1.Micro	&	25	&	1 	&	6.79 \\
19   &  AWS  	&	T1.Micro	&	10	&	5 	&	7.19 \\
20   &  AWS  	&	M1.Large	&	25	&	1 	&	7.52 \\
21   &  FTP  	&	x4150	&	1	&	50 	&	8.20 \\
22   &  FTP  	&	M1.Large	&	5	&	10 	&	8.64 \\
23   &  FTP  	&	M1.Large	&	5	&	5 	&	9.33 \\
24   &  AWS  	&	M1.Large	&	10	&	5 	&	12.25 \\
25   &  AWS  	&	T1.Micro	&	25	&	5 	&	18.42 \\
26   &  FTP  	&	T1.Micro	&	50	&	1 	&	27.04.83 \\
27   &  FTP  	&	M1.Large	&	50	&	1 	&	28.37 \\
28   &  AWS  	&	M1.Large	&	25	&	5 	&	28.83 \\
29   &  FTP  	&	T1.Micro	&	50	&	10 	&	30.87 \\
30   &  FTP  	&	T1.Micro	&	50	&	5 	&	39.00 \\
31   &  AWS  	&	M1.Large	&	50	&	5 	&	53.11 \\
32   &  FTP  	&	T1.Micro	&	100	&	1 	&	55.66 \\
33   &  FTP  	&	M1.Large	&	100	&	1 	&	57.82 \\
34   &  FTP  	&	T1.Micro	&	100	&	10 	&	72.11 \\
35   &  FTP  	&	T1.Micro	&	100	&	5 	&	78.26 \\
36   &  FTP  	&	M1.Large	&	  50	&	10 	&	99.42 \\
37   &  FTP  	&	M1.Large	&	100	&	5 	&	189.20 \\
38   &  FTP  	&	M1.Large	&	100	&	10 	&	191.31 \\
39   &  FTP  	&	M1.XLarge&	100	&	10 	&	193.28 \\
40   &  FTP  	&	C1.XLarge&	100	&	100	&	212.27 \\
41   &  FTP  	&	M1.XLarge&	100	&	50 	&	223.10 \\
42   &  FTP  	&	C1.XLarge&	100	&	100 	&	233.79 \\
43   &  FTP/AWS  	&	C1.2XLarge&	70	&	20 	&	332.93 \\

\bottomrule
\caption{Big Picture processing rate summary data} 
\label{tab:longtable2}
\end{longtable}

\begin{table}[htbp]
\centering
  \begin{tabular} {ccc}
  \toprule

    \multicolumn{1}{b{4cm}}{\begin{center}Num. Instances\end{center}} &
   \multicolumn{1}{b{4cm}}{\begin{center}Web Servers.\end{center}} &
   \multicolumn{1}{b{4cm}}{\begin{center}Files per Second\end{center}} \\

\addlinespace[-5mm]

  \midrule
  1 	& 1AWS   &  1470 \\
  1 	& 1AWS   &  1560 \\
  1 	& 1AWS   &  1733 \\
  1 	& 1AWS   &  1971 \\
  1 	& FTP   &  2295 \\
  1 	& FTP   &  2446 \\
  5 	& 1AWS   &  7694 \\
  5 	& 1AWS   &  7845 \\
  5 	& FTP   &  10366 \\
  5 	& FTP   &  11197 \\
  10 	& 1AWS   &  14635 \\
  10 	& 1AWS   &  14702 \\
  50 	& FTP   &  105423 \\
  50 	& FTP   &  119318 \\
  100 	& 1AWS   &  129333 \\
  100 	& HEANET   &  133612 \\
  100 	& FTP   &  186777 \\
  100 	& FTP   &  227039 \\
  100 	& ALL   &  227514 \\
  100 	& FTP   &  229577 \\

  \bottomrule
\hline
  \end{tabular}
  \caption{Files Processed by varying number of M1.Large instances }

\label{tab:pearsons1}
\end{table}





